\documentclass[11pt,letterpaper]{article}
\usepackage[margin=1in]{geometry}
\usepackage{amsmath,amssymb,amsfonts,amsbsy,amsthm,mathtools}
\usepackage{bm,bbm,mathrsfs,dsfont}
\usepackage{graphicx,subcaption}
\usepackage{multirow,booktabs,enumerate,enumitem,verbatim}
\usepackage{xcolor}
\usepackage{algorithm,algorithmic}
\usepackage{needspace}
\usepackage[authoryear,round]{natbib}
\bibpunct[, ]{(}{)}{;}{a}{,}{,}

\usepackage[colorlinks=true,citecolor=blue,linkcolor=black,urlcolor=black]{hyperref}
\hypersetup{pdftitle={Subgradient Methods for Nonsmooth Sample Average Approximation}}
\allowdisplaybreaks[1]
\theoremstyle{plain}
\newtheorem{theorem}{Theorem}
\newtheorem{lemma}{Lemma}
\newtheorem{proposition}{Proposition}
\newtheorem{corollary}{Corollary}

\newtheorem{remark}{Remark}
\newtheorem{example}{Example}
\newtheorem{condition}{Condition}
\newtheorem{conditionse}{Condition}

\newtheorem{assumption}{Assumption}
\newtheorem{assumptionvariant}{Assumption}
\newcommand{\proxassumptionnumber}{}
\newenvironment{assumptionint}
  {\begin{assumptionvariant}}
  {\end{assumptionvariant}}

\let\widebar\relax
\providecommand{\argmin}{}

\newcommand{\Norm}[1]{\left\Vert #1\right\Vert}			
\newcommand{\vect}[1]{\boldsymbol{#1}}				

\newcommand{\tp}{{\rm T}}						
\newcommand{\diff}{\mathrm{d}}					
\renewcommand{\argmin}[1]{\mathop{\rm{argmin}}_{#1}}	
											
\newcommand*\widebar[1]{\,						
	\hbox{%
   		\kern0.01em%
		\vbox{%
			\hrule height 0.5pt		
			\kern0.33ex			
			\hbox{%
				\kern-0.1em		
				\ensuremath{#1}%
				\kern-0.1em		
			}%
		}%
	}%
\,}%
\let\hat\widehat
\let\tilde\widetilde
\let\bar\widebar

\newcommand{\mcA}{{\mathcal A}}
\newcommand{\mcB}{{\mathcal B}}					

\newcommand{\mcE}{{\mathcal E}}

\newcommand{\mcL}{{\mathcal L}}					

\newcommand{\mcN}{{\mathcal N}}					

\newcommand{\mcT}{{\mathcal T}}

\newcommand{\mcV}{{\mathcal V}}

\newcommand{\mbB}{{\mathbb B}}					

\newcommand{\mbE}{{\mathbb E}}					

\newcommand{\mbI}{{\mathbb I}}					

\newcommand{\mbP}{{\mathbb P}}					
\newcommand{\mbR}{{\mathbb R}}					

\newcommand{\mfN}{{\mathfrak N}}					

\DeclareMathAlphabet\EuScriptBF{U}{eus}{b}{n}

\newcommand{\supp}{\mathrm{supp}}	
\newcommand{\Thetar}{\Theta_{\mathrm{res}}}	
\newcommand{\infh}{\mathrm{inf},h}	
\newcommand{\inth}{\mathrm{int},h}	

\definecolor{DSgray}{cmyk}{0,1,0,0}

\definecolor{scolor}{cmyk}{0.5,2,0,0}

\begin{document}
\begin{center}
\resizebox{\linewidth}{!}{\Large\bfseries Subgradient Methods for Nonsmooth Sample Average Approximation}\par
\end{center}
\vspace{0.6em}

\begin{center}
    \normalsize
    Xi Chen\textsuperscript{1}
    \qquad
    Ouqi Fu\textsuperscript{2}
    \qquad
    Jiyuan Tu\textsuperscript{3}
    \qquad
    Yichen Zhang\textsuperscript{4}

    \vspace{0.6em}

    \small
    \textsuperscript{1}Stern School of Business, New York University, New York, USA

    \textsuperscript{2}School of Mathematics,
    Shanghai University of Finance and Economics, Shanghai, China

    \textsuperscript{3}School of Statistics and Data Science,
    Shanghai University of Finance and Economics, Shanghai, China

    \textsuperscript{4}Department of Quantitative Methods, Purdue University, West Lafayette, IN, USA

    \vspace{0.4em}

\end{center}

\vspace{1em}
\begingroup\small
\noindent We study direct subgradient methods for nonsmooth sample average approximation (SAA). Subgradient descent and Nesterov acceleration with constant step sizes attain polynomial and geometric rates matching their smooth counterparts. After $O(\log n)$ full-batch updates, where $n$ is the sample size, the computed solution has the same first-order expansion and limiting distribution as an SAA solution, even when the iterates do not converge. A bound on subgradient increments transfers population regularity to the sample objective, controls the algorithmic trajectory, and identifies the leading random term. The bound covers convex piecewise linear scenario costs, including two-stage stochastic linear programs with fixed recourse. 
Moreau smoothing displaces the population solution at first order and changes the limiting distribution of the computed solution. An explicit shift of the residual removes the first-order displacement and improves the expansion remainder. 
For sparse decisions, restricted increment bounds give convergence, support recovery, and expansions for proximal subgradient descent and iterative hard thresholding.\par
\vspace{6pt}
\par
\endgroup
\vspace{12pt}

\section{Introduction}

Sample average approximation (SAA) replaces an expected cost by an average over observed scenarios. The accuracy of a computed solution depends on both sampling error and optimization error. We study how quickly direct subgradient methods reduce the optimization error to the scale needed for inference when the scenario costs are nonsmooth.

Averaging piecewise linear costs against a distribution with a bounded conditional density produces a smooth expected objective. We show how its local strong convexity speeds computation on the nonsmooth sample objective and determines the sampling distribution of the computed solution.

Our main probabilistic result relates subgradient increments to the distance between decisions. Consider $\mcL(\vect\theta)=n^{-1}\sum_i f(Y_i-\vect Z_i^{\tp}\vect\theta)+\vect b^{\tp}\vect\theta$ on $\Theta\subseteq\mbR^p$, where $f$ is convex and Lipschitz. Let $\bar{\mcL}$ be its expected objective, write $\partial\mcL$ for a selected subgradient, and set $a_n=p\log n/n$. We prove
\[
\big|\partial\mcL(\vect\theta_1)-\partial\mcL(\vect\theta_2)-\nabla\bar{\mcL}(\vect\theta_1)+\nabla\bar{\mcL}(\vect\theta_2)\big|_2\leq C\{\sqrt{a_n|\vect\theta_1-\vect\theta_2|_2}+a_n\}
\]
with high probability, simultaneously for every pair $(\vect\theta_1,\vect\theta_2)\in\Theta\times\Theta$. A pinball example shows that the smoothness approximation scale $a_n$ is necessary up to a logarithmic factor. The assumptions control the conditional density of $Y$ and projections of $\vect Z$, without a moment condition on $Y$. The same bound extends to finite maxima of random affine costs, including two-stage stochastic linear programs with fixed recourse.

The increment bound controls successive iterates, extrapolated points, and the SAA solution on a common event, although all depend on the same observations. At distances larger than $a_n$, it transfers smoothness and strong convexity from the expected objective to approximate properties of the SAA objective. Near the population solution, the distance dependence gives a linear subgradient approximation that identifies the random term governing inference.

For Lipschitz convex objectives, subgradient descent with step sizes proportional to $t^{-\zeta}$, $1/2<\zeta<1$, attains objective error $O(T^{\zeta-1})$ for a weighted average. Under strong convexity, diminishing steps and averaging give $O(T^{-1})$ \citep{julien_etal.2012arXiv}. Our increment bound gives three phases of faster convergence for SAA. In Phase 1, subgradient descent and Nesterov acceleration attain objective rates $T^{-1}$ and $T^{-2}$ and enter a region of approximate strong convexity. In Phase 2, their errors decrease toward approximation floors at rates $\exp(-cT/\kappa)$ and $\exp(-cT/\sqrt\kappa)$, where $\kappa$ is the ratio of the smoothness and strong convexity parameters. A one-dimensional example shows that, from almost every initialization, subgradient descent cannot reduce its error below order $\epsilon_1$, the smoothness approximation scale. Phase 3 separates the remaining optimization error from the leading random term. Acceleration improves the geometric rate while preserving the expansion remainder.

These convergence bounds complement analyses based on linear growth and sharpness \citep{polyak.1969,davis_etal.2018jota,ding_wang.2023arXiv}. The two-phase method of \citet{shen_li_etal.2025aos} uses decreasing step sizes initially and a constant step size in its second phase. Our methods use a constant step size throughout and yield a distributional approximation of the computed solution. Beyond SAA, bounded local subgradient variation and quadratic growth give geometric convergence of subgradient descent and restarted Nesterov acceleration to neighborhoods whose radii have the order of the variation scale.

The iterates need not converge to a minimizer to support inference. After $O(\log n)$ updates, their optimization error is $O(a_n)$, and the computed solution has the same first-order expansion as an SAA solution, with remainder $O_P\{(p/n)^{3/4}(\log n)^{1/2}+a_n\}$. First-order equivalence holds under weaker dimension growth conditions than those needed for a Gaussian limit. The same analysis gives bounds on the population optimality gap and the sample complexity of an $\varepsilon$-optimal decision. Each update evaluates $n$ scenario subgradients, giving a total of $O(n\log n)$ scenario evaluations.

Moreau smoothing can change the computed solution's limiting distribution through a drift $\sqrt n h\,\vect v^{\tp}\vect B_\omega$ in a unit contrast $\vect v$. We identify the first-order population displacement coefficient $\vect B_\omega$ for losses with two affine pieces and general prox-functions. We derive a residual shift $hc_\omega$ that removes the first-order displacement. The expansion remainder then improves to $O_P\{(p/n)^{4/5}(\log n)^{2/5}\}$, with the limiting distribution preserved. Sample averaging keeps the smoothness and local strong convexity constants bounded as the bandwidth decreases to $Ca_n$, permitting constant step sizes and convergence to the smoothed SAA solution.

Sparse decisions lead to regularized SAA. Restricting the increments to sparse vectors gives an approximation scale $s\log p/n$ when the working sparsity is proportional to the population sparsity $s$. For proximal subgradient descent and iterative hard thresholding, this yields geometric convergence, support recovery, and expansions of the computed solution. On a recovered support or a deterministic retained set containing the population support, the expansions give coordinatewise normal limits after $O(\log n)$ updates.

We determine the iteration budget needed to retain the sampling distribution described by the asymptotic theory of SAA solutions \citep{king_rockafellar.1993mor,shapiro.1993mor,shapiro_etal.2021}. When fewer updates are used, the expansion also quantifies the remaining optimization error. Retrospective approximation and sample-allocation analyses use an inner solver's convergence rate to choose sampling and stopping schedules \citep{pasupathy.2010or,royset_szechtman.2013or}. We derive that rate for the nonsmooth SAA solver and relate its stopping accuracy to the solution expansion. The same result gives a full-batch counterpart to asymptotic normality for stochastic approximation with iterate averaging \citep{ruppert.1988,polyak1992acceleration,davis_drusvyatskiy_jiang.2024aos}. Here the increment bound controls dependence on the repeatedly used sample and gives an expansion of the final iterate.

The increment bound supplies both the optimization models and the probability control needed for this expansion. The objective recursions follow the inexact-oracle framework \citep{devolder_etal.2013,devolder_etal.2014mp}, whose accelerated bounds accumulate model error with iterations. Here long updates remove that error, and we prove containment of iterates and extrapolated points within the regions where the models hold. The resulting approximation scales also give high-probability values of the regularity parameters in universal-gradient and local-variation analyses \citep{nesterov.2015mp,diakonikolas_guzman.2026siopt}.

To pass from optimization bounds to an expansion, we use the decrease of the increment error with distance. Localization of objective deviations over near-optimal sets gives bounds on SAA solutions \citep{oliveira_thompson.2023mp1,oliveira_thompson.2023mp2}. Other nonsmooth SAA analyses give concentration of subdifferentials and Moreau envelopes \citep{davis_drusvyatskiy.2022mor}. In particular, \citet{ruan.2025mor} establishes uniform Hausdorff convergence of empirical subdifferentials. Subtracting two such approximation bounds retains their full error at nearby decisions. Our direct control of increments makes the error decrease as the decisions approach each other, which gives the smaller remainder in the computed solution's expansion.

\subsection{Organization and notation}

Section \ref{sec:scheme} states the regularity conditions. Section \ref{sec:method} gives the convergence bounds and their combination. Section \ref{sec:exp_model} verifies the conditions for SAA objectives and develops applications. Sections \ref{sec:smooth_tech} and \ref{sec:hd_pgd} treat smoothing and regularized SAA. 
Proofs are collected in the Appendix.

For a vector $\vect v\in\mbR^p$, we use $|\vect v|_2$, $|\vect v|_1$, and $|\vect v|_\infty$ for its Euclidean, $\ell_1$, and maximum norms. For an index set $I$, $\vect v_I$ is the subvector indexed by $I$, and $\vect v_{-l}$ omits coordinate $l$. For a matrix $\vect A$, $\Norm{\vect A}$ and $\Norm{\vect A}_\infty$ are the induced Euclidean and maximum norms. For a symmetric matrix, $\Lambda_{\max}(\vect A)$ and $\Lambda_{\min}(\vect A)$ are its largest and smallest eigenvalues. We write $\vect A_{I\times J}$ for the submatrix indexed by rows $I$ and columns $J$. The indicator $\mbI(A)$ equals one when the event or condition $A$ holds and zero otherwise. We write $\mbB(\vect\theta,r)$ for the closed Euclidean ball with center $\vect\theta$ and radius $r$. For a random variable $W$ and $q>0$, write $\|W\|_{\psi_q}=\inf\{c>0:\mbE\exp(|W/c|^q)\leq2\}$. For positive sequences, $a_n\asymp b_n$ means that their ratio is bounded above and away from zero. Unless stated otherwise, constants and the step sizes used in asymptotic iteration bounds are independent of $n$ and $p$.

\section{Problem and regularity conditions}\label{sec:scheme}

Let $\mcL$ be an SAA objective on the decision region $\Theta\subseteq\mbR^p$, and let $\bar{\mcL}$ be its expected objective. When individual losses are not integrable, use the \hyperref[rem:response_tails]{loss-difference definition} in Section \ref{sec:exp_model}. We write
\begin{equation}\label{eq:emp_loss}
\vect\theta^*\in\argmin{\vect\theta\in\Theta}\bar{\mcL}(\vect\theta),~~
\hat{\vect\theta}\in\argmin{\vect\theta\in\Theta}\mcL(\vect\theta)
\end{equation}
for the population solution and an SAA solution. For a computed solution $\tilde{\vect\theta}$, the differences $\tilde{\vect\theta}-\hat{\vect\theta}$ and $\hat{\vect\theta}-\vect\theta^*$ are the optimization error and sampling error, respectively. The sample remains fixed throughout the algorithm. We use $\partial\mcL(\vect\theta)$ for a measurable selection of a subgradient at $\vect\theta$. At a differentiable point it equals $\nabla\mcL(\vect\theta)$. All updates and derivatives are taken in the free decision coordinates. Affine feasible sets are parametrized by these coordinates.

We first state deterministic conditions under which subgradient descent and Nesterov acceleration inherit convergence properties of smooth optimization. Section \ref{sec:exp_model} verifies these conditions with high probability for a class of SAA objectives.

\subsection{Approximate smoothness and strong convexity}

\begin{condition}\label{cond:L_smooth}
For $L>0$ and $\epsilon_1\geq0$, the function $\mcL$ is $(L,\epsilon_1)$-smooth on a set $\mathcal U$ if, for all $\vect\theta_1,\vect\theta_2\in\mathcal U$,
\begin{equation*}
|\partial\mcL(\vect\theta_1)-\partial\mcL(\vect\theta_2)|_2
\leq L\max\{|\vect\theta_1-\vect\theta_2|_2,\epsilon_1\}.
\end{equation*}
\end{condition}

Condition \ref{cond:L_smooth} bounds subgradient increments by distance when that distance is at least $\epsilon_1$, and by $L\epsilon_1$ at smaller distances. This permits discontinuities in the selected subgradient while retaining the quadratic upper bound used in descent analysis. Every function with maximum local subgradient variation bounded by $V_{\epsilon_1}$ at radius $\epsilon_1>0$ on a convex region satisfies Condition \ref{cond:L_smooth} with $L=2V_{\epsilon_1}/\epsilon_1$. Partitioning a segment into pieces of length at most $\epsilon_1$ gives this implication from the $\mathrm{BVG}_{\max}$ class of \citet{diakonikolas_guzman.2026siopt}.

\begin{lemma}\label{prop:L-smooth}
Suppose Condition \ref{cond:L_smooth} holds for a finite convex function $\mcL$ on a convex region. For every pair $\vect\theta_1,\vect\theta_2$ in that region with $|\vect\theta_1-\vect\theta_2|_2\geq\epsilon_1$,
\begin{equation}\label{eq:smooth_cond}
\mcL(\vect\theta_1)-\mcL(\vect\theta_2)\leq\langle\partial\mcL(\vect\theta_2),\vect\theta_1-\vect\theta_2\rangle+L|\vect\theta_1-\vect\theta_2|_2^2.
\end{equation}
For $0<\eta\leq1/L$ and $\vect\theta,\vect\theta^+=\vect\theta-\eta\partial\mcL(\vect\theta)$ in the region,
\begin{equation}\label{eq:segment_descent}
\mcL(\vect\theta^+)\leq\mcL(\vect\theta)-\frac\eta2|\partial\mcL(\vect\theta)|_2^2+\frac{\eta L^2\epsilon_1^2}{2}.
\end{equation}
\end{lemma}

The descent bound \eqref{eq:segment_descent} applies at every update. For $\eta\leq1/(2L)$, an update of length at least $\epsilon_1$ removes the additive error by \eqref{eq:smooth_cond}.

\begin{condition}\label{cond:mu_convex}
For $\mu>0$ and $\epsilon_2\geq0$, the function $\mcL$ is $(\mu,\epsilon_2)$-strongly convex on a set $\mathcal U$ if
\begin{equation*}
\mcL(\vect\theta_1)-\mcL(\vect\theta_2)
-\langle\partial\mcL(\vect\theta_2),\vect\theta_1-\vect\theta_2\rangle
\geq\frac\mu4|\vect\theta_1-\vect\theta_2|_2^2
\end{equation*}
for all $\vect\theta_1,\vect\theta_2\in\mathcal U$ with $|\vect\theta_1-\vect\theta_2|_2\geq\epsilon_2$.
\end{condition}

Condition \ref{cond:mu_convex} requires a quadratic lower bound at distances of at least $\epsilon_2$. Section \ref{sec:exp_model} proves that the SAA objective satisfies this condition wherever the expected objective is strongly convex. Suppose both conditions hold on a convex region, $\mcL$ is finite and convex there, and its minimizer $\hat{\vect\theta}$ lies in the interior. For $|\vect\theta-\hat{\vect\theta}|_2\geq\max\{\epsilon_1,\sqrt2\epsilon_2\}$, Lemma \ref{lem:approx_subreg} gives
\begin{equation}\label{eq:approx_subreg_chain}
4L^2|\vect\theta-\hat{\vect\theta}|_2^2
\geq|\partial\mcL(\vect\theta)|_2^2
\geq\mu\{\mcL(\vect\theta)-\mcL(\hat{\vect\theta})\}
\geq\frac{\mu^2}{8}|\vect\theta-\hat{\vect\theta}|_2^2.
\end{equation}
This approximate form of metric subregularity \citep{Dontchev_Rockafellar.2004} lets a subgradient norm threshold control the objective gap and distance without knowing $\hat{\vect\theta}$.

\subsection{Local subgradient approximation}

\begin{condition}\label{cond:H_smooth}
The function $\mcL$ satisfies the local subgradient approximation with matrix $\vect H$, radius $\epsilon_3$, and subgradient remainder bound $r_n$ if $\vect H$ is symmetric positive definite and
\begin{equation*}
\sup_{|\vect\theta-\vect\theta^*|_2\leq\epsilon_3}
\left|\partial\mcL(\vect\theta)-\partial\mcL(\vect\theta^*)
-\vect H(\vect\theta-\vect\theta^*)\right|_2\leq r_n.
\end{equation*}
\end{condition}

Condition \ref{cond:H_smooth} approximates the subgradient increment by the linear map $\vect H(\vect\theta-\vect\theta^*)$. Typically, $\vect H=\nabla^2\bar{\mcL}(\vect\theta^*)$. When $r_n=o(\epsilon_3)$, this approximation gives a linear recursion for iterates near $\vect\theta^*$, up to a smaller remainder. Its solution identifies the contribution of the random subgradient $\partial\mcL(\vect\theta^*)$ and the remaining optimization error.

\subsection{A three-phase scheme}

The three phases use progressively finer consequences of the same subgradient increment bound, verified in Section \ref{sec:exp_model}. Phase 1 reaches a region where approximate strong convexity holds. Phase 2 uses this curvature to reduce the optimization error geometrically. Phase 3 reduces the transient term to the order of the subgradient remainder, identifying the leading random term of the computed solution.

Algorithm \ref{alg:three_stage} allows the update method and its parameters to vary between phases. Appendix \ref{app:phase_combination} gives complete procedures, iteration counts, and proofs of containment. The procedures use \eqref{eq:approx_subreg_chain} to set subgradient norm thresholds that ensure the required update lengths and distances before Phase 3 and determine entry into its local region.

\begin{algorithm}[H]
\caption{{\small Three-phase subgradient method}}\label{alg:three_stage}
\noindent\mbox{\textbf{Input:} $\vect\theta^{(0)}$, methods $\mcA_j$, and stopping criteria $\mathrm{Cri}_j$, $j=1,2,3$.}\par
\begin{algorithmic}[1]
\STATE Initialize $\tilde{\vect\theta}=\vect\theta^{(0)}$.
\FOR{$j=1,2,3$}
\WHILE{$\mathrm{Cri}_j$ is not met}
\STATE Apply the Phase $j$ update $\tilde{\vect\theta}^{+}=\mcA_j(\tilde{\vect\theta})$.
\STATE Set $\tilde{\vect\theta}=\tilde{\vect\theta}^{+}$.
\ENDWHILE
\ENDFOR
\end{algorithmic}
\textbf{Output:} Computed solution $\tilde{\vect\theta}$.
\end{algorithm}

\section{Subgradient methods}	\label{sec:method}

We establish phasewise convergence bounds for subgradient descent and Nesterov's accelerated subgradient method. Section \ref{sec:composite} combines the objective bounds to obtain geometric convergence with a constant step size throughout the computation. The proofs transfer the quadratic upper and lower models at the point pairs where each algorithm uses them. At distances above $\epsilon_1$ and $\epsilon_2$, these models have no additive error. At shorter distances, the potential weights determine how the errors in Lemma \ref{lem:oracle_bounds} in Appendix \ref{app:regularity} enter the final rate.

\subsection{Subgradient descent}	\label{sec:subgd}

\begin{algorithm}[H]
	\caption{{\small Subgradient descent}}
	\label{alg:sgd}
	\noindent\mbox{\textbf{Input:} Initial iterate $\vect\theta^{(0)}$, iterations $T$, and step size $\eta$.}\par
	\begin{algorithmic}[1]
		\FOR{$t=0,\dots,T-1$}
			\STATE Subgradient update:
				\begin{equation*}
					\vect{\theta}^{(t+1)} = \vect{\theta}^{(t)} - \eta\partial\mcL(\vect{\theta}^{(t)}).
				\end{equation*}
		\ENDFOR
	\end{algorithmic}
	\textbf{Output:}  Final iterate $\vect{\theta}^{(T)}$.
\end{algorithm}

Approximate smoothness gives the $T^{-1}$ objective rate in Phase 1, and approximate strong convexity gives geometric convergence in Phase 2. Theorem \ref{thm:subgd_phase1} also bounds the error when the updates continue below the approximation scales.

\begin{theorem}\label{thm:subgd_phase1}\label{thm:subgd_phase2}\label{thm:subgd_phase3}
Index each phase from $t=0$. For parts (i) and (ii), suppose $\mcL$ is convex and satisfies Condition \ref{cond:L_smooth} on a convex set $\Theta$ containing $\vect\theta^{(0)},\ldots,\vect\theta^{(T)}$ and $\hat{\vect\theta}$.
\begin{enumerate}[label=(\roman*)]
\item \textup{Phase 1.} Let $0<\eta\leq1/L$. Then
\begin{equation*}
\frac1T\sum_{t=1}^T\{\mcL(\vect\theta^{(t)})-\mcL(\hat{\vect\theta})\}
\leq\frac{|\vect\theta^{(0)}-\hat{\vect\theta}|_2^2}{2T\eta}+\frac{\eta L^2\epsilon_1^2}{2}.
\end{equation*}
If $\eta\leq1/(2L)$ and $|\vect\theta^{(t+1)}-\vect\theta^{(t)}|_2\geq\epsilon_1$ for $t=0,\ldots,T-1$, the objective values and distances to $\hat{\vect\theta}$ are nonincreasing, and
\begin{equation*}
\mcL(\vect\theta^{(T)})-\mcL(\hat{\vect\theta})\leq\frac{|\vect\theta^{(0)}-\hat{\vect\theta}|_2^2}{2T\eta}.
\end{equation*}
\item \textup{Phase 2.} Let $\Theta_{\rm sc}\subseteq\Theta$, and impose Condition \ref{cond:mu_convex} on that subset. Assume $\vect\theta^{(t)},\hat{\vect\theta}\in\Theta_{\rm sc}$ for $t=0,\ldots,T-1$. For $0<\eta\leq1/L$ and $0<\mu\leq2L$,
\begin{equation*}
\mcL(\vect\theta^{(T)})-\mcL(\hat{\vect\theta})\leq\{1-\mu(\eta-L\eta^2/2)\}^T\{\mcL(\vect\theta^{(0)})-\mcL(\hat{\vect\theta})\}+\frac{\mu\epsilon_2^2}{4}+\frac{L\epsilon_1^2}{2\mu(\eta-L\eta^2/2)}.
\end{equation*}
For any step size satisfying $0<\mu(\eta-\eta^2L)\leq1$, if $|\vect\theta^{(t+1)}-\vect\theta^{(t)}|_2\geq\epsilon_1$ and $|\vect\theta^{(t)}-\hat{\vect\theta}|_2\geq\epsilon_2$ for $t=0,\ldots,T-1$, then
\begin{equation*}
\mcL(\vect\theta^{(T)})-\mcL(\hat{\vect\theta})\leq\{1-\mu(\eta-\eta^2L)\}^T\{\mcL(\vect\theta^{(0)})-\mcL(\hat{\vect\theta})\}.
\end{equation*}
\item \textup{Phase 3.} Suppose the local subgradient approximation in Condition \ref{cond:H_smooth} holds with remainder $r_n$, and $m_H\mbI\preceq\vect{H}\preceq M_H\mbI$ for constants $0<m_H\leq M_H$. Assume $|\vect{\theta}^{(0)}-\vect{\theta}^*|_2\leq\epsilon_3/2$, $|\partial\mcL(\vect{\theta}^*)|_2+r_n\leq m_H\epsilon_3/2$, and $0<\eta\leq M_H^{-1}$. Then $|\vect{\theta}^{(t)}-\vect{\theta}^*|_2\leq\epsilon_3/2$ for every $t\geq0$, and
\begin{equation}\label{eq:bahadur_sgd}
\vect{\theta}^{(T)}-\vect{\theta}^*={}-\vect{H}^{-1}\partial\mcL(\vect{\theta}^*)+(\mbI-\eta\vect{H})^T\big\{\vect{\theta}^{(0)}-\vect{\theta}^*+\vect{H}^{-1}\partial\mcL(\vect{\theta}^*)\big\}+\vect{R}_T,
\end{equation}
where $|\vect R_T|_2\leq r_n/m_H$.
\end{enumerate}
\end{theorem}

In \eqref{eq:bahadur_sgd}, the term $-\vect H^{-1}\partial\mcL(\vect\theta^*)$ approximates the sampling error. The transient term depends on initialization and decreases geometrically. Condition \ref{cond:H_smooth} controls the expansion remainder. Section \ref{sec:exp_model} determines the iteration count needed to match the first-order expansion of an SAA solution. The expansion is often called a Bahadur representation in statistics \citep{he_shao.1996aos}.

Projected subgradient descent and projected Nesterov acceleration attain the Phase 1 and geometric Phase 2 bounds on any closed convex decision region. Corollary \ref{cor:projected_subgradient} in Appendix \ref{app:methods} proves these bounds and, for an interior population solution, the Phase 3 expansions \eqref{eq:bahadur_sgd} and \eqref{eq:ngd_bahadur_optimal}.

\subsection{Nesterov's accelerated subgradient method}	\label{sec:nasubgd}

Algorithm \ref{alg:ngd} combines a subgradient step with Nesterov's extrapolation \citep{nesterov1983method}.

\begin{algorithm}[H]
	\caption{{\small Nesterov's accelerated subgradient method}}
	\label{alg:ngd}
	\noindent\mbox{\textbf{Input:} Initial iterate $\vect\theta^{(0)}$, iterations $T$, step size $\eta$, and momentum $\gamma_t$.}\par
	\begin{algorithmic}[1]
		\STATE Initialize $\vect\beta^{(0)}=\vect\theta^{(0)}$.
		\FOR{$t=0,\dots,T-1$}
			\STATE Subgradient update:
				\begin{equation*}
					\vect{\theta}^{(t+1)} = \vect{\beta}^{(t)} - \eta\partial\mcL(\vect{\beta}^{(t)}).
				\end{equation*}
			\STATE Extrapolation:
				\begin{equation*}
					\vect{\beta}^{(t+1)} = (1+\gamma_t)\vect{\theta}^{(t+1)}-\gamma_t\vect{\theta}^{(t)}.
				\end{equation*}
		\ENDFOR
	\end{algorithmic}
	\textbf{Output:}  Final iterate $\vect{\theta}^{(T)}$.
\end{algorithm}

\begin{theorem}\label{thm:nagd_phase1}\label{thm:nagd_phase2}\label{thm:nagd_phase3}
Index each phase from $t=0$. For parts (i) and (ii), suppose $\mcL$ is convex on a convex set $\Theta$ containing $\vect\theta^{(0)},\ldots,\vect\theta^{(T)}$, $\vect\beta^{(0)},\ldots,\vect\beta^{(T-1)}$, and $\hat{\vect\theta}$. Take $\eta>0$.
\begin{enumerate}[label=(\roman*)]
\item \textup{Phase 1.} Let $\gamma_t=(\lambda_t-1)/\lambda_{t+1}$, where $\lambda_0=1$ and $\lambda_{t+1}=(1+\sqrt{1+4\lambda_t^2})/2$. Suppose $\delta_t\geq0$ satisfies
\[
\mcL(\vect\theta^{(t+1)})\leq\mcL(\vect\beta^{(t)})-\frac\eta2|\partial\mcL(\vect\beta^{(t)})|_2^2+\delta_t.
\]
Then
\begin{equation}\label{eq:nag_short_step_errors}
\mcL(\vect\theta^{(T)})-\mcL(\hat{\vect\theta})
\leq\frac{|\vect\theta^{(0)}-\hat{\vect\theta}|_2^2/(2\eta)+\sum_{t=0}^{T-1}\lambda_t^2\delta_t}{\lambda_{T-1}^2}.
\end{equation}
Under Condition \ref{cond:L_smooth} and $\eta\leq1/(2L)$, take $\delta_t=\eta L^2\epsilon_1^2\mbI\{|\vect\theta^{(t+1)}-\vect\beta^{(t)}|_2<\epsilon_1\}/2$. This gives
\[
\mcL(\vect\theta^{(T)})-\mcL(\hat{\vect\theta})\leq\frac{2|\vect\theta^{(0)}-\hat{\vect\theta}|_2^2}{\eta T^2}+\frac{T\eta L^2\epsilon_1^2}{2}.
\]
If $|\vect\theta^{(t+1)}-\vect\beta^{(t)}|_2\geq\epsilon_1$ for $t=0,\ldots,T-1$, then
\begin{equation*}
\mcL(\vect\theta^{(T)})-\mcL(\hat{\vect\theta})\leq\frac{2|\vect\theta^{(0)}-\hat{\vect\theta}|_2^2}{\eta T^2}.
\end{equation*}
\item \textup{Phase 2.} Suppose Condition \ref{cond:L_smooth} holds on $\Theta$ and $\eta\leq1/L$. Let $\Theta_{\rm sc}\subseteq\Theta$, and impose Condition \ref{cond:mu_convex} on that subset. Assume $\vect\beta^{(t)},\hat{\vect\theta}\in\Theta_{\rm sc}$ for $t=0,\ldots,T-1$ and $\gamma_t\equiv\gamma\in[0,1)$. For $\alpha=\min\{(1-\gamma)/(1+\gamma),(\eta\mu/2)(1+\gamma)/(1-\gamma)\}$,
\begin{equation*}
\mcL(\vect\theta^{(T)})-\mcL(\hat{\vect\theta})\leq(1-\alpha)^T\Big\{\mcL(\vect\theta^{(0)})-\mcL(\hat{\vect\theta})+\frac{\alpha^2}{2\eta}|\vect\theta^{(0)}-\hat{\vect\theta}|_2^2\Big\}+\frac{\eta L^2\epsilon_1^2}{2\alpha}+\frac{\mu\epsilon_2^2}{4}.
\end{equation*}
If $\eta\leq1/(2L)$, $|\vect\theta^{(t+1)}-\vect\beta^{(t)}|_2\geq\epsilon_1$, and $|\vect\beta^{(t)}-\hat{\vect\theta}|_2\geq\epsilon_2$ for $t=0,\ldots,T-1$, then
\begin{equation*}
\mcL(\vect\theta^{(T)})-\mcL(\hat{\vect\theta})\leq(1-\alpha)^T\left\{\mcL(\vect\theta^{(0)})-\mcL(\hat{\vect\theta})+\frac{\alpha^2}{2\eta}|\vect\theta^{(0)}-\hat{\vect\theta}|_2^2\right\}.
\end{equation*}
\item \textup{Phase 3.} Suppose Condition \ref{cond:H_smooth} holds with remainder $r_n$, and $m_H\mbI\preceq\vect H\preceq M_H\mbI$ for $0<m_H\leq M_H$. Let $0<\eta\leq M_H^{-1}$ and $\gamma=(1-\sqrt{\eta m_H})/(1+\sqrt{\eta m_H})$. Restart Algorithm \ref{alg:ngd} with $\gamma_t\equiv\gamma$. If $(1+2\gamma)(|\vect\theta^{(0)}-\vect\theta^*+\vect H^{-1}\partial\mcL(\vect\theta^*)|_2+r_n/m_H)+|\vect H^{-1}\partial\mcL(\vect\theta^*)|_2\leq\epsilon_3$, then every iterate and extrapolated point lies in $\mbB(\vect\theta^*,\epsilon_3)$. For every $T\geq1$,
\begin{equation}\label{eq:ngd_bahadur_optimal}
\vect\theta^{(T)}-\vect\theta^*=-\vect H^{-1}\partial\mcL(\vect\theta^*)+\vect P_T\big\{\vect\theta^{(0)}-\vect\theta^*+\vect H^{-1}\partial\mcL(\vect\theta^*)\big\}+\vect R_T,
\end{equation}
where $\vect P_T$ is a matrix polynomial in $\vect H$ satisfying $\Norm{\vect P_T}\leq(1+T\sqrt{\eta m_H})(1-\sqrt{\eta m_H})^T\leq2\exp(-T\sqrt{\eta m_H}/2)$, and $|\vect R_T|_2\leq r_n/m_H$.
\end{enumerate}
\end{theorem}

For the bound without additive terms in part (ii), the proof uses approximate smoothness on $(\vect\theta^{(t+1)},\vect\beta^{(t)})$ and approximate strong convexity on $(\vect\beta^{(t)},\hat{\vect\theta})$, at distances of at least $\epsilon_1$ and $\epsilon_2$, respectively. For the bounds with additive terms, take $0<\mu\leq2L$ and $\eta=1/L$. Nesterov acceleration with $\gamma=(\sqrt{2L}-\sqrt\mu)/(\sqrt{2L}+\sqrt\mu)$ gives factor $1-\sqrt{\mu/(2L)}$, compared with $1-\mu/(2L)$ for subgradient descent in Theorem \ref{thm:subgd_phase2}(ii). The Nesterov factor is smaller when $\mu<2L$, and both factors are zero at $\mu=2L$.

With $\eta=1/M_H$, part (iii) gives the accelerated geometric factor $1-\sqrt{m_H/M_H}$ and preserves the sampling term and remainder bound $r_n/m_H$ of subgradient descent. Consequently, acceleration reduces the condition-number dependence of the Phase 3 iteration count from $M_H/m_H$ to $\sqrt{M_H/m_H}$.

Proposition \ref{prop:nag_fixed_momentum} in Appendix \ref{app:methods} treats an arbitrary prescribed momentum under a smaller step-size bound and compares its geometric factor with subgradient descent at the same step size. Corollary \ref{cor:heavy_ball} in Appendix \ref{app:other_updates} gives geometric convergence and a Phase 3 expansion for heavy-ball updates.

\subsection{Convergence with a constant step size}\label{sec:composite}

The phase bounds combine into an optimization guarantee for every later iterate. The next corollary proves entry into the region of approximate strong convexity and containment after entry, including updates shorter than $\epsilon_1$. Section \ref{sec:exp_model} uses the resulting distance bound to compare the computed and SAA solutions.

\begin{corollary}\label{cor:absorption}
Suppose $\mcL$ is finite and convex on $\mbR^p$ and has a minimizer $\hat{\vect\theta}$. Let $R_0\geq|\vect\theta^{(0)}-\hat{\vect\theta}|_2$, $0<\mu\leq2L$, $\kappa=L/\mu$, and $0<\eta\leq1/(2L)$. Suppose Condition \ref{cond:L_smooth} holds on $\mbB(\hat{\vect\theta},2\max\{R_0,R\}+2\epsilon_1)$ and Condition \ref{cond:mu_convex} holds on $\mbB(\hat{\vect\theta},R)$, where $R>0$ and $R\geq\max\{4\epsilon_2,16\epsilon_1/(\eta\mu)\}$. Let $T_1=\lfloor8R_0^2/(\eta\mu R^2)\rfloor+1$.
\begin{enumerate}[label=(\roman*)]
\item \textup{Subgradient descent.} The iterates of Algorithm \ref{alg:sgd} satisfy $|\vect\theta^{(t)}-\hat{\vect\theta}|_2\leq\max\{R_0,R\}$ for every $t\geq0$, and
\begin{equation}\label{eq:absorb_subgd}
|\vect\theta^{(t)}-\hat{\vect\theta}|_2\leq\left(1-\frac{3\eta\mu}{4}\right)^{(t-T_1)/2}\frac R{\sqrt2}+\sqrt{\epsilon_2^2+\frac{4\eta L^2}{3\mu}\epsilon_1^2},~~t\geq T_1.
\end{equation}
\item \textup{Nesterov acceleration.} Suppose additionally that $R\geq64(\epsilon_2+\kappa\epsilon_1)$. Let $N_1=\max\{1,\lceil\sqrt{512/(\eta\mu)}\,R_0/R\rceil\}$. Run Algorithm \ref{alg:ngd} with the momentum sequence in Theorem \ref{thm:nagd_phase1}(i) for at most $N_1$ updates, stopping before an update if $\eta|\partial\mcL(\vect\beta^{(t)})|_2<\epsilon_1$. Restart at this extrapolated point if the test is met, or at $\vect\theta^{(N_1)}$ otherwise, using step size $\eta$ and constant momentum $\gamma=(1-\sqrt{\eta\mu/2})/(1+\sqrt{\eta\mu/2})$. Indexing the restarted iterates from $t=0$, they satisfy
\begin{equation}\label{eq:absorb_nagd}
|\vect\theta^{(t)}-\hat{\vect\theta}|_2\leq\left(1-\sqrt{\frac{\eta\mu}{2}}\right)^{t/2}\frac R3+\sqrt2\epsilon_2+2\kappa\left(\frac{\eta\mu}{2}\right)^{1/4}\epsilon_1,~~t\geq0.
\end{equation}
Every Phase 1 iterate and extrapolated point lies in $\mbB(\hat{\vect\theta},R_0)$. Every iterate and extrapolated point after the restart lies in $\mbB(\hat{\vect\theta},R)$.
\end{enumerate}
\end{corollary}

At $\eta=1/(2L)$, parts (i) and (ii) give radii $O(\epsilon_2+\kappa^{1/2}\epsilon_1)$ and $O(\epsilon_2+\kappa^{3/4}\epsilon_1)$, respectively. Acceleration improves the condition-number dependence of the iteration count from $\kappa$ to $\sqrt\kappa$ at the cost of a factor $\kappa^{1/4}$ in the $\epsilon_1$ term. The $\epsilon_1$ term also follows from the accelerated inexact-oracle objective floor $O(\sqrt\kappa\,\delta)$ in \citet{devolder_etal.2013} and quadratic growth when $\epsilon_2=0$ and $\delta=L\epsilon_1^2/2$.

The $\epsilon_1$ dependence in \eqref{eq:absorb_subgd} is necessary even when $\epsilon_2=0$. At $\eta=1/(2L)$, Appendix \ref{app:methods} gives an example with limiting distance at least $\epsilon_1/14$ for every subgradient selection and Lebesgue-almost every initialization. The upper and lower radii therefore differ by at most a factor of order $\sqrt\kappa$.

Corollary \ref{cor:bvg_growth} in Appendix \ref{app:methods} extends the analysis of the $\mathrm{BVG}_{\max}$ class of \citet{diakonikolas_guzman.2026siopt} to geometric convergence under quadratic growth. Subgradient descent and restarted Nesterov acceleration attain the two rates above, to radii $O(\sqrt\kappa r)$ and $O(\kappa^{3/4}r)$, where $r$ is the variation scale. The lower bound makes the dependence on $r$ sharp for subgradient descent when $\kappa$ is bounded.

\section{Subgradient increments and applications}\label{sec:exp_model}

The subgradient increment bound gives optimization rates and a sampling expansion for the computed solution. We first treat residual costs,
\begin{equation}\label{eq:loss_expres}
\mcL(\vect\theta)=\frac1n\sum_{i=1}^n f(Y_i-\vect Z_i^{\tp}\vect\theta)+\vect b^{\tp}\vect\theta,\qquad \vect\theta\in\Theta\subseteq\mbR^p,
\end{equation}
\phantomsection\label{rem:response_tails}
where $(Y_i,\vect Z_i)$ are independent copies of $(Y,\vect Z)$ and $\vect b$ is deterministic. For a reference decision $\vect\theta_{\rm ref}\in\Theta$, define $\bar{\mcL}(\vect\theta)=\mbE\{f(Y-\vect Z^{\tp}\vect\theta)-f(Y-\vect Z^{\tp}\vect\theta_{\rm ref})\}+\vect b^{\tp}(\vect\theta-\vect\theta_{\rm ref})$. The difference inside the expectation is bounded in absolute value by $C_f|\vect Z^{\tp}(\vect\theta-\vect\theta_{\rm ref})|$ and is therefore integrable without a moment condition on $Y$. Reference subtraction preserves objective differences, subgradients, and updates.

\begin{assumption}\label{cond:residual_theta}
The decision region $\Theta$ is convex and satisfies $\sup_{\vect\theta\in\Theta}|\vect\theta|_2\leq C_\Theta$.
\end{assumption}

\begin{assumption}\label{ass:residual_distribution}
\begin{enumerate}[label=\textup{(\roman*)},ref=\theassumption(\roman*),leftmargin=*]
\item\label{cond:residual_density} The conditional distribution of $Y$ given $\vect Z$ has a density $\rho(y\mid\vect Z)$ with $\sup_y\rho(y\mid\vect Z)\leq C_\rho$ almost surely.
\item\label{cond:residual_cov} The covariates satisfy $\sup_{|\vect v|_2=1}\|\vect v^{\tp}\vect Z\|_{\psi_2}\leq C_Z$.
\end{enumerate}
\end{assumption}

\begin{assumption}\label{cond:discont}
The function $f$ is finite and convex on $\mbR$, and every subgradient of $f$ has absolute value at most $C_f$.
\end{assumption}

We assume $a_n\to0$. Assumption \ref{cond:discont} includes piecewise linear, Huber, and smooth convex Lipschitz losses. Write $g=f'_+$ and let $\nu=\diff g$ be its Stieltjes measure, with $\nu(\mbR)\leq2C_f$.

\begin{theorem}\label{thm:Lip_subgrad}
Suppose Assumptions \ref{cond:residual_theta}--\ref{cond:discont} hold. For each deterministic $\vect\theta_0\in\Theta$, there are $C,\gamma>0$ and an event of probability at least $1-Cn^{-\gamma p}$ on which
\begin{equation}\label{eq:lip_subgrad}
\Big|\partial\mcL(\vect\theta_1)-\partial\mcL(\vect\theta_2)-\nabla\bar{\mcL}(\vect\theta_1)+\nabla\bar{\mcL}(\vect\theta_2)\Big|_2\leq C\{\sqrt{|\vect\theta_1-\vect\theta_2|_2a_n}+a_n\}
\end{equation}
for every $\vect\theta_1,\vect\theta_2\in\Theta$ and every selection of scenario subgradients, and
\[
|\partial\mcL(\vect\theta_0)-\nabla\bar{\mcL}(\vect\theta_0)|_2\leq C\sqrt{a_n}.
\]
\end{theorem}

The error decreases as the decisions approach each other. Uniformity over $\Theta$ allows the same bound to follow the iterates from initialization to the region used for inference. At a common decision, \eqref{eq:lip_subgrad} bounds the difference between any two subgradients by $Ca_n$, controlling differences between the scenario-subgradient choices used by an implementation. Here $a_n = p \log n/n$ is the empirical complexity scale arising from the Bernstein and covering arguments in the proof of Theorem \ref{thm:Lip_subgrad}.

\begin{remark}\label{rem:func_approx}
Uniform objective approximation alone does not identify the solution's expansion. Even if a uniform error $\delta_n\asymp\sqrt{p/n}$ is available, population strong convexity gives only an $O\{(p/n)^{1/4}\}$ minimizer bound, above the sampling scale $\sqrt{p/n}$. This square-root dependence is attained by $\varphi(x)=x^2/2$ and $\varphi_a(x)=\max\{x^2/2,a^2/2-ax\}$, whose uniform difference is $a^2$ and whose minimizers differ by $(\sqrt2-1)a$ for $a>0$. For decisions $\vect\theta_1,\vect\theta_2$ at distance $d=|\vect\theta_1-\vect\theta_2|_2$, the subgradient increment error is bounded by $C\{\sqrt{da_n}+a_n\}$. This bound decreases to $Ca_n$ as $d$ shrinks, enabling the population Hessian approximation to identify the leading random term.
\end{remark}

\begin{corollary}\label{thm:H_smooth}
Under the assumptions of Theorem \ref{thm:Lip_subgrad}, the following conclusions hold on its event.
\begin{itemize}
\item[1)] The SAA objective satisfies Condition \ref{cond:L_smooth} on $\Theta$ with $L=L_0$ and $\epsilon_1=C_1a_n$.
\item[2)] If $\bar{\mcL}$ is $\mu_0$-strongly convex on a convex subset $\Theta_{\rm sc}\subseteq\Theta$, then Condition \ref{cond:mu_convex} holds for pairs in that subset with $\mu=\mu_0$ and $\epsilon_2=C_2a_n$.
\item[3)] Suppose the population Hessian is Lipschitz in operator norm on a neighborhood of $\vect\theta^*$, as when $\rho$ has a bounded continuous first derivative. If $\vect H=\nabla^2\bar{\mcL}(\vect\theta^*)$ is positive definite, then, for every $C_3>0$ and all $\vect\theta$ in the ball $|\vect\theta-\vect\theta^*|_2\leq\epsilon_3=C_3\sqrt{a_n}$ contained in that neighborhood and in $\Theta$,
\[
\Big|\partial\mcL(\vect\theta)-\partial\mcL(\vect\theta^*)-\vect H(\vect\theta-\vect\theta^*)\Big|_2\leq C_4a_n^{3/4}.
\]
Thus Condition \ref{cond:H_smooth} holds with subgradient remainder $r_n=C_4a_n^{3/4}$.
\end{itemize}
\end{corollary}

A positive lower eigenvalue bound at $\vect\theta^*$ and the Hessian Lipschitz property give population strong convexity on a neighborhood of $\vect\theta^*$. With $\epsilon_1,\epsilon_2\asymp a_n$, both radii in Corollary \ref{cor:absorption} are $O(a_n)$, below the sampling scale $\sqrt{p/n}$ when $p(\log n)^2=o(n)$. Thus the extra $\kappa^{1/4}$ factor from acceleration does not alter the first-order expansion in the next theorem.

\begin{theorem}\label{cor:model_algorithm}
Suppose the assumptions of Theorem \ref{thm:Lip_subgrad} hold, $a_n\to0$, $\rho$ has a bounded continuous first derivative, and $\vect\theta^*$ is an interior population minimizer with $m_H\mbI\preceq\vect H=\nabla^2\bar{\mcL}(\vect\theta^*)\preceq M_H\mbI$, where $0<m_H\leq M_H<\infty$. Use $L=L_0$, $\mu=m_H/2$, $\epsilon_j=C_ja_n$ for $j=1,2$, and $0<\eta\leq1/(2L)$.

Let $\Theta$ be closed with $\mbB(\vect\theta^*,2R)\subseteq\Theta$ for some $R>0$. From any $\vect\theta^{(0)}\in\Theta$, use either projected procedure in Corollary \ref{cor:projected_subgradient}\ref{part:projected_saa}. Let $\vect\theta^{(t)}$ denote the iterates, counting all phases.
\begin{enumerate}[label=\textup{(\roman*)},leftmargin=*]
\item There are $c,C,\gamma>0$ such that, for sufficiently large $n$, there is an event of probability at least $1-Cn^{-\gamma p}$ on which an SAA solution $\hat{\vect\theta}$ exists and the following bounds hold simultaneously over $\vect\theta^{(0)}\in\Theta$ and all selections of scenario subgradients.
\[
\sup_{t\geq\lceil c\log n\rceil}|\vect\theta^{(t)}-\hat{\vect\theta}|_2\leq Ca_n,\quad
\max\left\{\bar{\mcL}(\hat{\vect\theta})-\bar{\mcL}(\vect\theta^*),\ \sup_{t\geq\lceil c\log n\rceil}\{\bar{\mcL}(\vect\theta^{(t)})-\bar{\mcL}(\vect\theta^*)\}\right\}\leq Ca_n.
\]
\item Write $\vect\psi_i=\vect b-\vect Z_i f'_+(Y_i-\vect Z_i^{\tp}\vect\theta^*)$. Every output $\tilde{\vect\theta}=\vect\theta^{(T)}$ with $T\geq\lceil c\log n\rceil$ satisfies
\[
\tilde{\vect\theta}-\vect\theta^*=-\vect H^{-1}\frac1n\sum_{i=1}^n\vect\psi_i+O_P\!\left\{(p/n)^{3/4}\sqrt{\log n}+a_n\right\}.
\]
If $p\log n=o(\sqrt n)$, then $\sqrt n\,|\tilde{\vect\theta}-\hat{\vect\theta}|_2=o_P(1)$.
\item For a deterministic unit vector $\vect v_n$, let $\sigma_n^2=\operatorname{Var}(\vect v_n^{\tp}\vect H^{-1}\vect\psi_i)$. If $\inf_n\sigma_n>0$ and $p=o\{n^{1/3}(\log n)^{-2/3}\}$, then
\[
\frac{\sqrt n\,\vect v_n^{\tp}(\tilde{\vect\theta}-\vect\theta^*)}{\sigma_n}\xrightarrow{d}\mcN(0,1),
\qquad
n\{\bar{\mcL}(\tilde{\vect\theta})-\bar{\mcL}(\hat{\vect\theta})\}=O_P(p^{3/2}\log n/\sqrt n)=o_P(1).
\]
\end{enumerate}
\end{theorem}

Appendix \ref{app:phase_combination} gives the \hyperref[app:unprojected_saa]{unprojected extension} of Theorem \ref{cor:model_algorithm}. Under their respective region and initialization requirements, the complete procedures there and the interior switches in Corollary \ref{cor:projected_subgradient}\ref{part:projected_inference} attain the same expansion, normal limit, and population-gap conclusions, with Phase 3 counts $\lceil c(\eta_3m_H)^{-1}\log n\rceil$ for subgradient descent and $\lceil c(\eta_Nm_H)^{-1/2}\log n\rceil$ for Nesterov acceleration, where $\eta_3$ and $\eta_N$ are the Phase 3 step sizes. Corollary \ref{cor:subweibull}\ref{cor:subweibull_polynomial} in Appendix \ref{app:heavier_tails} treats directional sub-Weibull tails and preserves the dimension exponent in the normal limit, with additional logarithmic factors.

For $0<\varepsilon\leq1$ and sufficiently large $n$, Theorem \ref{cor:model_algorithm}(i) gives $\bar{\mcL}(\vect\theta^{(T)})-\bar{\mcL}(\vect\theta^*)\leq\varepsilon$ with probability at least $1-Cn^{-\gamma p}$ when $n\geq C(p/\varepsilon)\log(ep/\varepsilon)$ and $T\geq\lceil c\log n\rceil$.

\begin{remark}\label{rem:inference_without_convergence}
When $p(\log n)^2=o(n)$, further iterations after $\lceil c\log n\rceil$ updates do not improve the first-order statistical accuracy of the computed solution. Theorem \ref{cor:model_algorithm} confines every later iterate to an $O(a_n)$ neighborhood of the SAA solution, below both the sampling scale and the expansion remainder order, for every subgradient selection. Additional updates therefore retain the same leading sampling error, as well as the normal limit and population-gap equivalence under part \textup{(iii)}. The pinball trajectories in Proposition \ref{prop:selection_step_lower} show that fluctuations persist, but they need not vanish to attain SAA accuracy.
\end{remark}

\begin{proposition}\label{prop:smoothness_scale_lower}
Consider \eqref{eq:loss_expres} with $f(x)=cx\{\tau-\mbI(x\leq0)\}$, where $c>0$ and $\tau\in(0,1]$. Suppose the matrix with rows $\vect Z_i^{\tp}$ has rank $p$, there is an index $l$ with $Z_{il}=1$ for every $i$, and $\mcL$ attains its minimum on $\mbR^p$. There is a minimizer $\hat{\vect\theta}$ such that every pair $(L,\epsilon_1)$ satisfying Condition \ref{cond:L_smooth} on a neighborhood of $\hat{\vect\theta}$ obeys
\[
L\epsilon_1\geq\frac{cp}{n}.
\]
\end{proposition}
Thus, for $L$ of constant order, the scale $p\log n/n$ differs from the necessary scale in this class by at most a logarithmic factor. For pinball SAA, Proposition \ref{prop:selection_step_lower} in Appendix \ref{app:increments} gives constant-step trajectory lower bounds simultaneously for every scenario-subgradient selection. 

The subgradient increment bound also applies to scenario costs whose active affine piece changes with the decision.

\begin{corollary}\label{cor:max_affine}
Let $\Theta\subseteq\mbR^p$ be convex and bounded, $\xi$ be a random scenario whose cost has $K$ affine pieces with intercepts $\alpha_k(\xi)$ and slopes $\vect\beta_k(\xi)\in\mbR^p$, and let $\xi_1,\ldots,\xi_n$ be independent copies of $\xi$ with cost 
\[
Q(\vect\theta,\xi)=\max_{1\leq k\leq K}\{\alpha_k(\xi)+\vect\beta_k^{\tp}(\xi)\vect\theta\},\qquad
\mcL(\vect\theta)=\frac1n\sum_{i=1}^nQ(\vect\theta,\xi_i)+\vect b^{\tp}\vect\theta.
\]
Suppose $\sup_{k,|\vect v|_2=1}\|\vect v^{\tp}\vect\beta_k\|_{\psi_2}\leq C_\beta$ and, for each $j<k$, the difference $\alpha_j-\alpha_k$ conditional on $\vect\beta_j-\vect\beta_k$ has a density bounded by $C_\rho$. Define $\bar{\mcL}$ through expected loss differences. Then the increment and pointwise bounds in Theorem \ref{thm:Lip_subgrad} and parts 1 and 2 of Corollary \ref{thm:H_smooth} hold, with constants depending on $K$, for every selection from the convex hull of active slopes. If the population Hessian is Lipschitz near an interior minimizer $\vect\theta^*$ and $m_H\mbI\preceq\nabla^2\bar{\mcL}(\vect\theta^*)\preceq M_H\mbI$, the procedures in Theorem \ref{cor:model_algorithm} and Appendix \ref{app:phase_combination}, with their region and initialization requirements, give all conclusions of Theorem \ref{cor:model_algorithm} with score $\vect\psi_i=\vect b+\vect\beta_{k_i^*}(\xi_i)$, where $k_i^*$ maximizes the affine pieces at $\vect\theta^*$.
\end{corollary}

For a two-stage stochastic linear program with fixed recourse matrix $\vect W$ and cost $\vect q$, the recourse value is
\[
Q(\vect\theta,\xi)=\min\{\vect q^{\tp}\vect y:\vect W\vect y=\vect d(\xi)-\vect T(\xi)\vect\theta,\ \vect y\geq\vect0\}.
\]
If this value is finite and the dual feasible polytope has vertices $\vect\pi_1,\ldots,\vect\pi_K$, linear programming duality gives $\alpha_k=\vect\pi_k^{\tp}\vect d(\xi)$ and $\vect\beta_k=-\vect T(\xi)^{\tp}\vect\pi_k$. The corollary then gives the increment bound and optimization rates from density conditions on the scenario differences. The constants depend on the number of dual vertices.

\subsection{Applications}

The examples below give ordering rules, hedging, and classification. Appendices \ref{app:rank} and \ref{app:smoothing_applications} treat rank regression.

\begin{example}\label{exp:newsvendor}\label{exp:qr}
A decision maker uses demand scenarios to compute a linear ordering rule. For demand $D$ and features $\vect Z$, consider the rule $\vect Z^{\tp}\vect\beta$. For underage and overage costs $c_u,c_o>0$, the SAA objective is
\[
\mcL_{\rm NV}(\vect\beta)=\frac1n\sum_{i=1}^n[c_u(D_i-\vect Z_i^{\tp}\vect\beta)_++c_o(\vect Z_i^{\tp}\vect\beta-D_i)_+].
\]
This includes the decision rules in \citet{ban_rudin.2019or} and the sparse inventory policies in \citet{chang_etal.2026or}. Its scenario cost equals $(c_u+c_o)$ times the pinball loss used in quantile regression, $f(x)=x\{\tau-\mbI(x\leq0)\}$, with $\tau=c_u/(c_u+c_o)$ \citep{koenker.2005}. The population solution minimizes expected cost differences over linear ordering rules, including when the conditional demand quantile is nonlinear.

Suppose $\vect Z$ satisfies Assumption \ref{cond:residual_cov}, and the conditional demand density is bounded with a bounded first derivative. At an interior population solution $\vect\beta^*$,
\[
\vect H=(c_u+c_o)\mbE[\rho_D(\vect Z^{\tp}\vect\beta^*\mid\vect Z)\vect Z\vect Z^{\tp}].
\]
Positive lower bounds on this density and on $\Lambda_{\min}\{\mbE[\vect Z\vect Z^{\tp}]\}$ give the Hessian condition in Theorem \ref{cor:model_algorithm}. Starting from any feasible ordering rule in a closed convex decision region containing a ball about $\vect\beta^*$, the projected methods attain the SAA expansion and normal limit after $O(\log n)$ updates. Evaluating scenario subgradients costs $O(np\log n)$ operations in total, nearly linear in $n$, without requiring an exact solution of the SAA linear program. Their later iterates remain within $Ca_n$ of the SAA solution. Nonnegativity and capacity constraints are therefore inactive at those iterates.
\end{example}

\begin{example}\label{exp:cvar}
Let $B$ be an exogenous loss with $\mbE B_+<\infty$ and $\vect Z$ a vector of hedge payoffs. Positions $\vect w$ leave a loss $B-\vect Z^{\tp}\vect w$. At confidence level $\alpha_{\rm CVaR}\in(0,1)$, the Rockafellar--Uryasev SAA objective is \citep{rockafellar_uryasev.2000jor}
\[
\mcL_{\rm CVaR}(a,\vect w)=a+\frac1{n(1-\alpha_{\rm CVaR})}\sum_{i=1}^n(B_i-\vect Z_i^{\tp}\vect w-a)_+.
\]
Minimizing this objective gives the hedge and its risk threshold. It has the form \eqref{eq:loss_expres} with response $B$, covariate vector $\vect V=(\vect Z^{\tp},1)^{\tp}$, decision $(a, \vect w)$, and affine coefficient $(\vect0^{\tp},1)^{\tp}$. The affine term cancels in subgradient increments.

Suppose $\vect V$ satisfies the directional sub-Gaussian bound and the conditional density of $B$ given $\vect Z$ is bounded with a bounded first derivative. At an interior population solution $(a^*,\vect w^*)$,
\[
\vect H_{\rm CVaR}=\frac1{1-\alpha_{\rm CVaR}}\mbE[\rho_B(\vect Z^{\tp}\vect w^*+a^*\mid\vect Z)\vect V\vect V^{\tp}].
\]
Positive lower bounds on this density and on $\Lambda_{\min}\{\mbE[\vect V\vect V^{\tp}]\}$ give the Hessian condition in Theorem \ref{cor:model_algorithm}, whose score here is $\vect\psi_i=(\vect0^{\tp},1)^{\tp}-\mbI(B_i>\vect Z_i^{\tp}\vect w^*+a^*)\vect V_i/(1-\alpha_{\rm CVaR})$.
After $O(\log n)$ updates, the computed hedge and threshold have the same limiting distribution as the SAA decision, with covariance $\vect H_{\rm CVaR}^{-1}\operatorname{Var}(\vect\psi_i)\vect H_{\rm CVaR}^{-1}$ in fixed dimension. For growing dimension, Theorem \ref{cor:model_algorithm} gives the normal limit of decision contrasts. The conditional expected-shortfall formulation in \citet{he_tan_zhou.2023jrssb} provides a related regression approach to tail-risk decisions.
\end{example}

\begin{example}\label{exp:svm}
For a class label $Y\in\{-1,1\}$ and covariates $\vect Z$, the linear support vector machine uses the hinge objective \citep{koo_etal.2008jmlr}
\[
\mcL_{\rm SVM}(\vect\theta)=\frac1n\sum_{i=1}^n(1-\vect X_i^{\tp}\vect\theta)_+,~~\vect X_i=Y_i\vect Z_i.
\]
This is an index objective $n^{-1}\sum_i f(\vect X_i^{\tp}\vect\theta)$ with $f(t)=(1-t)_+$ and every coordinate of $\vect\theta$ varying. On a bounded convex region with $|\theta_1|\geq c_0>0$, Proposition \ref{prop:index_objectives} in Appendix \ref{app:index_models} gives the increment bound when $X_1$ given $\vect X_{-1}$ has a bounded density and $\vect X$ satisfies the directional sub-Gaussian bound, since $\nu=\delta_1$ has compact support. Under the Hessian and initialization conditions of Theorem \ref{cor:model_algorithm} or its \hyperref[app:unprojected_saa]{unprojected extension}, the expansion holds after $O(\log n)$ updates with $\vect\psi_i=-\vect X_i\mbI(\vect X_i^{\tp}\vect\theta^*<1)$, and the normal limit holds under its dimension and variance conditions.
\end{example}

\section{Smoothing}\label{sec:smooth_tech}

Direct subgradient updates already yield the SAA expansion without smoothing. Proposition \ref{prop:selection_step_lower} gives a lower bound on persistent fluctuations of nonsmooth iterates. Smoothing sets $\epsilon_1=\epsilon_2=0$, and Corollary \ref{cor:absorption} then gives geometric convergence to the smoothed SAA solution. Its effect on inference depends on the population displacement. Infimal convolution has displacement $O(h)$, with a leading coefficient determined by the prox-function. An explicit shift of the residual removes its first-order term. The expansion remainder then improves, with the limiting distribution preserved.

The corresponding results for integral convolution, together with its definition and relation to existing inference methods, are given in Appendix \ref{app:smoothing}.

\subsection{Convolution and regularity}\label{sec:inf_conv}\label{sec:int_conv}

For a strongly convex prox-function $\omega$, define the scalar infimal convolution
\begin{equation}\label{eq:infimum_conv}
f_h(x)=\inf_{y\in\mbR}\{f(y)+h\omega((x-y)/h)\}.
\end{equation}
The choice $\omega(u)=u^2/2$ gives the Moreau envelope \citep{moreau1965proximite}. The Fenchel-dual representation of \eqref{eq:infimum_conv} is a scalar instance of Nesterov smoothing \citep{nesterov2005smooth}. We analyze both the regularity and the population displacement for general prox-functions $\omega$, including nonquadratic choices. Write
\[
\mcL_h(\vect\theta)=\frac1n\sum_{i=1}^n f_h(Y_i-\vect Z_i^{\tp}\vect\theta)+\vect b^{\tp}\vect\theta,
\]
and let $\bar{\mcL}_h$ be its expected objective, defined through loss differences as in Section \ref{sec:exp_model}. Each update evaluates a scalar proximal map, which is explicit for piecewise linear losses.
\begin{assumption}\label{cond:prox_fun}\xdef\proxassumptionnumber{\theassumption}
The prox-function $\omega$ is twice continuously differentiable, with second derivative bounded above and bounded away from zero, and $\omega''$ is Lipschitz continuous. The function $f$ is $C^3$ on the intervals separated by a finite number of breakpoints, with second and third derivatives bounded on those intervals. The conditional density in Assumption \ref{cond:residual_density} is continuously differentiable with bounded first derivative.
\end{assumption}

Assumption \ref{cond:prox_fun} includes piecewise linear and Huber losses. The convolution acts on a scalar scenario loss and remains one-dimensional as $p$ grows, whereas \citet{diakonikolas_guzman.2026siopt} smooth the $p$-dimensional decision vector.

\begin{theorem}\label{thm:smoothing_increments}\label{thm:infconv_Lip_subgrad}\label{thm:intconv_Lip_subgrad}
Suppose Assumptions \ref{cond:residual_theta}--\ref{cond:prox_fun} hold. Let $Aa_n\leq h\leq h_0$, for sufficiently large $A$ and sufficiently small $h_0$. With probability at least $1-Cn^{-\gamma p}$,
\[
\begin{aligned}
&\Big|\nabla\mcL_h(\vect\theta_1)-\nabla\mcL_h(\vect\theta_2)-\nabla\bar{\mcL}_h(\vect\theta_1)+\nabla\bar{\mcL}_h(\vect\theta_2)\Big|_2\\
&\qquad\leq C\min\{\sqrt{|\vect\theta_1-\vect\theta_2|_2a_n},\ |\vect\theta_1-\vect\theta_2|_2\sqrt{a_n/h}\},
\end{aligned}
\]
uniformly over $\vect\theta_1,\vect\theta_2\in\Theta$.
\end{theorem}

The bound is linear in distance below the bandwidth and retains the square-root bound above it. Lemma \ref{lem:inf_Hess_conc} in Appendix \ref{app:smoothing} bounds the Hessian fluctuation by $O(\sqrt{a_n/h}+a_n/h)$.

\begin{corollary}\label{cor:smoothing_regularity}\label{thm:inf_conv_cond}\label{thm:int_conv_cond1}
Under the assumptions of Theorem \ref{thm:smoothing_increments}, the following hold with probability at least $1-Cn^{-\gamma p}$.
\begin{itemize}
\item[1)] Condition \ref{cond:L_smooth} holds for $\mcL_h$ with $\epsilon_1=0$ and a constant $L_0$.
\item[2)] If $\bar{\mcL}$ is $\mu_0$-strongly convex on a convex set $\Theta_{\rm sc}\subseteq\Theta$, then Condition \ref{cond:mu_convex} holds for $\mcL_h$ on that set with $\mu=\mu_0$ and $\epsilon_2=0$.
\item[3)] Suppose $\vect H=\nabla^2\bar{\mcL}(\vect\theta^*)$ is positive definite and $\mbB(\vect\theta^*,C_3\sqrt{a_n})\subseteq\Theta$. Then uniformly on that ball,
\begin{equation}\label{eq:H_smooth}
\Big|\nabla\mcL_h(\vect\theta)-\nabla\mcL_h(\vect\theta^*)-\vect H(\vect\theta-\vect\theta^*)\Big|_2 \leq C_4\{\min(a_n/\sqrt h,a_n^{3/4})+h\sqrt{a_n}\}.
\end{equation}
\end{itemize}
The density derivative bound gives the population Hessian Lipschitz property used in part 3.
\end{corollary}

The $O(1/h)$ smoothness bound for an individual smoothed loss also enters homotopy smoothing methods \citep{xu_yan_etal.2016nips}. Combined with an $O(h)$ function approximation error, it gives the accelerated bound $O(h)+O(1/(hT^2))$, which becomes $O(T^{-1})$ after balancing $h$ with $T$ \citep{beck_teboulle.2012siopt}. Under sample averaging, the smoothed SAA objective instead permits the Phase 1 optimization rate $O(T^{-2})$ at each bandwidth $h\geq Aa_n$, with a step size independent of $h$.

The threshold is necessary up to a logarithmic factor. For infimal convolution of the pinball loss, the smoothed second derivative equals $\omega''(x/h)/h$ within its transition interval. An intercept shift of size $O(h)$ from the minimizer in Proposition \ref{prop:smoothness_scale_lower} places at least $p$ residuals inside this interval. A smoothness bound of constant order on a fixed neighborhood of that minimizer therefore requires $h\gtrsim p/n$, as shown in Appendix \ref{app:infimal_convolution}.

\subsection{Population displacement and inference}\label{sec:smoothing_inference}

Smoothing can displace the population minimizer. For infimal convolution the displacement is $O(h)$, and Proposition \ref{prop:infimal_bias} identifies its leading term. Uniform function approximation bounds control objective values in smoothing analyses \citep{beck_teboulle.2012siopt}, whereas inference for $\vect\theta^*$ depends on this displacement.

For a loss with one breakpoint $k$ and affine pieces of slopes $a<b$, let $\omega^*$ be the convex conjugate of $\omega$ and define
\[
c_\omega=\frac{\omega^*(b)-\omega^*(a)}{b-a},\qquad
\vect B_\omega=-(b-a)c_\omega\vect H^{-1}\mbE[\vect Z\rho(k+\vect Z^{\tp}\vect\theta^*\mid\vect Z)].
\]

\begin{proposition}\label{prop:infimal_bias}
Suppose the model and population assumptions of Theorem \ref{cor:model_algorithm} and Assumption \ref{cond:prox_fun} hold, and $f$ has one breakpoint $k$, with affine pieces of slopes $a<b$. For sufficiently small $h$, the infimal population minimizer is unique, lies in the interior of $\Theta$, and satisfies
\[
\vect\theta^*_h-\vect\theta^*=h\vect B_\omega+O(h^2).
\]
\end{proposition}

For the losses in Proposition \ref{prop:infimal_bias}, define the shifted objective
\begin{equation}\label{eq:shifted_objective}
\mcL^{\rm sh}_h(\vect\theta)=\frac1n\sum_{i=1}^n f_h(Y_i-\vect Z_i^{\tp}\vect\theta+hc_\omega)+\vect b^{\tp}\vect\theta.
\end{equation}
Let $U_0$ be uniform on $[a,b]$, independently of $(Y,\vect Z)$, and put $W=(\omega')^{-1}(U_0)$. The following identity explains how the residual shift removes the first-order displacement.

\begin{proposition}\label{prop:infimal_correction}
Under the assumptions of Proposition \ref{prop:infimal_bias}, $\mbE W=c_\omega$ and, for every $r\in\mbR$ and $h>0$,
\begin{equation}\label{eq:infimal_shift_identity}
f_h(r+hc_\omega)=\mbE[f\{r-h(W-c_\omega)\}]+h\{ac_\omega-\omega^*(a)\}.
\end{equation}
For sufficiently small $h$, the expected objective $\bar{\mcL}^{\rm sh}_h$ has a unique minimizer $\vect\theta^{*,\rm sh}_h$ in the interior of $\Theta$, with
\[
|\vect\theta^{*,\rm sh}_h-\vect\theta^*|_2\leq Ch^2.
\]
For $\omega(u)=u^2/2$, the variable $W-c_\omega$ is uniform on $[-(b-a)/2,(b-a)/2]$, and the additive constant in \eqref{eq:infimal_shift_identity} is $hab/2$.
\end{proposition}

The perturbation $W-c_\omega$ has mean zero, which removes the first-order term in the population gradient. For Moreau smoothing, the shifted loss therefore has the same gradient as integral convolution with the uniform kernel on $[-(b-a)/2,(b-a)/2]$. The scalar proximal formula gives this gradient without computing a kernel integral.

\begin{remark}\label{rem:infimal_drift}
Without the residual shift, infimal smoothing can alter first-order inference for the original solution $\vect\theta^*$. At the $n^{-1/2}$ scale, its effect on a contrast is determined by the drift $\sqrt n h\,\vect v_n^{\tp}\vect B_\omega$. A nonnegligible drift alters the limiting distribution even after the optimization error becomes negligible. The residual shift in Proposition \ref{prop:infimal_correction} removes this first-order bias. Under the normal-limit conditions of Corollary \ref{cor:smoothing_algorithm}, the remaining $O(h^2)$ displacement is negligible at this scale, giving a centered normal limit at $\vect\theta^*$ with the original score variance.

For Moreau smoothing, $c_\omega=(a+b)/2$. For unit pinball loss at level $\tau$, the intercept displacement is $-(2\tau-1)h/2+O(h^2)$, including the scalar location model with $Z=1$, and the residual shift is $h(2\tau-1)/2$.

For the CVaR objective in Example \ref{exp:cvar}, the risk threshold has displacement $-h/\{2(1-\alpha_{\rm CVaR})\}+O(h^2)$, while the hedge weights have displacement $O(h^2)$. The threshold displacement is downward, with a first-order coefficient increasing in magnitude as $\alpha_{\rm CVaR}\to1$. The residual shift is $h/\{2(1-\alpha_{\rm CVaR})\}$.
\end{remark}

The Moreau displacement also depends on whether a linear term is placed in $f$ or in $\vect b^{\tp}\vect\theta$. To see this for scalar VaR estimation without hedge payoffs, write $\alpha=\alpha_{\rm CVaR}$, $q=(1-\alpha)^{-1}$, and $f_\alpha(r)=r\{\alpha-\mbI(r\leq0)\}$. The identity $qr_+=qf_\alpha(r)+r$ makes the CVaR objective and the scaled pinball objective differ only by the sample average of $B_i$. They therefore have the same minimizer. Their Moreau coefficients are $q/2$ and $q(2\alpha-1)/2$, respectively, which differ by one. For unit pinball loss the coefficient is $(2\alpha-1)/2$, and comparison at the same $h$ also changes the loss scaling. More generally, for $\tilde f(r)=f(r)+\ell r$, the quadratic prox-function gives
\[
\tilde f_h(r)=f_h(r-h\ell)+\ell r-h\ell^2/2.
\]
Adding $\ell r$ to the loss increases $c_\omega$ by $\ell$. The corresponding residual shift in Proposition \ref{prop:infimal_correction} removes the first-order displacement for either placement of the linear term.

The $O(h^2)$ displacement in Proposition \ref{prop:infimal_correction} and the localized increment term $\sqrt{pa_n/(nh)}$ yield the bandwidth below. The unshifted expansion is centered at $\vect\theta^*_h$, and the shifted expansion is centered at $\vect\theta^*$.

\begin{corollary}\label{cor:smoothing_algorithm}\label{cor:infimal_centered}
Impose the model and population assumptions of Theorem \ref{cor:model_algorithm} and Assumption \ref{cond:prox_fun}. Suppose $a_n\to0$ and choose $h\asymp\max\{a_n,(p/n)^{2/5}(\log n)^{1/5}\}$ with $h\geq Aa_n$ for sufficiently large $A$. Use a method from Theorem \ref{cor:model_algorithm} or its \hyperref[app:unprojected_saa]{unprojected extension}, with its initialization and region requirements, step size $0<\eta\leq1/(2L_0)$, and $\epsilon_1=\epsilon_2=0$. For either complete procedure use the \hyperref[app:phase_refinement]{local refinement in Phase 3} in Appendix \ref{app:phase_combination}. After $\lceil c\log n\rceil$ updates for sufficiently large $c$, the computed solution has distance $O_P(n^{-1})$ from the smoothed SAA solution and satisfies
\[
\tilde{\vect\theta}_h-\vect\theta^*_h=-\vect H^{-1}\{\nabla\mcL_h(\vect\theta^*_h)-\nabla\bar{\mcL}_h(\vect\theta^*_h)\}+O_P\Big\{(p/n)^{4/5}(\log n)^{2/5}+a_n^2\Big\}.
\]
If $p(\log n)^{4/3}=o(n)$, then $h\asymp(p/n)^{2/5}(\log n)^{1/5}$ and the remainder is $O_P\{(p/n)^{4/5}(\log n)^{2/5}\}$. For any deterministic unit contrast $\vect v_n$ whose unsmoothed score variance $\sigma_n^2=\operatorname{Var}(\vect v_n^{\tp}\vect H^{-1}\vect\psi_i)$ is bounded below, $p=o\{n^{3/8}(\log n)^{-1/2}\}$ gives
\[
\frac{\sqrt n\,\vect v_n^{\tp}(\tilde{\vect\theta}_h-\vect\theta^*_h)}{\sigma_n}\xrightarrow{d}\mcN(0,1).
\]
For an output from the shifted objective in Proposition \ref{prop:infimal_correction}, the preceding conclusions hold with $\mcL^{\rm sh}_h$ in place of $\mcL_h$ and $\vect\theta^*$ in place of $\vect\theta^*_h$.

For the unshifted objective, put $d_n=\sqrt n\,\vect v_n^{\tp}(\vect\theta^*_h-\vect\theta^*)/\sigma_n$. The standardized contrast centered at $\vect\theta^*$ has limiting distribution $\mcN(\lambda,1)$ if $d_n\to\lambda$, retains the displayed limit if $d_n\to0$, and diverges in absolute value in probability if $|d_n|\to\infty$. For the losses in Proposition \ref{prop:infimal_bias},
\[
d_n=\sqrt n h\,\vect v_n^{\tp}\vect B_\omega/\sigma_n+o(1)
\]
under the dimension condition for the normal limit.
\end{corollary}

The iteration bounds and expansions extend to the index objectives covered by Proposition \ref{lem:index_smoothing} in Appendix \ref{app:smoothing_applications}.

Smoothing serves two distinct purposes at two bandwidths. At $h=Aa_n$ with sufficiently large $A$, it gives exact geometric convergence. At this bandwidth, the population displacement is $o(n^{-1/2})$ when $p\log n=o(\sqrt n)$. At $h\asymp(p/n)^{2/5}(\log n)^{1/5}$, the shifted objective improves the expansion remainder, with the expansion centered at $\vect\theta^*$. Without the shift, the Moreau drift diverges in absolute value under the normal-limit conditions of Corollary \ref{cor:smoothing_algorithm} if $|\vect v_n^{\tp}\vect B_\omega|/\sigma_n$ stays bounded away from zero.

\section{Regularized SAA}	\label{sec:hd_pgd}

When $p>n$, strong convexity of the expected objective need not carry over to the SAA objective. Sparsity allows us to control objective changes along the directions used by sparse decisions. Write $S=\supp(\vect\theta^*)$ and $s=|S|\ll p$. In this section, $\hat{\vect{\theta}}$ denotes a solution of the regularized SAA problem
\[
\hat{\vect\theta}\in\argmin{\vect\theta\in\Theta}F(\vect\theta),\qquad
F(\vect\theta)=\mcL(\vect\theta)+\lambda r(\vect\theta),
\]
where $r$ is a penalty and $\lambda>0$. Throughout this section, $\mcL$ is convex. We extend the increment bound to sparse directions and use the resulting regularity to analyze proximal subgradient descent and iterative hard thresholding. Their local expansions give inference on retained sets containing $S$.


\subsection{Restricted regularity conditions}	\label{sec:hd_early}

Restricting the increment bound to sparse decisions makes its approximation scales depend on sparsity, with only logarithmic dependence on $p$. We use $\Thetar=\{\vect\theta\in\Theta:|\vect\theta|_0\leq\tilde s\}$, where $s\leq\tilde s\asymp s$. For $\vect a\in\mbR^p$, define $|\vect a|_{(k)}=\max_{|J|\leq k}|\vect a_J|_2$, the Euclidean norm of its $k$ largest coordinates in magnitude.

\begin{conditionse}\label{cond:hdL_smooth}
The SAA objective $\mcL$ is $(L,\epsilon_1)$-restricted smooth on $\Thetar$ if, for every $\vect\theta_1,\vect\theta_2\in\Thetar$,
\[
|\partial\mcL(\vect\theta_1)-\partial\mcL(\vect\theta_2)|_{(2\tilde s)}
\leq L\max\{|\vect\theta_1-\vect\theta_2|_2,\epsilon_1\}.
\]
\end{conditionse}

The norm in Condition \ref{cond:hdL_smooth} pairs with a $2\tilde s$-sparse decision difference. A maximum-norm increment bound with constant $L_0$ implies this condition with $L=\sqrt{2\tilde s}L_0$.

\begin{conditionse}	\label{cond:hdmu_convex}
	The SAA objective $\mcL$ is $(\mu,\epsilon_2)$-restricted strongly convex on $\Theta_{\mathrm{sc}}\subseteq\Theta$ if, for every $\vect{\theta}_1,\vect{\theta}_2\in\Theta_{\mathrm{sc}}\cap\Thetar$,
	\begin{equation*}	
		\mcL(\vect{\theta}_1) - \mcL(\vect{\theta}_2) - \langle \partial\mcL(\vect{\theta}_2),\vect{\theta}_1-\vect{\theta}_2 \rangle \geq \frac{\mu}{4}|\vect{\theta}_1-\vect{\theta}_2|_2^2,~~\text{ for }|\vect{\theta}_1-\vect{\theta}_2|_2\geq \epsilon_2.
	\end{equation*}
\end{conditionse}

Condition \ref{cond:hdmu_convex} gives quadratic growth between sparse vectors separated by at least $\epsilon_2$. The distance cutoff makes this condition applicable to piecewise linear SAA objectives. Together with the restricted increment bounds, it extends the use of restricted strong convexity in recovery analysis \citep{negahban_etal.2009nips,Wang_He.2024et,wang_peng_etal.2020jasa} and fast optimization \citep{agarwal_etal.2012aos} to expansions of subgradient iterates. Phase 3 uses a subgradient approximation near $\vect\theta^*$.

\begin{conditionse}	\label{cond:hdH_smooth}
	The SAA objective $\mcL$ admits a restricted local subgradient approximation at $\vect{\theta}^*$ with matrix $\vect{H}$, radius $\epsilon_3$, and subgradient remainder bound $r_n$ if, for every $\vect{\theta}\in\Thetar$,
	\begin{align*}
		\Big|\partial\mcL(\vect{\theta}) - \partial\mcL(\vect{\theta}^*) - \vect{H}(\vect{\theta} - \vect{\theta}^*) \Big|_{\infty} \leq r_n,~~&\text{for }|\vect{\theta} - \vect{\theta}^*|_2\leq\epsilon_3.
	\end{align*}
	The matrix $\vect H$ is symmetric and satisfies $m_H\mbI\preceq\vect H\preceq M_H\mbI$ for constants $0<m_H\leq M_H<\infty$.
\end{conditionse}

\begin{theorem}\label{thm:hdLip_subgrad}
Consider the residual objective \eqref{eq:loss_expres} with i.i.d. observations and $p>n\geq2$, and write $a_s=\tilde s\log p/n$. Suppose $\Theta$ is convex with $\sup_{\vect\theta\in\Theta}|\vect\theta|_2\leq C_\Theta$, the conditional density of $Y$ given $\vect Z$ is bounded by $C_\rho$, and $f$ is finite, convex, and Lipschitz.
\begin{itemize}
\item[1)] If $|\vect Z|_\infty\leq C_Z$ almost surely and $\Lambda_{\max}\{\mbE(\vect Z\vect Z^{\tp})\}=O(1)$, then, with probability at least $1-Cp^{-\gamma\tilde s}$,
\begin{equation}\label{eq:hdlip_subgrad}
\Big|\partial\mcL(\vect\theta_1)-\partial\mcL(\vect\theta_2)
-\nabla\bar{\mcL}(\vect\theta_1)+\nabla\bar{\mcL}(\vect\theta_2)\Big|_\infty
\leq C\{\sqrt{|\vect\theta_1-\vect\theta_2|_2a_s}+a_s\}.
\end{equation}
\item[2)] If $\sup_{|\vect v|_2=1,\,|\vect v|_0\leq2\tilde s}\|\vect v^{\tp}\vect Z\|_{\psi_2}\leq C_Z$, then \eqref{eq:hdlip_subgrad} holds with $|\cdot|_\infty$ replaced by $|\cdot|_{(2\tilde s)}$, with the same probability bound.
\end{itemize}
Both increment bounds hold for some $C,\gamma>0$, uniformly over $\vect\theta_1,\vect\theta_2\in\Theta$ with $|\supp(\vect\theta_1)\cup\supp(\vect\theta_2)|\leq2\tilde s$, for every selection of scenario subgradients. Under either set of assumptions, at $\vect\theta^*\in\Theta$, the bound $|\partial\mcL(\vect\theta^*)-\nabla\bar{\mcL}(\vect\theta^*)|_\infty\leq C\sqrt{\log p/n}$ holds with probability at least $1-Cp^{-\gamma}$. The increment and pointwise bounds hold jointly with probability at least $1-Cp^{-\gamma}$.
\end{theorem}

These bounds require no response moment condition. Corollary \ref{cor:sparse_index} in Appendix \ref{app:regularized} gives these results for index objectives.

The restricted norm controls objective changes without an additional factor $\sqrt{\tilde s}$. To verify the local approximation, assume that the expected objective is differentiable near $\vect\theta^*$ and, for some matrix $\vect H$,
\begin{equation}\label{eq:hd_taylor}
\big|\nabla\bar{\mcL}(\vect\theta)-\nabla\bar{\mcL}(\vect\theta^*)-\vect H(\vect\theta-\vect\theta^*)\big|_\infty
\leq C|\vect\theta-\vect\theta^*|_2^2
\end{equation}
for sparse $\vect\theta$ with $|\vect\theta-\vect\theta^*|_2\leq\epsilon_3$. Under Theorem \ref{thm:hdLip_subgrad}(2), a bounded derivative of the conditional density of $Y$ given $\vect Z$ gives this Taylor bound and its version in $|\cdot|_{(2\tilde s)}$ used for inference below.

\begin{corollary}\label{thm:hdH_smooth}
Under the assumptions of Theorem \ref{thm:hdLip_subgrad}(2), suppose $a_s=o(1)$. On its increment event, the following hold.
\begin{itemize}
\item[1)] Condition \ref{cond:hdL_smooth} holds with $L=L_0$ and $\epsilon_1=C_1a_s$.
\item[2)] If $\bar{\mcL}$ is $\mu_0$-strongly convex on a convex set $\Theta_{\mathrm{sc}}\subseteq\Theta$, then Condition \ref{cond:hdmu_convex} holds with $\mu=\mu_0$ and $\epsilon_2=C_2a_s$.
\item[3)] If \eqref{eq:hd_taylor} holds and $\vect H$ satisfies the spectral bounds in Condition \ref{cond:hdH_smooth}, then that condition holds with $\epsilon_3=C_3\sqrt{a_s}$ and $r_n=C_4a_s^{3/4}$.
\end{itemize}
Under the bounded-coordinate assumptions of Theorem \ref{thm:hdLip_subgrad}, these conclusions hold with $L=\sqrt{2\tilde s}L_0$ and $\epsilon_2=C_2\tilde s a_s$. The values of $\epsilon_1,\epsilon_3,r_n$ are unchanged.
\end{corollary}

The bound $L=L_0$ permits step sizes independent of $s$ in Phases 1 and 2.

\subsection{Proximal subgradient descent}	\label{sec:psubgd}

We use proximal subgradient descent (PGD) with the $\ell_1$-penalty in Phases 1 and 2 and the $\ell_0$-penalty in Phase 3. The $\ell_1$-penalty preserves convexity of the regularized objective and gives soft thresholding. Under the sufficient condition in Appendix \ref{app:proximal_updates}, soft thresholding preserves a working set of size at most $\tilde s$. The $\ell_0$-penalty gives hard thresholding, which removes small coordinates without shrinking the retained coordinates. On the recovered support $S$, the updates coincide with unpenalized subgradient descent and yield the expansion in Theorem \ref{thm:pgd_phases}(iii).

\begin{algorithm}[H]
	\caption{{\small Proximal subgradient descent}}
	\label{alg:psubgd}
	\noindent\mbox{\textbf{Input:} Initial iterate $\vect\theta^{(0)}$, iterations $T$, step size $\eta$, and penalty weight $\lambda$.}\par
	\begin{algorithmic}[1]
		\FOR{$t=0,\dots,T-1$}
			\STATE Compute the proximal subgradient update
		\begin{equation}	\label{eq:pgd_iterate}
                    \vect{\theta}^{(t+1)} = \argmin{\vect{\theta}}\Big\{  \big\langle\partial\mcL(\vect{\theta}^{(t)}),\vect{\theta} - \vect{\theta}^{(t)}\big\rangle + \lambda r(\vect{\theta}) + \frac{1}{2\eta}|\vect{\theta} - \vect{\theta}^{(t)}|_2^2\Big\}.
		\end{equation}
					\ENDFOR
	\end{algorithmic}
	\textbf{Output:}  Final iterate $\vect{\theta}^{(T)}$.
\end{algorithm}

In Phase 3, the $\ell_0$-penalty thresholds the coordinates of $\vect{\theta}-\eta\partial\mcL(\vect{\theta})$ at $\sqrt{2\eta\lambda}$. Let $b_n=|\partial\mcL(\vect{\theta}^*)|_\infty$ and use $r_n$ from Condition \ref{cond:hdH_smooth}. For a step size $\eta>0$, write $q=\|\mbI-\eta\vect{H}\|_\infty$. For $d>0$, define $\mcB_d=\{\vect{\theta}\in\mbR^p:\ |\vect{\theta}|_0\leq\tilde{s},\ |\vect{\theta}-\vect{\theta}^*|_\infty\leq d\}$.

\begin{theorem}\label{thm:pgd_phases}\label{thm:early_pgd}\label{thm:pgd_phase2}\label{thm:post_pgd}
Consider Algorithm \ref{alg:psubgd}, with $t=0$ denoting the first iterate of each phase.
\begin{enumerate}[label=\textup{(\roman*)},leftmargin=*]
\item \textup{Phase 1.} Under Condition \ref{cond:hdL_smooth}, let $r(\vect\theta)=|\vect\theta|_1$ and let $\hat{\vect\theta}\in\Thetar$ minimize $F$. Suppose $\vect\theta^{(t)}\in\Thetar$ for $0\leq t\leq T$ and $0<\eta\leq1/(2L)$. Then
\[
\frac1T\sum_{t=1}^T\{F(\vect\theta^{(t)})-F(\hat{\vect\theta})\}
\leq\frac{|\vect\theta^{(0)}-\hat{\vect\theta}|_2^2}{2T\eta}+\frac{\eta L^2\epsilon_1^2}{2}.
\]
If $|\vect\theta^{(t+1)}-\vect\theta^{(t)}|_2\geq\epsilon_1$ for $0\leq t<T$, then
\[
F(\vect\theta^{(T)})-F(\hat{\vect\theta})\leq\frac{|\vect\theta^{(0)}-\hat{\vect\theta}|_2^2}{2T\eta}.
\]
\item \textup{Phase 2.} Under Conditions \ref{cond:hdL_smooth} and \ref{cond:hdmu_convex}, let $r(\vect\theta)=|\vect\theta|_1$ and let $\hat{\vect\theta}\in\Theta_{\mathrm{sc}}\cap\Thetar$ minimize $F$. Suppose $\vect\theta^{(t)}\in\Theta_{\mathrm{sc}}\cap\Thetar$ for $0\leq t\leq T$. For $0<\eta\leq1/(2L)$, define $\alpha=\min\{\eta\mu/2,1\}$. Then
\[
|\vect\theta^{(T)}-\hat{\vect\theta}|_2^2
\leq(1-\alpha)^T|\vect\theta^{(0)}-\hat{\vect\theta}|_2^2+\epsilon_2^2+\frac{\eta^2L^2\epsilon_1^2}{\alpha}.
\]
If $|\vect\theta^{(t+1)}-\vect\theta^{(t)}|_2\geq\epsilon_1$ and $|\vect\theta^{(t)}-\hat{\vect\theta}|_2\geq\epsilon_2$ for $0\leq t<T$, both additive terms can be omitted.
\item \textup{Phase 3.} Assume Condition \ref{cond:hdH_smooth} holds. Suppose that $\mcB_d\subseteq\Theta$, $\sqrt{\tilde{s}}d\leq\epsilon_3$, and $q<1$. Assume $\vect{\theta}^{(0)}\in\mcB_d$, $\eta(b_n+r_n)\leq(1-q)d$, and $d<\sqrt{2\eta\lambda}<\min_{j\in S}|\theta_j^*|-d$.
Then the $\ell_0$-penalized proximal subgradient iterates satisfy $\supp(\vect{\theta}^{(t)})=S$ for every $t\geq1$. For $T\geq1$,
\begin{equation}\label{eq:Bahadur}
\Big|\{\vect{\theta}^{(T)}-\vect{\theta}^*\}_{S}
+\vect{H}_{S\times S}^{-1}\{\partial\mcL(\vect{\theta}^*)\}_{S}\Big|_\infty
\leq q^{T-1}\Big(d+\frac{\eta b_n}{1-q}\Big)+\frac{\eta r_n}{1-q}.
\end{equation}
\end{enumerate}
\end{theorem}

The $O(T^{-1})$ objective bound in Theorem \ref{thm:pgd_phases}(i) matches the smooth convex rate, improving on the $O(T^{-1/2})$ guarantee for nonsmooth proximal subgradient descent \citep[Theorem~9.26]{beck.2017}.

A positive step size with $q<1$ exists precisely when $\vect H$ is strictly diagonally dominant. Theorem \ref{thm:iht_phases}(iv) gives containment and an expansion without diagonal dominance for IHT on an unchanged retained set containing $S$. The admissible step sizes and the inverse bound for $q<1$ are given in the \hyperref[app:maximum_norm_contraction]{local recursion calculation} in Appendix \ref{app:regularized}. After support recovery, the leading term in \eqref{eq:Bahadur} is the same as if $S$ were known. For $q\leq q_0<1$, a logarithmic number of updates reduces the optimization term to order $r_n$ when $d/r_n$ grows at most polynomially in $n$. This adds a distributional approximation of the computed solution to geometric-convergence and recovery guarantees for $\ell_1$-penalized methods \citep{xiao_zhang.2013siamjo} and folded concave penalties \citep{wang_liu_zhang.2014}.

Appendix \ref{app:proximal_updates} gives the penalty calibration for Phase 3. Under the proximal-map condition stated there, support recovery and the expansion also hold for SCAD, MCP, and capped-$\ell_1$ penalties.

\subsection{Iterative hard thresholding}	\label{sec:hd_iht}

Iterative hard thresholding (IHT) preserves sparsity by retaining the $\tilde{s}$ largest coordinates in magnitude after each subgradient step \citep{blumensath.2009acha, jain_etal.2014nips}. Algorithm \ref{alg:iht} is the proximal subgradient update \eqref{eq:pgd_iterate} with a penalty equal to zero on vectors with at most $\tilde s$ nonzero coordinates and $+\infty$ otherwise. Its Phase 2 and Phase 3 bounds use the population solution $\vect\theta^*$ as the reference point.

\begin{algorithm}[H]
	\caption{{\small Iterative hard thresholding}}
	\label{alg:iht}
	\noindent\mbox{\textbf{Input:} Initial iterate $\vect\theta^{(0)}$, iterations $T$, step size $\eta$, and sparsity level $\tilde s$.}\par
	\begin{algorithmic}[1]
		\FOR{$t=0,\dots,T-1$}
			\STATE Compute the IHT update
				\begin{equation*}
					\vect{\theta}^{(t+1)} = \argmin{|\vect{\theta}|_0\leq\tilde{s}}\Big\{  \big\langle\partial\mcL(\vect{\theta}^{(t)}),\vect{\theta} - \vect{\theta}^{(t)}\big\rangle + \frac{1}{2\eta}|\vect{\theta} - \vect{\theta}^{(t)}|_2^2\Big\}.
				\end{equation*}
		\ENDFOR
	\end{algorithmic}
	\textbf{Output:}  Final iterate $\vect{\theta}^{(T)}$.
\end{algorithm}

\begin{theorem}\label{thm:iht_phases}\label{thm:early_iht}\label{thm:iht_phase2}\label{thm:post_iht}\label{cor:stable_iht_spectral}
Consider Algorithm \ref{alg:iht}, with $t=0$ denoting the first iterate of each phase. For part \textup{(iii)}, assume Condition \ref{cond:hdH_smooth} holds, $\tilde s\geq s$, $\mcB_d\subseteq\Theta$, $\sqrt{\tilde s}d\leq\epsilon_3$, $q<1$, $\vect\theta^{(0)}\in\mcB_d$, and $\eta(b_n+r_n)\leq(1-q)d$.
\begin{enumerate}[label=\textup{(\roman*)},leftmargin=*]
\item \textup{Phase 1.} Under Condition \ref{cond:hdL_smooth}, suppose $\vect\theta^{(0)}\in\Thetar$, $\vect\theta^{(t)}\in\Theta$ for $1\leq t\leq T$, and $\inf_{\vect\theta\in\Thetar}\mcL(\vect\theta)>-\infty$. If $0<\eta\leq1/(4L)$, then
\[
\min_{0\leq t<T}|\vect\theta^{(t+1)}-\vect\theta^{(t)}|_2^2
\leq\frac{4\eta}{T}\{\mcL(\vect\theta^{(0)})-\inf_{\vect\theta\in\Thetar}\mcL(\vect\theta)\}+4\eta^2L^2\epsilon_1^2.
\]
The final term can be omitted if $|\vect\theta^{(t+1)}-\vect\theta^{(t)}|_2\geq\epsilon_1$ for $0\leq t<T$.
\item \textup{Phase 2.} Under Conditions \ref{cond:hdL_smooth} and \ref{cond:hdmu_convex}, suppose $\vect\theta^*\in\Theta_{\mathrm{sc}}$ and $\vect\theta^{(t)}\in\Theta_{\mathrm{sc}}\cap\Thetar$ with $S\subseteq\supp(\vect\theta^{(t)})$ for $0\leq t\leq T$. For $0<\eta\leq1/(2L)$, define $\alpha=\min\{\eta\mu/2,1\}$. Then
\[
\mcL(\vect\theta^{(T)})-\mcL(\vect\theta^*)
\leq(1-\alpha)^T\{\mcL(\vect\theta^{(0)})-\mcL(\vect\theta^*)\}
+\frac{\mu\epsilon_2^2}{4}+\frac{\eta L^2\epsilon_1^2}{2\alpha}.
\]
If $|\vect\theta^{(t+1)}-\vect\theta^{(t)}|_2\geq\epsilon_1$ and $|\vect\theta^{(t)}-\vect\theta^*|_2\geq\epsilon_2$ for $0\leq t<T$, both additive terms can be omitted.
\item \textup{Phase 3 support retention.} If $\min_{j\in S}|\theta_j^*|>2d$, then $\vect{\theta}^{(t)}\in\mcB_d$ and $S\subseteq\supp(\vect{\theta}^{(t)})$ for every $t\geq0$. Moreover,
\[
|\vect{\theta}^{(T)}-\vect{\theta}^*|_\infty
\leq q^T|\vect{\theta}^{(0)}-\vect{\theta}^*|_\infty
+\frac{\eta(1-q^T)}{1-q}(b_n+r_n).
\]
If $\tilde{s}=s$, then $\supp(\vect{\theta}^{(t)})=S$ for every $t\geq0$.
\item \textup{Phase 3 on a retained set.}
Under the assumptions of Theorem \ref{thm:hdLip_subgrad}(2), assume \eqref{eq:hd_taylor} with $|\cdot|_\infty$ replaced by $|\cdot|_{(2\tilde s)}$, $\nabla\bar{\mcL}(\vect\theta^*)=\vect0$, and $\vect H=\nabla^2\bar{\mcL}(\vect\theta^*)$ with $m_H\mbI\preceq\vect H\preceq M_H\mbI$. Assume $a_s\to0$, take $0<\eta\leq M_H^{-1}$, and set $\epsilon_3=C_3\sqrt{a_s}$ for sufficiently large $C_3$. Let $\widehat J\supseteq S$ be a random retained set of size $\tilde s$. Suppose that, on an event $\mcE_n$ of probability at least $1-\delta_n$, where $\delta_n\to0$, the next $T$ IHT updates retain $\widehat J$, $\supp(\vect\theta^{(0)})\subseteq\widehat J$, $|\vect\theta^{(0)}-\vect\theta^*|_2\leq\epsilon_3$, and
\[
\{\vect\theta:\supp(\vect\theta)\subseteq\widehat J,\ |\vect\theta-\vect\theta^*|_2\leq\epsilon_3\}\subseteq\Theta.
\]
Write $\vect e_t=\{\vect\theta^{(t)}-\vect\theta^*\}_{\widehat J}$, $\vect g_{\widehat J}=\{\partial\mcL(\vect\theta^*)\}_{\widehat J}$, and $q_2=1-\eta m_H/2$, which accounts for the subgradient remainder in the containment bound. For sufficiently large $n$, with probability at least $1-\delta_n-Cp^{-\gamma}$,
\[
|\vect e_t|_2\leq q_2^t|\vect e_0|_2+\frac{2(1-q_2^t)}{m_H}\{|\vect g_{\widehat J}|_2+Ca_s\}\leq\epsilon_3,\qquad 0\leq t\leq T,
\]
and
\[
\left|\vect e_T+\vect H_{\widehat J\times\widehat J}^{-1}\vect g_{\widehat J}\right|_2\leq C\{(1-\eta m_H)^T\sqrt{a_s}+a_s^{3/4}\}.
\]
\end{enumerate}
\end{theorem}

Theorem \ref{thm:iht_phases}(ii) parallels the second-phase result of \citet{shen_li_etal.2025aos}, whose first phase uses decreasing step sizes. For Euclidean localization, Corollary \ref{cor:iht_warm_start} in Appendix \ref{app:regularized} uses overspecified working sparsity to give geometric convergence to an $O(\sqrt{a_s})$ neighborhood. On an unchanged retained set containing $S$, part \textup{(iv)} also dispenses with a minimum-coefficient condition.

\subsection{Sampling distribution and stopping rules}\label{sec:hd_stopping}

For inference on a retained set $J\supseteq S$, use $\vect\psi_i=\vect b-\vect Z_i f'_+(Y_i-\vect Z_i^{\tp}\vect\theta^*)$ and write $\sigma_{J,j}^2=\operatorname{Var}(\vect e_{j,J}^{\tp}\vect H_{J\times J}^{-1}\vect\psi_{i,J})$, where $\vect e_{j,J}$ is the coordinate vector for $j$ in $J$.

\begin{corollary}\label{cor:hd_normality}
Under the assumptions of Theorem \ref{thm:hdLip_subgrad}(2), impose the population assumptions of Theorem \ref{thm:iht_phases}(iv) and suppose $a_s=o(1)$. Use $\epsilon_3=C_3\sqrt{a_s}$ and $r_n=C_4a_s^{3/4}$ from Corollary \ref{thm:hdH_smooth}. Let $J=J_n\supseteq S$ be deterministic. For IHT, suppose the localization conditions of Theorem \ref{thm:iht_phases}(iv) or Proposition \ref{prop:stable_iht_inference}\ref{part:stable_iht_expansion} hold with probability tending to one for a retained set $\widehat J$ satisfying $\Pr(\widehat J=J)\to1$. For PGD, take $J=S$ and suppose the conditions of Theorem \ref{thm:pgd_phases}(iii) hold with probability tending to one. Under the conditions of Theorem \ref{thm:pgd_phases}(iii) or Proposition \ref{prop:stable_iht_inference}\ref{part:stable_iht_expansion}, take $q\leq q_0<1$ and $T\geq\lceil c\log n\rceil$. Under those of Theorem \ref{thm:iht_phases}(iv), take $T\geq\lceil c(\eta m_H)^{-1}\log n\rceil$. For sufficiently large $c$, the output satisfies
\[
\{\vect\theta^{(T)}-\vect\theta^*\}_J
=-\vect H_{J\times J}^{-1}\frac1n\sum_{i=1}^n\vect\psi_{i,J}
+O_P\left\{(|J|/n)^{3/4}\sqrt{\log n}+|J|\log n/n\right\}
\]
in Euclidean norm. For a deterministic $j\in J$, if $\sigma_{J,j}\geq c_0>0$ and $|J|=o\{n^{1/3}(\log n)^{-2/3}\}$, then
\[
\frac{\sqrt n(\theta_j^{(T)}-\theta_j^*)}{\sigma_{J,j}}\xrightarrow{d}\mcN(0,1).
\]
\end{corollary}

The expansion depends on $|J|$ and $n$, while the ambient dimension $p$ enters localization through $a_s$. Taking $J=S$ gives inference after support recovery. Proposition \ref{prop:stable_iht_inference} in Appendix \ref{app:regularized} gives simultaneous confidence statements when the retained set remains random. Corollary \ref{cor:sparse_index} extends these results to index objectives, including sparse SVM with a bounded density derivative.

Update lengths give stopping rules for reaching the local neighborhoods. For an update from $\vect\theta$ to $\vect\theta^+$, write $k=|\vect\theta^+-\vect\theta|_2$ and $G=k/\eta+L\max\{k,\epsilon_1\}$. Under the conditions of Proposition \ref{prop:restricted_stopping}(i)--(ii) in Appendix \ref{app:regularized}, the stopping tests are
\[
\begin{aligned}
k+\max\{4G/\mu,\epsilon_2\}&\leq\varrho&&\text{for PGD},\\
k+\max\{2(G+\sqrt{\tilde s}b_n)/\mu,\epsilon_2\}&\leq\varrho&&\text{for IHT}.
\end{aligned}
\]
These tests give $|\vect\theta-\hat{\vect\theta}|_2\leq\varrho$ for PGD with any proper convex penalty and $|\vect\theta-\vect\theta^*|_2\leq\varrho$ for IHT. For PGD, choose $\varrho+|\hat{\vect\theta}-\vect\theta^*|_2\leq\epsilon_3$ to include the sampling error. For IHT, choose $\varrho\leq\epsilon_3$. Proposition \ref{prop:restricted_stopping} also gives the lower bounds on $\varrho$ and the coordinatewise tests for entering $\mcB_d$.

%

\section{Conclusion}\label{sec:conclude}

The increment bound connects population regularity to convergence and inference for a computed SAA decision. It gives a first-order expansion after $O(\log n)$ subgradient or Nesterov updates even when the iterates do not converge. Projection permits any feasible initialization in a closed convex region containing a neighborhood of the population solution. Smoothing yields convergence to the smoothed minimizer, and its population displacement determines the effect on inference. The first-order Moreau bias appears as a deterministic drift, and an explicit shift of the residual removes this bias and improves the remainder without changing the limiting distribution. Restricted increments give expansions for sparse decisions, including IHT on retained sets under spectral Hessian bounds.

The accelerated procedure uses the strong convexity parameter, approximation scales, and radius ratios to choose momentum and phase transitions. Adaptive restart \citep{donoghue_candes.2015focm}, local curvature estimation \citep{Jensen_etal.2012bitnm}, and progress-based step search \citep{jin_scheinberg_xie.2024siopt} provide possible approaches to selecting these quantities. Applying these methods under our local regularity conditions requires containment and stopping guarantees that preserve the expansion. Nonsmooth nonconvex problems, including robust phase retrieval and models with rectified linear units, call for an extension of the local analysis. Combining increment control with sharpness may give fast convergence under linear objective growth. The expansions also suggest bootstrap inference based on computed solutions, building on smoothing-based bootstrap methods for median regression \citep{horowitz.1998ecomc}. Whether these iteration counts also preserve bootstrap validity across resampled objectives remains open.

\clearpage
\appendix
\section{Optimization under approximate regularity}\label{app:regularity}\label{app:methods}

\subsection{Oracle inequalities}

\begin{lemma}\label{lem:oracle_bounds}
Let $\mcL$ be finite and convex on a convex region. Under Condition \ref{cond:L_smooth}, every pair $\vect\theta_1,\vect\theta_2$ in the region satisfies
\begin{equation}\label{eq:oracle_upper}
0\leq\mcL(\vect\theta_1)-\mcL(\vect\theta_2)-\langle\partial\mcL(\vect\theta_2),\vect\theta_1-\vect\theta_2\rangle\leq\frac L2|\vect\theta_1-\vect\theta_2|_2^2+\frac L2\epsilon_1^2.
\end{equation}
If $\hat{\vect\theta}$ minimizes $\mcL$ over the region and lies in its interior, the same condition gives $|\partial\mcL(\hat{\vect\theta})|_2\leq L\epsilon_1$. If Condition \ref{cond:mu_convex} holds on a subset of the region, every pair in that subset satisfies
\begin{equation}\label{eq:oracle_lower}
\mcL(\vect\theta_1)-\mcL(\vect\theta_2)-\langle\partial\mcL(\vect\theta_2),\vect\theta_1-\vect\theta_2\rangle\geq\frac\mu4|\vect\theta_1-\vect\theta_2|_2^2-\frac\mu4\epsilon_2^2.
\end{equation}
\end{lemma}

\begin{proof}[Proof of Lemmas \ref{prop:L-smooth} and \ref{lem:oracle_bounds}]
Write $\vect d=\vect\theta_1-\vect\theta_2$ and $d=|\vect d|_2$. The result is immediate when $d=0$. Convexity gives the lower bound in \eqref{eq:oracle_upper}. The restriction of $\mcL$ to the segment is absolutely continuous, with derivative $\langle\partial\mcL(\vect\theta_2+t\vect d),\vect d\rangle$ almost everywhere. Hence
\begin{align*}
\mcL(\vect\theta_1)-\mcL(\vect\theta_2)-\langle\partial\mcL(\vect\theta_2),\vect d\rangle
&\leq Ld\int_0^1\max\{td,\epsilon_1\}\diff t
\leq\frac L2(d^2+\epsilon_1^2).
\end{align*}
The integral equals $L\epsilon_1d$ for $d\leq\epsilon_1$ and $L(d^2+\epsilon_1^2)/2$ for $d>\epsilon_1$. Since $\epsilon_1d\leq(d^2+\epsilon_1^2)/2$, this proves \eqref{eq:oracle_upper}. If $d\geq\epsilon_1$, the same bound is at most $Ld^2$, proving \eqref{eq:smooth_cond}.

For the subgradient update in the statement, write $s=|\vect\theta^+-\vect\theta|_2$ and take $0<\eta\leq1/L$. If $s\leq\epsilon_1$, the exact segment bound and $L\epsilon_1s\leq s^2/(2\eta)+\eta L^2\epsilon_1^2/2$ give \eqref{eq:segment_descent}. If $s\geq\epsilon_1$, then
\[
\frac L2(s^2+\epsilon_1^2)-\frac{s^2}{2\eta}-\frac{\eta L^2\epsilon_1^2}{2}
=-\frac{(1-\eta L)(s^2-\eta L\epsilon_1^2)}{2\eta}\leq0,
\]
which proves the same bound. When $\eta\leq1/(2L)$ and $s\geq\epsilon_1$, the segment remainder is at most $Ls^2\leq s^2/(2\eta)$, and the additive term can be omitted.

For \eqref{eq:oracle_lower}, Condition \ref{cond:mu_convex} gives the bound when $d\geq\epsilon_2$. When $d<\epsilon_2$, its right-hand side is nonpositive, and convexity suffices.

For the bound at the minimizer, use the directional limiting argument of \citet[Lemma~D.5]{lam_wang_wu_zhang.2022arxiv}. Let $\vect g=\partial\mcL(\hat{\vect\theta})$. If $\vect g\neq\vect0$, take $\vect u=-\vect g/|\vect g|_2$. Since $\hat{\vect\theta}$ lies in the interior, $\hat{\vect\theta}+s\vect u$ belongs to the region for sufficiently small $s>0$. Convexity and minimality give $\langle\partial\mcL(\hat{\vect\theta}+s\vect u),\vect u\rangle\geq0$, and Condition \ref{cond:L_smooth} implies $|\vect g|_2\leq L\max\{s,\epsilon_1\}$. Letting $s\downarrow0$ proves $|\vect g|_2\leq L\epsilon_1$. The claim is immediate when $\vect g=\vect0$.

\end{proof}

\begin{lemma}\label{lem:approx_subreg}
Suppose $\mcL$ is finite and convex on a convex set $\mathcal U$ and satisfies Conditions \ref{cond:L_smooth} and \ref{cond:mu_convex} there. Let $\hat{\vect\theta}$ minimize $\mcL$ over $\mathcal U$ and lie in its interior. For every $\vect\theta\in\mathcal U$,
\begin{equation}\label{eq:gap_growth}
|\vect\theta-\hat{\vect\theta}|_2^2\leq\frac4\mu\{\mcL(\vect\theta)-\mcL(\hat{\vect\theta})\}+\epsilon_2^2.
\end{equation}
If $|\vect\theta-\hat{\vect\theta}|_2\geq\max\{\epsilon_1,\sqrt2\epsilon_2\}$, the inequality chain \eqref{eq:approx_subreg_chain} also holds.
\end{lemma}

\begin{proof}[Proof of Lemma \ref{lem:approx_subreg}]
Write $d=|\vect\theta-\hat{\vect\theta}|_2$. If $d>\epsilon_2$, let $\vect\theta_s=\hat{\vect\theta}+s(\vect\theta-\hat{\vect\theta})$ for sufficiently small $s>0$. Convexity and minimality give $\langle\partial\mcL(\vect\theta_s),\vect\theta-\vect\theta_s\rangle\geq0$. Since $(1-s)d\geq\epsilon_2$, Condition \ref{cond:mu_convex} implies
\[
\mcL(\vect\theta)-\mcL(\hat{\vect\theta})\geq\mcL(\vect\theta)-\mcL(\vect\theta_s)\geq\frac\mu4(1-s)^2d^2.
\]
Letting $s\downarrow0$ gives $\mcL(\vect\theta)-\mcL(\hat{\vect\theta})\geq\mu d^2/4$. For $d\leq\epsilon_2$, minimality suffices for \eqref{eq:gap_growth}.

Lemma \ref{lem:oracle_bounds} gives $|\partial\mcL(\hat{\vect\theta})|_2\leq L\epsilon_1$. For $d\geq\epsilon_1$, Condition \ref{cond:L_smooth} therefore gives $|\partial\mcL(\vect\theta)|_2\leq Ld+L\epsilon_1\leq2Ld$. Applying Condition \ref{cond:mu_convex} at $\vect\theta$ with comparison point $\hat{\vect\theta}$ and completing the square yields
\begin{equation*}
\mcL(\hat{\vect\theta})\geq\mcL(\vect\theta)+\langle\partial\mcL(\vect\theta),\hat{\vect\theta}-\vect\theta\rangle+\frac\mu4d^2\geq\mcL(\vect\theta)-\frac1\mu|\partial\mcL(\vect\theta)|_2^2.
\end{equation*}
Finally, \eqref{eq:gap_growth} and $d^2\geq2\epsilon_2^2$ give $\mcL(\vect\theta)-\mcL(\hat{\vect\theta})\geq\mu d^2/8$, completing the chain.
\end{proof}

\begin{lemma}\label{eq:lambda_bound}
The sequence in Theorem \ref{thm:nagd_phase1}(i) satisfies $\lambda_t\geq(t+1)/2$.
\end{lemma}

\begin{proof}
The initial value is $\lambda_0=1$, and the recursion gives $\lambda_{t+1}\geq\lambda_t+1/2$. Induction proves the bound.
\end{proof}

Let $0<m_H\leq M_H$, $m_H\mbI\preceq\vect H\preceq M_H\mbI$, $\gamma\in(0,1)$, and $0<\eta\leq\{(1-\gamma)/(1+\gamma)\}^2M_H^{-1}$. Write $\vect M=\mbI-\eta\vect H$ and $\vect z_0=\vect\theta^{(0)}-\vect\theta^*+\vect H^{-1}\partial\mcL(\vect\theta^*)$. Define $\vect\rho_1,\vect\rho_2=\{(1+\gamma)\vect M\pm\sqrt{(1+\gamma)^2\vect M^2-4\gamma\vect M}\}/2$, $q_\rho=\Norm{\vect\rho_1}$, $d_\rho=\Lambda_{\min}(\vect\rho_1-\vect\rho_2)$, and
\[
A_\rho=\max\left\{m_H^{-1},\eta\min\left\{\frac1{(1-q_\rho)^2},\frac{q_\rho}{d_\rho(1-q_\rho)}\right\}\right\},
\]
with the convention $1/0=\infty$. Let $\vect P_0=\mbI$, $\vect P_1=\vect M$, and $\vect P_{t+1}=(1+\gamma)\vect M\vect P_t-\gamma\vect M\vect P_{t-1}$.

\begin{proposition}\label{prop:nag_fixed_momentum}
Suppose Condition \ref{cond:H_smooth} holds with remainder $r_n$. Restart Algorithm \ref{alg:ngd} with step size $\eta$ and constant momentum $\gamma_t\equiv\gamma$. Assume $|\vect\theta^{(0)}-\vect\theta^*|_2\leq\epsilon_3/[2(1+2\gamma)]$ and $A_\rho\{|\partial\mcL(\vect\theta^*)|_2+r_n\}\leq\epsilon_3/[2(1+2\gamma)]$. Then $|\vect\theta^{(t)}-\vect\theta^*|_2\leq\epsilon_3/(1+2\gamma)$ and $|\vect\beta^{(t)}-\vect\theta^*|_2\leq\epsilon_3$ for every $t\geq0$, and
\begin{equation}\label{eq:ngd_bahadur}
\vect\theta^{(T)}-\vect\theta^*=-\vect H^{-1}\partial\mcL(\vect\theta^*)+\vect P_T\vect z_0+\vect R_T,
\end{equation}
where $\Norm{\vect P_T}\leq\min\{1,(T+1)q_\rho^T,2q_\rho^T/d_\rho\}$ and $|\vect R_T|_2\leq A_\rho r_n$. The geometric factor satisfies $q_\rho<\Norm{\mbI-\eta\vect H}$.
\end{proposition}

\begin{proof}[Proof of Proposition \ref{prop:nag_fixed_momentum}]
\noindent\textbf{Step 1. The homogeneous recursion.}
All matrices below are functions of $\vect H$ and commute. The step-size inequality gives $\vect0\prec\vect\rho_2\preceq\vect\rho_1\prec\vect M\prec\mbI$. Define
\begin{equation*}
 \vect K_j=\sum_{k=0}^j\vect\rho_1^k\vect\rho_2^{j-k}\quad(j\geq0),~~\vect K_{-1}=\vect0.
\end{equation*}
This formula includes coincident roots. If $d_\rho>0$, it also equals $(\vect\rho_1^{j+1}-\vect\rho_2^{j+1})(\vect\rho_1-\vect\rho_2)^{-1}$. The matrices are positive semidefinite, and
\begin{equation*}
 \Norm{\vect K_j}\leq\min\{(j+1)q_\rho^j,q_\rho^{j+1}/d_\rho\},~~\eta\sum_{j=0}^\infty\Norm{\vect K_j}\leq A_\rho,~~\eta\sum_{j=0}^\infty\vect K_j=\vect H^{-1}.
\end{equation*}
The two norm bounds follow from the power sum and, when $d_\rho>0$, the difference formula. Summing them gives the bound by $A_\rho$. The last identity follows from $(\mbI-\vect\rho_1)(\mbI-\vect\rho_2)=\eta\vect H$. The polynomial recursion gives
\begin{equation*}
 \vect P_t=\vect K_t-\gamma\vect M\vect K_{t-1},~~\vect P_t-\vect P_{t+1}=\eta\vect H\vect K_t.
\end{equation*}
Since $q_\rho<1$, the first identity gives $\vect P_t\to\vect0$. The second then implies $\vect0\preceq\vect P_t\preceq\mbI$. Also $\vect P_t\preceq\vect K_t$, and $\Norm{\vect P_t}\leq\min\{1,(t+1)q_\rho^t,2q_\rho^t/d_\rho\}$. The convention for $d_\rho=0$ retains the finite bounds from the power sum at coincident roots.

\noindent\textbf{Step 2. Containment of the iterates and extrapolates.}
For this proof, define $\vect\theta^{(-1)}=\vect\theta^{(0)}$. Write $\vect g^*=\partial\mcL(\vect\theta^*)$ and $\vect x_t=\vect\theta^{(t)}-\vect\theta^*$. If $\vect\beta^{(j)}$ lies in the radius-$\epsilon_3$ ball, define
\begin{equation*}
 \vect e_j=\partial\mcL(\vect\beta^{(j)})-\vect g^*-\vect H(\vect\beta^{(j)}-\vect\theta^*),~~|\vect e_j|_2\leq r_n.
\end{equation*}
For every index up to which these bounds hold, the restarted recursion has solution
\begin{equation*}
 \vect x_t=\vect P_t\vect x_0-\eta\sum_{j=0}^{t-1}\vect K_{t-1-j}(\vect g^*+\vect e_j).
\end{equation*}
Consequently $|\vect x_t|_2\leq|\vect x_0|_2+A_\rho(|\vect g^*|_2+r_n)\leq\epsilon_3/(1+2\gamma)$. The restart gives the same bound at $t=-1$. Hence
\begin{equation*}
 |\vect\beta^{(t)}-\vect\theta^*|_2
 \leq(1+\gamma)|\vect x_t|_2+\gamma|\vect x_{t-1}|_2\leq\epsilon_3.
\end{equation*}
Starting with $\vect\beta^{(0)}=\vect\theta^{(0)}$, induction validates the recursion at every update.

\noindent\textbf{Step 3. The output expansion.}
The identities in Step 1 give $\eta\sum_{j=0}^{T-1}\vect K_j=(\mbI-\vect P_T)\vect H^{-1}$. Separating the constant input $\vect g^*$ therefore yields
\begin{equation*}
 \vect x_T=-\vect H^{-1}\vect g^*+\vect P_T(\vect x_0+\vect H^{-1}\vect g^*)-\eta\sum_{j=0}^{T-1}\vect K_{T-1-j}\vect e_j.
\end{equation*}
The last term has norm at most $A_\rho r_n$, and the bound for $\vect P_T$ gives the stated transient bound, including coincident roots. The strict inequality $\vect\rho_1\prec\vect M$ gives the stated rate comparison at a common step size.
\end{proof}

\begin{lemma}\label{lem:eigen_bound}
Let $0<m_H\leq M_H$, $m_H\mbI\preceq\vect H\preceq M_H\mbI$, $0<\eta\leq M_H^{-1}$, and $\gamma\in[0,1)$. Write $\vect M=\mbI-\eta\vect H$. For an eigenvalue $m$ of $\vect M$, let $r(\gamma,m)$ denote the largest modulus of the roots of $x^2-(1+\gamma)mx+\gamma m=0$, and define $q(\eta,\gamma)=\max_{m\in\operatorname{spec}(\vect M)}r(\gamma,m)$.
\begin{enumerate}[label=(\roman*)]
\item If $\eta\leq\{(1-\gamma)/(1+\gamma)\}^2M_H^{-1}$, the matrices $\vect\rho_1,\vect\rho_2$ given by the root formula preceding Proposition \ref{prop:nag_fixed_momentum} satisfy
\begin{equation*}
\gamma\mbI\preceq\vect\rho_2\preceq\frac{2\gamma}{1+\gamma}\mbI\preceq\frac{1+\gamma}{2}\vect M\preceq\vect\rho_1\preceq\vect M.
\end{equation*}
Consequently $q(\eta,\gamma)=\Norm{\vect\rho_1}\leq1-\eta m_H$.
\item If $\gamma=(1-\sqrt{\eta m_H})/(1+\sqrt{\eta m_H})$, then $q(\eta,\gamma)\leq1-\sqrt{\eta m_H}$. If $m_H=\Lambda_{\min}(\vect H)$, equality holds and this momentum minimizes $q(\eta,\gamma)$ over $\gamma\in[0,1)$.
\end{enumerate}
\end{lemma}

\begin{proof}[Proof of Lemma \ref{lem:eigen_bound}]
All matrices are functions of $\vect H$ and commute. It therefore suffices to compare scalar eigenvalues. For part (i), $m\geq4\gamma/(1+\gamma)^2$ gives real nonnegative roots. The case $\gamma=0$ is immediate. If $\gamma>0$, the root formula and $m\leq1$ give
\begin{equation*}
\frac{1+\gamma}{2}m\leq x_1\leq m,~~
x_2=\frac{2\gamma m}{(1+\gamma)m+\sqrt{(1+\gamma)^2m^2-4\gamma m}},~~
\gamma\leq x_2\leq\frac{2\gamma}{1+\gamma}.
\end{equation*}
The lower bound on $m$ completes the displayed chain.

For part (ii), write $a=\sqrt{\eta m_H}$. If $a=1$, then $\vect M=\vect0$ and the conclusion follows. Otherwise $\gamma=(1-a)/(1+a)$ and $4\gamma/(1+\gamma)^2=1-a^2$. For $0<m<1-a^2$, the roots are complex conjugates with modulus $\sqrt{\gamma m}<1-a$. At $m=1-a^2$, both roots equal $1-a$, and at $m=0$ they are zero. This proves the spectral bound and equality when $m_H=\Lambda_{\min}(\vect H)$. To verify optimality in the latter case, fix $m_*=1-\eta m_H$. Below the stated momentum, its larger real root decreases with $\gamma$, since implicit differentiation gives $\partial x_1/\partial\gamma=m_*(x_1-1)/\{2x_1-(1+\gamma)m_*\}<0$. Above that momentum, its root modulus is $\sqrt{\gamma m_*}$ and increases with $\gamma$. The bound already proved holds simultaneously for every eigenvalue, establishing optimality.

\end{proof}

\subsection{Proofs of the convergence bounds}

\begin{proof}[Proof of Theorem \ref{thm:subgd_phase1}(i)]
Write $\Delta_t=\mcL(\vect\theta^{(t)})-\mcL(\hat{\vect\theta})$ and $\vect g_t=\partial\mcL(\vect\theta^{(t)})$. Convexity gives
\begin{equation}\label{eq:gd_thetathat_conv}
\Delta_t\leq\langle\vect g_t,\vect\theta^{(t)}-\hat{\vect\theta}\rangle.
\end{equation}
Combining this inequality with \eqref{eq:segment_descent} and expanding the squared distance along $\vect\theta^{(t+1)}=\vect\theta^{(t)}-\eta\vect g_t$ yields
\begin{equation}\label{eq:gd_distance_step}
\Delta_{t+1}\leq\frac{|\vect\theta^{(t)}-\hat{\vect\theta}|_2^2-|\vect\theta^{(t+1)}-\hat{\vect\theta}|_2^2}{2\eta}+\frac{\eta L^2\epsilon_1^2}{2}.
\end{equation}
Summing over $t=0,\ldots,T-1$ proves the averaged bound. If every update has length at least $\epsilon_1$, \eqref{eq:smooth_cond} gives
\begin{equation}\label{eq:gd_thetatt1_smooth}
\mcL(\vect\theta^{(t+1)})-\mcL(\vect\theta^{(t)})\leq-(\eta-\eta^2L)|\vect g_t|_2^2.
\end{equation}
For $\eta\leq1/(2L)$, this proves objective descent and removes the additive term from \eqref{eq:gd_distance_step}. Nonnegativity of $\Delta_{t+1}$ then makes the distances nonincreasing. Summation and objective descent give $\Delta_T\leq T^{-1}\sum_{t=1}^T\Delta_t\leq|\vect\theta^{(0)}-\hat{\vect\theta}|_2^2/(2\eta T)$.
\end{proof}

\begin{proof}[Proof of Theorem \ref{thm:subgd_phase2}(ii)]
Let $\Delta_t=\mcL(\vect\theta^{(t)})-\mcL(\hat{\vect\theta})$ and $\vect g_t=\partial\mcL(\vect\theta^{(t)})$. Inequality \eqref{eq:oracle_lower} and completion of the square give
\[
\Delta_t\leq|\vect g_t|_2|\vect\theta^{(t)}-\hat{\vect\theta}|_2-\frac\mu4|\vect\theta^{(t)}-\hat{\vect\theta}|_2^2+\frac\mu4\epsilon_2^2\leq\frac{|\vect g_t|_2^2}{\mu}+\frac\mu4\epsilon_2^2.
\]
With $c_\eta=\mu(\eta-L\eta^2/2)$, applying \eqref{eq:oracle_upper} along the update therefore yields
\[
\Delta_{t+1}\leq(1-c_\eta)\Delta_t+\frac{c_\eta\mu\epsilon_2^2}{4}+\frac L2\epsilon_1^2.
\]
Here $0<c_\eta\leq\mu/(2L)\leq1$. Summing the geometric recursion gives the bound with additive terms. Under the two distance conditions, Condition \ref{cond:mu_convex} gives the first display without its $\epsilon_2$ term, hence $|\vect g_t|_2^2\geq\mu\Delta_t$. Equation \eqref{eq:gd_thetatt1_smooth} then gives $\Delta_{t+1}\leq\{1-\mu(\eta-\eta^2L)\}\Delta_t$. Iteration proves the remaining bound whenever $0<\mu(\eta-\eta^2L)\leq1$.
\end{proof}

\begin{proof}[Proof of Theorem \ref{thm:subgd_phase3}(iii)]
Write $\vect{g}^*=\partial\mcL(\vect{\theta}^*)$ and $\vect{M}=\mbI-\eta\vect{H}$. If $|\vect{\theta}^{(t)}-\vect{\theta}^*|_2\leq\epsilon_3/2$, the local subgradient approximation gives
\begin{equation*}
\vect{\theta}^{(t+1)}-\vect{\theta}^*=\vect{M}(\vect{\theta}^{(t)}-\vect{\theta}^*)-\eta\vect{g}^*-\eta\vect{e}_t,~~|\vect{e}_t|_2\leq r_n.
\end{equation*}
Whenever $|\vect{\theta}^{(t)}-\vect{\theta}^*|_2\leq\epsilon_3/2$, the bound $\Norm{\vect{M}}\leq1-\eta m_H$ implies
\begin{equation*}
|\vect{\theta}^{(t+1)}-\vect{\theta}^*|_2\leq(1-\eta m_H)\epsilon_3/2+\eta(|\vect{g}^*|_2+r_n)\leq\epsilon_3/2.
\end{equation*}
Induction keeps every iterate in this ball. Solving the recursion yields \eqref{eq:bahadur_sgd} with
\begin{equation*}
\vect{R}_T=-\eta\sum_{t=0}^{T-1}\vect{M}^{T-1-t}\vect{e}_t,~~|\vect{R}_T|_2\leq\eta r_n\sum_{j=0}^{T-1}(1-\eta m_H)^j\leq r_n/m_H.
\end{equation*}

\end{proof}

\begin{proof}[Proof of Theorem \ref{thm:nagd_phase1}(i)]
For this proof, define $\lambda_{-1}=0$, $\gamma_{-1}=-1$, and $\vect\theta^{(-1)}=\vect\theta^{(0)}$. Write $\Delta_t=\mcL(\vect\theta^{(t)})-\mcL(\hat{\vect\theta})$, $\vect g_t=\partial\mcL(\vect\beta^{(t)})$, and $\vect u_t=\lambda_{t-1}\vect\theta^{(t)}-(\lambda_{t-1}-1)\vect\theta^{(t-1)}$. Convexity gives
\begin{align}
\mcL(\vect\beta^{(t)})-\mcL(\hat{\vect\theta})&\leq\langle\vect g_t,\vect\beta^{(t)}-\hat{\vect\theta}\rangle,\label{eq:nagd_thetahatbeta_conv}\\
\mcL(\vect\beta^{(t)})-\mcL(\vect\theta^{(t)})&\leq\langle\vect g_t,\vect\beta^{(t)}-\vect\theta^{(t)}\rangle.\label{eq:nagd_thetatbeta_conv}
\end{align}
The momentum and update identities imply $\lambda_t(\lambda_t-1)=\lambda_{t-1}^2$, $\lambda_t\vect\beta^{(t)}-(\lambda_t-1)\vect\theta^{(t)}=\vect u_t$, and $\vect u_{t+1}=\vect u_t-\eta\lambda_t\vect g_t$. Weight the two convexity bounds by $\lambda_t$ and $\lambda_t(\lambda_t-1)$. Combining them with the one-step model multiplied by $\lambda_t^2$ and completing the square gives
\begin{equation}\label{eq:nag_phase1_potential_step}
\lambda_t^2\Delta_{t+1}-\lambda_{t-1}^2\Delta_t
\leq\frac{|\vect u_t-\hat{\vect\theta}|_2^2-|\vect u_{t+1}-\hat{\vect\theta}|_2^2}{2\eta}+\lambda_t^2\delta_t.
\end{equation}
Since $\vect u_0=\vect\theta^{(0)}$, summation proves \eqref{eq:nag_short_step_errors}. Under Condition \ref{cond:L_smooth} and $\eta\leq1/(2L)$, \eqref{eq:segment_descent} supplies the stated $\delta_t$. Monotonicity of $\lambda_t$ gives $\sum_{t<T}\lambda_t^2\delta_t/\lambda_{T-1}^2\leq T\eta L^2\epsilon_1^2/2$, and Lemma \ref{eq:lambda_bound} gives the $T^{-2}$ term. For an update of length at least $\epsilon_1$, \eqref{eq:smooth_cond} gives
\begin{equation}\label{eq:nagd_thetat1beta_smooth}
\mcL(\vect\theta^{(t+1)})\leq\mcL(\vect\beta^{(t)})-(\eta-\eta^2L)|\vect g_t|_2^2\leq\mcL(\vect\beta^{(t)})-\frac\eta2|\vect g_t|_2^2.
\end{equation}
Thus all $\delta_t$ vanish when every update is long enough, proving the error-free objective bound.

Before the first update shorter than $\epsilon_1$, \eqref{eq:nag_phase1_potential_step} and the momentum identities give, with $d_0=|\vect\theta^{(0)}-\hat{\vect\theta}|_2$,
\begin{equation}\label{eq:nag_phase1_containment}
\begin{gathered}
\lambda_{t-1}^2\{\mcL(\vect\theta^{(t)})-\mcL(\hat{\vect\theta})\}+\frac{|\vect u_t-\hat{\vect\theta}|_2^2}{2\eta}\leq\frac{d_0^2}{2\eta},\\
\vect\theta^{(t)}=\left(1-\frac1{\lambda_{t-1}}\right)\vect\theta^{(t-1)}+\frac{\vect u_t}{\lambda_{t-1}}\quad(t\geq1),\\
\vect\beta^{(t)}=\left(1-\frac1{\lambda_t}\right)\vect\theta^{(t)}+\frac{\vect u_t}{\lambda_t}.
\end{gathered}
\end{equation}
Since $\lambda_t\geq1$ and $\vect u_0=\vect\theta^{(0)}$, the last two identities are convex combinations. Nonnegativity of the objective gap and induction therefore keep $\vect u_t$, $\vect\theta^{(t)}$, and $\vect\beta^{(t)}$ within distance $d_0$ of $\hat{\vect\theta}$ whenever the preceding updates satisfy the distance and domain conditions.
\end{proof}

\begin{proof}[Proof of Theorem \ref{thm:nagd_phase2}(ii)]
Define $\vect\theta^{(-1)}=\vect\theta^{(0)}$, $\vect g_t=\partial\mcL(\vect\beta^{(t)})$, and $\Delta_t=\mcL(\vect\theta^{(t)})-\mcL(\hat{\vect\theta})$. Convexity and \eqref{eq:oracle_lower} give
\begin{align}
\mcL(\vect\beta^{(t)})&\leq\mcL(\vect\theta^{(t)})+\langle\vect g_t,\vect\beta^{(t)}-\vect\theta^{(t)}\rangle,\label{eq:thetatt1_convex}\\
\mcL(\vect\beta^{(t)})&\leq\mcL(\hat{\vect\theta})+\langle\vect g_t,\vect\beta^{(t)}-\hat{\vect\theta}\rangle-\frac\mu4|\vect\beta^{(t)}-\hat{\vect\theta}|_2^2+\frac\mu4\epsilon_2^2.\label{eq:thetatthat_strconvex}
\end{align}
Set $\vect r^{(0)}=\alpha(\vect\theta^{(0)}-\hat{\vect\theta})$, $\vect w^{(t)}=\vect\beta^{(t)}-(1-\alpha)\vect\theta^{(t)}-\alpha\hat{\vect\theta}$, and $\vect r^{(t+1)}=\vect w^{(t)}-\eta\vect g_t=\vect\theta^{(t+1)}-(1-\alpha)\vect\theta^{(t)}-\alpha\hat{\vect\theta}$. Weight the two inequalities by $1-\alpha$ and $\alpha$. Combining them with \eqref{eq:segment_descent} and completing the square gives
\begin{equation}\label{eq:ngd_geom_term2}
\Delta_{t+1}+\frac{|\vect r^{(t+1)}|_2^2}{2\eta}
\leq(1-\alpha)\Delta_t+\frac{|\vect w^{(t)}|_2^2}{2\eta}-\frac{\alpha\mu}{4}|\vect\beta^{(t)}-\hat{\vect\theta}|_2^2+\frac{\eta L^2\epsilon_1^2}{2}+\frac{\alpha\mu\epsilon_2^2}{4}.
\end{equation}
For $c=(1-\gamma)/(1+\gamma)$, direct expansion gives
	\begin{align*}
		&\frac{1}{2\eta}|\vect{w}^{(t)}|_2^2 - \frac{\alpha\mu}{4}|\hat{\vect{\theta}}-\vect{\beta}^{(t)}|_2^2 - \frac{1-\alpha}{2\eta}|\vect{r}^{(t)}|_2^2\\
		=&-\frac{1}{2\eta}\Big\{(c-\alpha)|\vect{r}^{(t)} + \gamma(\vect{\theta}^{(t)} - \vect{\theta}^{(t-1)})|_2^2+\alpha\gamma^2(1-\alpha)|\vect{\theta}^{(t)} - \vect{\theta}^{(t-1)}|_2^2\\
		&+\frac{\alpha}{2}(\mu\eta -2\alpha c)|\vect{\beta}^{(t)} - \hat{\vect{\theta}}|_2^2\Big\}\leq 0.
	\end{align*}
The last inequality uses $\alpha\leq c$ and $2\alpha c\leq\eta\mu$. Hence the potential
\begin{equation}\label{eq:nag_phase2_potential}
\Phi_t=\Delta_t+\frac{|\vect r^{(t)}|_2^2}{2\eta}
\end{equation}
satisfies
\begin{equation}\label{eq:nag_phase2_potential_step}
\Phi_{t+1}\leq(1-\alpha)\Phi_t+\frac{\eta L^2\epsilon_1^2}{2}+\frac{\alpha\mu\epsilon_2^2}{4}.
\end{equation}
Since $0<\alpha\leq1$ and $\Phi_0=\Delta_0+\alpha^2|\vect\theta^{(0)}-\hat{\vect\theta}|_2^2/(2\eta)$, summing this recursion proves the bound with additive terms. For $\eta\leq1/(2L)$ and under the two distance conditions, Condition \ref{cond:mu_convex} removes the $\epsilon_2$ term and \eqref{eq:nagd_thetat1beta_smooth} removes the $\epsilon_1$ term. The same square completion then gives \eqref{eq:nag_phase2_potential_step} without either additive term and proves the remaining bound.
\end{proof}

\begin{proof}[Proof of Theorem \ref{thm:nagd_phase3}(iii)]
For this proof, define $\vect\theta^{(-1)}=\vect\theta^{(0)}$. Write $\vect g^*=\partial\mcL(\vect\theta^*)$, $\vect M=\mbI-\eta\vect H$, and $\vect z_0=\vect\theta^{(0)}-\vect\theta^*+\vect H^{-1}\vect g^*$. Define $\vect P_0=\mbI$, $\vect P_1=\vect M$, and $\vect P_{t+1}=(1+\gamma)\vect M\vect P_t-\gamma\vect M\vect P_{t-1}$.
Write $a=\sqrt{\eta m_H}$ and $\rho=1-a$. Let $\vect K_{-1}=\vect0$, $\vect K_0=\mbI$, and $\vect K_{j+1}=(1+\gamma)\vect M\vect K_j-\gamma\vect M\vect K_{j-1}$. Then $\vect P_t=\vect K_t-\gamma\vect M\vect K_{t-1}$. Diagonalizing the symmetric matrix $\vect H$ reduces both sequences to scalar polynomials. For an eigenvalue $m\in[0,1-a^2]$ of $\vect M$, the characteristic roots $x_1,x_2$ satisfy $|x_i|\leq\rho$ by Lemma \ref{lem:eigen_bound}(ii). The polynomial $K_j$ therefore satisfies
\begin{equation*}
K_j(m)=\sum_{k=0}^j x_1^kx_2^{j-k},~~|K_j(m)|\leq(j+1)\rho^j,
\end{equation*}
including complex and repeated roots. Consequently $\Norm{\vect K_j}\leq(j+1)\rho^j$ and $\eta\sum_{j\geq0}\Norm{\vect K_j}\leq\eta/(1-\rho)^2=1/m_H$.

We next bound the homogeneous polynomial without separating the roots. If $0<a<1$ and $0<m<1-a^2$, write the roots as $r\exp(\pm\mathrm i\phi)$, where $r=\sqrt{\gamma m}\leq\rho$ and $\cos\phi=(1+\gamma)m/(2r)$. Since $(1-\gamma)/(1+\gamma)=a$,
\begin{equation*}
P_t(m)=r^t\left\{\cos(t\phi)+a\cos\phi\frac{\sin(t\phi)}{\sin\phi}\right\}.
\end{equation*}
The inequality $|\sin(t\phi)|\leq t\sin\phi$ gives $|P_t(m)|\leq(1+at)\rho^t$. At $m=1-a^2$, the double root is $\rho$ and $P_t(m)=(1+at)\rho^t$. At $m=0$, $P_t(m)=0$ for $t\geq1$. When $a=1$, $\vect M=\vect0$ and the same bounds hold directly. Hence
\begin{equation*}
\Norm{\vect P_t}\leq(1+at)\rho^t\leq1,~~\Norm{\vect P_t}\leq(1+at)e^{-at}\leq2e^{-at/2}.
\end{equation*}
The bound by $1$ follows from $1+at\leq(1-a)^{-t}$ for $0<a<1$.

Write $\vect w_t=\vect\theta^{(t)}-\vect\theta^*+\vect H^{-1}\vect g^*$. While the extrapolated points lie in the local ball, define $\vect e_t=\partial\mcL(\vect\beta^{(t)})-\vect g^*-\vect H(\vect\beta^{(t)}-\vect\theta^*)$. Condition \ref{cond:H_smooth} gives
\begin{equation*}
\vect w_{t+1}=(1+\gamma)\vect M\vect w_t-\gamma\vect M\vect w_{t-1}-\eta\vect e_t,~~|\vect e_t|_2\leq r_n,
\end{equation*}
with $\vect w_{-1}=\vect w_0=\vect z_0$. Solving this recursion yields
\begin{equation*}
\vect w_t=\vect P_t\vect z_0-\eta\sum_{j=0}^{t-1}\vect K_{t-1-j}\vect e_j,~~|\vect w_t|_2\leq|\vect z_0|_2+r_n/m_H.
\end{equation*}
Let $C=|\vect z_0|_2+r_n/m_H$. Initially $|\vect w_{-1}|_2=|\vect w_0|_2\leq C$. If the errors through index $t-1$ are bounded by $r_n$, the solution formula gives $|\vect w_t|_2\leq C$. Since $\vect\beta^{(t)}-\vect\theta^*=(1+\gamma)\vect w_t-\gamma\vect w_{t-1}-\vect H^{-1}\vect g^*$, the entry condition in Theorem \ref{thm:nagd_phase3}(iii) gives $|\vect\beta^{(t)}-\vect\theta^*|_2\leq (1+2\gamma)C+|\vect H^{-1}\vect g^*|_2\leq\epsilon_3$ and validates the bound on $\vect e_t$. Induction therefore covers every $t\geq0$. Also $|\vect\theta^{(t)}-\vect\theta^*|_2\leq C+|\vect H^{-1}\vect g^*|_2\leq\epsilon_3$. The solution formula and the bounds for $\vect P_t$ and $\vect K_j$ give \eqref{eq:ngd_bahadur_optimal}.
\end{proof}

\begin{proof}[Proof of Corollary \ref{cor:absorption}]
Write $\Delta(\vect\theta)=\mcL(\vect\theta)-\mcL(\hat{\vect\theta})$, $d(\vect\theta)=|\vect\theta-\hat{\vect\theta}|_2$, and $R_1=2\max\{R_0,R\}+2\epsilon_1$. Lemma \ref{lem:oracle_bounds} gives $|\partial\mcL(\hat{\vect\theta})|_2\leq L\epsilon_1$, hence
\[
|\partial\mcL(\vect\theta)|_2\leq L\{d(\vect\theta)+2\epsilon_1\},~~\vect\theta\in\mbB(\hat{\vect\theta},R_1).
\]
For $\vect\theta\in\mbB(\hat{\vect\theta},R)$, \eqref{eq:oracle_lower} and \eqref{eq:gap_growth} give
\begin{equation}\label{eq:absorb_gap_distance}
|\partial\mcL(\vect\theta)|_2^2\geq\mu\{\Delta(\vect\theta)-\mu\epsilon_2^2/4\},~~
d(\vect\theta)^2\leq\frac4\mu\Delta(\vect\theta)+\epsilon_2^2.
\end{equation}
The first inequality follows by completing the square in $|\partial\mcL(\vect\theta)|_2d-\mu d^2/4$.

Set $b=\mu R^2/16$. Since $R\geq4\epsilon_2$, \eqref{eq:absorb_gap_distance} gives $\Delta\geq b$ on the sphere $d=R$. Convexity along rays from $\hat{\vect\theta}$ then gives $\Delta(\vect\theta)\geq\mu R d(\vect\theta)/16$ for $d(\vect\theta)>R$. Thus $\{\Delta<b\}\subseteq\mbB(\hat{\vect\theta},R)$. Every point in the smoothness region with $\Delta\geq b$ has a subgradient step of length at least $\epsilon_1$. Inside the radius-$R$ ball, \eqref{eq:absorb_gap_distance} gives $|\partial\mcL|_2\geq\mu R/(4\sqrt2)$, and outside it convexity gives $|\partial\mcL|_2\geq\Delta/d\geq\mu R/16$. The condition $R\geq16\epsilon_1/(\eta\mu)$ suffices in both cases.

\noindent\textbf{Part (i).}
Let $\tau=\inf\{t:\Delta(\vect\theta^{(t)})<b\}$ and $q=1-3\eta\mu/4$. Before $\tau$, suppose $d(\vect\theta^{(t)})\leq R_0$. The proposed next iterate has distance at most $3R_0/2+\epsilon_1\leq R_1$, and the step is long enough for the error-free bound in Theorem \ref{thm:subgd_phase1}(i). Its distance to $\hat{\vect\theta}$ therefore cannot increase. Induction gives $\Delta(\vect\theta^{(t)})\leq R_0^2/(2\eta t)$ up to $\tau$. Hence $T_1>8R_0^2/(\eta\mu R^2)$ implies $\tau\leq T_1$.

For a current point $\vect\theta$ with $d(\vect\theta)\leq R$, write $\vect g=\partial\mcL(\vect\theta)$. The auxiliary point $\vect\theta-\vect g/L$ lies in $\mbB(\hat{\vect\theta},R_1)$ since its distance is at most $2R+2\epsilon_1$. Minimality and \eqref{eq:oracle_upper} applied to this step yield $|\vect g|_2^2\leq2L\Delta(\vect\theta)+L^2\epsilon_1^2$. Applying \eqref{eq:oracle_lower} with comparison point $\hat{\vect\theta}$ and then \eqref{eq:gap_growth} gives
\begin{align*}
d(\vect\theta-\eta\vect g)^2
&\leq\left(1-\frac{\eta\mu}{2}\right)d(\vect\theta)^2-\eta\Delta(\vect\theta)+\frac{\eta\mu}{2}\epsilon_2^2+\eta^2L^2\epsilon_1^2\\
&\leq q\,d(\vect\theta)^2+(1-q)\left(\epsilon_2^2+\frac{4\eta L^2}{3\mu}\epsilon_1^2\right).
\end{align*}
The first inequality uses $2\eta(1-\eta L)\geq\eta$. The radius assumptions imply $R^2\geq\epsilon_2^2+4\eta L^2\epsilon_1^2/(3\mu)$, making this ball invariant. At entry, \eqref{eq:absorb_gap_distance} gives $d(\vect\theta^{(\tau)})^2\leq R^2/4+\epsilon_2^2\leq R^2/2$. Iterating the distance recursion therefore gives
\[
d(\vect\theta^{(t)})^2\leq q^{t-\tau}\frac{R^2}{2}+\epsilon_2^2+\frac{4\eta L^2}{3\mu}\epsilon_1^2,~~t\geq\tau.
\]
Taking square roots and using $\tau\leq T_1$ proves \eqref{eq:absorb_subgd} and the distance bound.

\noindent\textbf{Part (ii).}
During the initial Nesterov phase, every executed update has length at least $\epsilon_1$. The potential and convex-combination identities in \eqref{eq:nag_phase1_containment} keep the completed iterates and the next extrapolated point within distance $R_0$ of $\hat{\vect\theta}$. The proposed next iterate lies within distance $3R_0/2+\epsilon_1\leq R_1$. Induction therefore validates each application of the error-free potential bound until the short-step test stops this phase or all $N_1$ updates are completed.

Let $\bar{\vect\theta}$ be the point at which the constant-momentum phase starts. If the short-step test is met, $|\partial\mcL(\bar{\vect\theta})|_2<\epsilon_1/\eta\leq\mu R/16$. The lower bound outside the radius-$R$ ball proved above places $\bar{\vect\theta}$ inside that ball. Completing the square in \eqref{eq:oracle_lower} and then applying \eqref{eq:gap_growth} give
\[
\Delta(\bar{\vect\theta})\leq\frac{|\partial\mcL(\bar{\vect\theta})|_2^2}{\mu}+\frac\mu4\epsilon_2^2
\leq\frac{65\mu R^2}{16384},\qquad
d(\bar{\vect\theta})^2\leq\frac{264R^2}{16384}.
\]
If all $N_1$ updates are completed, Theorem \ref{thm:nagd_phase1}(i) gives instead
\[
\Delta(\bar{\vect\theta})\leq\frac{2R_0^2}{\eta N_1^2}\leq\frac{\mu R^2}{256},\qquad
d(\bar{\vect\theta})^2\leq\frac{65R^2}{4096}.
\]
Here the objective bound first places $\bar{\vect\theta}$ inside the radius-$R$ ball, and \eqref{eq:gap_growth} gives the distance bound. The additional radius assumption gives $\epsilon_2\leq R/64$ in both cases.

Reset the time index at the restart with $\vect\theta^{(0)}=\bar{\vect\theta}$, and define $\vect\theta^{(-1)}=\vect\theta^{(0)}$ for this proof. Let $\alpha=\sqrt{\eta\mu/2}$ and use the potential from the proof of Theorem \ref{thm:nagd_phase2}(ii),
\[
\Phi_t=\Delta(\vect\theta^{(t)})+\frac1{2\eta}\big|\vect\theta^{(t)}-(1-\alpha)\vect\theta^{(t-1)}-\alpha\hat{\vect\theta}\big|_2^2.
\]
Since $\alpha^2/(2\eta)=\mu/4$, its initial value and forcing bound satisfy
\[
\Phi_0\leq\frac{131\mu R^2}{16384}<\frac{\mu R^2}{120},~~
E_N=\frac{\eta L^2\epsilon_1^2}{2\alpha}+\frac{\mu\epsilon_2^2}4
=\mu\alpha\kappa^2\epsilon_1^2+\frac\mu4\epsilon_2^2\leq\frac{\mu R^2}{4096}.
\]
The last inequality uses $\alpha\leq1$ and $\epsilon_2+\kappa\epsilon_1\leq R/64$.

We prove $\Phi_t\leq(1-\alpha)^t\Phi_0+E_N<\mu R^2/100$ while verifying the points used by the potential recursion. If the bound holds through time $t$, the sublevel-set inclusion places $\vect\theta^{(t)}$ in the radius-$R$ ball. Inequality \eqref{eq:gap_growth} then gives $d(\vect\theta^{(t)})^2<R^2/16$. The same holds at $t-1$, including the initial index $-1$ by the restart. Thus $|\vect\beta^{(t)}-\hat{\vect\theta}|_2\leq3R/4<R$, and the proposed $\vect\theta^{(t+1)}$ lies within distance $3R/2+\epsilon_1\leq R_1$. The one-step potential bound is therefore valid and gives
\[
\Phi_{t+1}\leq(1-\alpha)\Phi_t+\eta L^2\epsilon_1^2/2+\alpha\mu\epsilon_2^2/4
=(1-\alpha)\Phi_t+\alpha E_N.
\]
This closes the induction. Finally, \eqref{eq:gap_growth} gives
\[
d(\vect\theta^{(t)})^2\leq\frac1{30}(1-\alpha)^tR^2+2\epsilon_2^2+4\alpha\kappa^2\epsilon_1^2.
\]
Taking square roots and using $1/\sqrt{30}<1/3$ proves \eqref{eq:absorb_nagd}.
\end{proof}

\begin{proof}[Proof of the radius lower bound following Corollary \ref{cor:absorption}]
For $0<\mu\leq L/2$ and $\epsilon_1>0$, consider $\mcL(\vect\theta)=a|\vect\theta|_1+(\mu/2)|\vect\theta|_2^2$, where $a=(L-\mu)\epsilon_1/(2\sqrt p)$. Its minimizer is $\hat{\vect\theta}=\vect0$. Use step size $\eta=1/(2L)$.
For every subgradient selection, $|\partial\mcL(\vect\theta_1)-\partial\mcL(\vect\theta_2)|_2\leq\mu|\vect\theta_1-\vect\theta_2|_2+2a\sqrt p=\mu|\vect\theta_1-\vect\theta_2|_2+(L-\mu)\epsilon_1$, which verifies Condition \ref{cond:L_smooth}. The quadratic term verifies Condition \ref{cond:mu_convex} with $\epsilon_2=0$. Away from zero, each coordinate evolves by $x\mapsto(1-\eta\mu)x-\eta a\operatorname{sign}(x)$. Both affine branches have nonzero slope. Thus at most $2^t$ scalar initial values first reach zero at time $t$, and their union over $t\geq0$ is countable. The initial vectors having a coordinate in this set form a Lebesgue-null set common to all selections. Outside it, every iterate has nonzero coordinates, hence a unique subgradient, and all selections generate the same trajectory. The coordinate magnitudes evolve by $u\mapsto|(1-\eta\mu)u-\eta a|$, a contraction with fixed point $\eta a/(2-\eta\mu)$. Each magnitude therefore converges to this value, and $|\vect\theta^{(t)}-\hat{\vect\theta}|_2$ converges to $\sqrt p\eta a/(2-\eta\mu)=(L-\mu)\epsilon_1/\{2(4L-\mu)\}$. Since $0<\mu\leq L/2$, this limit is at least $\epsilon_1/14$.
\end{proof}

\subsection{Complete procedures}\label{app:phase_combination}

\phantomsection\label{app:unprojected_saa}
Under the model, population, and step-size assumptions of Theorem \ref{cor:model_algorithm}, suppose $|\vect\theta^{(0)}-\vect\theta^*|_2\leq D$ and $\mbB(\vect\theta^*,2D+R)\subseteq\Theta$ for $D>0$ and sufficiently small $R>0$. Run either unprojected method in Corollary \ref{cor:absorption} with radius $R/4$ and initialization bound $R_0=D+R/8$, counting all phases. All three conclusions of Theorem \ref{cor:model_algorithm} hold for these iterates, uniformly over the stated initial decisions and all scenario-subgradient selections.

\begin{proof}[Proof of the unprojected extension]
The localization argument in the proof of Theorem \ref{cor:model_algorithm} gives $\delta_n=|\hat{\vect\theta}-\vect\theta^*|_2\leq R/8$ on the common increment event for sufficiently large $n$. For sufficiently small $R$, the same argument verifies Conditions \ref{cond:L_smooth} and \ref{cond:mu_convex} on $\Theta$ and $\mbB(\vect\theta^*,R)$, respectively. With $R_{\rm abs}=R/4$ and $R_0=D+R/8$, the two regions needed by Corollary \ref{cor:absorption} satisfy
\[
\mbB(\hat{\vect\theta},R_{\rm abs})\subseteq\mbB(\vect\theta^*,R),~~\mbB(\hat{\vect\theta},2\max\{R_0,R_{\rm abs}\}+2\epsilon_1)\subseteq\mbB(\vect\theta^*,2D+R)
\]
for sufficiently large $n$. For the second inclusion, use
$\delta_n+2\max\{D+R/8,R/4\}+2\epsilon_1\leq2D+R$.
Also $R_{\rm abs}\geq\max\{64(\epsilon_2+\kappa\epsilon_1),16\epsilon_1/(\eta\mu)\}$ eventually, where $\kappa=L/\mu$. The initialization bound is valid and Phase 1 takes a bounded number of updates. Since $a_n\geq n^{-1}$ eventually, Corollary \ref{cor:absorption} gives $|\vect\theta^{(t)}-\hat{\vect\theta}|_2\leq Ca_n$ for every $t\geq\lceil c\log n\rceil$. Combining this bound with the SAA expansion and population-gap calculation in the proof of Theorem \ref{cor:model_algorithm} proves the extension.
\end{proof}

The following corollaries combine the phase bounds with subgradient stopping rules. They provide explicit iteration counts and a first-order expansion of the computed solution.

For subgradient descent with $D,R>0$ and eigenvalue bounds $0<m_H\leq M_H$, define
\begin{equation*}
 \eta=\frac1{2L},~~ \eta_3=M_H^{-1},~~
 \tau_n=\frac{\mu\epsilon_3}{8},~~
 \epsilon_0=\max\{\epsilon_1,\sqrt2\epsilon_2\},~~ q=1-\frac{\mu}{4L}.
\end{equation*}
Write $\log_+(x)=\max\{0,\log x\}$. For $r_n>0$, define the iteration counts
\begin{equation*}
 T_1=\left\lceil\frac{8(D+R/4)^2}{\eta\mu R^2}\right\rceil,~~
 T_2=\left\lceil\frac{\log_+(2L^2R^2/\tau_n^2)}{-\log q}\right\rceil,~~
 K=\left\lceil\frac{1}{\eta_3m_H}\log\left(\frac{m_H\epsilon_3}{r_n}\right)\right\rceil.
\end{equation*}
Phase 1 uses step size $\eta$ for at most $T_1$ updates. If $|\partial\mcL(\vect{\theta})|_2\leq\tau_n$ at any check, the algorithm proceeds directly to Phase 3. Otherwise Phase 2 continues with the same step size until this inequality holds. Phase 3 uses step size $\eta_3$ for $K$ updates.

\begin{corollary}\label{thm:full_algorithm}
Suppose an event $\mcE_n$ has probability at least $1-\delta_n$. Let $\mathcal K\subseteq\Theta$ be convex and contain the ball of radius $2D+R$ about $\vect\theta^*$. On $\mcE_n$, assume that $\mcL$ is finite and convex on $\mbR^p$, Condition \ref{cond:L_smooth} holds for pairs in $\mathcal K$, and Condition \ref{cond:mu_convex} holds for pairs in the radius-$R$ ball about $\vect{\theta}^*$. Suppose the local subgradient approximation in Condition \ref{cond:H_smooth} holds with remainder bound $r_n$, and
\begin{gather*}
 m_H\mbI\preceq\vect{H}\preceq M_H\mbI,~~ 0<\mu\leq2L,~~ 0<\epsilon_3\leq R/2,~~
 |\vect{\theta}^{(0)}-\vect{\theta}^*|_2\leq D,\\
 \epsilon_0\leq\frac{\tau_n}{4L},~~ |\partial\mcL(\vect{\theta}^*)|_2\leq\min\{\mu/16,m_H/4\}\epsilon_3,~~ r_n\leq m_H\epsilon_3/4.
\end{gather*}
On $\mcE_n$, every iterate remains in $\mathcal K$, Phase 2 stops within $T_2$ updates, and Phase 3 starts within distance $\epsilon_3/2$ of $\vect{\theta}^*$. After at most $T_1+T_2+K$ updates, the computed solution and the SAA solution satisfy
\begin{equation}\label{eq:complete_expansions}
 \tilde{\vect{\theta}}-\vect{\theta}^*=-\vect{H}^{-1}\partial\mcL(\vect{\theta}^*)+\vect{R}_n,~~
 \hat{\vect{\theta}}-\vect{\theta}^*=-\vect{H}^{-1}\partial\mcL(\vect{\theta}^*)+\hat{\vect{R}}_n,
\end{equation}
where $|\vect R_n|_2\leq2r_n/m_H$ and $|\hat{\vect R}_n|_2\leq r_n/m_H$. Consequently, $|\tilde{\vect{\theta}}-\hat{\vect{\theta}}|_2\leq3r_n/m_H$.
\end{corollary}

\begin{proof}[Proof of Corollary \ref{thm:full_algorithm}]
\noindent\textbf{Step 1. Localization of the SAA solution.}
Work on $\mcE_n$ and write $\vect{g}(\vect{\theta})=\partial\mcL(\vect{\theta})$ and $s_n=|\vect{g}(\vect{\theta}^*)|_2$. We first locate the SAA solution. For each unit vector $\vect{u}$, Condition \ref{cond:mu_convex} gives
\begin{equation*}
\mcL(\vect{\theta}^*+R\vect{u})-\mcL(\vect{\theta}^*)\geq-s_nR+\frac{\mu R^2}{4}\geq\frac{\mu R^2}{8}.
\end{equation*}
For $t>R$, convexity along the same direction implies
\begin{equation*}
\mcL(\vect{\theta}^*+t\vect{u})-\mcL(\vect{\theta}^*)\geq\frac{t}{R}\{\mcL(\vect{\theta}^*+R\vect{u})-\mcL(\vect{\theta}^*)\}\geq\frac{\mu Rt}{8}.
\end{equation*}
The SAA objective is therefore coercive, and every SAA solution lies inside the ball of radius $R$. Since $\mcL(\hat{\vect{\theta}})\leq\mcL(\vect{\theta}^*)$, Condition \ref{cond:mu_convex} gives
\begin{equation*}
|\hat{\vect{\theta}}-\vect{\theta}^*|_2\leq\max\{\epsilon_2,4s_n/\mu\}\leq R/4.
\end{equation*}

The assumptions also bound this distance by $\epsilon_3/4$. Hence every global minimizer belongs to $\Theta$, and the global minimizers are exactly the SAA solutions in \eqref{eq:emp_loss}.

Since $\hat{\vect\theta}$ lies in the interior of $\mbB(\vect\theta^*,2D+R)$, Lemma \ref{lem:oracle_bounds} gives $|\vect g(\hat{\vect\theta})|_2\leq L\epsilon_1$.

\noindent\textbf{Step 2. Feasible iterates and entry into Phase 2.}
Initially $|\vect\theta^{(0)}-\hat{\vect\theta}|_2\leq D+R/4$. Suppose this bound holds at a current iterate with $|\vect g(\vect\theta^{(t)})|_2>\tau_n$. Condition \ref{cond:L_smooth} and $|\vect g(\hat{\vect\theta})|_2\leq L\epsilon_1$ give $|\vect g(\vect\theta^{(t)})|_2\leq L\{|\vect\theta^{(t)}-\hat{\vect\theta}|_2+2\epsilon_1\}$. The proposed update therefore satisfies
\[
|\vect\theta^{(t+1)}-\vect\theta^*|_2\leq\frac32(D+R/4)+\epsilon_1+R/4\leq2D+R,
\]
since the assumptions imply $\epsilon_1\leq R/32$. The update lies in $\mathcal K$ and has length greater than $\eta\tau_n\geq\epsilon_1$. Equations \eqref{eq:gd_distance_step} and \eqref{eq:gd_thetatt1_smooth}, applied to this update with comparator $\hat{\vect\theta}$, give nonincreasing distance and objective values, with no additive error. Induction validates every update before Phase 3. If all $T_1$ updates are used, the sum of \eqref{eq:gd_distance_step} gives
\[
\mcL(\vect\theta^{(T_1)})-\mcL(\hat{\vect\theta})\leq\frac{(D+R/4)^2}{2\eta T_1}\leq\frac{\mu R^2}{16}.
\]
The lower bound in Step 1 places this iterate inside the radius-$R$ ball, and objective descent keeps Phase 2 there.

\noindent\textbf{Step 3. The Phase 2 stopping time.}
At a Phase 2 iterate with subgradient norm greater than $\tau_n$, write $d=|\vect{\theta}-\hat{\vect{\theta}}|_2$ and $\Delta=\mcL(\vect{\theta})-\mcL(\hat{\vect{\theta}})$. If $d<\epsilon_0$, approximate smoothness would give $|\vect{g}(\vect{\theta})|_2\leq2L\epsilon_0\leq\tau_n/2$. Hence $d\geq\epsilon_0$. Applying Lemma \ref{lem:approx_subreg} on $\mbB(\vect\theta^*,R)$ gives
\begin{equation*}
\mu\Delta\leq|\vect{g}(\vect{\theta})|_2^2\leq\frac{32L^2}{\mu}\Delta.
\end{equation*}
The first inequality and the descent bound contract $\Delta$ by $q$ at each Phase 2 update. After $T_2$ updates, the objective gap is at most $\mu\tau_n^2/(32L^2)$. If the subgradient norm still exceeded $\tau_n$, the second inequality would give $|\vect g(\vect\theta)|_2^2\leq\tau_n^2$, a contradiction. This proves the bound on the Phase 2 stopping time.

\noindent\textbf{Step 4. Entry into Phase 3 and the output expansion.}
We now locate the iterate that satisfies $|\vect{g}(\vect{\theta})|_2\leq\tau_n$. This argument also covers a direct transition from Phase 1 to Phase 3. Applying Condition \ref{cond:mu_convex} in both orders between $\vect{\theta}^*+R\vect{u}$ and $\vect{\theta}^*$ yields
\begin{equation*}
\langle\vect{g}(\vect{\theta}^*+R\vect{u}),\vect{u}\rangle\geq\frac{\mu R}{2}-s_n>\tau_n.
\end{equation*}
Monotonicity of subgradients gives the same lower bound at $\vect{\theta}^*+t\vect{u}$ for $t>R$. An iterate meeting the stopping condition therefore lies inside the ball of radius $R$. Applying Condition \ref{cond:mu_convex} in both orders between this iterate and $\vect{\theta}^*$ gives
\begin{equation*}
|\vect{\theta}-\vect{\theta}^*|_2\leq\max\{\epsilon_2,2(\tau_n+s_n)/\mu\}\leq\epsilon_3/2.
\end{equation*}
Since $s_n+r_n\leq m_H\epsilon_3/2$, Theorem \ref{thm:subgd_phase3}(iii) keeps every Phase 3 iterate in the required neighborhood. The choice of $K$ gives $(1-\eta_3m_H)^K\leq\exp(-\eta_3m_HK)\leq r_n/(m_H\epsilon_3)$. The transient term consequently has norm at most $3r_n/(4m_H)$, and the expansion remainder has norm at most $r_n/m_H$. Their sum is bounded by $2r_n/m_H$.

\noindent\textbf{Step 5. Accuracy relative to the SAA solution.}
The same approximation gives the expansion of the SAA solution. Its localization bound above implies $|\hat{\vect\theta}-\vect\theta^*|_2\leq\epsilon_3/4$. Define $\vect A=\vect g(\vect\theta^*)+\vect H(\hat{\vect\theta}-\vect\theta^*)$. If $\vect A\ne\vect0$, let $\vect u=\vect A/|\vect A|_2$. For sufficiently small $t>0$, the point $\hat{\vect\theta}-t\vect u$ lies in the ball of radius $\epsilon_3$. The subgradient inequality and minimality of $\hat{\vect\theta}$ give
\begin{equation*}
0\geq\langle\vect g(\hat{\vect\theta}-t\vect u),\vect u\rangle\geq|\vect A|_2-tM_H-r_n.
\end{equation*}
Letting $t\downarrow0$ yields $|\vect A|_2\leq r_n$, which also holds when $\vect A=\vect0$. Thus $\hat{\vect R}_n=\vect H^{-1}\vect A$ has norm at most $r_n/m_H$. Subtracting the two first-order expansions gives $|\tilde{\vect\theta}-\hat{\vect\theta}|_2\leq3r_n/m_H$. All conclusions hold on $\mcE_n$, and their joint probability is at least $1-\delta_n$.
\end{proof}

For Nesterov acceleration, fix $0<\eta_N\leq M_H^{-1}$ and let $\gamma_3=(1-\sqrt{\eta_Nm_H})/(1+\sqrt{\eta_Nm_H})$. Define the remaining algorithm parameters by
\begin{equation*}
 \alpha_2=\sqrt{\frac{\mu}{4L}},~~ \gamma_2=\frac{1-\alpha_2}{1+\alpha_2},~~
 c_2=1+2\gamma_2,~~ c_3=1+2\gamma_3,~~ \tau_N=\frac{\mu\epsilon_3}{8c_3}.
\end{equation*}
The iteration counts are
\begin{equation*}
 N_1=\left\lceil\frac{16c_2(D+R/4)}{R\sqrt{\eta\mu}}\right\rceil,~~
 N_2=1+\left\lceil\frac{\log_+(4L^2R^2/\tau_N^2)}{-\log(1-\alpha_2)}\right\rceil,~~
 N_3=\left\lceil\frac{2}{\sqrt{\eta_Nm_H}}\log_+\left(\frac{2m_H\epsilon_3}{r_n}\right)\right\rceil.
\end{equation*}
Phase 1 runs Algorithm \ref{alg:ngd} with $\eta=1/(2L)$ and the momentum sequence in Theorem \ref{thm:nagd_phase1}(i) for at most $N_1$ updates. Before each update, check whether $|\partial\mcL(\vect\beta^{(t)})|_2\leq\tau_N$. If so, proceed directly to Phase 3 at this extrapolated point. Otherwise, after $N_1$ updates restart Nesterov's method at $\vect\theta^{(N_1)}$ with $\gamma_2$ and the same $\eta$ in Phase 2, using the same norm test at $\vect\beta^{(t)}$. Phase 3 restarts at the extrapolated point meeting the test, uses $\gamma_3,\eta_N$, and runs $N_3$ updates.

\begin{corollary}\label{thm:accelerated_algorithm}
Suppose the assumptions of Corollary \ref{thm:full_algorithm} hold. In addition, assume
\begin{equation*}
 \epsilon_0\leq\min\Big\{\frac{\tau_N}{4Lc_2},\frac{R}{8c_2}\Big\},~~
 |\partial\mcL(\vect\theta^*)|_2\leq
 \min\Big\{\frac{\mu}{16c_3},\frac{m_H}{4(c_3+1)}\Big\}\epsilon_3,
 ~~ r_n\leq\frac{m_H\epsilon_3}{4c_3}.
\end{equation*}
On $\mcE_n$, every Phase 1 iterate and extrapolated point remains in $\mathcal K$. Phase 2 remains in the strongly convex region and stops within $N_2$ updates. Phase 3 starts within distance $\epsilon_3/(2c_3)$ of $\vect\theta^*$, and every subsequent extrapolated iterate remains in the radius-$\epsilon_3$ ball. After at most $N_1+N_2+N_3$ updates, the computed solution satisfies \eqref{eq:complete_expansions} with $|\vect R_n|_2\leq2r_n/m_H$ and $|\tilde{\vect\theta}-\hat{\vect\theta}|_2\leq3r_n/m_H$.
\end{corollary}

\begin{proof}[Proof of Corollary \ref{thm:accelerated_algorithm}]
Work on $\mcE_n$ and write $\vect g=\partial\mcL$ and $s_n=|\vect g(\vect\theta^*)|_2$.

\noindent\textbf{Step 1. Entry into Phase 2.}
The localization and selected-subgradient arguments in the proof of Corollary \ref{thm:full_algorithm} give
\begin{equation*}
 |\hat{\vect\theta}-\vect\theta^*|_2\leq\max\{\epsilon_2,4s_n/\mu\}\leq\epsilon_3/4\leq R/8,~~|\vect g(\hat{\vect\theta})|_2\leq L\epsilon_1.
\end{equation*}
Write $d_0=|\vect\theta^{(0)}-\hat{\vect\theta}|_2\leq D+R/4$. The Phase 1 threshold obeys $\eta\tau_N\geq\epsilon_1$, and every executed update has length at least $\epsilon_1$. The potential and convex-combination identities in \eqref{eq:nag_phase1_containment} keep every completed iterate and the next extrapolated point within distance $d_0$ of $\hat{\vect\theta}$. The proposed next iterate satisfies
\[
|\vect\theta^{(t+1)}-\vect\theta^*|_2
\leq\frac32d_0+\epsilon_1+|\hat{\vect\theta}-\vect\theta^*|_2
\leq2D+R.
\]
It therefore lies in $\mathcal K$, validating the next potential inequality. Induction proves Phase 1 containment before the norm test stops the phase or all $N_1$ updates are completed. If Phase 1 does not stop early, Theorem \ref{thm:nagd_phase1}(i) gives
\begin{equation*}
 \mcL(\vect\theta^{(N_1)})-\mcL(\hat{\vect\theta})\leq\frac{2(D+R/4)^2}{\eta N_1^2}\leq\frac{\mu R^2}{128c_2^2}.
\end{equation*}
This iterate lies in the radius-$R$ ball, since every point outside that ball has objective gap at least $\mu R^2/8$.

\noindent\textbf{Step 2. Containment during accelerated Phase 2.}
Index Phase 2 from zero and restart with $\vect\theta^{(-1)}=\vect\theta^{(0)}$. Use the potential \eqref{eq:nag_phase2_potential} with $\alpha=\alpha_2$ and $\eta=1/(2L)$, and denote it by $E_t$. Write $\Delta_t=\mcL(\vect\theta^{(t)})-\mcL(\hat{\vect\theta})$.
For a point in the radius-$R$ ball, \eqref{eq:gap_growth} and $\epsilon_0\geq\sqrt2\epsilon_2$ give
\begin{equation*}
 |\vect\theta-\hat{\vect\theta}|_2^2\leq\max\{\epsilon_0^2,8[\mcL(\vect\theta)-\mcL(\hat{\vect\theta})]/\mu\}.
\end{equation*}
Since $\alpha_2^2/(2\eta)=\mu/4$, the initial potential satisfies
\begin{equation*}
 E_0\leq3\Delta_0+\frac\mu4\epsilon_0^2
 \leq\frac{7\mu R^2}{256c_2^2}<\frac{\mu R^2}{32c_2^2}.
\end{equation*}
We prove inductively that every Phase 2 iterate lies in the radius-$R$ ball and $E_t\leq(1-\alpha_2)^tE_0$ until the norm test stops the phase. The bound holds at initialization. If it holds at two consecutive iterates, the last two displays give
\begin{equation*}
 |\vect\theta^{(t)}-\hat{\vect\theta}|_2,
 |\vect\theta^{(t-1)}-\hat{\vect\theta}|_2\leq\frac{R}{2c_2},~~|\vect\beta^{(t)}-\hat{\vect\theta}|_2\leq\frac R2.
\end{equation*}
The same conclusion at $t=0$ follows from the restart. Thus $\vect\beta^{(t)}$ lies in the radius-$R$ ball. Approximate smoothness bounds the next proposed iterate by
\begin{equation*}
 |\vect\theta^{(t+1)}-\vect\theta^*|_2
 \leq\frac32|\vect\beta^{(t)}-\hat{\vect\theta}|_2+\epsilon_1+|\hat{\vect\theta}-\vect\theta^*|_2
 \leq\frac{7R}{8}+\epsilon_1\leq R.
\end{equation*}
If the phase has not stopped, $|\vect g(\vect\beta^{(t)})|_2>\tau_N$ implies $|\vect\beta^{(t)}-\hat{\vect\theta}|_2\geq\epsilon_0$ by the selected-subgradient bound and approximate smoothness. Its update length is at least $\eta\tau_N\geq\epsilon_1$. All pairs used in Theorem \ref{thm:nagd_phase2}(ii) therefore lie in their required regions before its potential inequality is applied. Equation \eqref{eq:nag_phase2_potential_step}, with $\alpha=\alpha_2$ and both additive terms absent under these distance conditions, gives $E_{t+1}\leq(1-\alpha_2)E_t$, completing the induction.

\noindent\textbf{Step 3. The norm stopping rule.}
After $N_2-1$ updates, the potential is at most $\mu\tau_N^2/(128L^2c_2^2)$. If another update is needed, the same bound holds for both $E_{N_2-1}$ and $E_{N_2}$. The growth inequality in Step 2 and $\epsilon_0\leq\tau_N/(4Lc_2)$ then give
\begin{equation*}
 |\vect\theta^{(N_2)}-\hat{\vect\theta}|_2,
 |\vect\theta^{(N_2-1)}-\hat{\vect\theta}|_2\leq\frac{\tau_N}{4Lc_2},~~|\vect\beta^{(N_2)}-\hat{\vect\theta}|_2\leq\frac{\tau_N}{4L}.
\end{equation*}
Consequently $|\vect g(\vect\beta^{(N_2)})|_2\leq\tau_N/4+L\epsilon_1\leq\tau_N/2$. Hence the phase stops within $N_2$ updates.

\noindent\textbf{Step 4. Entry into Phase 3 and the final error.}
Step 4 of the proof of Corollary \ref{thm:full_algorithm} applies to the stopping point in either Phase 1 or Phase 2 and yields
\begin{equation*}
 |\vect\theta^{(0)}-\vect\theta^*|_2
 \leq\max\{\epsilon_2,2(\tau_N+s_n)/\mu\}\leq\frac{\epsilon_3}{2c_3}.
\end{equation*}
Define $\vect z_0=\vect\theta^{(0)}-\vect\theta^*+\vect H^{-1}\vect g(\vect\theta^*)$. The score and remainder bounds give
\[
c_3\big(|\vect z_0|_2+r_n/m_H\big)+|\vect H^{-1}\vect g(\vect\theta^*)|_2
\leq\frac{\epsilon_3}{2}+\frac{(c_3+1)s_n+c_3r_n}{m_H}\leq\epsilon_3.
\]
Theorem \ref{thm:nagd_phase3}(iii) therefore proves containment. Also $|\vect z_0|_2\leq\epsilon_3$, and the choice of $N_3$ gives $2\exp\{-\sqrt{\eta_Nm_H}N_3/2\}|\vect z_0|_2\leq r_n/m_H$. Adding the local remainder gives $|\vect R_n|_2\leq2r_n/m_H$. The SAA expansion in Corollary \ref{thm:full_algorithm} then gives $|\tilde{\vect\theta}-\hat{\vect\theta}|_2\leq3r_n/m_H$ on $\mcE_n$.
\end{proof}

Acceleration reduces the Phase 1 count from order $(L/\mu)(1+D/R)^2$ to $\sqrt{L/\mu}(1+D/R)$ and the Phase 2 count from order $(L/\mu)\log(R/\tau_n)$ to $\sqrt{L/\mu}\log(R/\tau_N)$. At $\eta_N=M_H^{-1}$, Phase 3 takes $O\{\sqrt{M_H/m_H}\log(m_H\epsilon_3/r_n)\}$ updates with the same expansion remainder bound as subgradient descent. Polynomially decreasing approximation scales give $O(\log n)$ updates. Each update evaluates $n$ scenario subgradients.

\phantomsection\label{app:phase_refinement}
Condition \ref{cond:H_smooth} and the entry conditions of the complete procedures give containment in $\mbB(\vect\theta^*,\epsilon_3)$ for every Phase 3 iterate and extrapolated point by Theorems \ref{thm:subgd_phase3}(iii) and \ref{thm:nagd_phase3}(iii). The SAA solution also lies in this ball. Write $\vect g=\partial\mcL$. Suppose that, on this ball, for some $0\leq\delta\leq m_H/64$ and $\zeta_n\geq0$,
\begin{equation}\label{eq:local_phase3_error}
|\vect g(\vect\theta)-\vect H(\vect\theta-\hat{\vect\theta})|_2
\leq\delta|\vect\theta-\hat{\vect\theta}|_2+\zeta_n.
\end{equation}
Index Phase 3 from zero and use the restarted momentum in the accelerated procedure. Then
\begin{equation}\label{eq:local_phase3_comparison}
|\vect\theta^{(t)}-\hat{\vect\theta}|_2
\leq4q_{\rm loc}^t|\vect\theta^{(0)}-\hat{\vect\theta}|_2+\frac{2\zeta_n}{m_H},\qquad
q_{\rm loc}=\begin{cases}
1-\eta_3m_H/2,&\text{subgradient descent},\\
1-\sqrt{\eta_Nm_H}/2,&\text{Nesterov acceleration}.
\end{cases}
\end{equation}
For subgradient descent, set $\vect d_t=\vect\theta^{(t)}-\hat{\vect\theta}$. Since $0<\eta_3\leq M_H^{-1}$, \eqref{eq:local_phase3_error} gives
\[
|\vect d_{t+1}|_2\leq(1-\eta_3m_H+\eta_3\delta)|\vect d_t|_2+\eta_3\zeta_n
\leq q_{\rm loc}|\vect d_t|_2+\eta_3\zeta_n.
\]
Summing this recursion proves \eqref{eq:local_phase3_comparison}.

For Nesterov acceleration, let $\vect d_{-1}=\vect d_0$, $a=\sqrt{\eta_Nm_H}$, $\rho=1-a$, and $q_{\rm loc}=1-a/2$. Use the polynomials $\vect P_t$ and $\vect K_j$ from the proof of Theorem \ref{thm:nagd_phase3}(iii), with $\eta_N$ in place of $\eta$. Their bounds imply
\[
\|\vect P_t\|\leq2q_{\rm loc}^t,\qquad
\eta_N\sum_{j\geq0}\|\vect K_j\|\leq\frac1{m_H},\qquad
\eta_N\sum_{j\geq0}\|\vect K_j\|q_{\rm loc}^{-j-1}
\leq\frac{\eta_Nq_{\rm loc}}{(q_{\rm loc}-\rho)^2}\leq\frac4{m_H}.
\]
Here the first bound follows from $(1+at)(\rho/q_{\rm loc})^t\leq(1+at)e^{-at/2}\leq2$. Let $\vect b_t=\vect\beta^{(t)}-\hat{\vect\theta}=(1+\gamma_3)\vect d_t-\gamma_3\vect d_{t-1}$ and $\vect e_t=\vect g(\vect\beta^{(t)})-\vect H\vect b_t$. The local recursion has solution
\[
\vect d_t=\vect P_t\vect d_0-\eta_N\sum_{j=0}^{t-1}\vect K_{t-1-j}\vect e_j,\qquad
|\vect e_j|_2\leq\delta|\vect b_j|_2+\zeta_n.
\]
Put $D_0=|\vect d_0|_2$ and $B=2\zeta_n/m_H$. If $|\vect d_j|_2\leq4q_{\rm loc}^jD_0+B$ up to index $t-1$, then $q_{\rm loc}\geq1/2$ and $0\leq\gamma_3<1$ give $|\vect b_j|_2\leq16q_{\rm loc}^jD_0+3B$. The three polynomial bounds therefore yield
\[
|\vect d_t|_2\leq\left(2+\frac{64\delta}{m_H}\right)q_{\rm loc}^tD_0
+\frac{3\delta B+\zeta_n}{m_H}
\leq4q_{\rm loc}^tD_0+B.
\]
The initialization satisfies the same bound, including the convention $\vect d_{-1}=\vect d_0$. Induction proves \eqref{eq:local_phase3_comparison} for every $t$.

Under the model and population assumptions of Theorem \ref{cor:model_algorithm} and the region and initialization conditions in the \hyperref[app:unprojected_saa]{unprojected extension}, choose $\epsilon_3=C_3\sqrt{a_n}$ and $r_n=C_4a_n^{3/4}$. For sufficiently large $C_3$, the score bound in the proof of that theorem verifies the score inequalities in Corollaries \ref{thm:full_algorithm} and \ref{thm:accelerated_algorithm}. Their remaining scale inequalities follow from $\epsilon_0=O(a_n)=o(\epsilon_3)$ and $r_n/\epsilon_3\to0$. The ball inclusion verifies their regions and gives containment for the Phase 3 iterates and extrapolated points. The increment bound in Theorem \ref{thm:Lip_subgrad}, applied to $\vect\theta$ and $\hat{\vect\theta}$, and the Lipschitz population Hessian give, with $d=|\vect\theta-\hat{\vect\theta}|_2$,
\[
|\vect g(\vect\theta)-\vect H(\vect\theta-\hat{\vect\theta})|_2
\leq|\vect g(\hat{\vect\theta})|_2+C\{\sqrt{da_n}+a_n+\epsilon_3d\}.
\]
Lemma \ref{lem:oracle_bounds} bounds $|\vect g(\hat{\vect\theta})|_2$ by $L\epsilon_1=O(a_n)$. Young's inequality bounds $C\sqrt{da_n}$ by $\delta d/2+C_{\delta}a_n$, and $C\epsilon_3\leq\delta/2$ for sufficiently large $n$. Taking $\delta=m_H/64$ thus verifies \eqref{eq:local_phase3_error} with $\zeta_n=O(a_n)$. Equation \eqref{eq:local_phase3_comparison} gives the $O(a_n)$ optimization error with the Phase 3 counts stated after Theorem \ref{cor:model_algorithm}. At $\eta_3=\eta_N=M_H^{-1}$, these counts have condition-number dependence $M_H/m_H$ and $\sqrt{M_H/m_H}$, respectively.

The projected procedures admit the interior switches in Corollary \ref{cor:projected_subgradient}\ref{part:projected_interior_switch}, with the SAA calibration in part \ref{part:projected_saa}. Combining each optimization bound with the SAA expansion in the proof of Theorem \ref{cor:model_algorithm} gives its parts \textup{(ii)} and \textup{(iii)} and the population-gap bound in part \textup{(i)} for the terminal outputs.

\subsection{Other update rules}\label{app:other_updates}

Let $\mathcal C\subseteq\mbR^p$ be nonempty, closed, and convex, and let $\Pi_{\mathcal C}$ be its Euclidean projection. Projected subgradient descent starts at $\vect x_0\in\mathcal C$ and uses $\vect x_{t+1}=\Pi_{\mathcal C}\{\vect x_t-\eta\partial\mcL(\vect x_t)\}$. The accelerated method starts a block at $\vect x_0=\vect z_0\in\mathcal C$, sets $\lambda_0=1$ and $\lambda_{t+1}=(1+\sqrt{1+4\lambda_t^2})/2$, and uses
\begin{equation}\label{eq:feasible_nag_updates}
\begin{aligned}
\vect y_t&=(1-\lambda_t^{-1})\vect x_t+\lambda_t^{-1}\vect z_t,\\
\vect z_{t+1}&=\Pi_{\mathcal C}\{\vect z_t-\eta\lambda_t\partial\mcL(\vect y_t)\},\\
\vect x_{t+1}&=(1-\lambda_t^{-1})\vect x_t+\lambda_t^{-1}\vect z_{t+1}.
\end{aligned}
\end{equation}
All three sequences are feasible. A restart sets $\lambda_0=1$ and both initial sequences to the preceding block output.

\begin{corollary}\label{cor:projected_subgradient}
Let $\mcL$ be finite and convex on $\mathcal C$. For parts \textup{(i)} and \textup{(ii)}, suppose $\hat{\vect\theta}\in\argmin{\vect\theta\in\mathcal C}\mcL(\vect\theta)$ exists, impose Condition \ref{cond:L_smooth} on $\mathcal C$, and take $0<\eta\leq1/(2L)$. Write $\Delta(\vect x)=\mcL(\vect x)-\mcL(\hat{\vect\theta})$, $D_0=|\vect x_0-\hat{\vect\theta}|_2$, and $\delta=\eta L^2\epsilon_1^2/2$.
\begin{enumerate}[label=\textup{(\roman*)},ref=(\roman*),leftmargin=*]
\item\label{part:projected_convex} Let $T\geq1$.
\begin{enumerate}[label=\textup{(\alph*)},ref=\theenumi(\alph*),leftmargin=*]
\item\label{part:projected_convex_gd} Projected subgradient descent satisfies
\[
\frac1T\sum_{t=1}^T\Delta(\vect x_t)\leq\frac{D_0^2}{2\eta T}+\delta.
\]
If $|\vect x_{t+1}-\vect x_t|_2\geq\epsilon_1$ for $0\leq t<T$, its objective values are nonincreasing and $\Delta(\vect x_T)\leq D_0^2/(2\eta T)$.
\item\label{part:projected_convex_nag} For a block of $T$ accelerated updates, put $\delta_t=\delta\mbI\{|\vect x_{t+1}-\vect y_t|_2<\epsilon_1\}$. Then
\begin{equation}\label{eq:feasible_nag_phase1}
\Delta(\vect x_T)\leq\frac{D_0^2/(2\eta)+\sum_{t=0}^{T-1}\lambda_t^2\delta_t}{\lambda_{T-1}^2}
\leq\frac{2D_0^2}{\eta T^2}+T\delta.
\end{equation}
Every $\vect x_t,\vect z_t$ for $0\leq t\leq T$ and $\vect y_t$ for $0\leq t<T$ has distance at most $\{D_0^2+2\eta\delta T^3\}^{1/2}$ from $\hat{\vect\theta}$. If every actual displacement $|\vect x_{t+1}-\vect y_t|_2$ is at least $\epsilon_1$, the radius is $D_0$ and the additive objective error vanishes.
\end{enumerate}
\item\label{part:projected_geometric}
\begin{enumerate}[label=\textup{(\alph*)},ref=\theenumi(\alph*),leftmargin=*]
\item\label{part:projected_geometric_gd} Suppose Condition \ref{cond:mu_convex} holds on $\mathcal C$ and $\alpha=\eta\mu/2\leq1$. Projected subgradient descent satisfies
\[
\begin{aligned}
|\vect x_T-\hat{\vect\theta}|_2^2&\leq(1-\alpha)^TD_0^2+F_G,
&F_G&=\epsilon_2^2+\frac{2\eta\delta}{\alpha},\\
\Delta(\vect x_T)&\leq(1-\alpha)^T\Delta(\vect x_0)+\frac{\mu\epsilon_2^2}{4}+\frac\delta\alpha.
\end{aligned}
\]
If $|\vect x_{t+1}-\vect x_t|_2\geq\epsilon_1$ and $|\vect x_t-\hat{\vect\theta}|_2\geq\epsilon_2$ for $0\leq t<T$, both additive errors vanish.
\item\label{part:projected_geometric_nag} Suppose Condition \ref{cond:mu_convex} holds on $\mathcal C$ and $\eta\mu\leq1$. Set $K=\lceil\sqrt{32/(\eta\mu)}\rceil$ and $E=\epsilon_2^2+4K\delta/\mu$. Restart the accelerated method every $K$ updates, write $\bar{\vect\theta}_j$ for the output of $j$ blocks, and put $D_j=|\bar{\vect\theta}_j-\hat{\vect\theta}|_2$. Then
\begin{equation}\label{eq:feasible_nag_restart}
D_j^2\leq4^{-j}D_0^2+\frac43(1-4^{-j})E,\qquad j\geq0,
\end{equation}
and $\Delta(\bar{\vect\theta}_j)\leq(\mu/4)4^{-j}D_0^2+\mu E/3$ for $j\geq1$. Block $j$ starts at $\bar{\vect\theta}_j$. The points in this block satisfy
\begin{equation}\label{eq:feasible_nag_global_containment}
|\vect x_t-\hat{\vect\theta}|_2^2,\ |\vect y_t-\hat{\vect\theta}|_2^2,\ |\vect z_t-\hat{\vect\theta}|_2^2
\leq4^{-j}D_0^2+26E,
\end{equation}
for $\vect x_t,\vect z_t$ with $0\leq t\leq K$ and $\vect y_t$ with $0\leq t<K$.
\item\label{part:projected_local_entry} Suppose Condition \ref{cond:mu_convex} holds only on $\mathcal C\cap\mbB(\hat{\vect\theta},R)$ for $R>0$. For projected subgradient descent, if $\alpha\leq1$, $F_G\leq R^2/4$, and $D_0\leq R/2$, all iterates stay within distance $R/2$ and the bounds in part \textup{(ii)(a)} hold. For the accelerated method, retain $\eta\mu\leq1$ and assume
\begin{equation}\label{eq:feasible_nag_local_accuracy}
\epsilon_2^2+\frac{2K\delta}{\mu}\leq\frac{R^2}{256}.
\end{equation}
If $D_0\leq R/2$, all points in each block remain in $\mathcal C\cap\mbB(\hat{\vect\theta},R)$ and the bounds in part \textup{(ii)(b)} hold.

For entry from any feasible point at distance at most $R_0$, assume $\epsilon_2\leq R/16$. Projected subgradient descent first runs $N_G=\max\{1,\lceil64R_0^2/(\eta\mu R^2)\rceil\}$ updates and restarts from their average. If $\delta\leq\mu R^2/128$, this average lies within distance $R/2$. The accelerated method first runs one block of $N=\max\{1,\lceil16R_0/(R\sqrt{\eta\mu})\rceil\}$ updates. If
\begin{equation}\label{eq:feasible_nag_entry_accuracy}
N\delta\leq\frac{\mu R^2}{128},
\end{equation}
its output lies within distance $R/2$. All three sequences in this initial block have distance at most $\{R_0^2+2\eta\delta N^3\}^{1/2}$ from $\hat{\vect\theta}$. Each method then continues under its local conditions above.
\end{enumerate}
\item\label{part:projected_inference}
\begin{enumerate}[label=\textup{(\alph*)},ref=\theenumi(\alph*),leftmargin=*]
\item\label{part:projected_interior_switch} Suppose Condition \ref{cond:H_smooth} holds on $\mbB(\vect\theta^*,\epsilon_3)\subseteq\mathcal C$ and $m_H\mbI\preceq\vect H\preceq M_H\mbI$. Under the initialization and score conditions of Theorem \ref{thm:subgd_phase3}(iii), projected subgradient descent with $0<\eta_H\leq M_H^{-1}$ has inactive projections and satisfies \eqref{eq:bahadur_sgd}.

For Nesterov acceleration, let $\hat{\vect\theta}$ be a constrained minimizer and put $\vect w=\vect H^{-1}\partial\mcL(\vect\theta^*)$, $b_*=|\hat{\vect\theta}-\vect\theta^*+\vect w|_2$, and $\gamma_H=(1-\sqrt{\eta_Hm_H})/(1+\sqrt{\eta_Hm_H})$, where $0<\eta_H\leq M_H^{-1}$. Choose $\rho>0$ with
\begin{equation}\label{eq:feasible_nag_switch}
(1+2\gamma_H)(\rho+b_*+r_n/m_H)+|\vect w|_2\leq\epsilon_3.
\end{equation}
From any feasible point within distance $\rho$ of $\hat{\vect\theta}$, restart Algorithm \ref{alg:ngd} with $\eta_H,\gamma_H$. Its iterates and extrapolated points remain in $\mbB(\vect\theta^*,\epsilon_3)$ and satisfy \eqref{eq:ngd_bahadur_optimal}. Projection onto $\mathcal C$ leaves these updates unchanged. The block outputs in part \textup{(ii)(b)} or \textup{(ii)(c)} meet this distance requirement when $\sqrt{4E/3}\leq\rho/2$ and $2^{-j}D_0\leq\rho/2$.
\item\label{part:projected_saa} Suppose the assumptions of Theorem \ref{thm:Lip_subgrad} or Proposition \ref{prop:index_objectives} hold with $a_n\to0$, and take $\mathcal C=\Theta$ closed and bounded. Let $\vect\theta^*$ be an interior population minimizer with a Lipschitz population Hessian on a neighborhood and $m_H\mbI\preceq\vect H=\nabla^2\bar{\mcL}(\vect\theta^*)\preceq M_H\mbI$. Choose $R>0$ sufficiently small that $\mbB(\vect\theta^*,2R)\subseteq\Theta$ and the population Hessian is at least $m_H\mbI/2$ there. Use $L=L_0$, $\mu=m_H/2$, and $\epsilon_j=C_ja_n$ from Corollary \ref{thm:H_smooth}, increasing $L$ if needed so that $\mu\leq2L$. Take $0<\eta\leq1/(2L)$ and $R_0\geq\operatorname{diam}(\Theta)$. Use either entry and continuation procedure in part \textup{(ii)(c)}.

For sufficiently large $n$, there are $c,C,\gamma>0$ and one event of probability at least $1-Cn^{-\gamma p}$ on which an SAA solution lies in the interior of $\Theta$ and the following bounds hold for every initial decision in $\Theta$ and every selection of scenario subgradients. Every decision after $\lceil c\log n\rceil$ updates has distance at most $Ca_n$ from the SAA solution, including all three accelerated sequences. The population optimality gaps of the SAA solution and these decisions are at most $Ca_n$. The expansion and distributional conclusions of Theorem \ref{cor:model_algorithm} consequently hold under their dimension and variance conditions. Switching to either interior Phase 3 in part \textup{(iii)(a)} preserves these conclusions after another $O(\log n)$ updates, with the optimization error given by \eqref{eq:local_phase3_comparison}.
\end{enumerate}
\end{enumerate}
\end{corollary}

\begin{proof}
Write $\vect g=\partial\mcL$. For $\vect z^+=\Pi_{\mathcal C}(\vect z-a\vect g(\vect y))$, projection gives, for every $\vect v\in\mathcal C$,
\[
\langle\vect g(\vect y),\vect z^+-\vect v\rangle
\leq\frac{|\vect z-\vect v|_2^2-|\vect z^+-\vect v|_2^2-|\vect z^+-\vect z|_2^2}{2a}.
\]
The segment bound in the proof of Lemma \ref{prop:L-smooth} and $\eta L\leq1/2$ give $L\max\{k,\epsilon_1\}k-k^2/(2\eta)\leq\delta\mbI\{k<\epsilon_1\}$ for every feasible displacement of length $k$. For $k<\epsilon_1$, this follows by completing the square, and for $k\geq\epsilon_1$ the left side is nonpositive.

\emph{Part (i).} For a projected subgradient update, use $\vect y=\vect z=\vect x$, $a=\eta$, and $\vect v=\hat{\vect\theta}$ in the projection inequality. Convexity and the remainder bound give
\[
\Delta(\vect x^+)\leq\frac{|\vect x-\hat{\vect\theta}|_2^2-|\vect x^+-\hat{\vect\theta}|_2^2}{2\eta}+\delta.
\]
Summation proves part \textup{(i)(a)}. If each update length is at least $\epsilon_1$, the same calculation has zero error. The comparison $\vect v=\vect x$ gives monotonicity and hence the last-iterate bound.

For acceleration, feasibility follows from \eqref{eq:feasible_nag_updates}. Set $\vect g_t=\vect g(\vect y_t)$, $k_t=|\vect x_{t+1}-\vect y_t|_2$, and $\lambda_{-1}=0$. Weight the convexity inequalities at $\vect x_t$ and $\hat{\vect\theta}$ by $\lambda_t(\lambda_t-1)$ and $\lambda_t$, respectively. Applying the projection inequality with $a=\eta\lambda_t$, $\vect z=\vect z_t$, and $\vect v=\hat{\vect\theta}$, and using $|\vect z_{t+1}-\vect z_t|_2=\lambda_tk_t$ and $\lambda_t(\lambda_t-1)=\lambda_{t-1}^2$, gives
\begin{equation}\label{eq:feasible_nag_potential}
\lambda_t^2\Delta(\vect x_{t+1})+\frac{|\vect z_{t+1}-\hat{\vect\theta}|_2^2}{2\eta}
\leq\lambda_{t-1}^2\Delta(\vect x_t)+\frac{|\vect z_t-\hat{\vect\theta}|_2^2}{2\eta}+\lambda_t^2\delta_t.
\end{equation}
Summation and $\lambda_{T-1}\geq T/2$ prove \eqref{eq:feasible_nag_phase1}. Since $\Delta\geq0$, each auxiliary squared distance is at most $D_0^2+2\eta\sum_{s<t}\lambda_s^2\delta_s$. The bounds $\lambda_s\leq s+1$ and $\sum_{s<T}(s+1)^2\leq T^3$ prove the asserted radius for $\vect z_t$. Each $\vect x_t$ is a convex combination of preceding $\vect x$ and $\vect z$ points, as is $\vect y_t$. This proves their radii as well. If all $k_t\geq\epsilon_1$, each $\delta_t=0$ and all distances are at most $D_0$.

\emph{Part (ii).} The proof of \eqref{eq:gap_growth} applies to a constrained minimizer. Indeed, for $\vect x_s=\hat{\vect\theta}+s(\vect x-\hat{\vect\theta})$, convexity and constrained minimality give $\langle\vect g(\vect x_s),\vect x-\vect x_s\rangle\geq0$. If $d=|\vect x-\hat{\vect\theta}|_2>\epsilon_2$, Condition \ref{cond:mu_convex} at $(\vect x,\vect x_s)$ and $s\downarrow0$ give $\Delta(\vect x)\geq\mu d^2/4$. Together with minimality for $d\leq\epsilon_2$, this gives
\begin{equation}\label{eq:feasible_nag_qg}
d^2\leq4\Delta(\vect x)/\mu+\epsilon_2^2.
\end{equation}

For projected subgradient descent, retain the lower-model term \eqref{eq:oracle_lower} at $(\hat{\vect\theta},\vect x)$ in the calculation for part \textup{(i)(a)}. Since $\Delta(\vect x^+)\geq0$ and $\alpha=\eta\mu/2$, this gives
\[
|\vect x^+-\hat{\vect\theta}|_2^2\leq(1-\alpha)|\vect x-\hat{\vect\theta}|_2^2+\alpha\epsilon_2^2+2\eta\delta.
\]
For the objective bound, the projected update minimizes $Q_{\vect x}(\vect v)=\langle\vect g(\vect x),\vect v-\vect x\rangle+|\vect v-\vect x|_2^2/(2\eta)$ over $\mathcal C$. Compare its value with $\vect v=(1-\alpha)\vect x+\alpha\hat{\vect\theta}$. Since $\alpha=\eta\mu/2$, the squared-distance terms cancel and
\[
\Delta(\vect x^+)\leq(1-\alpha)\Delta(\vect x)+\frac{\alpha\mu\epsilon_2^2}{4}+\delta.
\]
Iteration proves part \textup{(ii)(a)}. The stated distance cutoffs remove the two error terms. Both one-step inequalities use strong convexity only at the pair $(\hat{\vect\theta},\vect x)$. If $|\vect x-\hat{\vect\theta}|_2\leq R/2$ and $F_G\leq R^2/4$, the distance recursion therefore proves local containment before invoking strong convexity at the next point. This proves the projected subgradient conclusion in part \textup{(ii)(c)}.

For an accelerated block of length $K$, part \textup{(i)(b)} and \eqref{eq:feasible_nag_qg} give
\[
D_{j+1}^2\leq\frac8{\eta\mu K^2}D_j^2+\frac{4K\delta}{\mu}+\epsilon_2^2\leq\frac14D_j^2+E.
\]
Iteration proves \eqref{eq:feasible_nag_restart}. Since $2/(\eta K^2)\leq\mu/16$, for $j\geq1$,
\[
\Delta(\bar{\vect\theta}_j)\leq\frac\mu{16}D_{j-1}^2+K\delta\leq\frac\mu4\,4^{-j}D_0^2+\frac\mu{12}E+K\delta\leq\frac\mu4\,4^{-j}D_0^2+\frac\mu3E.
\]
Since $\eta\mu K^2\leq(\sqrt{32}+1)^2<49$, we have $2\eta\delta K^3\leq49E/2$. The bound on squared distances within each block in part \textup{(i)(b)} is therefore $D_j^2+2\eta\delta K^3\leq4^{-j}D_0^2+26E$, proving part \textup{(ii)(b)}.

For local containment, put $H=\epsilon_2^2+2K\delta/\mu\leq R^2/256$. A block starting within distance $R/2$ has every inner squared distance at most $R^2/4+49H\leq113R^2/256<R^2$ by part \textup{(i)(b)}, which uses only approximate smoothness on $\mathcal C$. Applying \eqref{eq:feasible_nag_qg} to its output then gives $D_{j+1}^2\leq R^2/16+E\leq R^2/16+2H\leq9R^2/128<R^2/4$. Induction proves the accelerated conclusions in part \textup{(ii)(c)}.

For entry, \eqref{eq:feasible_nag_qg} and $\epsilon_2\leq R/16$ give $\Delta(\vect x)>\mu R^2/8$ on the feasible radius-$R$ sphere. Convexity along segments to $\hat{\vect\theta}$ gives the same lower bound at every feasible point outside that sphere. A feasible point of gap at most $\mu R^2/64$ must therefore lie inside it, where \eqref{eq:feasible_nag_qg} gives squared distance at most $R^2/16+\epsilon_2^2<R^2/4$. For projected subgradient descent, its feasible average satisfies
\[
\Delta\left(\frac1{N_G}\sum_{t=1}^{N_G}\vect x_t\right)\leq\frac{R_0^2}{2\eta N_G}+\delta\leq\frac{\mu R^2}{64}.
\]
For acceleration, \eqref{eq:feasible_nag_phase1} gives $\Delta(\vect x_N)\leq2R_0^2/(\eta N^2)+N\delta\leq\mu R^2/64$. The initial accelerated radius is the bound in part \textup{(i)(b)}. This proves both entry procedures.

\emph{Part (iii).} For projected subgradient descent in part \textup{(iii)(a)}, Theorem \ref{thm:subgd_phase3}(iii) places each unprojected proposal in $\mbB(\vect\theta^*,\epsilon_3/2)\subseteq\mathcal C$. Each projection is therefore inactive, proving \eqref{eq:bahadur_sgd}. For the accelerated switch from $\vect x$, the bound $|\vect x-\hat{\vect\theta}|_2\leq\rho$ gives $|\vect x-\vect\theta^*+\vect w|_2\leq\rho+b_*$. The block bound \eqref{eq:feasible_nag_restart} implies $|\bar{\vect\theta}_j-\hat{\vect\theta}|_2\leq\rho$ under the two stated block conditions. Condition \eqref{eq:feasible_nag_switch} verifies the initialization in Theorem \ref{thm:nagd_phase3}(iii), which proves the expansion and containment of all iterates and extrapolated points. Projection onto $\mathcal C$ leaves each proposed update unchanged.

For part \textup{(iii)(b)}, work on the joint increment and pointwise event in Theorem \ref{thm:Lip_subgrad} or Proposition \ref{prop:index_objectives}. The population first-order condition and pointwise score bound give $s_n=|\vect g(\vect\theta^*)|_2\leq C\sqrt{a_n}$. Corollary \ref{thm:H_smooth} gives approximate smoothness on $\Theta$ and approximate strong convexity on $\mbB(\vect\theta^*,2R)$. The model objective is continuous. Hence compactness of $\Theta$ gives a minimizer. On the radius-$R/2$ sphere about $\vect\theta^*$,
\[
\mcL(\vect\theta^*+R\vect u/2)-\mcL(\vect\theta^*)\geq-s_nR/2+\mu R^2/16\geq\mu R^2/32
\]
for every unit $\vect u$ and sufficiently large $n$. Convexity along feasible segments excludes every point outside this sphere from the minimizer set. Each SAA minimizer therefore lies in the interior of $\Theta$ and satisfies
\[
\delta_n:=|\hat{\vect\theta}-\vect\theta^*|_2\leq\max\{\epsilon_2,4s_n/\mu\}=O(\sqrt{a_n}).
\]
Eventually $\delta_n\leq R/8$. Hence $\mbB(\hat{\vect\theta},R)\subseteq\mbB(\vect\theta^*,2R)\subseteq\Theta$ and \eqref{eq:feasible_nag_qg} holds throughout $\mbB(\hat{\vect\theta},R)$.
Since $R_0\geq\operatorname{diam}(\Theta)$, every initialization has distance at most $R_0$. The entry lengths $N_G,N$ and restart length $K$ have constant order, while $\delta=O(a_n^2)$ and $\epsilon_2=O(a_n)$. All local and entry accuracy conditions in part \textup{(ii)(c)} hold eventually. That part therefore proves entry and containment for both methods. Their distance floors are $F_G=O(a_n^2)$ and $E=O(a_n^2)$. The geometric bounds give distance $O(a_n)$ after $O(\log n)$ updates for all later projected subgradient iterates and all points in the accelerated blocks. The constants are uniform over initial decisions, and the event controls all scenario-subgradient choices.

Population smoothness and stationarity bound the population gap by $C|\vect x-\vect\theta^*|_2^2$. The localization and optimization bounds therefore give the asserted gaps on the same event. At $\vect\theta^*$, the scenario subgradient is unique almost surely, and $\vect g(\vect\theta^*)=n^{-1}\sum_i\vect\psi_i$ has mean zero and second moment at most $Cp/n$. Hence $\delta_n=O_P(\sqrt{p/n}+a_n)$. Interiority, Lemma \ref{lem:oracle_bounds}, the increment bound, and the population Taylor expansion give
\[
|\vect H(\hat{\vect\theta}-\vect\theta^*)+\vect g(\vect\theta^*)|_2
\leq L\epsilon_1+C\{\sqrt{\delta_n a_n}+a_n+\delta_n^2\}
=O_P\!\left\{(p/n)^{3/4}\sqrt{\log n}+a_n\right\}.
\]
Adding the $O(a_n)$ optimization error proves the computed-solution expansion. The first-order equivalence conclusions follow from $a_n=o(\sqrt{p/n})$ when $p=o\{n(\log n)^{-2}\}$ and $\sqrt n a_n\to0$ when $p\log n=o(\sqrt n)$. Every deterministic unit projection of $\vect H^{-1}\vect\psi_i$ has bounded third absolute moment. Thus the variance lower bound gives Lyapunov's condition, and $p=o\{n^{1/3}(\log n)^{-2/3}\}$ makes the expansion remainder $o_P(n^{-1/2})$. Slutsky's theorem proves the normal limit. Finally, population smoothness gives
\[
|\bar{\mcL}(\tilde{\vect\theta})-\bar{\mcL}(\hat{\vect\theta})|
\leq C\{\delta_n|\tilde{\vect\theta}-\hat{\vect\theta}|_2+|\tilde{\vect\theta}-\hat{\vect\theta}|_2^2\}
=O_P\{a_n\sqrt{p/n}+a_n^2\}.
\]
Multiplication by $n$ gives $O_P(p^{3/2}\log n/\sqrt n)=o_P(1)$ under the same dimension condition.

For the interior switch, choose $\epsilon_3=C_3\sqrt{a_n}$ with sufficiently large $C_3$. The local approximation has remainder $r_n=C_4a_n^{3/4}$. The score and localization bounds verify the projected subgradient initialization in part \textup{(iii)(a)} after reaching optimization error $O(a_n)$. For acceleration, $\vect w=\vect H^{-1}\vect g(\vect\theta^*)$ satisfies $|\vect w|_2\leq C\sqrt{a_n}$, and approximation at the interior minimizer gives $b_*\leq(L\epsilon_1+r_n)/m_H=o(\epsilon_3)$. Choose $\rho=c_3\epsilon_3$ for small $c_3>0$ and increase $C_3$ to verify \eqref{eq:feasible_nag_switch}. Theorems \ref{thm:subgd_phase3}(iii) and \ref{thm:nagd_phase3}(iii) make subsequent projections inactive. Finally, the increment bound applied to $\vect\theta$ and $\hat{\vect\theta}$ verifies \eqref{eq:local_phase3_error} with $\zeta_n=O(a_n)$, as in Appendix \ref{app:phase_combination}. Its local comparison \eqref{eq:local_phase3_comparison} gives optimization error $O(a_n)$ after another $O(\log n)$ updates. This completes part \textup{(iii)}.
\end{proof}

\begin{corollary}\label{cor:heavy_ball}
Let $\mcL$ be finite and convex on $\mbR^p$, let $\hat{\vect\theta}\in\argmin{\vect\theta\in\mbR^p}\mcL(\vect\theta)$, and consider the heavy-ball updates
\begin{equation}\label{eq:heavy_ball_update}
\vect\theta^{(t+1)}=\vect\theta^{(t)}-\eta\partial\mcL(\vect\theta^{(t)})+\beta(\vect\theta^{(t)}-\vect\theta^{(t-1)}),\qquad \vect\theta^{(-1)}=\vect\theta^{(0)}.
\end{equation}
\begin{enumerate}[label=\textup{(\roman*)},leftmargin=*]
\item Suppose Conditions \ref{cond:L_smooth} and \ref{cond:mu_convex} hold on $\mbR^p$. Take $\beta=1/2$ and $0<\eta\leq1/(4L)$. Write
\[
E_t=\mcL(\vect\theta^{(t)})-\mcL(\hat{\vect\theta})+\frac{|\vect\theta^{(t)}-\vect\theta^{(t-1)}|_2^2}{4\eta},\qquad q_C=\frac{\mu\eta}{2(1+\mu\eta)},\qquad B_C=\frac{\mu\epsilon_2^2}{4}+\frac{2L^2(1+\mu\eta)}{\mu}\epsilon_1^2.
\]
For every $T\geq0$,
\begin{equation}\label{eq:heavy_ball_global}
E_T\leq(1-q_C)^TE_0+\{1-(1-q_C)^T\}B_C,\qquad |\vect\theta^{(T)}-\hat{\vect\theta}|_2^2\leq\frac4\mu E_T+\epsilon_2^2.
\end{equation}
\item Suppose Conditions \ref{cond:L_smooth} and \ref{cond:mu_convex} hold on $\mbB(\hat{\vect\theta},R)$ for $R>0$, and take $\vect\theta^{(0)}\in\mbB(\hat{\vect\theta},R)$. Use the parameters and quantities in part \textup{(i)}. Set $r=\{4\max\{E_0,B_C\}/\mu+\epsilon_2^2\}^{1/2}$. If
\begin{equation}\label{eq:heavy_ball_local_entry}
2r+\eta L\{\max\{r,\epsilon_1\}+\epsilon_1\}\leq R,
\end{equation}
then $|\vect\theta^{(t)}-\hat{\vect\theta}|_2\leq r$ for every $t\geq0$, and \eqref{eq:heavy_ball_global} holds.
\item Suppose Condition \ref{cond:H_smooth} holds with remainder $r_n$, and $m_H\mbI\preceq\vect H\preceq M_H\mbI$ for $0<m_H\leq M_H$. Restart \eqref{eq:heavy_ball_update} with
\[
\eta=\frac4{(\sqrt{M_H}+\sqrt{m_H})^2},\qquad q_H=\frac{\sqrt{M_H}-\sqrt{m_H}}{\sqrt{M_H}+\sqrt{m_H}},\qquad \beta=q_H^2,\qquad A_H=\frac{1+q_H}{1-q_H}.
\]
Write $\vect g^*=\partial\mcL(\vect\theta^*)$ and $\vect z_0=\vect\theta^{(0)}-\vect\theta^*+\vect H^{-1}\vect g^*$. If
\begin{equation}\label{eq:heavy_ball_phase3_entry}
A_H|\vect\theta^{(0)}-\vect\theta^*|_2+\frac{|\vect g^*|_2+r_n}{m_H}\leq\epsilon_3,
\end{equation}
then $|\vect\theta^{(t)}-\vect\theta^*|_2\leq\epsilon_3$ for every $t\geq0$. Define $\vect P_0=\mbI$, $\vect P_1=\mbI-\eta\vect H$, and $\vect P_{t+1}=\{(1+\beta)\mbI-\eta\vect H\}\vect P_t-\beta\vect P_{t-1}$. For every $T\geq1$,
\begin{equation}\label{eq:heavy_ball_expansion}
\begin{split}
\vect\theta^{(T)}-\vect\theta^*&=-\vect H^{-1}\vect g^*+\vect P_T\vect z_0+\vect R_T,\qquad |\vect R_T|_2\leq r_n/m_H,\\
\Norm{\vect P_T}&\leq\min\big\{A_H,\{1+(1+q_H)T\}q_H^T,2A_H\exp\{-(1-q_H)T/2\}\big\}.
\end{split}
\end{equation}
\end{enumerate}
\end{corollary}

\begin{proof}
Write $\vect v_t=\vect\theta^{(t)}-\vect\theta^{(t-1)}$, $\vect g_t=\partial\mcL(\vect\theta^{(t)})$, and $\Delta_t=\mcL(\vect\theta^{(t)})-\mcL(\hat{\vect\theta})$. The segment bound in the proof of Lemma \ref{prop:L-smooth} and Young's inequality bound the linearization remainder for an update of length $k$ by $L\max\{k,\epsilon_1\}k\leq k^2/(4\eta)+\eta L^2\epsilon_1^2$ when $\eta\leq1/(4L)$. With $\beta=1/2$, the update gives $\eta\vect g_t=\vect v_t/2-\vect v_{t+1}$. Adding the difference of the kinetic terms to the upper bound for $\Delta_{t+1}-\Delta_t$ yields
\begin{equation}\label{eq:heavy_ball_energy_descent}
E_{t+1}\leq E_t-\frac{|\vect v_{t+1}|_2^2+|\vect v_t-\vect v_{t+1}|_2^2}{4\eta}+\eta L^2\epsilon_1^2.
\end{equation}
The lower bound \eqref{eq:oracle_lower} with comparison point $\hat{\vect\theta}$ and a square completion give $\Delta_t\leq|\vect g_t|_2^2/\mu+\mu\epsilon_2^2/4$. Since $2\eta\vect g_t=(\vect v_t-\vect v_{t+1})-\vect v_{t+1}$,
\[
E_t\leq\frac{1+\mu\eta}{2\mu\eta^2}\big\{|\vect v_{t+1}|_2^2+|\vect v_t-\vect v_{t+1}|_2^2\big\}+\frac{\mu\epsilon_2^2}{4}.
\]
Substitution into \eqref{eq:heavy_ball_energy_descent} gives $E_{t+1}\leq(1-q_C)E_t+q_CB_C$. Iteration proves the first bound in \eqref{eq:heavy_ball_global}. Applying \eqref{eq:gap_growth} and using $\Delta_T\leq E_T$ proves its distance bound.

For part \textup{(ii)}, write $B=\max\{E_0,B_C\}$. The initial point belongs to the radius-$R$ ball, and \eqref{eq:gap_growth} gives $|\vect\theta^{(0)}-\hat{\vect\theta}|_2\leq r$. The restart gives this bound at $t=-1$. Suppose the bounds $E_t\leq B$ and $|\vect\theta^{(j)}-\hat{\vect\theta}|_2\leq r$ hold through index $t$. Lemma \ref{lem:oracle_bounds} bounds $|\partial\mcL(\hat{\vect\theta})|_2$ by $L\epsilon_1$. Hence Condition \ref{cond:L_smooth} gives $|\vect g_t|_2\leq L\{\max\{r,\epsilon_1\}+\epsilon_1\}$. The proposed update satisfies
\[
|\vect\theta^{(t+1)}-\hat{\vect\theta}|_2\leq\tfrac32|\vect\theta^{(t)}-\hat{\vect\theta}|_2+\tfrac12|\vect\theta^{(t-1)}-\hat{\vect\theta}|_2+\eta|\vect g_t|_2\leq R.
\]
Its segment with the current point therefore lies in the radius-$R$ ball. The calculation for part \textup{(i)} gives $E_{t+1}\leq(1-q_C)B+q_CB_C\leq B$, and \eqref{eq:gap_growth} then gives $|\vect\theta^{(t+1)}-\hat{\vect\theta}|_2\leq r$. Induction proves containment and validates \eqref{eq:heavy_ball_global} at every update.

For part \textup{(iii)}, let $\vect A=(1+\beta)\mbI-\eta\vect H$, define $\vect K_{-1}=\vect0$, $\vect K_0=\mbI$, and set $\vect K_{j+1}=\vect A\vect K_j-\beta\vect K_{j-1}$. For an eigenvalue $\lambda\in[m_H,M_H]$ of $\vect H$, the scalar coefficient $1+q_H^2-\eta\lambda$ belongs to $[-2q_H,2q_H]$. Its characteristic roots therefore have modulus $q_H$. If these roots are $z_1,z_2$, the polynomial $K_j(\lambda)=\sum_{k=0}^jz_1^kz_2^{j-k}$ satisfies $|K_j(\lambda)|\leq(j+1)q_H^j$. Diagonalization of $\vect H$ and summation give
\begin{equation}\label{eq:heavy_ball_kernel}
\Norm{\vect K_j}\leq(j+1)q_H^j,\qquad \eta\sum_{j=0}^{\infty}\Norm{\vect K_j}\leq\frac{\eta}{(1-q_H)^2}=\frac1{m_H}.
\end{equation}
The polynomial sequence in the statement satisfies $\vect P_t=\vect K_t-\beta\vect K_{t-1}$. For $q_H>0$ and $t\geq1$, this identity gives $\Norm{\vect P_t}\leq\{1+(1+q_H)t\}q_H^t$. Since $(t+1)q_H^t\leq\sum_{j=0}^tq_H^j$, it also gives $\Norm{\vect P_t}\leq A_H$. With $x=(1-q_H)t$, the inequality $(1+x)e^{-x}\leq2e^{-x/2}$ gives the exponential bound in \eqref{eq:heavy_ball_expansion}. If $q_H=0$, then $\vect H=m_H\mbI$, $\vect A=\vect0$, $\vect K_j=\vect0$ for $j\geq1$, and $\vect P_t=\vect0$ for $t\geq1$. The bounds follow directly in this case.

Write $\vect x_t=\vect\theta^{(t)}-\vect\theta^*$. While $|\vect x_t|_2\leq\epsilon_3$, Condition \ref{cond:H_smooth} gives $\vect u_t=\partial\mcL(\vect\theta^{(t)})-\vect g^*-\vect H\vect x_t$ with $|\vect u_t|_2\leq r_n$. The restarted recursion has the solution
\[
\vect x_t=\vect P_t\vect x_0-\eta\sum_{j=0}^{t-1}\vect K_{t-1-j}(\vect g^*+\vect u_j).
\]
The entry condition implies $|\vect x_0|_2\leq\epsilon_3$. If the error bounds hold through index $t-1$, the solution and \eqref{eq:heavy_ball_kernel} give $|\vect x_t|_2\leq A_H|\vect x_0|_2+(|\vect g^*|_2+r_n)/m_H\leq\epsilon_3$. Induction therefore validates every update. The polynomial identity $\vect P_t-\vect P_{t+1}=\eta\vect H\vect K_t$ gives $\eta\sum_{j=0}^{T-1}\vect K_j=(\mbI-\vect P_T)\vect H^{-1}$. Separating the constant term $\vect g^*$ proves \eqref{eq:heavy_ball_expansion} with $\vect R_T=-\eta\sum_{j=0}^{T-1}\vect K_{T-1-j}\vect u_j$. Its norm is bounded by $r_n/m_H$ through \eqref{eq:heavy_ball_kernel}.
\end{proof}

\subsection{Functions with bounded local subgradient variation}

\begin{corollary}\label{cor:bvg_growth}
Let $\mcL$ be finite and convex on $\mbR^p$, with a minimizer $\hat{\vect\theta}$. Suppose that, for $r,V_r>0$, its selected subgradient satisfies $|\partial\mcL(\vect x)-\partial\mcL(\vect y)|_2\leq V_r$ whenever $|\vect x-\vect y|_2\leq r$, and that
\[
\mcL(\vect\theta)-\mcL(\hat{\vect\theta})\geq\frac\mu2|\vect\theta-\hat{\vect\theta}|_2^2,\qquad \vect\theta\in\mbR^p.
\]
Set $L=2V_r/r$, $\kappa=L/\mu\geq1$, and $\eta=1/(2L)$.
\begin{enumerate}[label=\textup{(\roman*)},leftmargin=*]
\item Subgradient descent satisfies
\[
|\vect\theta^{(t)}-\hat{\vect\theta}|_2^2
\leq q^t|\vect\theta^{(0)}-\hat{\vect\theta}|_2^2+(1-q^t)\kappa r^2,
\qquad q=(1+1/(2\kappa))^{-1}.
\]
\item Restart the Phase 1 Nesterov method every $K=\lceil\sqrt{32\kappa}\rceil$ updates, and write $\vect z^{(j)}$ for the output after $j$ such blocks. Then
\[
|\vect z^{(j)}-\hat{\vect\theta}|_2^2
\leq4^{-j}|\vect z^{(0)}-\hat{\vect\theta}|_2^2+
\frac{2K\kappa}{3}(1-4^{-j})r^2.
\]
\end{enumerate}
Thus the distances decrease at rates $\exp(-ct/\kappa)$ and $\exp(-ct/\sqrt\kappa)$ to radii of orders $\sqrt\kappa r$ and $\kappa^{3/4}r$, respectively. The order $r$ is necessary for constant-step subgradient descent even when $\kappa$ is bounded.
\end{corollary}

\begin{proof}
Partitioning a segment into pieces of length at most $r$ gives Condition \ref{cond:L_smooth} with $\epsilon_1=r$ and $L=2V_r/r$. Write $D_t=|\vect\theta^{(t)}-\hat{\vect\theta}|_2$. The upper oracle bound \eqref{eq:oracle_upper}, convexity, and the subgradient update give
\[
2\eta\{\mcL(\vect\theta^{(t+1)})-\mcL(\hat{\vect\theta})\}
\leq D_t^2-D_{t+1}^2-\eta^2(1-\eta L)|\partial\mcL(\vect\theta^{(t)})|_2^2+\eta Lr^2.
\]
Quadratic growth and $1-\eta L\geq0$ imply $(1+\eta\mu)D_{t+1}^2\leq D_t^2+\eta Lr^2$. Summing this recursion proves part \textup{(i)}.

For one Nesterov block, Theorem \ref{thm:nagd_phase1}(i) and quadratic growth yield
\[
|\vect z^{(j+1)}-\hat{\vect\theta}|_2^2
\leq\frac{4}{\eta\mu K^2}|\vect z^{(j)}-\hat{\vect\theta}|_2^2+
\frac{K\eta L^2}{\mu}r^2
\leq\frac14|\vect z^{(j)}-\hat{\vect\theta}|_2^2+\frac{K\kappa}{2}r^2.
\]
Iteration proves part \textup{(ii)}. Since $K\asymp\sqrt\kappa$, the block count gives the stated rate in the total number of updates.

For the lower bound take $p=1$ and $\mcL(x)=r|x|/4+x^2/4$. Its local subgradient variation at radius $r$ is at most $V_r=r$, and its quadratic growth parameter can be chosen as $\mu=1/2$. Hence $L=2$, $\kappa=4$, and $\eta=1/4$. Except on the countable set of initial values whose trajectory reaches zero, every subgradient selection gives $x_{t+1}=7x_t/8-r\operatorname{sign}(x_t)/16$. The magnitudes converge to $r/30$ by the contraction $u\mapsto|7u/8-r/16|$.
\end{proof}

\section{Subgradient increment bounds}\label{app:increments}

\subsection{Auxiliary results}

We use Bernstein's inequality in its moment form \citep{vershynin.2018hdp}.

\begin{lemma}\label{lemma:cai_liu}
Let $W_1,\ldots,W_n$ be independent mean-zero random variables. Suppose that, for every integer $k\geq2$,
\[
\sum_{i=1}^n\mbE|W_i|^k\leq\frac{k!}{2}VB^{k-2}.
\]
Then, for every $t>0$,
\[
\mbP\left\{\left|\sum_{i=1}^nW_i\right|>\sqrt{2Vt}+2Bt\right\}\leq2e^{-t}.
\]
\end{lemma}

\begin{proof}
For $|\lambda|B<1$, independence, the power series for the exponential, and $\log(1+x)\leq x$ give
\[
\log\mbE\exp\left(\lambda\sum_iW_i\right)
\leq\sum_i\sum_{k\geq2}\frac{|\lambda|^k\mbE|W_i|^k}{k!}
\leq\frac{V\lambda^2}{2(1-B|\lambda|)}.
\]
Chernoff's bound with $\lambda=\sqrt{2t/V}/\{1+B\sqrt{2t/V}\}$ gives the upper-tail bound at $\sqrt{2Vt}+Bt$, and applying the same calculation to $-W_i$ gives the lower tail. The stated, slightly larger threshold follows. If $V=0$, all summands vanish almost surely.

\end{proof}

\subsection{Residual objectives}

\begin{proof}[Proof of Theorem \ref{thm:Lip_subgrad}]
Let $g=f'_+$ and let $\nu=\diff g$ be its Stieltjes measure. Convexity and the Lipschitz bound give $\nu\geq0$ and $\nu(\mbR)\leq2C_f$. Write $r_{\vect\theta}=Y-\vect Z^{\tp}\vect\theta$, $P_nq=n^{-1}\sum_iq(Y_i,\vect Z_i)$, and $Pq=\mbE[q(Y,\vect Z)]$. Conditional absolute continuity and dominated differentiation of the expected objective defined through loss differences give
\[
\nabla\bar{\mcL}(\vect\theta)=\vect b-P[\vect Zg(r_{\vect\theta})].
\]
For unit vectors $\vect u,\vect v$ and every integer $k\geq1$, the directional sub-Gaussian assumption gives
\[
P|\vect v^{\tp}\vect Z|^k\leq(C\sqrt{k})^k,~~P[|\vect v^{\tp}\vect Z|^k|\vect u^{\tp}\vect Z|]\leq C^{k+1}k^{k/2}.
\]

For $d=|\vect\theta_1-\vect\theta_2|_2>0$, write $\vect\theta_1-\vect\theta_2=d\vect u$. For every $t\in\mbR$, conditioning on $\vect Z$ gives
\[
P\big[|\vect v^{\tp}\vect Z|^k\mbI\{t\text{ lies between }r_{\vect\theta_1}\text{ and }r_{\vect\theta_2}\}\big] \leq C_{\rho}d\,P[|\vect v^{\tp}\vect Z|^k|\vect u^{\tp}\vect Z|] \leq d\,k!C^k,~~k\geq2.
\]
Since $|g(x)-g(y)|\leq2C_f$, integration with respect to $\nu$ gives, for
$W_{\vect v}=\vect v^{\tp}\vect Z\{g(r_{\vect\theta_1})-g(r_{\vect\theta_2})\}$,
\begin{equation}\label{eq:grad_diff_term1}
P|W_{\vect v}|\leq Cd,~~P|W_{\vect v}|^k\leq d\,k!C^k\quad(k\geq2).
\end{equation}
The same bounds, with larger constants, hold after subtracting the mean. Lemma \ref{lemma:cai_liu} therefore yields
\begin{equation}\label{eq:grad_diff_term2}
\mbP\{|(P_n-P)W_{\vect v}|>C(\sqrt{dt/n}+t/n)\}\leq2e^{-t}.
\end{equation}

We now extend the fixed-pair bound to the full decision region. Let $\delta=n^{-12}$, take a $\delta$-net $\mfN$ of $\Theta$, and take a $1/4$-net $\mcV$ of the unit sphere. Their logarithmic cardinalities are at most $Cp\log n$ and $Cp$. A sphere net and the sub-Gaussian tail bound imply, for a sufficiently large $C_B$,
\[
\mbP\{\max_{i\leq n}|\vect Z_i|_2>B_n\}\leq n^{-\gamma p},~~B_n=C_B\sqrt{p\log n}.
\]
Increasing $C_B$ if needed also gives
$\sup_{\vect v\in\mcV}P[|\vect v^{\tp}\vect Z|\mbI\{|\vect Z|_2>B_n\}]\leq n^{-10}$.
Here $p\log n\leq n$ eventually, since $a_n\to0$.

For each $\vect\theta$, choose $\tilde{\vect\theta}\in\mfN$ within $\delta$. On the displayed sample event, the difference between their residuals is at most $\delta B_n$. With $w_n=2\delta B_n$, monotonicity bounds every subgradient selection by the strict bracket
\[
B_{\tilde{\vect\theta}}=g(r_{\tilde{\vect\theta}}+w_n)-f'_-(r_{\tilde{\vect\theta}}-w_n).
\]
For each $\vect v\in\mcV$, the nonnegative variable $|\vect v^{\tp}\vect Z|B_{\tilde{\vect\theta}}$ has mean at most $Cw_n$ and $k$th moment at most $w_n k!C^k$. These assertions follow by integrating
$P(|r_{\tilde{\vect\theta}}-t|\leq w_n\mid\vect Z)\leq2C_{\rho}w_n$
against $\nu$. Bernstein's inequality and a union bound over $\mfN\times\mcV$ therefore give
\begin{equation}\label{eq:inf_conv_cond1_term4}
\max_{\tilde{\vect\theta}\in\mfN,\,\vect v\in\mcV}
(P_n+P)[|\vect v^{\tp}\vect Z|B_{\tilde{\vect\theta}}]\leq C(w_n+a_n)\leq Ca_n
\end{equation}
with failure probability at most $Cn^{-\gamma p}$. At a common decision $\vect\theta$, the difference between any two scenario-subgradient choices is bounded by the same strict bracket. Each projection of the resulting difference between subgradients of $\mcL$ is therefore bounded by $P_n[|\vect v^{\tp}\vect Z|B_{\tilde{\vect\theta}}]\leq Ca_n$. The sphere-net inequality gives the Euclidean bound simultaneously for all $\vect\theta\in\Theta$ and all choices.

The population part outside $|\vect Z|_2\leq B_n$ contributes at most $Cn^{-10}$. Consequently,
\begin{equation}\label{eq:inf_conv_cond1_term3}
|(P_n-P)\vect v^{\tp}\vect Z\{g(r_{\vect\theta_1})-g(r_{\vect\theta_2})\}| \leq |(P_n-P)\vect v^{\tp}\vect Z\{g(r_{\tilde{\vect\theta}_1})-g(r_{\tilde{\vect\theta}_2})\}|+Ca_n
\end{equation}
for all $\vect v\in\mcV$. Applying \eqref{eq:grad_diff_term2} to every net pair with $t=Ap\log n$, where $A$ exceeds the cardinality constants, gives the right side $C\{\sqrt{(d+2\delta)a_n}+a_n\}$. The sphere-net inequality $|\vect z|_2\leq2\max_{\vect v\in\mcV}|\vect v^{\tp}\vect z|$ proves \eqref{eq:lip_subgrad}. Summing the failure probabilities for truncation, brackets, and net pairs preserves the order $Cn^{-\gamma p}$. The truncation was used only for interpolation.

For the pointwise bound, each projection of $\vect Zg(r_{\vect\theta_0})$ is sub-Gaussian. Apply the scalar tail bound on $\mcV$ with $t=Ap\log n$, followed by the sphere-net inequality. This gives $C\sqrt{a_n}$ with the same failure order. Add that failure probability to the preceding event.
\end{proof}

\begin{proof}[Proof of Corollary \ref{thm:H_smooth}]
The crossing bound for the first moment in the proof of Theorem \ref{thm:Lip_subgrad}, integrated against $\nu$, gives $|\nabla\bar{\mcL}(\vect\theta_1)-\nabla\bar{\mcL}(\vect\theta_2)|_2\leq C|\vect\theta_1-\vect\theta_2|_2$.
On the event of Theorem \ref{thm:Lip_subgrad}, with $d=|\vect\theta_1-\vect\theta_2|_2$,
\begin{equation}\label{eq:lip_subgrad_repo}
|\partial\mcL(\vect\theta_1)-\partial\mcL(\vect\theta_2)-\nabla\bar{\mcL}(\vect\theta_1)+\nabla\bar{\mcL}(\vect\theta_2)|_2\leq C(\sqrt{da_n}+a_n).
\end{equation}
Adding the population bound and considering $d\geq a_n$ and $d<a_n$ proves part 1.

For part 2, integration along the segment in the convex region $\Theta_{\rm sc}$ gives
\[
\mcL(\vect\theta_1)-\mcL(\vect\theta_2)-\langle\partial\mcL(\vect\theta_2),\vect\theta_1-\vect\theta_2\rangle \geq\frac{\mu_0}{2}d^2-\frac{2C}{3}\sqrt{a_n}\,d^{3/2}-Ca_nd \geq\frac{\mu_0}{4}d^2
\]
when $d\geq C_2a_n$. This is Condition \ref{cond:mu_convex} with $\mu=\mu_0$.

For part 3, on $|\vect\theta-\vect\theta^*|_2\leq C_3\sqrt{a_n}$, \eqref{eq:lip_subgrad_repo} gives
\begin{equation}\label{eq:neighbor_grad_diff}
|\partial\mcL(\vect\theta)-\partial\mcL(\vect\theta^*)-\nabla\bar{\mcL}(\vect\theta)+\nabla\bar{\mcL}(\vect\theta^*)|_2\leq Ca_n^{3/4}.
\end{equation}
Taylor's formula bounds the remaining population increment by $C|\vect\theta-\vect\theta^*|_2^2\leq Ca_n$. Since $a_n\leq a_n^{3/4}$ eventually, part 3 follows.

To verify the sufficient density condition in part 3, suppose $\rho$ has a bounded continuous first derivative. Conditional integration gives
\[
\vect H(\vect\theta)=P\left[\vect Z\vect Z^{\tp}\int_{\mbR}\rho(t+\vect Z^{\tp}\vect\theta\mid\vect Z)\nu(\diff t)\right].
\]
For unit $\vect u,\vect v$,
\[
|\vect v^{\tp}\{\vect H(\vect\theta+d\vect u)-\vect H(\vect\theta)\}\vect v|
\leq Cd\,P[(\vect v^{\tp}\vect Z)^2|\vect u^{\tp}\vect Z|]\leq Cd.
\]
Hence the population Hessian is Lipschitz in operator norm. The \hyperref[rem:response_tails]{loss-difference bound} justifies dominated differentiation without a moment assumption on $Y$.

\end{proof}

\begin{proof}[Proof of Theorem \ref{cor:model_algorithm}]
Work on the event supplied by Theorem \ref{thm:Lip_subgrad} and write $\vect g(\vect\theta)=\partial\mcL(\vect\theta)$. The population Hessian is Lipschitz by assumption. For sufficiently small $R$, its smallest eigenvalue is therefore at least $m_H/2$ throughout $\mbB(\vect\theta^*,R)$. The population first-order condition and the pointwise bound give $s_n=|\vect g(\vect\theta^*)|_2\leq C\sqrt{a_n}$. Corollary \ref{thm:H_smooth} gives approximate smoothness on $\Theta$ and approximate strong convexity on $\mbB(\vect\theta^*,R)$, with $L=L_0$, $\mu=m_H/2$, and $\epsilon_j=C_ja_n$ for $j=1,2$. We may increase $L$ so that $\mu\leq2L$.

For every unit vector $\vect u$ and sufficiently large $n$,
\[
\mcL(\vect\theta^*+R\vect u)-\mcL(\vect\theta^*)\geq-s_nR+\frac{\mu R^2}4\geq\frac{\mu R^2}8.
\]
Convexity along rays therefore makes the finite convex objective in \eqref{eq:loss_expres} coercive, with every minimizer inside $\mbB(\vect\theta^*,R)$. Approximate strong convexity and optimality then give
\[
\delta_n:=|\hat{\vect\theta}-\vect\theta^*|_2\leq\max\{\epsilon_2,4s_n/\mu\}=O(\sqrt{a_n}).
\]
Corollary \ref{cor:projected_subgradient}\ref{part:projected_saa} gives geometric convergence to an $O(a_n)$ neighborhood of $\hat{\vect\theta}$ from every initial decision in $\Theta$. Since $a_n\geq n^{-1}$ eventually, the geometric terms are at most $a_n$ for every $t\geq\lceil c\log n\rceil$ when $c$ is sufficiently large.
The population gradient vanishes at $\vect\theta^*$ and is Lipschitz. Hence its objective gap is at most $C|\vect\theta-\vect\theta^*|_2^2$. The localization bound and the uniform optimization bound prove both asserted population-gap bounds on the same event. This event controls all scenario-subgradient choices simultaneously. The deterministic arguments therefore also cover choices depending on the preceding iterates.

At the deterministic decision $\vect\theta^*$, the scenario loss is differentiable almost surely. Thus $\vect g(\vect\theta^*)=n^{-1}\sum_i\vect\psi_i$, with $P\vect\psi_i=\vect0$. The directional sub-Gaussian assumption and $|f'_+|\leq C_f$ give $\mbE|\vect\psi_i|_2^2\leq Cp$, and independence yields $\mbE|\vect g(\vect\theta^*)|_2^2\leq Cp/n$. Consequently $s_n=O_P(\sqrt{p/n})$, and the preceding localization inequality sharpens to $\delta_n=O_P(\sqrt{p/n}+a_n)$. The uniform increment bound and the Lipschitz population Hessian apply at the random minimizer. Lemma \ref{lem:oracle_bounds} gives $|\vect g(\hat{\vect\theta})|_2\leq L\epsilon_1$. Hence
\[
\big|\vect H(\hat{\vect\theta}-\vect\theta^*)+\vect g(\vect\theta^*)\big|_2
\leq L\epsilon_1+C\{\sqrt{\delta_n a_n}+a_n+\delta_n^2\}
=O_P\!\left\{(p/n)^{3/4}\sqrt{\log n}+a_n\right\}.
\]
Combining this expansion with the $O(a_n)$ optimization bound proves the output expansion. When $p=o\{n(\log n)^{-2}\}$, the $a_n$ term is smaller than $(p/n)^{3/4}\sqrt{\log n}$.

For every deterministic unit $\vect v_n$, the third absolute moment of $\vect v_n^{\tp}\vect H^{-1}\vect\psi_i$ is bounded. Indeed, $|f'_+|\leq C_f$, $\|\vect H^{-1}\|\leq m_H^{-1}$, and Assumption \ref{cond:residual_cov} bounds every covariate projection. The population first-order condition gives $\mbE\vect\psi_i=\vect0$. The lower bound on $\sigma_n$ therefore gives Lyapunov's condition. If $p=o\{n^{1/3}(\log n)^{-2/3}\}$, then both $\sqrt n\,(p/n)^{3/4}\sqrt{\log n}$ and $\sqrt n\,a_n$ vanish. The expansion and Slutsky's theorem prove the normal limit. For fixed $p$, a score law and $\vect H$ that do not depend on $n$ give the vector limit $\mcN\{\vect0,\vect H^{-1}\mbE[\vect\psi_i\vect\psi_i^{\tp}]\vect H^{-1}\}$ by the multivariate central limit theorem.

Population smoothness and $\nabla\bar{\mcL}(\vect\theta^*)=\vect0$ give
\[
|\bar{\mcL}(\tilde{\vect\theta})-\bar{\mcL}(\hat{\vect\theta})|
\leq C\{|\hat{\vect\theta}-\vect\theta^*|_2|\tilde{\vect\theta}-\hat{\vect\theta}|_2+|\tilde{\vect\theta}-\hat{\vect\theta}|_2^2\}
=O_P\{a_n\sqrt{p/n}+a_n^2\}.
\]
Under the dimension condition for the normal limit, $a_n=o(\sqrt{p/n})$ and $na_n\sqrt{p/n}=p^{3/2}\log n/\sqrt n\to0$, proving the asserted equivalence.
The sample-complexity consequence follows from part \textup{(i)}, since $n\geq A(p/\varepsilon)\log(ep/\varepsilon)$ gives $Cp\log n/n\leq\varepsilon$ for sufficiently large $A$.
\end{proof}

\begin{proof}[Proof of Proposition \ref{prop:smoothness_scale_lower}]
Write $r_i(\vect\theta)=Y_i-\vect Z_i^{\tp}\vect\theta$. Choose a minimizer $\hat{\vect\theta}$ for which the rows with $r_i(\hat{\vect\theta})=0$ span a space of largest possible dimension, and let $A$ denote their index set. We first show that this dimension is $p$. Otherwise choose a nonzero $\vect v$ orthogonal to these rows. For small $|s|$, the active residuals remain zero and all other residuals keep their signs at $\hat{\vect\theta}+s\vect v$. The objective is affine on this segment, and minimality at its midpoint makes its slope zero. Full column rank ensures that some inactive residual changes along $\vect v$. Choose the sign of $\vect v$ so that a residual first reaches zero at a finite positive $s$. The objective remains affine, hence constant, up to that point. This gives another minimizer with all previous active rows and a new row not in their span, a contradiction. Therefore $|A|\geq p$.

Let $l$ be the coordinate with $Z_{il}=1$ for every $i$, and let $\vect e_l$ be its unit vector. For sufficiently small $\delta>0$, residuals outside $A$ keep their signs at $\hat{\vect\theta}\pm\delta\vect e_l$, while those in $A$ equal $\mp\delta$. Both points are differentiable, and
\[
\nabla\mcL(\hat{\vect\theta}+\delta\vect e_l)-\nabla\mcL(\hat{\vect\theta}-\delta\vect e_l)=\frac cn\sum_{i\in A}\vect Z_i.
\]
Its $l$th coordinate is $c|A|/n\geq cp/n$. Condition \ref{cond:L_smooth} therefore gives $cp/n\leq L\max\{2\delta,\epsilon_1\}$. Letting $\delta\downarrow0$ proves the claim.
\end{proof}

\begin{proposition}\label{prop:selection_step_lower}
Suppose the observations are independent and the conditional distribution of $Y_i$ given $\vect Z_i$ has a density for every $i$. Consider \eqref{eq:loss_expres} on $\mbR^p$ with a finite convex piecewise linear loss $f$ having finitely many pieces. Fix a deterministic initialization $\vect\theta^{(0)}$ and a constant step size $\eta>0$. With probability one, every selection of scenario subgradients generates the same subgradient descent trajectory, and no iterate has a residual at a breakpoint of $f$. On this event, the following bound holds for every $t\geq0$ and every minimizer $\hat{\vect\theta}$.
If $f(x)=cx\{\tau-\mbI(x\leq0)\}$, where $c>0$ and $\tau\in(0,1]$, and some coordinate $l$ satisfies $Z_{il}=1$ for every $i$ and $n(\tau-b_l/c)\notin\mathbb Z$, then
\[
|\vect\theta^{(t)}-\hat{\vect\theta}|_2+|\vect\theta^{(t+1)}-\hat{\vect\theta}|_2\geq\frac{\eta c}{n}\min_{k\in\mathbb Z}|n(\tau-b_l/c)-k|.
\]
For $f(x)=|x|$, $\vect b=\vect0$, $Z_{ik}\in\{-1,1\}$ for every $i,k$, and odd $n$, the left side is at least $\eta\sqrt p/n$.
\end{proposition}

\begin{proof}
Condition on $\vect Z_1,\ldots,\vect Z_n$. Independence implies that each $Y_i$ still has a conditional density. Before the first time that a residual reaches a breakpoint, every scenario subgradient belongs to the finite set of slopes of $f$. At a first hitting time $t$, the iterate therefore belongs to a finite set determined by the design, $\vect\theta^{(0)}$, and $\eta$. A hit would require $Y_i$ to equal $\vect Z_i^{\tp}\vect\theta$ plus a breakpoint for one of these finitely many points. Each equality has conditional probability zero. Taking the countable union over $t$ excludes every first hit on one event, simultaneously for every selection. Hence all selections yield the same iterates and unique scenario subgradients.

For the pinball loss, let $K_t$ be the number of negative residuals at $\vect\theta^{(t)}$. The $l$th coordinate of the objective gradient is $(c/n)\{K_t-n(\tau-b_l/c)\}$, hence each step has length at least $(\eta c/n)\min_{k\in\mathbb Z}|n(\tau-b_l/c)-k|$. For the absolute loss, every gradient coordinate is a sum of $n$ numbers in $\{-1,1\}$ divided by $n$. Its absolute value is at least $1/n$ when $n$ is odd, giving step length at least $\eta\sqrt p/n$. The triangle inequality gives both conclusions.
\end{proof}

\subsection{Index objectives}\label{app:index_models}
The index formulation allows every coefficient of a random scenario vector to vary,
\begin{equation}\label{eq:index_loss}
\mcL(\vect\theta)=\frac1n\sum_{i=1}^nf(\vect X_i^{\tp}\vect\theta)+\vect b^{\tp}\vect\theta.
\end{equation}
Let $\vect X_i$ be independent copies of $\vect X$.

\begin{assumption}\label{cond:theta}
The decision region $\Theta$ is convex and satisfies $\sup_{\vect\theta\in\Theta}|\vect\theta|_2\leq C_{\Theta}$ and $\inf_{\vect\theta\in\Theta}|\theta_1|\geq c_0>0$.
\end{assumption}

\begin{assumption}\label{ass:distribution}
\begin{enumerate}[label=\textup{(\roman*)},ref=\theassumption(\roman*),leftmargin=*]
\item\label{cond:density}
The conditional distribution of $X_1$ given $\vect X_{-1}$ has a density $\rho_X(x\mid\vect X_{-1})$ with $\sup_x\rho_X(x\mid\vect X_{-1})\leq C_{\rho}$ almost surely.
\item\label{cond:cov}
The vector $\vect X$ is sub-Gaussian with $\sup_{|\vect v|_2=1}\|\vect v^{\tp}\vect X\|_{\psi_2}\leq C_X$.
\end{enumerate}
\end{assumption}

\begin{proposition}\label{prop:index_objectives}
Suppose Assumptions \ref{cond:discont}, \ref{cond:theta}, and \ref{ass:distribution} hold and at least one of the following conditions holds.
\begin{enumerate}[label=\textup{(\alph*)},leftmargin=*]
\item $|\vect X|_2\leq C_X$ almost surely.
\item The measure $\nu$ has compact support.
\item The conditional density $\rho_X$ has a bounded continuous first derivative.
\end{enumerate}
For each deterministic $\vect\theta_0\in\Theta$, there are $C,\gamma>0$ and an event of probability at least $1-Cn^{-\gamma p}$ on which the following bounds hold. For every $\vect\theta_1,\vect\theta_2\in\Theta$ and every selection of scenario subgradients,
\begin{equation}\label{eq:index_lip_subgrad}
\Big|\partial\mcL(\vect\theta_1)-\partial\mcL(\vect\theta_2)-\nabla\bar{\mcL}(\vect\theta_1)+\nabla\bar{\mcL}(\vect\theta_2)\Big|_2 \leq C\{\sqrt{|\vect\theta_1-\vect\theta_2|_2a_n}+a_n\}.
\end{equation}
At $\vect\theta_0$, $|\partial\mcL(\vect\theta_0)-\nabla\bar{\mcL}(\vect\theta_0)|_2\leq C\sqrt{a_n}$. Parts 1 and 2 of Corollary \ref{thm:H_smooth} follow. In cases \textup{(a)} and \textup{(b)}, a bounded continuous first derivative of $\rho_X$ also gives a Lipschitz population Hessian. If the population Hessian is Lipschitz near an interior minimizer $\vect\theta^*$ and $m_H\mbI\preceq\nabla^2\bar{\mcL}(\vect\theta^*)\preceq M_H\mbI$, part 3 of Corollary \ref{thm:H_smooth} holds. The procedures in Theorem \ref{cor:model_algorithm} and Appendix \ref{app:phase_combination}, with their region and initialization requirements, give all its conclusions with $\vect\psi_i=\vect b+\vect X_i f'_+(\vect X_i^{\tp}\vect\theta^*)$.
\end{proposition}

\begin{proof}[Proof of Proposition \ref{prop:index_objectives}]\phantomsection\label{app:index_objectives}
Write $g=f'_+$, $\nu=\diff g$, and $\vect U=\vect X_{-1}$, and let $C_\Theta=\sup_{\vect\theta\in\Theta}|\vect\theta|_2$. Case \textup{(b)} includes every piecewise affine loss with finitely many breakpoints. Cases \textup{(a)} and \textup{(c)} allow arbitrary finite convex Lipschitz losses. The Lipschitz bound on $f$ and the sub-Gaussian assumption make $f(\vect X^{\tp}\vect\theta)$ integrable.

The curvature measure satisfies $\nu\geq0$ and $\nu(\mbR)\leq2C_f$. For $c\in\mbR$, define
\[
a_{\vect\theta}(c)=\frac{c-\vect U^{\tp}\vect\theta_{-1}}{\theta_1},\qquad
\vect x_{\vect\theta}(c)=(a_{\vect\theta}(c),\vect U^{\tp})^{\tp},\qquad
\rho_{\vect\theta}(c\mid\vect U)=\frac{\rho_X(a_{\vect\theta}(c)\mid\vect U)}{|\theta_1|}.
\]
Convexity of $\Theta$ fixes the sign of $\theta_1$. In case \textup{(a)}, only $|c|\leq C_XC_\Theta$ can contribute, and all projections of $\vect x_{\vect\theta}(c)$ on this interval are bounded. In the sub-Gaussian cases, those projections have $k$th moments at most $C^k(|c|+\sqrt k)^k$. This suffices on every compact level interval, including a neighborhood of the support of $\nu$ in case \textup{(b)}.

In case \textup{(c)} we need a bound at every level. The marginal density $p_{\vect\theta}(c)=\mbE\rho_{\vect\theta}(c\mid\vect U)$ is uniformly bounded and Lipschitz, while $\vect X^{\tp}\vect\theta$ is uniformly sub-Gaussian. Thus
\begin{equation}\label{eq:svm_section_density_tail}
p_{\vect\theta}(c)\leq C\exp(-c_1c^2),\qquad c\in\mbR,\quad\vect\theta\in\Theta.
\end{equation}
Indeed, if $L_1>0$ bounds the density Lipschitz constants, density at least $p_{\vect\theta}(c)/2$ persists on an interval of radius $p_{\vect\theta}(c)/(2L_1)$ about $c$. Its probability is at least $p_{\vect\theta}(c)^2/(2L_1)$, and for large $|c|$ it lies in the sub-Gaussian tail $|\vect X^{\tp}\vect\theta|\geq|c|/2$. This proves the display. For unit $\vect u,\vect v$ and nonnegative integers $j,k$ with $m=j+k\geq1$, Cauchy--Schwarz and $\rho_{\vect\theta}\leq C$ now give
\[
\mbE[\rho_{\vect\theta}(c\mid\vect U)|\vect v^{\tp}\vect x_{\vect\theta}(c)|^j|\vect u^{\tp}\vect x_{\vect\theta}(c)|^k]
\quad\leq C^{m+1}e^{-c_1c^2/2}(|c|+\sqrt m)^m\leq C^{m+1}m^{m/2}.
\]
In particular, the section moments
\begin{equation}\label{eq:svm_section_moments}
\begin{aligned}
\mbE[\rho_{\vect\theta}(c\mid\vect U)|\vect v^{\tp}\vect x_{\vect\theta}(c)|^{2k}]&\leq k!C^k,\\
\mbE[\rho_{\vect\theta}(c\mid\vect U)|\vect v^{\tp}\vect x_{\vect\theta}(c)|^k|\vect u^{\tp}\vect x_{\vect\theta}(c)|]&\leq k!C^k
\end{aligned}
\end{equation}
hold uniformly at all levels in cases \textup{(a)} and \textup{(c)}, and on every compact level interval in case \textup{(b)}.

Write $\vect\theta_1-\vect\theta_0=d\vect u$ and $\vect\theta_s=\vect\theta_0+sd\vect u$. Conditional on $\vect U$, the threshold $a_{\vect\theta_s}(c)$ is monotone in $s$, with derivative of absolute value $d|\vect u^{\tp}\vect x_{\vect\theta_s}(c)|/|\theta_{s,1}|$. Integrating along this interval gives
\begin{equation}\label{eq:svm_all_level_crossing}
\begin{aligned}
&\mbE[|\vect v^{\tp}\vect X|^k|\mbI(\vect X^{\tp}\vect\theta_1<c)-\mbI(\vect X^{\tp}\vect\theta_0<c)|]\\
&\quad=d\int_0^1\mbE[\rho_{\vect\theta_s}(c\mid\vect U)|\vect v^{\tp}\vect x_{\vect\theta_s}(c)|^k|\vect u^{\tp}\vect x_{\vect\theta_s}(c)|]\diff s\leq d\,k!C^k.
\end{aligned}
\end{equation}
Integration over $[c-w,c+w]$ similarly gives
\begin{equation}\label{eq:svm_all_level_slabs}
\mbE[|\vect v^{\tp}\vect X|^k\mbI\{|\vect X^{\tp}\vect\theta-c|\leq w\}]\leq wC^{k+1}k^{k/2}.
\end{equation}
In case \textup{(b)} take $c$ in the support of $\nu$ and $0<w\leq1$. The other two cases allow every $c$ and $w>0$. Integrating the crossing bound against $\nu$ gives, for $W_{\vect v}=\vect v^{\tp}\vect X\{g(\vect X^{\tp}\vect\theta_1)-g(\vect X^{\tp}\vect\theta_0)\}$,
\[
\mbE|W_{\vect v}|\leq Cd,\qquad \mbE|W_{\vect v}|^k\leq d\,k!C^k,\quad k\geq2.
\]
The same bounds hold after subtracting the mean, with larger constants. Write $P_n$ for the empirical average and $P$ for expectation under the law of $\vect X$. Lemma \ref{lemma:cai_liu} yields
\[
\Pr\{|(P_n-P)W_{\vect v}|>C(\sqrt{dt/n}+t/n)\}\leq2e^{-t}.
\]

Let $\delta=n^{-12}$, let $\mfN$ be a $\delta$-net of $\Theta$, and let $\mcV$ be a $1/4$-net of the unit sphere. Their logarithmic cardinalities are at most $Cp\log n$ and $Cp$. The sub-Gaussian bound gives
\[
\Pr\{\max_{i\leq n}|\vect X_i|_2>B_n\}\leq n^{-\gamma p},\qquad B_n=C_B\sqrt{p\log n}.
\]
Choose $C_B$ sufficiently large that also $\sup_{\vect v\in\mcV}P[|\vect v^{\tp}\vect X|\mbI\{|\vect X|_2>B_n\}]\leq n^{-10}$. For each decision choose a representative $\tilde{\vect\theta}\in\mfN$ within $\delta$ and put $w_n=2\delta B_n$. On the sample norm event, monotonicity bounds the difference between its scenario subgradient and the derivative at the representative by
\[
B_{\tilde{\vect\theta}}=g(\vect X^{\tp}\tilde{\vect\theta}+w_n)-f'_-(\vect X^{\tp}\tilde{\vect\theta}-w_n).
\]
The slab bound integrated against $\nu$ gives mean at most $Cw_n$ and $k$th moment at most $w_n k!C^k$ for $|\vect v^{\tp}\vect X|B_{\tilde{\vect\theta}}$. In case \textup{(b)}, $w_n\leq1$ eventually and only the specified neighborhood of the support of $\nu$ is used. Bernstein's inequality and the union bound over $\mfN\times\mcV$ give
\[
\max_{\tilde{\vect\theta}\in\mfN,\,\vect v\in\mcV}(P_n+P)[|\vect v^{\tp}\vect X|B_{\tilde{\vect\theta}}]\leq C(w_n+a_n)\leq Ca_n
\]
with failure probability at most $Cn^{-\gamma p}$. The population tail outside the norm event contributes at most $Cn^{-10}$. Thus interpolation of each projected mean-zero increment from its two net representatives costs at most $Ca_n$. This also bounds the difference between any two scenario-subgradient selections at a common decision.

Apply the fixed-pair bound on $\mfN\times\mfN\times\mcV$ with $t=Ap\log n$, taking $A$ larger than the cardinality constants. The representative distance is at most $d+2\delta$. Adding the interpolation error and using the sphere-net inequality $|\vect z|_2\leq2\max_{\vect v\in\mcV}|\vect v^{\tp}\vect z|$ prove \eqref{eq:index_lip_subgrad}. Conditional absolute continuity and dominated differentiation give
\[
\nabla\bar{\mcL}(\vect\theta)=\vect b+P[\vect Xg(\vect X^{\tp}\vect\theta)].
\]
At a deterministic decision every scenario index has a density, and each projection of $\vect Xg(\vect X^{\tp}\vect\theta_0)$ is sub-Gaussian. A scalar tail bound on $\mcV$ gives the pointwise bound $C\sqrt{a_n}$. Adding the failure probabilities for truncation, brackets, net pairs, and the pointwise bound preserves the order $Cn^{-\gamma p}$.

We verify the asserted Hessian Lipschitz property in cases \textup{(a)} and \textup{(b)} under the density derivative bound. Conditional integration with the notation above gives
\[
\vect H(\vect\theta)=\mbE\int_{\mbR}\rho_{\vect\theta}(c\mid\vect U)\vect x_{\vect\theta}(c)\vect x_{\vect\theta}(c)^{\tp}\nu(\diff c).
\]
For bounded $\vect X$, only the compact interval $|c|\leq C_XC_\Theta$ contributes. For compactly supported $\nu$, the integral is already over a compact interval. On either interval, projections of $\vect x_{\vect\theta}(c)$ have bounded sub-Gaussian norms, and subtracting the defining fractions gives $\|a_{\vect\theta}(c)-a_{\vect\theta'}(c)\|_{\psi_2}\leq C|\vect\theta-\vect\theta'|_2$. The density derivative bound and H\"older's inequality applied to differences of the integrand matrices therefore yield $\|\vect H(\vect\theta)-\vect H(\vect\theta')\|\leq C|\vect\theta-\vect\theta'|_2$.
\end{proof}

\begin{proof}[Proof of the conclusions in Example \ref{exp:svm}]
Set $f(t)=(1-t)_+$ and $\vect b=\vect0$ in the general index formulation. Then $f'_+(t)=-\mbI(t<1)$ and $\nu=\delta_1$. Case \textup{(b)} therefore gives the increment and pointwise score bounds under the density assumption.

Under the additional density derivative bound, the compact support of $\nu$ gives the population Hessian Lipschitz property proved in Proposition \ref{prop:index_objectives}. The Hessian formula becomes
\[
\nabla^2\bar{\mcL}_{\rm SVM}(\vect\theta)
=\mbE\left[\frac{\rho_X(a_{\vect\theta}(1)\mid\vect U)}{|\theta_1|}
\vect x_{\vect\theta}(1)\vect x_{\vect\theta}(1)^{\tp}\right].
\]
The population score at the interior solution is $\vect\psi_i=-\vect X_i\mbI(\vect X_i^{\tp}\vect\theta^*<1)$ and has mean zero. Applying Theorem \ref{cor:model_algorithm} and its \hyperref[app:unprojected_saa]{unprojected extension} with this score gives the containment, expansion, and normal limits asserted in the example. Appendix \ref{app:phase_combination} gives the same conclusions for the complete procedures.
\end{proof}

\subsection{Piecewise linear costs}

\begin{proof}[Proof of Corollary \ref{cor:max_affine}]
Write $\Delta\alpha_{jk}=\alpha_j-\alpha_k$ and $\Delta\vect\beta_{jk}=\vect\beta_j-\vect\beta_k$. Each pair of affine pieces ties with probability zero at a deterministic decision. If the maximizing pieces at $\vect\theta_0$ and $\vect\theta_1$ differ, their difference changes sign between these decisions. For $\vect\theta_1-\vect\theta_0=d\vect u$ and any unit $\vect v$, the conditional density assumption in Corollary \ref{cor:max_affine} gives
\[
\begin{aligned}
&\mbE\left[|\vect v^{\tp}\Delta\vect\beta_{jk}|^q\mbI\{\text{pieces $j,k$ cross between $\vect\theta_0$ and $\vect\theta_1$}\}\right]\\
&\quad\leq C_\rho d\,\mbE[|\vect v^{\tp}\Delta\vect\beta_{jk}|^q|\vect u^{\tp}\Delta\vect\beta_{jk}|]
\leq d\,q!C^q,\qquad q\geq1.
\end{aligned}
\]
The absolute projected gradient increment is bounded by the sum of these weighted crossing indicators over $j<k$. Since there are at most $K(K-1)/2$ pairs, its first moment is at most $Cd$ and its $q$th moment at most $d\,q!C^q$. Thus the Bernstein bound in \eqref{eq:grad_diff_term2} holds for each pair of decisions and direction, with constants depending on $K$.

To cover ties and data-dependent decisions, use the decision and sphere nets from the proof of Theorem \ref{thm:Lip_subgrad}. On the sample event $\max_{i,k}|\vect\beta_{ik}|_2\leq B_n=C_B\sqrt{p\log n}$, every affine piece active at a decision within $\delta$ of a net point either agrees with its maximizing piece or belongs to a pair satisfying
\[
|\Delta\alpha_{jk}+\Delta\vect\beta_{jk}^{\tp}\tilde{\vect\theta}|\leq 2\delta B_n.
\]
For $w>0$, the conditional density assumption in Corollary \ref{cor:max_affine} gives
\[
\mbE[|\vect v^{\tp}\Delta\vect\beta_{jk}|^q\mbI\{|\Delta\alpha_{jk}+\Delta\vect\beta_{jk}^{\tp}\tilde{\vect\theta}|\leq w\}]
\leq 2C_\rho w\,\mbE|\vect v^{\tp}\Delta\vect\beta_{jk}|^q
\leq w\,q!C^q.
\]
Summing over pairs bounds every convex combination of active pieces by the same bracket. Its Bernstein bound and the union bound give interpolation error $Ca_n$. The sample norm tail, population tail, and net-pair bounds proceed exactly as in the proof of Theorem \ref{thm:Lip_subgrad}, establishing the increment bound for all subgradient choices. At a deterministic decision, each score projection is bounded in absolute value by $\max_k|\vect v^{\tp}\vect\beta_k|$ and is therefore sub-Gaussian. This gives the pointwise bound.

The crossing bound for the first moment proves that $\nabla\bar{\mcL}$ is Lipschitz. The segment integration and Taylor arguments in the proof of Corollary \ref{thm:H_smooth} then give its conclusions under the stated Hessian assumption. At an interior population solution, $\vect\psi_i$ has mean zero, second moment at most $Cp$, and bounded third moments in deterministic unit directions. The proof of Theorem \ref{cor:model_algorithm} consequently applies. Finally,
\[
|Q(\vect\theta,\xi)-Q(\vect\theta_{\rm ref},\xi)|\leq\max_k|\vect\beta_k^{\tp}(\vect\theta-\vect\theta_{\rm ref})|
\]
is integrable, which justifies the expected loss differences and their derivatives without offset moments.
For any pair $j<k$, replacing the conditional density assumption by Assumptions \ref{cond:theta} and \ref{ass:distribution} on $\vect W_{jk}=(\Delta\vect\beta_{jk}^{\tp},\Delta\alpha_{jk})^{\tp}$ and the region $\{(\vect\theta^{\tp},1)^{\tp}:\vect\theta\in\Theta\}$, after the same permutation of coordinates, gives the same conclusions. Apply \eqref{eq:svm_all_level_crossing} and \eqref{eq:svm_all_level_slabs} at level zero with test direction $(\vect v^{\tp},0)^{\tp}$ and displacement direction $(\vect u^{\tp},0)^{\tp}$. The positive-part loss has curvature measure $\delta_0$, and case \textup{(b)} of Proposition \ref{prop:index_objectives} gives the crossing and slab bounds used above.
\end{proof}

\subsection{Rank regression}\label{app:rank}

\begin{example}\label{exp:rank}
Rank regression uses pairwise residual differences \citep{jaeckel.1972aoms}, with objective
\[
\mcL_{\rm rank}(\vect\theta)=\frac1{n(n-1)}\sum_{i\ne j}|(Y_i-\vect Z_i^{\tp}\vect\theta)-(Y_j-\vect Z_j^{\tp}\vect\theta)|.
\]
This formulation is used in tuning-free sparse regression and can improve efficiency over single-quantile regression \citep{wang_peng_etal.2020jasa}. We use slope coordinates, since pairwise differences eliminate an intercept.

On a convex bounded decision region, under the distributional assumptions of Theorem \ref{thm:Lip_subgrad}, the pairwise objective satisfies its uniform increment and pointwise score bounds. If the conditional density derivative bound holds and the expected pairwise objective satisfies the Hessian, region, and initialization conditions of Theorem \ref{cor:model_algorithm} or its \hyperref[app:unprojected_saa]{unprojected extension}, the corresponding methods and complete procedures attain distance $O_P(a_n)$ from a minimizer after $O(\log n)$ updates using the pairwise objective. For independent copies $\vect X=(Y,\vect Z^{\tp})^{\tp}$ and $\vect X'=(Y',\vect Z'^{\tp})^{\tp}$, define the score kernel and its first-order projection by
\[
\vect k(\vect X,\vect X')=-(\vect Z-\vect Z')\operatorname{sign}\{Y-Y'-(\vect Z-\vect Z')^{\tp}\vect\theta^*\},~~
\vect\psi(\vect X)=2\mbE[\vect k(\vect X,\vect X')\mid\vect X].
\]
Their outputs satisfy
\[
\tilde{\vect\theta}-\vect\theta^*=-\vect H^{-1}\frac1n\sum_{i=1}^n\vect\psi(\vect X_i)+O_P\!\left\{(p/n)^{3/4}\sqrt{\log n}+a_n\right\}.
\]
The remainder reduces to $O_P\{(p/n)^{3/4}\sqrt{\log n}\}$ when $p=o\{n(\log n)^{-2}\}$. For any deterministic unit contrast with variance bounded below, $p=o\{n^{1/3}(\log n)^{-2/3}\}$ gives the normal limit in Theorem \ref{cor:model_algorithm}, with variance $\operatorname{Var}\{\vect v_n^{\tp}\vect H^{-1}\vect\psi(\vect X)\}$.
\end{example}

\begin{proof}[Proof of the rank-regression conclusions in Example \ref{exp:rank}]
Write $\Delta Y=Y-Y'$ and $\Delta\vect Z=\vect Z-\vect Z'$ for independent observations. The conditional density
\[
\rho_\Delta(t\mid\vect Z,\vect Z')=\int\rho(t+y\mid\vect Z)\rho(y\mid\vect Z')\diff y
\]
is bounded by $C_\rho$ and, under the density derivative bound, has derivative bounded by $C_{\rho,1}$. Every projection of $\Delta\vect Z$ has sub-Gaussian norm at most $2C_Z$. Thus an independent pair satisfies the crossing and strict-bracket moments in the proof of Theorem \ref{thm:Lip_subgrad}, with $f(x)=|x|$. Expected loss differences are defined as in Section \ref{sec:exp_model}.

To handle the dependent pairs, put $U_nQ=\binom n2^{-1}\sum_{i<j}Q(\vect X_i,\vect X_j)$, $PQ=\mbE Q(\vect X,\vect X')$, and $m=\lfloor n/2\rfloor$ for a symmetric scalar kernel $Q$. Counting pairs over permutations $\pi$, followed by Jensen's inequality and independence of disjoint pairs, gives
\[
\begin{aligned}
U_nQ&=\frac1{n!}\sum_\pi\frac1m\sum_{j=1}^mQ(\vect X_{\pi(2j-1)},\vect X_{\pi(2j)}),\\
\mbE\exp\{\lambda(U_nQ-PQ)\}
&\leq\left[\mbE\exp\{\lambda(Q(\vect X,\vect X')-PQ)/m\}\right]^m.
\end{aligned}
\]
Hence the independent-pair Bernstein bounds hold with $m$ in place of $n$. Nonsymmetric kernels are first symmetrized, preserving their expectations and moment bounds by convexity.

For projected score increments, the crossing moments $d\,k!C^k$ give deviation $C\{\sqrt{dt/n}+t/n\}$, where $d=|\vect\theta_1-\vect\theta_2|_2$. Use the decision and sphere nets in the proof of Theorem \ref{thm:Lip_subgrad}. On $\max_i|\vect Z_i|_2\leq C\sqrt{p\log n}$, pairwise differences have at most twice this norm and the strict brackets use twice the width. A bracket of width $w$ has independent-pair moments $w\,k!C^k$, and the corresponding population tail is $Cn^{-10}$. Applying the displayed permutation bound to these brackets and the net pairs at $t=Ap\log n$ gives \eqref{eq:lip_subgrad} for every decision pair and subgradient selection. Projected pairwise scores are sub-Gaussian, giving the pointwise bound $C\sqrt{a_n}$ by the permutation bound and sphere net. Summing the sample-norm, bracket, net-pair, and pointwise failure probabilities gives $Cn^{-\gamma p}$.

Conditional-density integration gives
\[
\vect H(\vect\theta)=2\mbE[\rho_\Delta(\Delta\vect Z^{\tp}\vect\theta\mid\vect Z,\vect Z')\Delta\vect Z\Delta\vect Z^{\tp}].
\]
The density derivative and directional moments bound its operator norm and Lipschitz constant. At deterministic $\vect\theta^*$ the score is unique almost surely, has mean zero, and has kernel second moment at most $Cp$. The permutation representation and Jensen's inequality give $\mbE|\partial\mcL_{\rm rank}(\vect\theta^*)|_2^2\leq Cp/n$. These bounds verify the localization and expansion inputs in the proof of Theorem \ref{cor:model_algorithm}. The methods in Theorem \ref{cor:model_algorithm} and Appendix \ref{app:phase_combination}, with their initialization requirements and Phase 3 counts, therefore give the claimed optimization accuracy and expansion with leading term $-\vect H^{-1}\partial\mcL_{\rm rank}(\vect\theta^*)$.

To identify this term, let $\vect k_1(\vect X)=\mbE[\vect k(\vect X,\vect X')\mid\vect X]$ and $\vect k_2(\vect X,\vect X')=\vect k(\vect X,\vect X')-\vect k_1(\vect X)-\vect k_1(\vect X')$. Since $\mbE[\vect k_2(\vect X,\vect X')\mid\vect X]=\vect0$, the Hoeffding decomposition and orthogonality of degenerate pairs, including pairs sharing one observation, give
\[
\begin{aligned}
\partial\mcL_{\rm rank}(\vect\theta^*)&=\frac1n\sum_{i=1}^n\vect\psi(\vect X_i)+\vect D_n,\qquad \vect D_n=U_n\vect k_2,\\
\mbE|\vect D_n|_2^2&=\frac{2}{n(n-1)}\mbE|\vect k_2(\vect X,\vect X')|_2^2\leq Cp/n^2.
\end{aligned}
\]
Here $|\vect k|_2\leq|\Delta\vect Z|_2$ and $\mbE|\Delta\vect Z|_2^2\leq Cp$. Thus $|\vect D_n|_2=O_P(\sqrt p/n)$ is negligible relative to the expansion remainder. Conditional Jensen's inequality makes $\vect v_n^{\tp}\vect H^{-1}\vect\psi(\vect X)$ sub-Gaussian. Its mean is zero, and a variance bounded below gives Lyapunov's condition. The dimension condition in the example makes the expansion remainder $o_P(n^{-1/2})$, proving the contrast limit by Slutsky's theorem. For fixed dimension, the multivariate CLT gives covariance $\vect H^{-1}\operatorname{Var}\{\vect\psi(\vect X)\}\vect H^{-1}$.
\end{proof}

\subsection{Heavier tails}\label{app:heavier_tails}

\begin{corollary}\label{cor:subweibull}
For parts \textup{(i)}--\textup{(iii)}, consider i.i.d. observations within each $n$ and either the residual objective \eqref{eq:loss_expres} under Assumptions \ref{cond:residual_theta}, \ref{cond:residual_density}, and \ref{cond:discont}, or the index objective \eqref{eq:index_loss} under Assumptions \ref{cond:discont}, \ref{cond:theta}, and \ref{cond:density}. Write $\vect V=\vect Z$ for the residual objective and $\vect V=\vect X$ for the index objective. In parts \textup{(i)} and \textup{(ii)}, the index objective also satisfies one of cases \textup{(a)}--\textup{(c)} in Proposition \ref{prop:index_objectives}, with its sub-Gaussian assumption replaced by the displayed tail assumption.
\begin{enumerate}[label=\textup{(\roman*)},ref=(\roman*),leftmargin=*]
\item\label{cor:subweibull_directional} Suppose $\sup_{|\vect v|_2=1}\|\vect v^{\tp}\vect V\|_{\psi_\alpha}\leq C_V$ for some $\alpha>0$. Put
\[
 \bar\alpha=\min\{\alpha,1\},\qquad
 u_n=\frac{(p\log n)^{1/\bar\alpha}}n,\qquad
 \pi_n=Cn^{-\gamma p}.
\]
\item\label{cor:subweibull_polynomial} Suppose $\sup_{|\vect v|_2=1}\|\vect v^{\tp}\vect V\|_{\psi_\alpha}\leq C_V$ for $0<\alpha<1$. For every $M>0$, choose $C_M$ sufficiently large and put
\[
 B_n=C_M\sqrt{p+\log n}\,(\log n)^{1/\alpha-1/2},\qquad
 u_n=B_na_n,\qquad
 \pi_n=Cn^{-M}+Cn^{-\gamma p}.
\]
\item\label{cor:subweibull_mixture} Suppose $(\vect V_i,R_i)$ are i.i.d. within each $n$, $R\geq1$, and, for some $\beta>0$,
\[
 \mbP(R>t)\leq C\exp(-ct^\beta),\qquad
 \sup_{|\vect v|_2=1}\|\vect v^{\tp}\vect V\mid R=r\|_{\psi_2}\leq Cr,
 \qquad r\geq1.
\]
For the residual objective, suppose the conditional density of $Y$ given $(\vect Z,R)$ is bounded. For the index objective, suppose the conditional density of $X_1$ given $(\vect X_{-1},R)$ is bounded and either $\nu$ has compact support or this conditional density has a bounded continuous first derivative. All conditional bounds are uniform in $r$. For every $M>0$, choose $C_M$ sufficiently large and put
\[
 \tau_n=(C_M\log n)^{1/\beta},\qquad
 u_n=\tau_na_n,\qquad
 \pi_n=Cn^{-M}+Cn^{-\gamma p}.
\]
\end{enumerate}
Under any of parts \textup{(i)}--\textup{(iii)}, suppose $u_n\to0$. For each deterministic $\vect\theta_0\in\Theta$, there is an event of probability at least $1-\pi_n$ on which the following bounds hold simultaneously for every endpoint pair in $\Theta$ and every selection of scenario subgradients.
\[
 \left|\partial\mcL(\vect\theta_1)-\partial\mcL(\vect\theta_2)-\nabla\bar{\mcL}(\vect\theta_1)+\nabla\bar{\mcL}(\vect\theta_2)\right|_2
 \leq C\{\sqrt{|\vect\theta_1-\vect\theta_2|_2a_n}+u_n\},
\]
\[
 |\partial\mcL(\vect\theta_0)-\nabla\bar{\mcL}(\vect\theta_0)|_2
 \leq C\{\sqrt{a_n}+u_n\}.
\]
Parts 1 and 2 of Corollary \ref{thm:H_smooth} hold with $\epsilon_j=C_ju_n$, $j=1,2$. Under the Hessian conditions of its part 3, if $\mbB(\vect\theta^*,C_3\sqrt{u_n})\subseteq\Theta$, that conclusion holds with radius $C_3\sqrt{u_n}$ and remainder $C_4u_n^{3/4}$. For the residual objective, a bounded continuous conditional density derivative gives the population Hessian Lipschitz property for every finite convex Lipschitz loss. For the index objective, this verification holds in cases \textup{(a)} and \textup{(b)} in parts \textup{(i)} and \textup{(ii)}, and with compactly supported $\nu$ in part \textup{(iii)}. In part \textup{(iii)}, these derivative bounds are conditional on $R$ and uniform in $r$.

Suppose the population Hessian is Lipschitz near an interior minimizer $\vect\theta^*$ and $m_H\mbI\preceq\vect H=\nabla^2\bar{\mcL}(\vect\theta^*)\preceq M_H\mbI$. Use a procedure from Theorem \ref{cor:model_algorithm} or its \hyperref[app:unprojected_saa]{unprojected extension}, with its region and initialization requirements and $\epsilon_j=C_ju_n$. On the same event, an SAA solution $\hat{\vect\theta}$ exists, every iterate after $\lceil c\log n\rceil$ updates has distance at most $Cu_n$ from it, and the population objective gaps of the SAA solution and these iterates are at most $Cu_n$. Every such output satisfies
\[
 \tilde{\vect\theta}-\vect\theta^*
 =-\vect H^{-1}\frac1n\sum_{i=1}^n\vect\psi_i
 +O_P\!\left\{(p/n)^{3/4}\sqrt{\log n}+u_n\right\},
\]
where $\vect\psi_i=\vect b-\vect Z_i f'_+(Y_i-\vect Z_i^{\tp}\vect\theta^*)$ for the residual objective and $\vect\psi_i=\vect b+\vect X_i f'_+(\vect X_i^{\tp}\vect\theta^*)$ for the index objective. The two complete procedures give the same bounds with their local Phase 3 refinement calibrated at $u_n$. If $\sqrt n\,u_n\to0$, then $\sqrt n\,|\tilde{\vect\theta}-\hat{\vect\theta}|_2=o_P(1)$. For a deterministic unit contrast whose variance in Theorem \ref{cor:model_algorithm} is bounded below, its normal limit holds under
\[
\begin{array}{ll}
 p=o\{n^{1/3}(\log n)^{-2/3}\},\quad \sqrt n\,u_n\to0,
   &\text{in part \textup{(i)}},\\[2pt]
 p=o\{n^{1/3}(\log n)^{-(2/\alpha+1)/3}\},
   &\text{in part \textup{(ii)}},\\[2pt]
 p=o\{n^{1/3}(\log n)^{-2/3}\},
   &\text{in part \textup{(iii)}}.
\end{array}
\]
In part \textup{(ii)}, $p\leq n^{1/3}$ also gives remainder $O_P\{(p/n)^{3/4}(\log n)^{1/\alpha+1/2}\}$. The mixtures in part \textup{(iii)} satisfy the directional $\psi_\alpha$ bound with $\alpha=2\beta/(\beta+2)$. For the index objective they include $\vect X=R\vect G$, where $\vect G\sim\mcN(\vect0,\vect\Sigma)$ is independent of $R$ and $c\mbI\preceq\vect\Sigma\preceq C\mbI$.

\begin{enumerate}[label=\textup{(\roman*)},ref=(\roman*),leftmargin=*,start=4]
\item\label{cor:subweibull_coordinate}\phantomsection\label{app:sparse_subweibull} Consider the residual objective \eqref{eq:loss_expres} with i.i.d. observations, $p>n\geq2$, a convex bounded region $\Theta$, a bounded conditional density of $Y$ given $\vect Z$, and a finite convex Lipschitz loss. Fix a deterministic $\vect\theta^*\in\Theta$ and suppose
\[
 \max_{j\leq p}\|Z_j\|_{\psi_\alpha}\leq C_X,\qquad
 \Lambda_{\max}\{\mbE(\vect Z\vect Z^{\tp})\}\leq C_\Sigma,
 \qquad \alpha>0.
\]
Write $m=\tilde s$ and use $a_s$ from Section \ref{sec:hd_pgd}. For every $\gamma>0$, choose $C_\tau$ sufficiently large and put $\tau=C_\tau\{\log(np)\}^{1/\alpha}$. For sufficiently large $n$, there is an event of probability at least $1-Cp^{-\gamma}$ on which the following conclusions hold.
\begin{enumerate}[label=\textup{(\alph*)},leftmargin=*]
\item Uniformly over endpoint pairs in $\Theta$ with joint support of size at most $2m$, and over every selection of scenario subgradients,
\[
 \left|\partial\mcL(\vect\theta_1)-\partial\mcL(\vect\theta_2)-\nabla\bar{\mcL}(\vect\theta_1)+\nabla\bar{\mcL}(\vect\theta_2)\right|_\infty
 \leq C\{\sqrt{Da_s}+\tau a_s\},\qquad
 D=|\vect\theta_1-\vect\theta_2|_2.
\]
At $\vect\theta^*$,
\[
 |\partial\mcL(\vect\theta^*)-\nabla\bar{\mcL}(\vect\theta^*)|_\infty
 \leq C\{\sqrt{\log p/n}+\tau\log p/n\}.
\]
\item Condition \ref{cond:hdL_smooth} holds with $L=C\sqrt m\,\tau^2$ and $\epsilon_1=Ca_s$. If $\bar{\mcL}$ is $\mu_0$-strongly convex on a convex set $\Theta_{\rm sc}\subseteq\Theta$, then Condition \ref{cond:hdmu_convex} holds there with $\mu=\mu_0$ and $\epsilon_2=Cm\tau^3a_s$.
\item Suppose $\vect\theta^*\in\Thetar$, $a_s\to0$, and, for all $\vect\theta\in\Thetar$ with $|\vect\theta-\vect\theta^*|_2\leq C_3\sqrt{a_s}$,
\[
 |\nabla\bar{\mcL}(\vect\theta)-\nabla\bar{\mcL}(\vect\theta^*)-\vect H(\vect\theta-\vect\theta^*)|_\infty
 \leq C_0|\vect\theta-\vect\theta^*|_2^2,
\]
where $m_H\mbI\preceq\vect H\preceq M_H\mbI$. Then Condition \ref{cond:hdH_smooth} holds with $\epsilon_3=C_3\sqrt{a_s}$ and $r_n=C\tau^{3/2}a_s^{3/4}$.
\end{enumerate}
For the index objective under Assumptions \ref{cond:discont}, \ref{cond:theta}, and \ref{cond:density}, impose the displayed tail and second-moment bounds on $\vect X$. Then part \textup{(a)} holds with increment bound $C\{\tau^{3/2}\sqrt{Da_s}+\tau a_s\}$, and parts \textup{(b)} and \textup{(c)} hold under their population assumptions. Substitution of these parameters into Theorems \ref{thm:pgd_phases} and \ref{thm:iht_phases}, under their remaining hypotheses, gives their optimization and expansion bounds.
\end{enumerate}
\end{corollary}

\begin{proof}
\emph{Part \textup{(i)}.} We give the index calculation first. The residual calculation uses the conditional density of $Y$ given $\vect Z$ in each crossing and slab bound. Write $g=f'_+$ and $\nu=\diff g$. The directional condition gives $\|\vect v^{\tp}\vect X\|_{L^q}\leq Cq^{1/\bar\alpha}$ for every unit $\vect v$. Use the reconstructed sections in the proof of Proposition \ref{prop:index_objectives}. On compact level intervals,
\[
 \|\vect v^{\tp}\vect x_{\vect\theta}(c)\|_{L^q}
 \leq C(|c|+q^{1/\bar\alpha}).
\]
In case \textup{(c)}, the marginal density is bounded and Lipschitz. Its value at level $c$ persists on an interval whose length is proportional to that value. The directional tail bound consequently gives $p_{\vect\theta}(c)\leq C\exp(-c_1|c|^{\bar\alpha})$. Cauchy--Schwarz then bounds a weighted section moment of total degree $q$ by
\[
 C^{q+1}e^{-c_1|c|^{\bar\alpha}/2}(|c|+q^{1/\bar\alpha})^q
 \leq C^{q+1}q^{q/\bar\alpha}.
\]
These bounds hold at every level in case \textup{(c)}, on the required compact level interval in case \textup{(b)}, and with bounded sections in case \textup{(a)}. For $W_{\vect v}=\vect v^{\tp}\vect X\{g(\vect X^{\tp}\vect\theta_1)-g(\vect X^{\tp}\vect\theta_0)\}$ and $d=|\vect\theta_1-\vect\theta_0|_2$, integration along the threshold crossing and against $\nu$ gives
\[
 \mbE|W_{\vect v}|\leq Cd,\qquad
 \mbE|W_{\vect v}|^k\leq dC^kk^{k/\bar\alpha},\quad k\geq2.
\]
For residual objectives, these inequalities follow by conditioning on $\vect Z$. A slab of width $w$ has the same moment bounds with $d$ replaced by $w$.

Write $Y_i=W_{i,\vect v}-\mbE W_{i,\vect v}$ and let $q\geq2$ be even. For $2\leq k\leq q$, the preceding moments imply
\[
 \mbE|Y_i|^k\leq\frac{k!}{2}V B_q^{k-2},\qquad
 V=Cd,\quad B_q=Cq^{1/\bar\alpha-1}.
\]
Indeed, $k^{k/\bar\alpha}/k!\leq C^kk^{k(1/\bar\alpha-1)}$, and the factor $k^{2(1/\bar\alpha-1)}$ is absorbed into $C^k$. Expansion of the $q$th moment and comparison of nonnegative power-series coefficients give
\[
 \mbE\left|\sum_iY_i\right|^q
 \leq q!\,[z^q]\exp\left\{\frac{nVz^2}{2(1-B_qz)}\right\}.
\]
If $V=0$, every summand vanishes almost surely. Otherwise, for $0<z<1/B_q$, the coefficient is at most $z^{-q}$ times the displayed exponential. Taking $z=\min\{\sqrt{q/(2nV)},1/(2B_q)\}$ gives
\[
 \left\|\sum_iY_i\right\|_{L^q}\leq C\{\sqrt{ndq}+q^{1/\bar\alpha}\}.
\]
Markov's inequality with even $q\in[t,t+2]$ gives, for $t\geq2$,
\[
 \mbP\{|(P_n-P)W_{\vect v}|>C(\sqrt{dt/n}+t^{1/\bar\alpha}/n)\}\leq2e^{-t}.
\]
The slab variables satisfy this bound, and the pointwise projected score satisfies it with $d$ replaced by one.

Take the decision and sphere nets from the proof of Theorem \ref{thm:Lip_subgrad}, with decision mesh $n^{-12}$ and logarithmic cardinalities $O(p\log n)$. A sphere net and the directional tail bound give
\[
 \mbP\{\max_{i\leq n}|\vect X_i|_2>C(p\log n)^{1/\bar\alpha}\}\leq n^{-\gamma p}.
\]
Increasing the constant makes the corresponding population tail in any unit projection negligible. Since $(p\log n)^{1/\bar\alpha}=nu_n=o(n)$, the strict-bracket width is $o(n^{-10})$. Applying the slab and pairwise bounds with $t=Ap\log n$ over the nets gives bracket error $Cu_n$. The representative pair has distance at most $d+2n^{-12}$. The sphere-net inequality proves the increment bound for all endpoint pairs and subgradient selections. Applying the score bound on the sphere net proves the pointwise claim.

\emph{Part \textup{(ii)}.} Let $\vect G\sim\mcN(\vect0,\mbI_p)$ be independent of $\vect X$ and let $G_1\sim\mcN(0,1)$. Gaussian integration and the directional moment bound give, for $q\geq2$,
\[
 \mbE|G_1|^q\,\mbE|\vect X|_2^q
 =\mbE|\vect G^{\tp}\vect X|^q
 \leq C^qq^{q/\alpha}\mbE|\vect G|_2^q.
\]
Since $\|G_1\|_{L^q}\geq c\sqrt q$ and $\||\vect G|_2\|_{L^q}\leq C\sqrt{p+q}$,
\[
 \||\vect X|_2\|_{L^q}\leq C\sqrt{p+q}\,q^{1/\alpha-1/2}.
\]
Markov's inequality with $q=C_1\log n$ and a union bound give $\mbP\{\max_i|\vect X_i|_2>B_n\}\leq Cn^{-M}$. Increasing $C_M$ and using bounded directional second moments also give
\[
 \sup_{|\vect v|_2=1}\mbE[|\vect v^{\tp}\vect X|\mbI\{|\vect X|_2>B_n\}]\leq Cn^{-M}.
\]
Set $\mcA_n=\{|\vect X|_2\leq B_n\}$ and weight each projected increment by $\mbI_{\mcA_n}$. Evaluate this indicator at the reconstructed vector in the crossing identity, taking expectations under the original law. The section moments from \hyperref[cor:subweibull_directional]{part \textup{(i)}} give crossing first moment and variance $Cd$ and slab first moment and variance $Cw$, while the envelopes are $CB_n$. Bernstein's inequality gives fixed-pair deviation $C\{\sqrt{dt/n}+B_nt/n\}$ and the analogous slab bound.

Apply the decision and sphere nets from part \textup{(i)} at $t=Ap\log n$. Their mesh $n^{-12}$ gives bracket width $2B_nn^{-12}=o(a_n)$, since $B_na_n\to0$, and interpolation error $CB_na_n=Cu_n$. On the sample norm event the weighted and original empirical fields agree. Their population difference is uniformly $Cn^{-M}$, which is absorbed into $u_n$ by increasing the tail exponent. The pointwise score has bounded projected variance and envelope $CB_n$, giving its bound by the same sphere-net Bernstein argument. The total failure probability is $Cn^{-M}+Cn^{-\gamma p}$. For the residual objective use $\vect Z$ and conditional response-density integration.

\emph{Part \textup{(iii)}.} Let $A_n=\{R\leq\tau_n\}$ and $P_\tau=P(\cdot\mid A_n)$. Conditional on $R=r$, the reconstructed sections satisfy
\[
 \|\vect v^{\tp}\vect x_{\vect\theta}(c)\|_{L^q(P\mid R=r)}\leq C(|c|+r\sqrt q).
\]
On compact level intervals this is at most $Cr\sqrt q$. Under the conditional density derivative bound, the density-tail argument in Proposition \ref{prop:index_objectives} gives $p_{\vect\theta,r}(c)\leq C\exp(-c_1c^2/r^2)$. Cauchy--Schwarz then bounds a weighted section moment of total degree $q$ by
\[
 C^{q+1}e^{-c_1c^2/(2r^2)}(|c|+r\sqrt q)^q\leq C^{q+1}r^q q^{q/2}.
\]
For $W_{\vect v}$ as in part \textup{(i)}, integration against $\nu$ gives
\[
 \mbE(|W_{\vect v}|\mid R=r)\leq Cdr^2,\qquad
 \mbE(|W_{\vect v}|^k\mid R=r)\leq d\,k!C^k r^{k+1},\quad k\geq2.
\]
For residual objectives, conditioning on $(\vect Z,R)$ and integrating the bounded response density gives the same inequalities without a response moment assumption. Every fixed moment of $R$ is finite, and $P(A_n)\geq1/2$ eventually. Hence
\[
 \mbE_\tau R^{k+1}\leq2\tau_n^{k-2}\mbE R^3,\qquad k\geq2.
\]
The centered moments satisfy Lemma \ref{lemma:cai_liu} with variance parameter $Cd$ and scale $C\tau_n$. Thus
\[
 \mbP_\tau\{|(P_n-P_\tau)W_{\vect v}|>C(\sqrt{dt/n}+\tau_nt/n)\}\leq2e^{-t}.
\]
A level slab of width $w$ has conditional $k$th moment at most $wk!C^kr^k$. The bound $\mbE_\tau R^k\leq2\tau_n^{k-2}\mbE R^2$ gives variance parameter $Cw$ and scale $C\tau_n$, and its first moment is at most $Cw\mbE R$. At a deterministic decision, the projected score has bounded variance and scale $C\tau_n$ by the same argument.

Under $P_\tau$, apply the nets from \hyperref[cor:subweibull_directional]{part \textup{(i)}} with norm cutoff $C\tau_n\sqrt{p\log n}$ and Bernstein scale $C\tau_n$. Conditional sub-Gaussian tails give failure $Cn^{-\gamma p}$ and a negligible population interpolation tail. The width $Cn^{-12}\tau_n\sqrt{p\log n}=o(a_n)$ and $t=Ap\log n$ give bracket error $C\tau_na_n$ and increment error $C\{\sqrt{da_n}+\tau_na_n\}$ for all endpoint pairs and selections, with conditional failure $Cn^{-\gamma p}$.

Choose $C_M$ sufficiently large that
\[
 n\mbP(R>\tau_n)\leq Cn^{-M},\qquad
 \mbE[R^j\mbI\{R>\tau_n\}]\leq Cn^{-M},\quad j=1,2,3.
\]
Conditional on $A_n$ for every observation, the observations remain i.i.d. with law $P_\tau$. The conditional density and derivative bounds are preserved, since, given the remaining covariates, they are averages of the bounds conditional on $R$. Taking a supremum over unit projections gives
\[
 \sup_{\vect\theta\in\Theta}|\nabla\bar{\mcL}_\tau(\vect\theta)-\nabla\bar{\mcL}(\vect\theta)|_2
 \leq C\mbE[R\mbI\{R>\tau_n\}]+C\mbP(R>\tau_n)\leq Cn^{-M}.
\]
A larger exponent in the tail estimate absorbs this difference into $\tau_na_n$. Removing the conditioning proves the increment and pointwise claims. Finally,
\[
 \|\vect v^{\tp}\vect V\|_{L^q}\leq C\sqrt q\,\|R\|_{L^q}\leq Cq^{1/2+1/\beta},\qquad q\geq2,
\]
which gives the asserted directional $\psi_{2\beta/(\beta+2)}$ bound. For $\vect X=R\vect G$, the Gaussian conditional density of $X_1$ given $(\vect X_{-1},R)$ is bounded by $C/R\leq C$ and its derivative by $C/R^2\leq C$. The covariance bounds also give the required conditional projection bound.

\emph{The consequences in parts \textup{(i)}--\textup{(iii)}.} The crossing first moment gives population Lipschitz continuity. The density derivative and moments of degree at most three give the Hessian Lipschitz verifications. The residual calculation uses only $\nu(\mbR)\leq2C_f$, while the index calculation uses bounded sections or compact support of $\nu$. In part \textup{(iii)}, integrate these bounds over $R$ using its finite moments. Since $a_n\leq u_n$, the proof of Corollary \ref{thm:H_smooth} gives $\epsilon_j=C_ju_n$. On a ball of radius $C_3\sqrt{u_n}$, the increment error is at most $Cu_n^{3/4}$ and the population Taylor error is at most $Cu_n$, proving its part 3.

At a stationary population solution the score has mean zero, second moment at most $Cp$, and bounded third moments in deterministic unit directions. The localization argument gives
\[
 \delta_n:=|\hat{\vect\theta}-\vect\theta^*|_2=O_P(\sqrt{p/n}+u_n).
\]
The increment bound at an SAA solution, Taylor's formula, and Lemma \ref{lem:oracle_bounds} give
\[
 |\vect H(\hat{\vect\theta}-\vect\theta^*)+\partial\mcL(\vect\theta^*)|_2
 \leq C\{\sqrt{\delta_na_n}+u_n+\delta_n^2\}
 =O_P\!\left\{(p/n)^{3/4}\sqrt{\log n}+u_n\right\}.
\]
Corollary \ref{cor:absorption} and the local recursions in Corollary \ref{cor:projected_subgradient}\ref{part:projected_local_entry}, calibrated at $u_n$, give the optimization bound for all later iterates. The local comparison \eqref{eq:local_phase3_comparison} gives the complete procedures' bound at their Phase 3 iteration rates. On the displayed event, the pointwise bound gives SAA localization $C\sqrt{u_n}$. Population smoothness and stationarity then bound the population objective gaps by $Cu_n$. Adding optimization error proves the output expansion, and multiplying that error by $\sqrt n$ proves first-order equivalence whenever $\sqrt n\,u_n\to0$.

For the normal limit, the variance lower bound and third moments give Lyapunov's condition. In part \textup{(i)}, the two stated dimension conditions make the remainder $o_P(n^{-1/2})$. In part \textup{(ii)}, if $p\geq\log n$,
\[
 \sqrt n\,u_n\leq C\sqrt{\frac{p^3(\log n)^{2/\alpha+1}}n}\longrightarrow0.
\]
For $p<\log n$, the same conclusion follows from a polylogarithmic bound. Its stated dimension condition also makes $(p/n)^{3/4}\sqrt{n\log n}\to0$. In part \textup{(iii)}, since $u_n=\tau_na_n$ and $\tau_n$ is a power of $\log n$, both remainder terms multiplied by $\sqrt n$ vanish under $p=o\{n^{1/3}(\log n)^{-2/3}\}$. Slutsky's theorem proves the limits. If $p\leq n^{1/3}$ and $p\geq\log n$ in part \textup{(ii)},
\[
 \frac{u_n}{(p/n)^{3/4}}
 \leq C\frac{p^{3/4}}{n^{1/4}}(\log n)^{1/\alpha+1/2}
 \leq C(\log n)^{1/\alpha+1/2}.
\]
For $p<\log n$, direct substitution gives the same bound eventually. This proves its additional remainder assertion.

\emph{Part \textup{(iv)}.} First consider the index objective. Let $g=f'_+$, $\nu=\diff g$, $\mcA_\tau=\{\vect x:|\vect x|_\infty\leq\tau\}$, and $\vect U=\vect X_{-1}$. All expectations use the original law with indicator weight $\mbI_{\mcA_\tau}$. For any prescribed $M>0$, increasing $C_\tau$ gives
\[
 \Pr\{\text{some }i\leq n:\vect X_i\notin\mcA_\tau\}\leq Cp^{-M},\qquad
 \max_j\mbE[|X_j|\mbI_{\mcA_\tau^c}]\leq Cp^{-M}.
\]
The first inequality follows from the coordinate tails and a union bound over $np$. The second follows from Cauchy--Schwarz, bounded second moments, and the same tail probability. Take $M>\max\{\gamma,12\}$.

Use the sections in Proposition \ref{prop:index_objectives}. When $\vect x_{\vect\theta}(c)\in\mcA_\tau$, every section coordinate is bounded by $\tau$. For every unit $\vect u$, the conditional density and second-moment bounds give, at every level $c$,
\[
\mbE[\rho_{\vect\theta}(c\mid\vect U)\mbI\{\vect x_{\vect\theta}(c)\in\mcA_\tau\}|\vect u^{\tp}\vect x_{\vect\theta}(c)|]\qquad\leq C\{|u_1|\tau+\mbE|\vect u_{-1}^{\tp}\vect U|\}\leq C\tau.
\]
For endpoints at distance $d$, apply the crossing formula along their segment with its unit displacement direction and the displayed indicator. Integrating against $\nu$ gives, for $W_j^\tau=X_j\mbI_{\mcA_\tau}\{g(\vect X^{\tp}\vect\theta_1)-g(\vect X^{\tp}\vect\theta_2)\}$,
\[
 \mbE|W_j^\tau|^k\leq dC^k\tau^{k+1},\quad k\geq2,\qquad
 |W_j^\tau|\leq2C_f\tau.
\]
The variance is at most $C\tau^3d$ and the centered envelope at most $4C_f\tau$. Bernstein's inequality gives fixed-pair deviation $C\{\tau^{3/2}\sqrt{dt/n}+\tau t/n\}$ with failure probability $2e^{-t}$.

Use the sparse decision nets in the bounded-coordinate proof of Theorem \ref{thm:hdLip_subgrad}(1), with mesh $p^{-12}$ on subspaces of dimension at most $2m$ and logarithmic cardinality $O(m\log p)$. On $\mcA_\tau$ the bracket width is $w=\tau\sqrt{2m}\,p^{-12}\leq a_s$ eventually. Conditional integration gives
\[
 \mbE\left[\{|X_j|\mbI_{\mcA_\tau}[g(\vect X^{\tp}\tilde{\vect\theta}+2w)-f'_-(\vect X^{\tp}\tilde{\vect\theta}-2w)]\}^k\right]
 \leq wC^k\tau^k.
\]
At $t=Am\log p$, the coordinate union bound and strict-bracket interpolation give error $C\tau a_s$ and cover every selection. Combining the fixed-pair bound with this interpolation proves the weighted increment bound. On $\{\vect X_i\in\mcA_\tau\text{ for all }i\}$ the weighted and original empirical fields agree, while their population difference is uniformly $Cp^{-M}$ and is absorbed into $\tau a_s$. Combining these losses gives failure probability $Cp^{-\gamma}$.

At a deterministic decision, each weighted score coordinate has variance at most $C_f^2\mbE X_j^2\leq C$ and envelope at most $C_f\tau$. Coordinatewise Bernstein bounds and the same population-tail comparison prove the pointwise claim. The first-moment crossing calculation gives a truncated population increment at most $C\tau^2D$ in maximum norm, while the full population increment differs by at most $Cp^{-M}$. Adding the empirical increment bound, separately for $D\geq a_s$ and $D<a_s$, gives
\[
 |\partial\mcL(\vect\theta_1)-\partial\mcL(\vect\theta_2)|_\infty
 \leq C\tau^2\max\{D,a_s\}.
\]
Conversion to $|\cdot|_{(2m)}$ proves \textup{(b)}'s smoothness assertion. Under population strong convexity, pairing the increment error with a sparse displacement and integrating gives
\[
 \frac{\mu_0}{2}D^2-C\sqrt m\,\tau^{3/2}\sqrt{a_s}\,D^{3/2}-C\sqrt m\,\tau a_sD.
\]
For $D\geq Cm\tau^3a_s$, this is at least $\mu_0D^2/4$, proving restricted strong convexity. Combining \textup{(a)} with the assumed Taylor bound on $D\leq C_3\sqrt{a_s}$ gives
\[
 C\{\tau^{3/2}\sqrt{Da_s}+\tau a_s+D^2\}\leq C_1\tau^{3/2}a_s^{3/4},
\]
which proves \textup{(c)}.

For residual objectives, conditional response-density integration and the tail and second-moment bounds give
\[
 \mbE[Z_j^2|\vect u^{\tp}\vect Z|]
 \leq(\mbE Z_j^4)^{1/2}(\mbE|\vect u^{\tp}\vect Z|^2)^{1/2}\leq C.
\]
The truncated crossing variance is therefore $Cd$, while its Bernstein scale remains $C\tau$. The net and tail argument just given proves the residual increment bound. Its first-moment population bound follows from $\mbE[|Z_j||\vect u^{\tp}\vect Z|]\leq C$, and the segment and Taylor calculations prove \textup{(b)} and \textup{(c)} with the stated parameters.
\end{proof}

\section{Smoothing}\label{app:smoothing}

Write $g=f'_+$ and $\nu=\diff g$, and let $P_n$ and $P$ denote the empirical average and expectation. Write $\circ\in\{{\rm inf},{\rm int}\}$ for infimal convolution in \eqref{eq:infimum_conv} and integral convolution in \eqref{eq:integral_conv}. In this appendix, the subscript ${\rm inf},h$ denotes the infimal convolution written with subscript $h$ in Section \ref{sec:smooth_tech}. For the residual objective, impose Assumptions \ref{cond:residual_theta}--\ref{cond:discont}, together with Assumption \ref{cond:prox_fun} for infimal convolution or Assumption \ref{cond:kernel_fun} for integral convolution. For the index objective, impose Assumptions \ref{cond:discont}, \ref{cond:theta}, and \ref{ass:distribution}, with a bounded continuous first derivative of $\rho_X$. Its infimal-convolution calculations use the conditions on $f$ and $\omega$ in Assumption \ref{cond:prox_fun}, and its integral-convolution calculations use the conditions on $K$ in Assumption \ref{cond:kernel_fun}.

For the common moment and concentration arguments, put $(Q_{\vect\theta},\vect V)=(Y-\vect Z^{\tp}\vect\theta,\vect Z)$ for the residual objective and $(Q_{\vect\theta},\vect V)=(\vect X^{\tp}\vect\theta,\vect X)$ for the index objective. Write $q_h=f_{\circ,h}^{\prime\prime}$ where this derivative exists. The scenario gradients have opposite signs, and both Hessians equal $q_h(Q_{\vect\theta})\vect V\vect V^{\tp}$.

Write $\vect U=\vect X_{-1}$ and define
\[
a_{\vect\theta}(c)=\frac{c-\vect U^{\tp}\vect\theta_{-1}}{\theta_1},\qquad
\vect x_{\vect\theta}(c)=(a_{\vect\theta}(c),\vect U^{\tp})^{\tp},\qquad
\rho_{\vect\theta}(c\mid\vect U)=\frac{\rho_X(a_{\vect\theta}(c)\mid\vect U)}{|\theta_1|}.
\]

Also write
\[
\vect A_{\vect\theta}(c)=\mbE[\rho_{\vect\theta}(c\mid\vect U)\vect x_{\vect\theta}(c)\vect x_{\vect\theta}(c)^{\tp}],\qquad
\vect B_{\vect\theta}(c)=\mbE[\vect X\mbI(\vect X^{\tp}\vect\theta\geq c)].
\]

\subsection{Infimal convolution}\label{app:infimal_convolution}

Fix constants $0<\mu_{\omega}\leq C_{\omega}<\infty$ such that $\mu_{\omega}\leq\omega''\leq C_{\omega}$. Write $y_h(x)$ for the unique minimizer in \eqref{eq:infimum_conv}. Its optimality equation is
\begin{equation}\label{eq:envlp_subgd}
\omega'((x-y_h(x))/h)\in\partial f(y_h(x)),~~f'_{\infh}(x)=\omega'((x-y_h(x))/h).
\end{equation}
Since $|\partial f|\leq C_f$ and $\omega''\geq\mu_{\omega}$, the quantity $(x-y_h(x))/h$ is bounded uniformly. In particular,
\begin{equation}\label{eq:yxtheta_def}
|y_h(x)-x|\leq Ch,~~|f'_{\infh}(x)|\leq C_f.
\end{equation}
Let $g_h=f'_{\infh}$ and $\nu_h=\diff g_h$. The derivative is nondecreasing, and $\nu_h(\mbR)=\nu(\mbR)\leq2C_f$. The inverse subgradient graphs satisfy
\begin{equation}\label{eq:inf_esteq}
g_h^{-1}(s)=g^{-1}(s)+h(\omega')^{-1}(s),~~g(-\infty)<s<g(+\infty),
\end{equation}
where generalized inverses can be used at flat portions. This follows from \eqref{eq:envlp_subgd}, or from $(f_{\infh})^*=f^*+h\omega^*$, where $*$ denotes convex conjugacy \citep{rockafellar1997convex}. Thus the measures $\nu_h$ and $\nu$ admit a coupling that moves mass by at most $Ch$.

\begin{lemma}\label{lem:non-exp}
The proximal minimizer satisfies $|y_h(x)-y_h(z)|\leq|x-z|$. Consequently, $g_h$ is nondecreasing and $C/h$-Lipschitz.
\end{lemma}

\begin{proof}
Monotonicity of $\partial f$ and \eqref{eq:envlp_subgd} give
\[
\{\omega'((x-y_h(x))/h)-\omega'((z-y_h(z))/h)\}\{y_h(x)-y_h(z)\}\geq0.
\]
Since $\omega'$ is strictly increasing, the differences $y_h(x)-y_h(z)$ and $x-z-y_h(x)+y_h(z)$ have the same sign. Both lie between zero and $x-z$ when $x\geq z$. This proves nonexpansiveness and, using $\omega''\leq C_{\omega}$ in \eqref{eq:envlp_subgd}, the Lipschitz bound for $g_h$.

\end{proof}

Let $q_h=g_h'$ wherever this derivative exists. Lemma \ref{lem:non-exp} gives
\begin{equation}\label{eq:inf_fpp_bound}
0\leq q_h\leq C/h,~~\int_{\mbR}q_h(t)\diff t\leq2C_f.
\end{equation}
Under Assumption \ref{cond:prox_fun}, $q_h$ is Lipschitz with constant $C/h^2$ between finitely many transition endpoints. Indeed, on a smooth piece, differentiation of \eqref{eq:envlp_subgd} gives
\[
q_h(x)=\frac{f''(y_h(x))\omega''((x-y_h(x))/h)}{\omega''((x-y_h(x))/h)+hf''(y_h(x))}.
\]
The assumed derivative bounds and $\omega''\geq\mu_{\omega}$ bound its derivative on that piece. If $y_h(x)$ equals a breakpoint $t_j$ whose one-sided derivatives are $m_{j,-},m_{j,+}$, then
\[
q_h(x)=h^{-1}\omega''((x-t_j)/h),~~t_j+h(\omega')^{-1}(m_{j,-})<x<t_j+h(\omega')^{-1}(m_{j,+}).
\]
Thus there are finitely many mapped boundaries, and the claimed piecewise Lipschitz constant follows. Define $q_h=0$ at those boundaries. Its values there do not affect any population expectation or any integral along a segment.

For the residual objective, the curvature coupling and conditional-density integration give
\begin{equation}\label{eq:prox_control}
\sup_{\vect\theta\in\Theta}|\nabla\bar{\mcL}_{\infh}(\vect\theta)-\nabla\bar{\mcL}(\vect\theta)|_2\leq Ch,\qquad
\sup_{\vect\theta\in\Theta}\|\nabla^2\bar{\mcL}_{\infh}(\vect\theta)-\nabla^2\bar{\mcL}(\vect\theta)\|\leq Ch.
\end{equation}
Each unit projection of the gradient difference is bounded by $Ch\nu(\mbR)\mbE|\vect v^{\tp}\vect Z|$. The Hessian difference is bounded by $Ch\nu(\mbR)\mbE(\vect v^{\tp}\vect Z)^2$, using the bounded first derivative of the conditional response density. This derivative bound also gives the population Hessian Lipschitz property.

For the index objective, the population gradient and Hessian are $\vect b+g(-\infty)\mbE\vect X+\int\vect B_{\vect\theta}(c)\nu(\diff c)$ and $\int\vect A_{\vect\theta}(c)\nu(\diff c)$. The all-level section moments in the proof of Proposition \ref{prop:index_objectives} bound $\vect A_{\vect\theta}(c)$, $\vect B_{\vect\theta}(c)$, and $\vect B'_{\vect\theta}(c)=-\mbE[\rho_{\vect\theta}(c\mid\vect U)\vect x_{\vect\theta}(c)]$ in operator or Euclidean norm by $C$. The curvature coupling in \eqref{eq:inf_esteq} therefore gives the gradient bound in \eqref{eq:prox_control}. For the Hessian bound, decompose $\nu$ into the bounded density $f''$ on its smooth pieces and the finitely many atoms at its breakpoints. On a smooth piece, the formula for $q_h$ above and $|y_h(x)-x|\leq Ch$ give $|q_h(x)-f''(x)|\leq Ch$ away from $Ch$-neighborhoods of the breakpoints. The component of $\nu_h$ transported from the density $f''$ has density at most $f''(y_h(x))\leq C$ on its mapped smooth pieces and zero on the intervals receiving an atom. Its contribution, and that of $f''$, in the breakpoint neighborhoods is $O(h)$ by the weighted slab bound \eqref{eq:svm_all_level_slabs}. Each atom is transported by at most $Ch$ near a fixed level, where $\vect A_{\vect\theta}(c)$ is Lipschitz in $c$ by the density derivative bound and the section moments. Its Hessian contribution consequently changes by $O(h)$. This proves \eqref{eq:prox_control} for the index objective. The original index Hessian is also Lipschitz in $\vect\theta$ under Assumption \ref{cond:prox_fun}. For its continuous-curvature part, bounded $f'''$ gives $C d\mbE[(\vect v^{\tp}\vect X)^2|\vect u^{\tp}\vect X|]$ off the finitely many breakpoint crossings, while \eqref{eq:svm_all_level_crossing} bounds their contributions by $Cd$. The atomic parts are Lipschitz by the compact-level calculation in the proof of Proposition \ref{prop:index_objectives}. Here $\vect u$ is the displacement direction, $\vect v$ is a unit test vector, and $d$ is the displacement length.

\begin{proposition}\label{lem:conv_popu_min}
Under the infimal-convolution assumptions specified above, suppose $\vect\theta^*$ is an interior population solution and $\bar{\mcL}$ is strongly convex on a ball about $\vect\theta^*$. For sufficiently small $h$, the smoothed population solution $\vect\theta^*_{\infh}$ is unique, lies in the interior of $\Theta$, and satisfies
\[
|\vect\theta^*_{\infh}-\vect\theta^*|_2\leq Ch.
\]
\end{proposition}

\begin{proof}[Proof of Proposition \ref{lem:conv_popu_min}]
Choose a ball $\mbB(\vect\theta^*,r)\subseteq\Theta$ on which the expected objective is strongly convex. By \eqref{eq:prox_control}, the smoothed expected Hessian is bounded below on this ball for small $h$. The original gradient points strictly outward on its boundary, and its perturbation has norm $O(h)$. Hence the smoothed gradient also points strictly outward there. Minimizing over the closed ball gives an interior stationary point. Convexity makes it a global minimizer, and strict convexity on the ball makes it the unique global minimizer. Its gradient equation and \eqref{eq:prox_control} give $|\vect\theta^*_{\infh}-\vect\theta^*|_2\leq C|\nabla\bar{\mcL}_{\infh}(\vect\theta^*)|_2\leq Ch$.

\end{proof}

\begin{lemma}\label{lem:inf_exp_lip}
Under the infimal-convolution assumptions specified above, $g_h$ is bounded and nondecreasing with total variation at most $2C_f$. For $d=|\vect\theta_1-\vect\theta_2|_2$ and every unit $\vect v$,
\[
P|\vect v^{\tp}\vect V\{g_h(Q_{\vect\theta_1})-g_h(Q_{\vect\theta_2})\}|^k\leq d\,k!C^k,\qquad k\geq2.
\]
The smoothed population gradients are Lipschitz with a constant independent of $h$.
\end{lemma}

\begin{proof}
For the index objective, integrate the all-level crossing bound \eqref{eq:svm_all_level_crossing} against $\nu_h$. For the residual objective, use conditional response-density integration as in \eqref{eq:grad_diff_term1}. Monotonicity and $\nu_h(\mbR)\leq2C_f$ give the displayed moments. Their first-moment version proves the population Lipschitz bound.
\end{proof}

\begin{lemma}\label{lem:inf_Hess_conc}
Under the infimal-convolution assumptions specified above,
\[
\sup_{\vect\theta\in\Theta}\|(P_n-P)[q_h(Q_{\vect\theta})\vect V\vect V^{\tp}]\|
\leq C\{\sqrt{a_n/h}+a_n/h\}
\]
with probability at least $1-Cn^{-\gamma p}$. The same assertion holds for integral convolution under Assumption \ref{cond:kernel_fun}, with $q_h=f''_{\inth}$ wherever the derivative exists.
\end{lemma}

\begin{proof}
First fix $\vect\theta$ and a unit vector $\vect v$. For the index objective, integrate the section moment \eqref{eq:svm_section_moments} against $q_h(c)\diff c$. For the residual objective, condition on $\vect Z$ and use its bounded response density. Together with \eqref{eq:inf_fpp_bound}, these give, for every $k\geq2$,
\[
P|q_h(Q_{\vect\theta})(\vect v^{\tp}\vect V)^2|^k\leq h^{1-k}k!C^k.
\]
After subtracting the mean, Lemma \ref{lemma:cai_liu} applies with variance constant $C/h$ and scale constant $C/h$. The fixed-pair deviation is therefore $C\{\sqrt{t/(nh)}+t/(nh)\}$ with failure probability $2e^{-t}$. 

Take a decision net with mesh $\delta=n^{-12}$ and a $1/4$-sphere net $\mcV$. Their combined logarithmic cardinality is $O(p\log n)$. On the sample event $\max_i|\vect V_i|_2\leq B_n=C_B\sqrt{p\log n}$ established above, let $w_n=2\delta B_n$. For a decision and its net representative, the piecewise Lipschitz description gives
\[
|q_h(Q_{\vect\theta})-q_h(Q_{\tilde{\vect\theta}})|
\leq Cw_n/h^2+(C/h)\sum_j\mbI\{|Q_{\tilde{\vect\theta}}-e_{j,h}|\leq w_n\},
\]
where the finite set $\{e_{j,h}\}$ consists of the mapped boundaries. The index slab bound \eqref{eq:svm_all_level_slabs}, or conditional response-density integration, gives
\[
P[|\vect v^{\tp}\vect V|^{2k}\mbI\{|Q_{\tilde{\vect\theta}}-e_{j,h}|\leq w_n\}]
\leq w_n k!C^k.
\]
Bernstein's inequality and the net union bound show that every empirical weighted slab average, and its expectation, is at most $C(w_n+a_n)$ with failure probability $Cn^{-\gamma p}$. A separate application to $(\vect v^{\tp}\vect V)^2$ gives $(P_n+P)(\vect v^{\tp}\vect V)^2\leq C$ for all $\vect v\in\mcV$. Thus the interpolation error for quadratic forms is at most
\[
Cw_n/h^2+C(w_n+a_n)/h+Cn^{-10}/h.
\]
The last term bounds the population contribution outside $|\vect V|_2\leq B_n$, by the fourth directional moment and the exponentially small norm tail. Since $h\geq Aa_n\geq A\log n/n$ and $p\log n\leq n$, the terms involving $w_n$ and $n^{-10}$ are negligible relative to $a_n/h$.

For integral convolution, the finite Stieltjes measure $\nu=\diff g$ gives
\[
q_h(x)=\int h^{-1}K((x-t)/h)\nu(\diff t),~~\|q_h\|_\infty\leq C/h,~~\int q_h\leq2C_f.
\]
The fixed-pair moment calculation is therefore identical. If $K$ is globally Lipschitz, then $\operatorname{Lip}(q_h)\leq C/h^2$, and interpolation uses only the $Cw_n/h^2$ term, without boundary slabs.

If $K$ has bounded variation, choose the right-continuous representative of $q_h$ and let $\lambda_h=\diff q_h$ be its signed Stieltjes measure. Convolution gives $|\lambda_h|(\mbR)\leq\nu(\mbR)\operatorname{TV}(K)/h\leq C/h$. For scalar indices or residuals within $w_n$, their curvature difference is bounded by
\[
A_{\tilde{\vect\theta}}=\int\mbI\{|Q_{\tilde{\vect\theta}}-t|\leq w_n\}\,|\lambda_h|(\diff t).
\]
The same slab bound and the directional moments give, for $k\geq1$,
\[
P[((\vect v^{\tp}\vect V)^2A_{\tilde{\vect\theta}})^k]
\leq w_n h^{-k}k!C^k.
\]
Bernstein's inequality and the same net union bound therefore bound the empirical bracket and its expectation by
$C\{w_n+\sqrt{w_na_n}+a_n\}/h\leq Ca_n/h$.
The population tail outside the sample event is at most $Cn^{-10}/h$. This proves the same interpolation bound without a Lipschitz assumption on $K$.
In both cases apply the fixed-pair deviation with $t=Ap\log n$ over the nets. The symmetric-matrix sphere-net inequality converts their quadratic-form bounds to the operator norm bound. Adding the finite-net, sample-tail, covariance, and bracket failure probabilities gives $Cn^{-\gamma p}$.

\end{proof}

\begin{proof}[Infimal-convolution case of Theorem \ref{thm:smoothing_increments}]
Write $d=|\vect\theta_1-\vect\theta_2|_2$. Lemma \ref{lem:inf_exp_lip} and the corresponding index or residual net proof give
\begin{equation}\label{eq:inf_conv_term1}
|\nabla\mcL_{\infh}(\vect\theta_1)-\nabla\mcL_{\infh}(\vect\theta_2)-\nabla\bar{\mcL}_{\infh}(\vect\theta_1)+\nabla\bar{\mcL}_{\infh}(\vect\theta_2)|_2
\leq C(\sqrt{da_n}+a_n).
\end{equation}
Alternatively, integrate the uniform Hessian bound in Lemma \ref{lem:inf_Hess_conc} along the segment, which lies in $\Theta$ by convexity. Absolute continuity of $g_h$ then gives
\begin{equation}\label{eq:inf_conv_term2}
|\nabla\mcL_{\infh}(\vect\theta_1)-\nabla\mcL_{\infh}(\vect\theta_2)-\nabla\bar{\mcL}_{\infh}(\vect\theta_1)+\nabla\bar{\mcL}_{\infh}(\vect\theta_2)|_2
\leq Cd\{\sqrt{a_n/h}+a_n/h\}.
\end{equation}
A segment lying in a mapped boundary contributes a constant scenario derivative, and the assigned value $q_h=0$ gives its correct contribution. At all other points the ordinary chain rule applies almost everywhere. For $d\leq h$, \eqref{eq:inf_conv_term2} gives $Cd\sqrt{a_n/h}$, since $a_n/h\leq1$. For $d\geq h\geq Aa_n$, the additive term in \eqref{eq:inf_conv_term1} is absorbed by $\sqrt{da_n}$. These two ranges prove the minimum bound on the joint event.

\end{proof}

\begin{proof}[Infimal-convolution case of Corollary \ref{cor:smoothing_regularity}]
The population gradient is Lipschitz by Lemma \ref{lem:inf_exp_lip}. Adding the linear bound in Theorem \ref{thm:infconv_Lip_subgrad} proves part 1. For part 2, consider any segment in $\Theta_{\rm sc}$ and its unit direction $\vect u$. Strong convexity gives a second directional derivative of the original expected objective of at least $\mu_0$ in the interior of this segment. Equation \eqref{eq:prox_control} makes the smoothed expected second directional derivative at least $3\mu_0/4$ for small $h_0$. Choose $A$ in $h\geq Aa_n$ large enough that Lemma \ref{lem:inf_Hess_conc} bounds its empirical deviation in this direction by $\mu_0/4$. The SAA second derivative along the segment is then at least $\mu_0/2$ almost everywhere. Integrating along the segment proves Condition \ref{cond:mu_convex} with $\mu=\mu_0$ and $\epsilon_2=0$.

For part 3, write $d=|\vect\theta-\vect\theta^*|_2\leq C_3\sqrt{a_n}$. Theorem \ref{thm:infconv_Lip_subgrad} bounds the mean-zero gradient increment by $C\min(a_n^{3/4},a_n/\sqrt h)$. The original population Hessian is Lipschitz, and \eqref{eq:prox_control} bounds the smoothing perturbation by $Ch$. Its difference from $\vect H$ along the segment is therefore at most $C(h+d)$. The remaining population increment is at most $C(hd+d^2)$. The $d^2=O(a_n)$ term is absorbed by the displayed minimum, since $h\leq h_0<1$. This proves part 3.

\end{proof}

\begin{proof}[Proof of Proposition \ref{prop:infimal_bias}]
Let $U_0$ be uniform on $[a,b]$ and put $W=(\omega')^{-1}(U_0)$. The variable $W$ is bounded and
\[
\mbE W=\frac1{b-a}\int_a^b(\omega')^{-1}(u)\diff u
=\frac{\omega^*(b)-\omega^*(a)}{b-a}=c_\omega.
\]
The proximal optimality equation gives $f'_{\infh}(k+hs)=\min\{b,\max\{a,\omega'(s)\}\}$. Hence $f'_{\infh}(x)=\mbE[f'_+(x-hW)]$, and matching the left affine tails gives
\[
f_{\infh}(x)=\mbE[f(x-hW)]+h\{ac_\omega-\omega^*(a)\}.
\]
Conditioning on $\vect Z$ and expanding the conditional response distribution gives
\[
\nabla\bar{\mcL}_{\infh}(\vect\theta^*)
=h(b-a)c_\omega\mbE[\vect Z\rho(k+\vect Z^{\tp}\vect\theta^*\mid\vect Z)]+O(h^2).
\]
The remainder has norm at most $Ch^2$ by the bounded density derivative and the directional moment bounds. Proposition \ref{lem:conv_popu_min} gives a unique interior population minimizer with displacement $O(h)$. Its gradient equation, \eqref{eq:prox_control}, and the population Hessian Lipschitz bound give the $h\vect B_\omega+O(h^2)$ displacement.

For the intercept calculation in Remark \ref{rem:infimal_drift}, $Z_l=1$ almost surely gives
\[
\vect H\vect e_l=(b-a)\mbE[\vect Z\rho(k+\vect Z^{\tp}\vect\theta^*\mid\vect Z)],\qquad
\vect B_\omega=-c_\omega\vect e_l.
\]
For Moreau smoothing, $c_\omega=(a+b)/2$. Substitution of the pinball slopes $\tau-1,\tau$ and the CVaR slopes $0,(1-\alpha_{\rm CVaR})^{-1}$ gives the coefficients in the remark.
\end{proof}

\begin{proof}[Proof of Proposition \ref{prop:infimal_correction}]
The proof of Proposition \ref{prop:infimal_bias} gives $\mbE W=c_\omega$. Substitution of $x=r+hc_\omega$ in its convolution identity proves \eqref{eq:infimal_shift_identity}. Write $V=W-c_\omega$ and $t_{\vect Z}=k+\vect Z^{\tp}\vect\theta^*$. The variable $V$ is bounded and has mean zero. Conditioning on $\vect Z$ and using $\nabla\bar{\mcL}(\vect\theta^*)=\vect0$ therefore gives
\[
\nabla\bar{\mcL}^{\rm sh}_{\infh}(\vect\theta^*)
=(b-a)\mbE\Big[\vect Z\,\mbE_V\int_0^{hV}\{\rho(t_{\vect Z}+s\mid\vect Z)-\rho(t_{\vect Z}\mid\vect Z)\}\diff s\Big].
\]
The bounded density derivative and $\sup_{|\vect v|_2=1}\mbE|\vect v^{\tp}\vect Z|\leq C$ bound the norm of this gradient by $Ch^2$. The conditional-density calculation in \eqref{eq:prox_control}, with the bounded shift $V$, gives an $O(h)$ Hessian perturbation. The boundary-gradient argument in the proof of Proposition \ref{lem:conv_popu_min} then gives a unique population minimizer in the interior of $\Theta$, at distance $O(h^2)$ from $\vect\theta^*$. For $\omega(u)=u^2/2$, we have $W=U_0$, $c_\omega=(a+b)/2$, and $ac_\omega-\omega^*(a)=ab/2$, proving the last assertion.
\end{proof}

\phantomsection\label{app:infimal_bandwidth}

For the bandwidth lower bound in Section \ref{sec:inf_conv}, consider the pinball objective in Proposition \ref{prop:smoothness_scale_lower}. Let $\hat{\vect\theta}$ be the minimizer supplied by that proposition, let $l$ be the intercept coordinate, and put $A=\{i:Y_i-\vect Z_i^{\tp}\hat{\vect\theta}=0\}$, with $|A|\geq p$. Write $a=c(\tau-1)$, $b=c\tau$, $s_0=(\omega')^{-1}\{(a+b)/2\}$, and $\tilde{\vect\theta}_h=\hat{\vect\theta}-hs_0\vect e_l$. At this decision, each scenario in $A$ has residual $hs_0$. Since $a<\omega'(s_0)<b$, it lies in the interior of the transition interval $h[(\omega')^{-1}(a),(\omega')^{-1}(b)]$, where $f'_{\infh}(x)=\omega'(x/h)$. Monotonicity makes every other scenario's contribution to the following directional increment nonnegative. Consequently,
\[
\liminf_{t\downarrow0}\frac{\nabla_l\mcL_{\infh}(\tilde{\vect\theta}_h+t\vect e_l)-\nabla_l\mcL_{\infh}(\tilde{\vect\theta}_h)}{t}
\geq\frac{|A|\omega''(s_0)}{nh}\geq\frac{p\mu_\omega}{nh}.
\]
The distance from $\tilde{\vect\theta}_h$ to $\hat{\vect\theta}$ is $h|s_0|$. Hence any Lipschitz bound on a fixed neighborhood of $\hat{\vect\theta}$ satisfies $L_h\geq p\mu_\omega/(nh)$ for sufficiently small $h$, proving the necessary bandwidth order for every prox-function in Assumption \ref{cond:prox_fun}.

\subsection{Integral convolution}\label{app:integral_convolution}
Integral convolution has been studied for inference and regularization in quantile regression. Bahadur representations and asymptotic inference are developed by \citet{fernandes_etal.2021jbes}, and nonasymptotic expansions, normal approximation, and bootstrap validity in growing dimension are established by \citet{he_pan_etal.2023joe}. Convolution smoothing is combined with concave regularization in \citet{tan_wang_zhou.2022jrssb} and with noisy iterative hard thresholding for sparse newsvendor policies in \citet{chang_etal.2026or}. Related smoothing-based refinement under memory constraints is studied by \citet{chen_etal.2019aos}. These works establish inference and regularization benefits of smoothing for quantile losses. The following results give the corresponding regularity, population displacement, and computed-solution expansions for the convex Lipschitz losses considered here.

For a probability kernel $K$, define the scalar integral convolution
\begin{equation}\label{eq:integral_conv}
f_{\inth}(x)=\int_{\mbR}h^{-1}K((x-y)/h)f(y)\diff y,
\end{equation}
and the objective $\mcL_{\inth}(\vect\theta)=n^{-1}\sum_{i=1}^nf_{\inth}(Y_i-\vect Z_i^{\tp}\vect\theta)+\vect b^{\tp}\vect\theta$ with expected objective $\bar{\mcL}_{\inth}$ defined through loss differences. Gaussian and uniform convolution give closed-form gradients for the piecewise linear losses in quantile regression and SVM. Without a closed form, the gradient $\mbE[f'(x-hV)]$, where $V$ has density $K$, requires numerical integration or sampling, and sampling turns a full-batch update into a stochastic update \citep{duchi_etal.2012siamjo}.
\begin{assumptionint}\label{cond:kernel_fun}
The kernel $K$ is a symmetric, bounded, nonnegative probability density with finite second moment, and is either globally Lipschitz continuous or of bounded variation. The conditional density in Assumption \ref{cond:residual_density} has a bounded continuous first derivative.
\end{assumptionint}
Assumption \ref{cond:kernel_fun} includes Gaussian and uniform kernels and permits every convex Lipschitz loss in Assumption \ref{cond:discont}, without additional piecewise smoothness.

\begin{corollary}\label{cor:integral_convolution}
Suppose Assumptions \ref{cond:residual_theta}--\ref{cond:discont} and \ref{cond:kernel_fun} hold. Theorem \ref{thm:smoothing_increments} and Corollary \ref{cor:smoothing_regularity} hold for $\mcL_{\inth}$ in place of $\mcL_{\infh}$, with Assumption \ref{cond:kernel_fun} in place of Assumption \ref{cond:prox_fun}. If $\vect\theta^*$ is an interior population solution and $\bar{\mcL}$ is strongly convex on a ball about it, then $|\vect\theta^*_{\inth}-\vect\theta^*|_2=O(h^2)$ for sufficiently small $h$. Under the hypotheses of Corollary \ref{cor:smoothing_algorithm}, with Assumption \ref{cond:kernel_fun} in place of Assumption \ref{cond:prox_fun}, its optimization bound, expansion, and normal limit hold with $\mcL_{\inth}$ in place of $\mcL_h$ and $\vect\theta^*$ in place of $\vect\theta^*_h$, without a residual shift.
\end{corollary}

The additional condition used for the integral index perturbation bounds is
\begin{equation}\label{eq:index_section_derivative}
\sup_{\vect\theta\in\Theta,\,|\vect v|_2=1}\int_{\mbR}\sup_{|s|\leq1}\mbE\Big[|\rho'_{\vect\theta}(c+s\mid\vect U)|\{\vect v^{\tp}\vect x_{\vect\theta}(c+s)\}^2\Big]\nu(\diff c)\leq C,
\end{equation}
where the prime denotes differentiation in $c$.

\begin{proposition}\label{lem:intm_dist}
Under the integral-convolution assumptions specified above, suppose $\vect\theta^*$ is an interior population solution and $\bar{\mcL}$ is strongly convex on a ball about $\vect\theta^*$. For an index objective impose \eqref{eq:index_section_derivative}. For sufficiently small $h$, the integral-convolution population minimizer is unique, lies in the interior of $\Theta$, and satisfies $|\vect\theta^*_{\inth}-\vect\theta^*|_2\leq Ch^2$. For the residual objective, no condition on the curvature support is needed.
\end{proposition}

\begin{proof}
Let $V$ have density $K$. For the index objective, differentiation of the conditional integrals gives
\[
\begin{aligned}
\vect B'_{\vect\theta}(c)&=-\mbE[\rho_{\vect\theta}(c\mid\vect U)\vect x_{\vect\theta}(c)],\\
\vect B''_{\vect\theta}(c)&=-\mbE[\rho'_{\vect\theta}(c\mid\vect U)\vect x_{\vect\theta}(c)+\rho_{\vect\theta}(c\mid\vect U)\vect e_1/\theta_1],\\
\vect A'_{\vect\theta}(c)&=\mbE[\rho'_{\vect\theta}(c\mid\vect U)\vect x_{\vect\theta}(c)\vect x_{\vect\theta}(c)^{\tp}
+\rho_{\vect\theta}(c\mid\vect U)\{\vect e_1\vect x_{\vect\theta}(c)^{\tp}+\vect x_{\vect\theta}(c)\vect e_1^{\tp}\}/\theta_1].
\end{aligned}
\]
Equation \eqref{eq:index_section_derivative}, Cauchy--Schwarz, and the section moments bound the integrals against $\nu$ of the local second derivative of $\vect B_{\vect\theta}$ and the local first derivative of $\vect A_{\vect\theta}$ by $C$. Their global norm bounds extend Taylor's inequalities to every shift, giving
\[
\begin{aligned}
\sup_{\vect\theta\in\Theta}\left|\int[\vect B_{\vect\theta}(c+s)-\vect B_{\vect\theta}(c)-s\vect B'_{\vect\theta}(c)]\nu(\diff c)\right|_2&\leq Cs^2,\\
\sup_{\vect\theta\in\Theta}\left\|\int[\vect A_{\vect\theta}(c+s)-\vect A_{\vect\theta}(c)]\nu(\diff c)\right\|&\leq C|s|.
\end{aligned}
\]
For $|s|\leq1$ these follow by integrating the displayed derivatives. For $|s|>1$, use the bounded norms of $\vect A$, $\vect B$, and $\vect B'$, and $\nu(\mbR)\leq2C_f$. The smoothed population gradient replaces $\vect B_{\vect\theta}(c)$ by $\mbE_V\vect B_{\vect\theta}(c+hV)$, and its Hessian makes the same replacement for $\vect A_{\vect\theta}$. Thus $\mbE V=0$ and $\mbE V^2<\infty$ give
\begin{equation}\label{eq:intc_convex}
\sup_{\vect\theta\in\Theta}|\nabla\bar{\mcL}_{\inth}(\vect\theta)-\nabla\bar{\mcL}(\vect\theta)|_2\leq Ch^2,
\end{equation}
\begin{equation}\label{eq:intc_stgconvex}
\sup_{\vect\theta\in\Theta}\|\nabla^2\bar{\mcL}_{\inth}(\vect\theta)-\nabla^2\bar{\mcL}(\vect\theta)\|\leq Ch.
\end{equation}

Here are the sufficient conditions for \eqref{eq:index_section_derivative}. If $\nu$ has compact support, the reconstructed projections have bounded second moments on its unit neighborhood and $\rho'_{\vect\theta}$ is bounded. If $\vect X$ is bounded, the conditional density and its derivative vanish outside the conditional support of $X_1$, including its boundary. Wherever the derivative contributes, $\vect x_{\vect\theta}(c)$ is therefore bounded. If $\rho_X''$ is bounded, nonnegativity and Taylor's inequality give $|\rho_X'|^2\leq C\rho_X$, and hence $|\rho'_{\vect\theta}|^2\leq C\rho_{\vect\theta}$. Cauchy--Schwarz and \eqref{eq:svm_section_density_tail} imply
\[
\mbE[|\rho'_{\vect\theta}(c\mid\vect U)|(\vect v^{\tp}\vect x_{\vect\theta}(c))^2]
\leq C\{\mbE\rho_{\vect\theta}(c\mid\vect U)\}^{1/2}\{\mbE(\vect v^{\tp}\vect x_{\vect\theta}(c))^4\}^{1/2}
\leq C(1+c^2)e^{-c_1c^2}\leq C.
\]
This also bounds the supremum over a unit level neighborhood, and integration against the finite measure $\nu$ verifies the condition.

For a residual objective, conditional integration gives the expected gradient through $g(-\infty)+\int\{1-F(c+\vect Z^{\tp}\vect\theta\mid\vect Z)\}\nu(\diff c)$. Taylor's inequality for $F$, the bounded response density derivative, and $\mbE V=0$ give \eqref{eq:intc_convex}. The density Lipschitz bound gives \eqref{eq:intc_stgconvex}. Each unit projection uses only $\mbE|\vect v^{\tp}\vect Z|$ or $\mbE(\vect v^{\tp}\vect Z)^2$, respectively, and no response moment is needed. Finally, the outward-gradient and strict-convexity argument in Proposition \ref{lem:conv_popu_min} gives the unique interior minimizer and its $O(h^2)$ displacement.
\end{proof}

Without \eqref{eq:index_section_derivative}, the index Hessian perturbation still tends uniformly to zero. Indeed, the section-density tail and the reconstructed fourth moments give $\|\vect A_{\vect\theta}(c)\|\leq Ce^{-c_1c^2}$, while the bounded density derivative gives $\|\vect A'_{\vect\theta}(c)\|\leq C(1+c^2)$. Splitting levels at $|c|=2|s|^{-1/4}$ yields $\sup_{\vect\theta,c}\|\vect A_{\vect\theta}(c+s)-\vect A_{\vect\theta}(c)\|\leq C\sqrt{|s|}$ for $|s|\leq1$, with a bounded extension for larger shifts. Consequently the Hessian perturbation is at most $C\sqrt h\,\mbE\sqrt{|V|}=o(1)$. This suffices for the ordinary strong convexity conclusion in Corollary \ref{cor:smoothing_regularity}, which does not impose \eqref{eq:index_section_derivative}.

\begin{lemma}\label{lem:intc_exp_lip}
Under the integral-convolution assumptions specified above, its smoothed derivative satisfies the moments in Lemma \ref{lem:inf_exp_lip}, and its population gradient is Lipschitz with a constant independent of $h$.
\end{lemma}

\begin{proof}
Convolution preserves monotonicity and the total variation bound of $g$. Integrate the index crossing bound or the residual conditional-density bound against the smoothed curvature measure. The first-moment calculation proves the population Lipschitz bound.
\end{proof}

\begin{proof}[Integral-convolution case of Theorem \ref{thm:smoothing_increments} and Corollary \ref{cor:smoothing_regularity}]
Write $d=|\vect\theta_1-\vect\theta_2|_2$. Lemma \ref{lem:intc_exp_lip} and the corresponding index or residual net argument give
\begin{equation}\label{eq:int_conv_term1}
|\nabla\mcL_{\inth}(\vect\theta_1)-\nabla\mcL_{\inth}(\vect\theta_2)-\nabla\bar{\mcL}_{\inth}(\vect\theta_1)+\nabla\bar{\mcL}_{\inth}(\vect\theta_2)|_2
\leq C(\sqrt{da_n}+a_n).
\end{equation}
For integral convolution, integrating the bound in Lemma \ref{lem:inf_Hess_conc} along the segment between the arguments gives
\begin{equation}\label{eq:int_conv_term2}
|\nabla\mcL_{\inth}(\vect\theta_1)-\nabla\mcL_{\inth}(\vect\theta_2)-\nabla\bar{\mcL}_{\inth}(\vect\theta_1)+\nabla\bar{\mcL}_{\inth}(\vect\theta_2)|_2
\leq Cd\{\sqrt{a_n/h}+a_n/h\}.
\end{equation}
For $d\leq h$, use \eqref{eq:int_conv_term2}. For $d\geq h\geq Aa_n$, absorb $a_n$ into the square-root term of \eqref{eq:int_conv_term1}. The sum of the two failure probabilities is $Cn^{-\gamma p}$, proving the theorem. Combining the population gradient Lipschitz bound in Lemma \ref{lem:intc_exp_lip} with \eqref{eq:int_conv_term2} proves part 1. For part 2, use the uniform $o(1)$ population Hessian perturbation proved above and the same Hessian concentration event along each segment in $\Theta_{\rm sc}$, as in the infimal-convolution case of Corollary \ref{cor:smoothing_regularity}. The second directional derivative along that segment is at least $\mu_0/2$ almost everywhere, and integration gives Condition \ref{cond:mu_convex} with $\mu=\mu_0$ and $\epsilon_2=0$. For part 3, the mean-zero increment on the radius $C_3\sqrt{a_n}$ is at most $C\min(a_n^{3/4},a_n/\sqrt h)$. Under the additional section-derivative condition for an index objective, or the response density derivative condition for a residual objective, equation \eqref{eq:intc_stgconvex} and the unsmoothed Hessian Lipschitz bound contribute at most $C(h\sqrt{a_n}+a_n)$. Absorb $a_n$ as in the infimal-convolution case of Corollary \ref{cor:smoothing_regularity} to obtain \eqref{eq:H_smooth} for integral convolution.

\end{proof}

For the bandwidth lower bound, consider the pinball objective in Proposition \ref{prop:smoothness_scale_lower} and a kernel satisfying Assumption \ref{cond:kernel_fun}, continuous at zero with $K(0)>0$. Let $\hat{\vect\theta}$ be the minimizer supplied by that proposition, write $r_i=Y_i-\vect Z_i^{\tp}\hat{\vect\theta}$, and put $A=\{i:r_i=0\}$, with $|A|\geq p$. Let $l$ be the intercept coordinate, with $Z_{il}=1$ for every $i$, and let $F_K(x)=\int_{-\infty}^xK(u)\diff u$. For $f(x)=cx\{\tau-\mbI(x\leq0)\}$, differentiation under the convolution gives $f'_{\inth}(x)=c\{\tau-1+F_K(x/h)\}$. Hence, for $t>0$,
\[
\begin{aligned}
&\frac{\nabla_l\mcL_{\inth}(\hat{\vect\theta}+t\vect e_l)-\nabla_l\mcL_{\inth}(\hat{\vect\theta})}{t}\\
&\quad=\frac{c}{nt}\sum_{i=1}^n\int_{(r_i-t)/h}^{r_i/h}K(u)\diff u
\geq\frac{c|A|}{nt}\int_{-t/h}^{0}K(u)\diff u.
\end{aligned}
\]
Every summand is nonnegative. Continuity of $K$ at zero therefore gives
\[
\liminf_{t\downarrow0}\frac{\nabla_l\mcL_{\inth}(\hat{\vect\theta}+t\vect e_l)-\nabla_l\mcL_{\inth}(\hat{\vect\theta})}{t}
\geq\frac{c|A|K(0)}{nh}\geq\frac{cpK(0)}{nh}.
\]
If the smoothed gradient is $L_h$-Lipschitz on a neighborhood of $\hat{\vect\theta}$, then $L_h\geq cpK(0)/(nh)$. Thus a constant-order bound on $L_h$ requires $h\gtrsim p/n$, compared with the sufficient scale $Aa_n=Ap\log n/n$.

\subsection{Proof of Corollary \ref{cor:smoothing_algorithm}}

\begin{proof}
Write $\mcL_h$ for the objective in question and $\vect\theta_h$ for its population minimizer. The population bounds give $|\vect\theta_h-\vect\theta^*|_2\leq Ch$ for the unshifted infimal objective and $Ch^2$ for integral convolution. Proposition \ref{prop:infimal_correction} gives $|\vect\theta_h-\vect\theta^*|_2\leq Ch^2$ for the shifted objective. This objective uses the translated loss $f(\cdot+hc_\omega)$, whose breakpoint is $k-hc_\omega$ and whose derivative bounds in Assumption \ref{cond:prox_fun} are unchanged. On a fixed ball about $\vect\theta^*$, the smoothed population and SAA objectives have strong convexity of constant order. At the deterministic minimizer $\vect\theta_h$, the expected score is zero. Boundedness of the smoothed scalar derivative and the directional moment bounds give its high-probability norm bound $C\sqrt{a_n}$ and $\mbE|\nabla\mcL_h(\vect\theta_h)|_2^2\leq Cp/n$. The boundary-gradient argument therefore gives a unique interior SAA minimizer $\hat{\vect\theta}_h$ with $|\hat{\vect\theta}_h-\vect\theta_h|_2=O_P(\sqrt{p/n})$. 

Since the population displacement tends to zero, the local strong-convexity ball about $\hat{\vect\theta}_h$ lies in the population curvature region for sufficiently large $n$. For the projected methods, the initialization distance is bounded by $\operatorname{diam}(\Theta)$. Corollary \ref{cor:projected_subgradient}\ref{part:projected_local_entry}, with $\epsilon_1=\epsilon_2=0$, gives both methods distance recursions with zero additive terms. Their initial phases take a bounded number of updates, and their geometric bounds give distance $O_P(n^{-1})$ after $\lceil c\log n\rceil$ updates from any permitted initialization. For an unprojected method, impose the ball-inclusion requirement in the \hyperref[app:unprojected_saa]{unprojected extension}. The vanishing population displacement then permits the region verification in Corollary \ref{cor:absorption}, whose bound with zero additive terms gives the same optimization accuracy.

For the complete procedures, center their Phase 3 neighborhood at $\vect\theta_h$, with radius $\epsilon_3=C_3\sqrt{a_n}$ and sufficiently large $C_3$. The score at $\vect\theta_h$ has mean zero and satisfies the preceding high-probability bound. The gradient increment relative to $\vect H$ has remainder $O(a_n^{3/4}+h\sqrt{a_n})=o(\epsilon_3)$ on this ball. This uses the population Hessian perturbation $O(h)$ and the unsmoothed Hessian Lipschitz property, since $|\vect\theta_h-\vect\theta^*|_2=O(h)$. The pointwise, region, and scale conditions in the complete procedures follow as in the proof of Theorem \ref{cor:model_algorithm}. On this ball, Lemma \ref{lem:inf_Hess_conc} and the population bounds give
\[
|\nabla\mcL_h(\vect\theta)-\vect H(\vect\theta-\hat{\vect\theta}_h)|_2
\leq C\{\sqrt{a_n/h}+a_n/h+h+\epsilon_3\}|\vect\theta-\hat{\vect\theta}_h|_2.
\]
Choose $A$ sufficiently large, then $n$ sufficiently large, so that this coefficient is at most $m_H/64$. The local comparison \eqref{eq:local_phase3_comparison}, with $\zeta_n=0$, reduces the optimization error to $O(n^{-1})$ in $O(\log n)$ updates for both procedures. For the unshifted infimal objective, the neighborhood is centered at $\vect\theta_h$, since its displacement can exceed $\sqrt{a_n}$.

Put $\vect\vartheta_h=\vect\theta_h$ for the unshifted infimal objective and $\vect\vartheta_h=\vect\theta^*$ for the shifted objective and for integral convolution. Independence gives a second moment at most $Cp/n$ for the score deviation from its expectation at this deterministic point. The expected score is zero for the unshifted infimal objective and has norm $O(h^2)$ for the shifted objective and for integral convolution. Strong convexity consequently gives
\[
\delta_{n,h}:=|\hat{\vect\theta}_h-\vect\vartheta_h|_2
=O_P(\sqrt{p/n}+h^2).
\]
The uniform linear increment bound and the population Hessian bounds imply
\[
|\vect H(\hat{\vect\theta}_h-\vect\vartheta_h)+\nabla\mcL_h(\vect\vartheta_h)|_2
\leq C\{\delta_{n,h}\sqrt{a_n/h}+h\delta_{n,h}+\delta_{n,h}^2\}
=O_P\{\sqrt{p a_n/(nh)}+h^2\}.
\]
Subtracting the expected score preserves this remainder. For the chosen bandwidth, $\sqrt{p a_n/(nh)}\leq C(p/n)^{4/5}(\log n)^{2/5}$ and $h^2\leq C\{(p/n)^{4/5}(\log n)^{2/5}+a_n^2\}$. The optimization error $O_P(n^{-1})$ is absorbed. If $p(\log n)^{4/3}=o(n)$, then $a_n/h_{\rm opt}\to0$ for $h_{\rm opt}=(p/n)^{2/5}(\log n)^{1/5}$ and $a_n^2=o(h_{\rm opt}^2)$, proving the sharper remainder.

To compare variances, write $\vect u_n=\vect H^{-1}\vect v_n$, which has bounded norm. The curvature coupling gives
\[
|f'_{\infh}(x)-g(x)|\leq\int\mbI\{|x-c|\leq Ch\}\nu(\diff c).
\]
For the index objective, the weighted slab bound makes its squared score difference in the direction $\vect u_n$ at most $Ch$. Moving from $\vect\theta_h$ to $\vect\theta^*$ adds at most $C|\vect\theta_h-\vect\theta^*|_2\leq Ch$ by the weighted crossing bound. For the shifted residual objective, \eqref{eq:infimal_shift_identity} gives the same bound with the bounded perturbation $W-c_\omega$. For integral convolution, Jensen's inequality and the level shift $hV$ give $Ch\mbE|V|$. Conditional response-density integration gives the same bounds for the residual objective. Subtracting means cannot increase these variances. The smoothed projected score variance therefore differs from the original $\sigma_n^2$ by $O(\sqrt h)$, and third absolute moments remain bounded. Lyapunov's theorem gives the normal limit for the score after subtracting its expectation. The condition $p=o\{n^{3/8}(\log n)^{-1/2}\}$ makes $\sqrt n$ times the expansion remainder vanish. Slutsky's theorem proves the limiting distribution. Centering the unshifted infimal output at $\vect\theta^*$ adds the exact deterministic drift $d_n$. If $d_n\to\lambda$, Slutsky's theorem gives $\mcN(\lambda,1)$, while $|d_n|\to\infty$ gives divergence in absolute value in probability. For the losses in Proposition \ref{prop:infimal_bias}, its displacement expansion gives $d_n=\sqrt n h\,\vect v_n^{\tp}\vect B_\omega/\sigma_n+O(\sqrt n h^2)$. The dimension condition for the normal limit makes $\sqrt n h^2=o(1)$, proving the drift expansion in Corollary \ref{cor:smoothing_algorithm}.
\end{proof}

\subsection{Other objectives}\label{app:smoothing_applications}

\begin{proposition}\label{lem:index_smoothing}
Consider the index objective \eqref{eq:index_loss} under Assumptions \ref{cond:discont}, \ref{cond:theta}, and \ref{ass:distribution}, and suppose $\rho_X$ has a bounded continuous first derivative. For infimal convolution impose the conditions on $f$ and $\omega$ in Assumption \ref{cond:prox_fun}, and for integral convolution impose the conditions on $K$ in Assumption \ref{cond:kernel_fun}. Then Theorem \ref{thm:smoothing_increments} and parts 1 and 2 of Corollary \ref{cor:smoothing_regularity} hold for the index objective.

For integral convolution additionally impose \eqref{eq:index_section_derivative}. Suppose the population Hessian is Lipschitz near $\vect\theta^*$ and $m_H\mbI\preceq\nabla^2\bar{\mcL}(\vect\theta^*)\preceq M_H\mbI$. Then part 3 of Corollary \ref{cor:smoothing_regularity} applies. The unshifted infimal conclusion of Corollary \ref{cor:smoothing_algorithm} and the integral conclusion of Corollary \ref{cor:integral_convolution} apply with their step sizes, bandwidths, and initialization requirements, using score $\vect\psi_i=\vect b+\vect X_i f'_+(\vect X_i^{\tp}\vect\theta^*)$. For infimal convolution these conclusions do not require \eqref{eq:index_section_derivative}. The integral condition holds when $\vect X$ is bounded or $\nu$ has compact support, under the first density derivative bound, and for arbitrary $\nu$ when $\rho_X$ has a bounded second derivative.

If $f$ has one breakpoint $k$ and affine pieces of slopes $a<b$, the expected index objective has an interior minimizer $\vect\theta^*$, and its Hessian $\vect H$ is positive definite and Lipschitz near that point, then
\[
\vect\theta^*_{\infh}-\vect\theta^*=h\vect B^{\rm idx}_\omega+O(h^2),\qquad
\vect B^{\rm idx}_\omega=(b-a)c_\omega\vect H^{-1}\mbE[\rho_{\vect\theta^*}(k\mid\vect U)\vect x_{\vect\theta^*}(k)],
\]
with $c_\omega$ as in Proposition \ref{prop:infimal_bias}. For $k\ne0$, $\vect B^{\rm idx}_\omega=(c_\omega/k)\vect\theta^*$. In particular, the Moreau envelope of the hinge loss has displacement $-h\vect\theta^*/2+O(h^2)$.
\end{proposition}

\begin{proof}[Proof of Proposition \ref{lem:index_smoothing}]
The index section moments and all-level crossing bounds in Proposition \ref{prop:index_objectives} verify Lemmas \ref{lem:inf_exp_lip}, \ref{lem:inf_Hess_conc}, and \ref{lem:intc_exp_lip}. Their proofs above therefore give the increment and Hessian concentration bounds for the index objective. Equations \eqref{eq:prox_control}, \eqref{eq:intc_convex}, and \eqref{eq:intc_stgconvex}, with the section-derivative condition in the integral case, give the population perturbation bounds. The sufficient conditions for \eqref{eq:index_section_derivative} were verified in the proof of Proposition \ref{lem:intm_dist}. These estimates prove the regularity conclusions and supply every input to the localization, optimization, expansion, and variance calculation in the proof of Corollary \ref{cor:smoothing_algorithm}. For infimal convolution, the expansion is centered at $\vect\theta^*_{\infh}$, and the unsmoothed index score gives the limiting variance.

For the displacement coefficient, use the bounded-shift convolution identity in the proof of Proposition \ref{prop:infimal_bias}. It gives
\[
\nabla\bar{\mcL}_{\infh}(\vect\theta)
=\vect b+a\mbE\vect X+(b-a)\mbE_W\vect B_{\vect\theta}(k+hW).
\]
The first two level derivatives of $\vect B_{\vect\theta}$ are bounded near the fixed level $k$ by the conditional density derivative and the section moments. Since $\mbE W=c_\omega$, Taylor expansion yields
\[
\nabla\bar{\mcL}_{\infh}(\vect\theta^*)
=-h(b-a)c_\omega\mbE[\rho_{\vect\theta^*}(k\mid\vect U)\vect x_{\vect\theta^*}(k)]+O(h^2).
\]
Proposition \ref{lem:conv_popu_min}, the Hessian perturbation \eqref{eq:prox_control}, and the gradient equation give the coefficient $\vect B^{\rm idx}_\omega$. Since $\vect x_{\vect\theta^*}(k)^{\tp}\vect\theta^*=k$, we have $\vect H\vect\theta^*=(b-a)k\mbE[\rho_{\vect\theta^*}(k\mid\vect U)\vect x_{\vect\theta^*}(k)]$. For $k\ne0$ this proves $\vect B^{\rm idx}_\omega=(c_\omega/k)\vect\theta^*$. The hinge loss has $a=-1$, $b=0$, and $k=1$. The quadratic prox-function gives $c_\omega=-1/2$.
\end{proof}

For SVM, $f(t)=(1-t)_+$ has curvature measure $\nu=\delta_1$. The density derivative bound in Proposition \ref{lem:index_smoothing} therefore verifies \eqref{eq:index_section_derivative}. For integral convolution, smoothing this hinge is equivalent to smoothing $x_+$ at $1-\vect X^{\tp}\vect\theta$, since $K$ is symmetric. For infimal convolution, the latter convention is the former with the reflected prox-function $u\mapsto\omega(-u)$, which satisfies the same assumptions. Both conventions therefore have the regularity and inference conclusions of that proposition, centered at the population minimizer of the chosen smoothed objective in the infimal case.

For rank regression, set $f(x)=|x|$, write $r_{ij}(\vect\theta)=Y_i-Y_j-(\vect Z_i-\vect Z_j)^{\tp}\vect\theta$, and define
\[
\mcL_{{\rm rank},\circ,h}(\vect\theta)=\frac1{n(n-1)}\sum_{i\ne j}f_{\circ,h}(r_{ij}(\vect\theta)),\qquad \circ\in\{{\rm inf},{\rm int}\}.
\]

\begin{proposition}\label{prop:smoothing_applications}
Impose the distributional assumptions in Example \ref{exp:rank}, including its density derivative bound, on a convex bounded decision region $\Theta$. Impose the conditions on the population minimizer and Hessian in Theorem \ref{cor:model_algorithm} on the pairwise expected objective, and assume $a_n\to0$. For infimal convolution impose the conditions on $\omega$ in Assumption \ref{cond:prox_fun}, and for integral convolution impose the conditions on $K$ in Assumption \ref{cond:kernel_fun}. Let $Aa_n\leq h\leq h_0$, with $A$ sufficiently large and $h_0$ sufficiently small. The conclusions of Theorem \ref{thm:smoothing_increments} and Corollary \ref{cor:smoothing_regularity} hold for the pairwise smoothed objective. In particular, its mean-zero gradient increments satisfy the same minimum bound uniformly over decision pairs, with probability at least $1-Cn^{-\gamma p}$, and its smoothness and strong convexity bounds have $\epsilon_1=\epsilon_2=0$. Its population minimizers are unique and lie in the interior of $\Theta$ for sufficiently small $h$, with displacements $O(h)$ and $O(h^2)$ for infimal and integral convolution, respectively.

With $h\asymp\max\{a_n,(p/n)^{2/5}(\log n)^{1/5}\}$ and $h\geq Aa_n$, the procedures in Corollary \ref{cor:smoothing_algorithm}, with their region and initialization requirements, produce an output after $O(\log n)$ updates with distance $O_P(n^{-1})$ from the smoothed SAA solution and its expansion with remainder $O_P\{(p/n)^{4/5}(\log n)^{2/5}+a_n^2\}$. The expansion is centered at $\vect\theta^*_{\infh}$ for infimal convolution and at $\vect\theta^*$ for integral convolution. For any deterministic unit contrast with unsmoothed rank-score variance bounded below, $p=o\{n^{3/8}(\log n)^{-1/2}\}$ gives the limiting distribution in that corollary, with the unsmoothed rank-regression variance in Example \ref{exp:rank}.
\end{proposition}

\begin{proof}
Let $\vect W=(Y,\vect Z^{\tp})^{\tp}$ and let $\vect W'$ be an independent copy. The proof of Example \ref{exp:rank} shows that $\Delta Y=Y-Y'$ conditional on $(\vect Z,\vect Z')$ has bounded density and bounded first density derivative, and every projection of $\Delta\vect Z=\vect Z-\vect Z'$ is sub-Gaussian. Since the ordered-pair objective includes both orientations, it equals the U-statistic with scalar loss $\ell_{\circ,h}(x)=\{f_{\circ,h}(x)+f_{\circ,h}(-x)\}/2$. For integral convolution, symmetry of $K$ gives $\ell_{\inth}=f_{\inth}$. For infimal convolution, its derivative is bounded and nondecreasing, and its curvature is the average of $f''_{\infh}(x)$ and $f''_{\infh}(-x)$. This curvature has bounded total mass, support in $[-Ch,Ch]$, and the piecewise Lipschitz property used in Lemma \ref{lem:inf_Hess_conc}. The absolute loss satisfies the piecewise derivative assumptions.

An independent pair therefore satisfies all fixed-decision, crossing, and slab moments in the residual proofs in Appendix \ref{app:smoothing}. The permutation moment-generating-function inequality in the proof of Example \ref{exp:rank}, applied to each projected gradient increment, quadratic form, and nonnegative interpolation bracket, gives the same Bernstein bounds with $\lfloor n/2\rfloor$ in place of $n$. On $\max_i|\vect Z_i|_2\leq C\sqrt{p\log n}$, pairwise design differences have at most twice this norm, and twice the bracket width suffices for interpolation. The same nets and union bounds give the increment bound and Hessian fluctuation $C\{\sqrt{a_n/h}+a_n/h\}$. The mean-zero pointwise gradient deviation has second moment at most $Cp/n$ by the permutation representation and Jensen's inequality. Conditional-density integration for the independent pair gives the population gradient perturbations $O(h)$ and $O(h^2)$ and Hessian perturbation $O(h)$. Symmetrization preserves these bounds. Thus the regularity and population-minimizer arguments in Appendix \ref{app:smoothing} apply.

Put $\vect\vartheta_h=\vect\theta^*_{\infh}$ for infimal convolution and $\vect\vartheta_h=\vect\theta^*$ for integral convolution. Define
\[
\begin{aligned}
\vect k_h(\vect W,\vect W')&=-\Delta\vect Z\,\ell'_{\circ,h}(\Delta Y-\Delta\vect Z^{\tp}\vect\vartheta_h),\\
\vect\psi_h(\vect W)&=2\{\mbE[\vect k_h(\vect W,\vect W')\mid\vect W]-\mbE\vect k_h(\vect W,\vect W')\}.
\end{aligned}
\]
This kernel is symmetric, since $\ell'_{\circ,h}$ is odd. The Hoeffding decomposition gives
\[
\nabla\mcL_{{\rm rank},\circ,h}(\vect\vartheta_h)-\nabla\bar{\mcL}_{{\rm rank},\circ,h}(\vect\vartheta_h)
=\frac1n\sum_{i=1}^n\vect\psi_h(\vect W_i)+\vect D_{n,h},\qquad \mbE|\vect D_{n,h}|_2^2\leq Cp/n^2.
\]
Orthogonality of distinct degenerate pairs gives the factor $2/\{n(n-1)\}$ multiplying their kernel second moment, which is at most $Cp$ by $|\ell'_{\circ,h}|\leq1$. Thus $|\vect D_{n,h}|_2=O_P(\sqrt p/n)$ is absorbed by the remainder.

The population and score bounds permit the localization argument in the proof of Corollary \ref{cor:smoothing_algorithm}. For the complete procedures, the Phase 3 neighborhood is centered at the corresponding smoothed population minimizer. The local comparison \eqref{eq:local_phase3_comparison} gives the stated optimization error. The same increment expansion then gives the remainder for both convolutions. Conditional-density integration for an independent pair bounds the projected second moment of the smoothed score difference by $Ch$, including the $O(h)$ movement of the infimal population minimizer. Conditional Jensen's inequality gives the same bound for the difference between $\vect\psi_h$ and the original influence vector in Example \ref{exp:rank}. Subtracting means preserves this bound, and third absolute moments remain bounded. Lyapunov's theorem and the remainder calculation prove the stated limiting distributions.
\end{proof}

\section{Regularized SAA}\label{app:regularized}

\subsection{Auxiliary results}

\begin{lemma}\label{prop:hdL-smooth}
Suppose $\mcL$ is convex and satisfies Condition \ref{cond:hdL_smooth}. Then, for every $\vect\theta_1,\vect\theta_2\in\Thetar$,
\begin{equation}\label{eq:HD_smooth_cond}
0\leq\mcL(\vect\theta_1)-\mcL(\vect\theta_2)-\langle\partial\mcL(\vect\theta_2),\vect\theta_1-\vect\theta_2\rangle \leq L\max\{|\vect\theta_1-\vect\theta_2|_2,\epsilon_1\}|\vect\theta_1-\vect\theta_2|_2.
\end{equation}
\end{lemma}

\begin{proof}[Proof of Lemma \ref{prop:hdL-smooth}]
Convexity, $|\vect\theta_1-\vect\theta_2|_0\leq2\tilde s$, and Condition \ref{cond:hdL_smooth} give
\[
\begin{aligned}
0&\leq\mcL(\vect\theta_1)-\mcL(\vect\theta_2)-\langle\partial\mcL(\vect\theta_2),\vect\theta_1-\vect\theta_2\rangle\\
&\leq\langle\partial\mcL(\vect\theta_1)-\partial\mcL(\vect\theta_2),\vect\theta_1-\vect\theta_2\rangle\\
&\leq L\max\{|\vect\theta_1-\vect\theta_2|_2,\epsilon_1\}|\vect\theta_1-\vect\theta_2|_2.
\end{aligned}
\]
\end{proof}

\begin{lemma}\label{lem:l0_pgd_onestep}
The update \eqref{eq:pgd_iterate} with the $\ell_0$-penalty $r(\vect\theta)=|\vect\theta|_0$ has the closed-form solution
\[
\theta_l^+=\theta_{\mcT,l}\mbI\{|\theta_{\mcT,l}|\geq\sqrt{2\eta\lambda}\},
\]
where $\vect\theta_{\mcT}=\vect\theta-\eta\partial\mcL(\vect\theta)$.
\end{lemma}

\begin{proof}[Proof of Lemma \ref{lem:l0_pgd_onestep}]
Multiplying the proximal objective by $\eta$ and removing terms independent of $\vect\beta$ gives $-\langle\vect\theta_{\mcT},\vect\beta\rangle+\eta\lambda|\vect\beta|_0+|\vect\beta|_2^2/2$, which separates across coordinates. For each $l$, choosing $\beta_l=0$ gives value zero in its coordinate contribution. Among nonzero choices, the minimum is attained at $\beta_l=\theta_{\mcT,l}$ when $\theta_{\mcT,l}\ne0$, with value $\eta\lambda-\theta_{\mcT,l}^2/2$. Therefore, one minimizer is
\[
\theta_l^+=\theta_{\mcT,l}\mbI\{|\theta_{\mcT,l}|\geq\sqrt{2\eta\lambda}\}.
\]
At equality either choice minimizes the coordinate contribution. This proves the lemma.

\end{proof}

\begin{lemma}\label{lem:supp_inclu}
Under the assumptions of Theorem \ref{thm:post_pgd}(iii), $\supp(\vect{\theta}^{(t)})=S$ for every $t\geq1$ and $|\vect{\theta}^{(t)}-\vect{\theta}^*|_\infty\leq d$ for every $t\geq0$.
\end{lemma}

\begin{proof}
The bound on the minimum nonzero coefficient gives $S\subseteq\supp(\vect{\theta}^{(0)})$. Suppose that $\vect{\theta}^{(t)}\in\mcB_d$ and $S\subseteq\supp(\vect{\theta}^{(t)})$. Its error has at most $\tilde{s}$ nonzero coordinates, hence $|\vect{\theta}^{(t)}-\vect{\theta}^*|_2\leq\sqrt{\tilde{s}}d\leq\epsilon_3$. Define the unthresholded update and the subgradient remainder by
\[
\vect{z}^{(t)}=\vect{\theta}^{(t)}-\eta\partial\mcL(\vect{\theta}^{(t)}),~~\vect{u}^{(t)}=\partial\mcL(\vect{\theta}^{(t)})-\partial\mcL(\vect{\theta}^*)-\vect{H}(\vect{\theta}^{(t)}-\vect{\theta}^*).
\]
The definitions of $q$, $b_n$, and $r_n$ give
\begin{equation*}
|\vect{z}^{(t)}-\vect{\theta}^*|_\infty \leq q|\vect{\theta}^{(t)}-\vect{\theta}^*|_\infty+\eta(b_n+r_n) \leq qd+\eta(b_n+r_n)\leq d.
\end{equation*}
For $j\notin S$, this gives $|z_j^{(t)}|\leq d<\sqrt{2\eta\lambda}$. For $j\in S$, it gives $|z_j^{(t)}|\geq|\theta_j^*|-d>\sqrt{2\eta\lambda}$. Lemma \ref{lem:l0_pgd_onestep} therefore gives $\supp(\vect{\theta}^{(t+1)})=S$ and $|\vect{\theta}^{(t+1)}-\vect{\theta}^*|_\infty\leq d$. Since $\mcB_d\subseteq\Theta$, this proves the induction.

\end{proof}

\subsection{Restricted subgradient increments}

\begin{proof}[Proof of Theorem \ref{thm:hdLip_subgrad}]
Let $g=f'_+$, let $\nu=\diff g$, and write $r_{\vect\theta}=Y-\vect Z^{\tp}\vect\theta$. Assumption \ref{cond:discont} gives $\nu\geq0$ and $\nu(\mbR)\leq2C_f$. Write $P_n$ and $P$ for the empirical average and expectation, respectively, and let $m=\tilde s$ and $a_s=m\log p/n$. The affine term cancels from every subgradient increment.

First consider the bounded-coordinate assumptions. For $d=|\vect\theta_1-\vect\theta_2|_2>0$, write $\vect\theta_1-\vect\theta_2=d\vect u$. At every level $t$, conditional-density integration gives
\[
P\{t\text{ lies between }r_{\vect\theta_1}\text{ and }r_{\vect\theta_2}\}
\leq C_\rho d\,P|\vect u^{\tp}\vect Z|\leq Cd.
\]
Integrating against $\nu$, and using $|g|\leq C_f$ and $|Z_j|\leq C_Z$, shows that
\[
W_j=Z_j\{g(r_{\vect\theta_1})-g(r_{\vect\theta_2})\}
\quad\text{satisfies}\quad P|W_j|^k\leq d\,k!C^k,~~k\geq2.
\]
Bernstein's inequality gives $|(P_n-P)W_j|\leq C(\sqrt{dt/n}+t/n)$ with failure probability $2e^{-t}$.

Take a $p^{-12}$-net of the bounded decision region on each coordinate subspace of dimension at most $2m$. The union has logarithmic cardinality at most $Cm\log p$. Every pair of $m$-sparse decisions belongs to one of these subspaces. The same is true of all pairs on the segment joining them. A decision and a net representative on the same subspace have residuals within $w=C_Z\sqrt{2m}\,p^{-12}$. Monotonicity bounds their selected-subgradient difference by
\[
B_{\tilde{\vect\theta}}=g(r_{\tilde{\vect\theta}}+2w)-f'_-(r_{\tilde{\vect\theta}}-2w).
\]
Conditional integration against $\nu$ gives $PB_{\tilde{\vect\theta}}^k\leq w k!C^k$. Applying Bernstein's inequality over the decision nets and coordinates yields
\[
\max_{\tilde{\vect\theta}}(P_n+P)B_{\tilde{\vect\theta}}\leq C(w+a_s)
\]
with failure probability $Cp^{-\gamma m}$. Since $w\leq Ca_s$, this bounds the interpolation error in maximum norm by $Ca_s$. Apply the fixed-pair bound with $t=Am\log p$ over all net pairs and coordinates. Approximating both endpoints proves the asserted maximum-norm increment bound. The strict brackets also cover arbitrary selections at nondifferentiable residuals.

Under the projection bound in Theorem \ref{thm:hdLip_subgrad}(2), test the increment against a unit vector $\vect v$ with at most $2m$ nonzero coordinates. If the two decisions have joint support of size at most $2m$, their displacement direction $\vect u$ has the same sparsity bound. The directional moment assumption and conditional-density integration imply, uniformly in the crossing level,
\[
P\big[|\vect v^{\tp}\vect Z|^k\mbI\{t\text{ lies between }r_{\vect\theta_1}\text{ and }r_{\vect\theta_2}\}\big]
\leq C_\rho d\,P[|\vect v^{\tp}\vect Z|^k|\vect u^{\tp}\vect Z|]
\leq d\,k!C^k.
\]
Thus the same fixed-pair Bernstein bound holds for $W_{\vect v}=\vect v^{\tp}\vect Z\{g(r_{\vect\theta_1})-g(r_{\vect\theta_2})\}$. Use the preceding decision nets and a $1/4$-net on the unit sphere of each coordinate subspace of dimension at most $2m$. The total logarithmic cardinality remains $Cm\log p$. A union bound over these sphere nets and the $n\leq p$ observations gives
\[
\max_{i\leq n}|\vect Z_i|_{(2m)}\leq B_p=C_B\sqrt{m\log p}
\]
with failure probability $Cp^{-\gamma m}$. Increasing $C_B$ also makes
$P[|\vect v^{\tp}\vect Z|\mbI\{|\vect Z|_{(2m)}>B_p\}]\leq p^{-10}$ uniformly over the test vectors.

On this event the residual interpolation width is $w=p^{-12}B_p$. The Stieltjes bracket has the moment bound
\[
P\big[\{|\vect v^{\tp}\vect Z|B_{\tilde{\vect\theta}}\}^k\big]\leq w k!C^k,~~k\geq1.
\]
Bernstein's inequality over the decision and sphere nets bounds its empirical average and expectation by $C(w+a_s)$. The population contribution outside the displayed sample event is at most $Cp^{-10}$. Hence interpolation again costs at most $Ca_s$. Applying the fixed-pair bound over the nets, followed by the sphere-net inequality on each test support, gives the stated bound in $|\cdot|_{(2m)}$. The proof applies to every pair whose joint support has size at most $2m$, including every pair along the segments used below.

Finally, at $\vect\theta^*$, each coordinate of $\vect Zg(r_{\vect\theta^*})$ is bounded under the first assumptions and sub-Gaussian under the second. A coordinate union bound gives $|\partial\mcL(\vect\theta^*)-\nabla\bar{\mcL}(\vect\theta^*)|_\infty\leq C\sqrt{\log p/n}$ with failure probability $Cp^{-\gamma}$. Intersecting this event with the increment event gives failure probability $Cp^{-\gamma}$, while the increment bound alone retains the sharper probability $Cp^{-\gamma m}$.

\end{proof}

\begin{proof}[Proof of Corollary \ref{thm:hdH_smooth}]
Write $m=\tilde s$ and $a_s=m\log p/n$. For unit $2m$-sparse vectors $\vect u,\vect v$, conditional-density integration and the sparse projection moments give
\[
P\left[|\vect v^{\tp}\vect Z|\,\big|f'_+(Y-\vect Z^{\tp}(\vect\theta+d\vect u))-f'_+(Y-\vect Z^{\tp}\vect\theta)\big|\right]\leq Cd.
\]
Thus the population gradient increment has $|\cdot|_{(2m)}$-norm at most $Cd$. Combining this bound with Theorem \ref{thm:hdLip_subgrad}, separately for $d\geq a_s$ and $d<a_s$, proves part 1.

For part 2, let $D=|\vect\theta_1-\vect\theta_2|_2$ and $\vect\theta_t=\vect\theta_2+t(\vect\theta_1-\vect\theta_2)$. This segment stays in $\Theta_{\rm sc}$, and every pair on it has joint support of size at most $2m$. Pairing the restricted empirical increment error with its sparse displacement and integrating along the segment gives
\[
\mcL(\vect\theta_1)-\mcL(\vect\theta_2)-\langle\partial\mcL(\vect\theta_2),\vect\theta_1-\vect\theta_2\rangle\geq\frac{\mu_0}{2}D^2-C\sqrt{a_s}D^{3/2}-Ca_sD\geq\frac{\mu_0}{4}D^2
\]
when $D\geq C_2a_s$, proving part 2.

For part 3, combine the empirical increment bound with the assumed Taylor bound. If $D=|\vect\theta-\vect\theta^*|_2\leq C_3\sqrt{a_s}$, then
\[
\big|\partial\mcL(\vect\theta)-\partial\mcL(\vect\theta^*)-\vect H(\vect\theta-\vect\theta^*)\big|_\infty\leq C\{\sqrt{Da_s}+a_s+D^2\}\leq C_4a_s^{3/4}.
\]
To verify the Taylor bound under a bounded conditional density derivative, let $\nu=\diff f'_+$. Conditional integration gives
\[
\vect H(\vect\theta)=P\left[\vect Z\vect Z^{\tp}\int\rho(t+\vect Z^{\tp}\vect\theta\mid\vect Z)\nu(\diff t)\right].
\]
The sparse projection moments imply $P[|\vect v^{\tp}\vect Z|\,|\vect Z^{\tp}(\vect\theta-\vect\theta^*)|^2]\leq C|\vect\theta-\vect\theta^*|_2^2$ for each unit $2m$-sparse $\vect v$. Taylor's inequality for the conditional distribution function therefore gives the population remainder in $|\cdot|_{(2m)}$.

Under the bounded-coordinate assumptions, the first-moment crossing bound gives a population maximum-norm increment at most $Cd$. Conversion to $|\cdot|_{(2m)}$ multiplies the smoothness constant by $\sqrt{2m}$. In the strong convexity calculation the scalar empirical error acquires the same factor, and $D\geq C_2ma_s$ suffices. The maximum-norm local approximation has the same remainder $C_4a_s^{3/4}$.
\end{proof}

\subsection{Proximal subgradient descent}\label{app:proximal_updates}

For Phases 1 and 2, suppose there is a set $J_0$ with $|J_0|\leq\tilde s$, $\supp(\vect\theta^{(0)})\subseteq J_0$, and $|\partial_j\mcL(\vect\theta)|\leq\lambda$ for every $j\notin J_0$ and every $\vect\theta\in\Theta$ supported on $J_0$. At each such iterate, soft thresholding keeps the next iterate supported on $J_0$.

\begin{proof}[Proof of Theorem \ref{thm:early_pgd}(i)]
Write $\vect x=\vect\theta^{(t)}$, $\vect x^+=\vect\theta^{(t+1)}$, and $\vect g=\partial\mcL(\vect x)$. The proximal objective is $1/\eta$-strongly convex, and its optimality condition gives, for every $\vect u\in\Thetar$,
\[
\langle\vect g,\vect x^+-\vect u\rangle+\lambda(|\vect x^+|_1-|\vect u|_1)
\leq\frac{|\vect x-\vect u|_2^2-|\vect x^+-\vect u|_2^2-|\vect x^+-\vect x|_2^2}{2\eta}.
\]
Inequality \eqref{eq:HD_smooth_cond} gives an upper remainder $L\max\{|\vect x^+-\vect x|_2,\epsilon_1\}|\vect x^+-\vect x|_2$. After subtracting $|\vect x^+-\vect x|_2^2/(2\eta)$, this remainder is nonpositive if $|\vect x^+-\vect x|_2\geq\epsilon_1$, since $\eta\leq1/(2L)$. Below $\epsilon_1$, completing the square bounds it by $\eta L^2\epsilon_1^2/2$. Thus
\begin{equation}\label{eq:Ftildetheta_thetap}
F(\vect x^+)-F(\vect u)\leq\frac{|\vect x-\vect u|_2^2-|\vect x^+-\vect u|_2^2}{2\eta}-D(\vect u,\vect x)+\frac{\eta L^2\epsilon_1^2}{2},
\end{equation}
where $D(\vect u,\vect x)=\mcL(\vect u)-\mcL(\vect x)-\langle\vect g,\vect u-\vect x\rangle\geq0$. Take $\vect u=\hat{\vect\theta}$ and sum over $t$ to obtain the average-objective bound. If all update lengths are at least $\epsilon_1$, the last term is absent. Taking $\vect u=\vect x$ then gives $F(\vect x^+)\leq F(\vect x)$. Hence the final objective gap is bounded by the average gap.

\end{proof}

\begin{proof}[Proof of Theorem \ref{thm:pgd_phase2}(ii)]
Let $\alpha=\min\{\eta\mu/2,1\}$. Convexity and Condition \ref{cond:hdmu_convex} give
\[
D(\hat{\vect\theta},\vect\theta^{(t)})
\geq\frac{\alpha}{2\eta}\big\{|\vect\theta^{(t)}-\hat{\vect\theta}|_2^2-\epsilon_2^2\big\}.
\]
Indeed, below $\epsilon_2$ the right side is nonpositive. Above it, use $\alpha/(2\eta)\leq\mu/4$. Since $F(\vect\theta^{(t+1)})\geq F(\hat{\vect\theta})$, \eqref{eq:Ftildetheta_thetap} gives
\[
|\vect\theta^{(t+1)}-\hat{\vect\theta}|_2^2
\leq(1-\alpha)|\vect\theta^{(t)}-\hat{\vect\theta}|_2^2+\alpha\epsilon_2^2+\eta^2L^2\epsilon_1^2.
\]
Iteration and summation of the geometric series prove the stated bound. When both distance conditions hold, the two additive terms are absent at every step, which proves the sharper conclusion.

\end{proof}

\begin{proof}[Proof of Theorem \ref{thm:post_pgd}(iii)]
By Lemma \ref{lem:supp_inclu}, the iterates have support $S$ for $t\geq1$. Using $\vect{u}^{(t)}$ from its proof, the update on $S$ is
\[
\{\vect{\theta}^{(t+1)}-\vect{\theta}^*\}_S
=(\mbI-\eta\vect{H}_{S\times S})\{\vect{\theta}^{(t)}-\vect{\theta}^*\}_S -\eta\{\partial\mcL(\vect{\theta}^*)\}_S-\eta\vect{u}^{(t)}_S,
\]
where $|\vect{u}^{(t)}|_\infty\leq r_n$. The principal submatrix satisfies $\|\mbI-\eta\vect{H}_{S\times S}\|_\infty\leq q<1$, and the Neumann series gives
\[
\|\vect{H}_{S\times S}^{-1}\|_\infty
\leq\eta\sum_{k=0}^{\infty}q^k=\frac{\eta}{1-q}.
\]
Adding $\vect{H}_{S\times S}^{-1}\{\partial\mcL(\vect{\theta}^*)\}_S$ to the recursion and iterating from $t=1$ gives
\begin{equation*}
\Big|\{\vect{\theta}^{(T)}-\vect{\theta}^*\}_S +\vect{H}_{S\times S}^{-1}\{\partial\mcL(\vect{\theta}^*)\}_S\Big|_\infty \leq q^{T-1}\Big(d+\frac{\eta b_n}{1-q}\Big) +\eta r_n\sum_{k=0}^{T-2}q^k,
\end{equation*}
with the sum understood as zero when $T=1$. This proves \eqref{eq:Bahadur}.

\end{proof}

Choosing $d\asymp b_n+r_n$ and $\lambda\asymp d^2$ gives the order $\lambda\asymp b_n^2$ when $r_n=O(b_n)$ \citep{bunea_etal.2007aos, chen_lee.2023joe}.

The support recovery and expansion also extend to SCAD, MCP, and capped-$\ell_1$ penalties. For a separable penalty $\lambda r(\vect\theta)=\sum_{j=1}^p p_\lambda(\theta_j)$, write $\mathcal P_\eta(u)=\argmin{v\in\mbR}\{p_\lambda(v)+(v-u)^2/(2\eta)\}$. Suppose $\mathcal P_\eta(u)=\{0\}$ for $|u|<\tau_0$ and $\mathcal P_\eta(u)=\{u\}$ for $|u|>\tau_1$, where $d<\tau_0\leq\tau_1<\min_{j\in S}|\theta_j^*|-d$. Replacing the $\ell_0$-threshold condition in Theorem \ref{thm:pgd_phases}(iii) by this separation gives the same support recovery after one update and expansion \eqref{eq:Bahadur}.

\begin{proof}[Proof of the penalty extension in Section \ref{sec:psubgd}]
The threshold separation gives $\min_{j\in S}|\theta_j^*|>2d$, hence $S\subseteq\supp(\vect\theta^{(0)})$. For $\vect\theta\in\mcB_d$ containing $S$ in its support, write $\vect z=\vect\theta-\eta\partial\mcL(\vect\theta)$. The neighborhood, contraction, and initialization conditions of Theorem \ref{thm:pgd_phases}(iii) give $|\vect z-\vect\theta^*|_\infty\leq qd+\eta(b_n+r_n)\leq d$. Thus $|z_j|\leq d<\tau_0$ for $j\notin S$, whereas $|z_j|\geq|\theta_j^*|-d>\tau_1$ for $j\in S$. Every coordinate proximal minimizer therefore equals zero outside $S$ and equals $z_j$ on $S$. The updated vector has support $S$ and remains in $\mcB_d$. The induction in Lemma \ref{lem:supp_inclu} and the unpenalized recursion in the proof of Theorem \ref{thm:pgd_phases}(iii) then give \eqref{eq:Bahadur}.
\end{proof}

\subsection{Iterative hard thresholding}

\begin{proof}[Proof of Theorem \ref{thm:early_iht}(i)]
Define $\vect g_t=\partial\mcL(\vect\theta^{(t)})$ and
$Q_t(\vect u)=\langle\vect g_t,\vect u-\vect\theta^{(t)}\rangle+(2\eta)^{-1}|\vect u-\vect\theta^{(t)}|_2^2$.
The IHT update minimizes $Q_t$ over all $\tilde s$-sparse vectors. Inequality \eqref{eq:HD_smooth_cond} yields
\begin{equation}\label{eq:iht_descent}
\mcL(\vect\theta^{(t+1)})\leq{}\mcL(\vect\theta^{(t)})+Q_t(\vect\theta^{(t+1)})-\frac{|\vect\theta^{(t+1)}-\vect\theta^{(t)}|_2^2}{2\eta} +L\max\{|\vect\theta^{(t+1)}-\vect\theta^{(t)}|_2,\epsilon_1\}|\vect\theta^{(t+1)}-\vect\theta^{(t)}|_2.
\end{equation}
Since $\eta\leq1/(4L)$, the last two terms are at most $-|\vect\theta^{(t+1)}-\vect\theta^{(t)}|_2^2/(4\eta)$ when the update length is at least $\epsilon_1$. Below $\epsilon_1$, completing the square gives the same bound with additive error $\eta L^2\epsilon_1^2$. Since $Q_t(\vect\theta^{(t+1)})\leq Q_t(\vect\theta^{(t)})=0$,
\[
\frac{|\vect\theta^{(t+1)}-\vect\theta^{(t)}|_2^2}{4\eta}
\leq\mcL(\vect\theta^{(t)})-\mcL(\vect\theta^{(t+1)})+\eta L^2\epsilon_1^2.
\]
Summing proves the claim. If every update length is at least $\epsilon_1$, the final term is absent.

\end{proof}

\begin{proof}[Proof of Theorem \ref{thm:iht_phase2}(ii)]
Let $\alpha=\min\{\eta\mu/2,1\}$, $\vect u_t=(1-\alpha)\vect\theta^{(t)}+\alpha\vect\theta^*$, and $Q_t(\vect u)=\langle\partial\mcL(\vect\theta^{(t)}),\vect u-\vect\theta^{(t)}\rangle+(2\eta)^{-1}|\vect u-\vect\theta^{(t)}|_2^2$. Since $S\subseteq\supp(\vect\theta^{(t)})$, the vector $\vect u_t$ is $\tilde s$-sparse. Comparison with $\vect u_t$ in the IHT update gives
\[
Q_t(\vect\theta^{(t+1)})\leq Q_t(\vect u_t)
=\alpha\langle\partial\mcL(\vect\theta^{(t)}),\vect\theta^*-\vect\theta^{(t)}\rangle
+\frac{\alpha^2}{2\eta}|\vect\theta^{(t)}-\vect\theta^*|_2^2.
\]
By convexity and Condition \ref{cond:hdmu_convex},
\[
\langle\partial\mcL(\vect\theta^{(t)}),\vect\theta^*-\vect\theta^{(t)}\rangle
\leq\mcL(\vect\theta^*)-\mcL(\vect\theta^{(t)})
-\frac\mu4|\vect\theta^{(t)}-\vect\theta^*|_2^2+\frac\mu4\epsilon_2^2.
\]
Write $\Delta_t=\mcL(\vect\theta^{(t)})-\mcL(\vect\theta^*)$. For $\eta\leq1/(2L)$, the last two terms in \eqref{eq:iht_descent} are bounded by $\eta L^2\epsilon_1^2/2$, as shown in the proof of Theorem \ref{thm:early_pgd}(i). Substituting the bound for $Q_t(\vect u_t)$ gives
\[
\Delta_{t+1}\leq(1-\alpha)\Delta_t
-\Big(\frac{\alpha\mu}{4}-\frac{\alpha^2}{2\eta}\Big)|\vect\theta^{(t)}-\vect\theta^*|_2^2
+\frac{\alpha\mu\epsilon_2^2}{4}+\frac{\eta L^2\epsilon_1^2}{2}.
\]
The squared-distance coefficient is nonnegative since $\alpha\leq\eta\mu/2$. Drop that term and iterate the resulting scalar inequality. The sum of the geometric weights is at most $1/\alpha$, which gives the remainder. Under the two distance conditions, both additive terms vanish before iteration.

\end{proof}

\begin{proof}[Proof of Theorem \ref{thm:post_iht}(iii)]
The bound on the minimum nonzero coefficient and $\vect{\theta}^{(0)}\in\mcB_d$ give $S\subseteq\supp(\vect{\theta}^{(0)})$. Suppose $\vect{\theta}^{(t)}\in\mcB_d$ and $S\subseteq\supp(\vect{\theta}^{(t)})$. The error has at most $\tilde{s}$ nonzero coordinates, hence $|\vect{\theta}^{(t)}-\vect{\theta}^*|_2\leq\sqrt{\tilde{s}}d\leq\epsilon_3$. Define $\vect{z}^{(t)}$ and $\vect{u}^{(t)}$ as in the proof of Lemma \ref{lem:supp_inclu}. Then
\[
|\vect{z}^{(t)}-\vect{\theta}^*|_\infty
\leq q|\vect{\theta}^{(t)}-\vect{\theta}^*|_\infty+\eta(b_n+r_n)
\leq d.
\]
For $j\in S$, the inequality $|z_j^{(t)}|\geq|\theta_j^*|-d>d$ holds. For $j\notin S$, $|z_j^{(t)}|\leq d$. Since $\tilde{s}\geq s$, hard thresholding retains every coordinate in $S$. It follows that
\[
|\vect{\theta}^{(t+1)}-\vect{\theta}^*|_\infty
\leq q|\vect{\theta}^{(t)}-\vect{\theta}^*|_\infty+\eta(b_n+r_n).
\]
The iterate $\vect{\theta}^{(t+1)}$ has at most $\tilde{s}$ nonzero coordinates and belongs to $\Theta$ by $\mcB_d\subseteq\Theta$. Induction gives the support inclusion, invariance of $\mcB_d$, and geometric bound. When $\tilde{s}=s$, every iterate has support $S$, including $t=0$.

To recover $S$ when $s$ is unknown, IHT can instead be followed by $\ell_0$-penalized PGD. Under the conditions of Theorem \ref{thm:iht_phases}(iii), the next unthresholded iterate has maximum-norm error at most $d$. Every threshold satisfying $d<\sqrt{2\eta\lambda}<\min_{j\in S}|\theta_j^*|-d$ removes all coordinates outside $S$ and retains all coordinates in $S$. The theorem's minimum-coefficient condition makes this interval nonempty.

\end{proof}

\begin{proof}[Proof of Theorem \ref{thm:iht_phases}(iv)]
The joint increment and score event in Theorem \ref{thm:hdLip_subgrad} has probability at least $1-Cp^{-\gamma}$. On this event, every retained set $J\supseteq S$ of size $\tilde s$ satisfies $|\vect g_J|_2\leq\sqrt{\tilde s}\,|\partial\mcL(\vect\theta^*)|_\infty\leq C_g\sqrt{a_s}$. The assumed population Taylor bound gives its Euclidean remainder on each retained set. Consequently, for every $\vect\theta\in\Theta$ supported on $J$ with $D=|\vect\theta-\vect\theta^*|_2\leq\epsilon_3$,
\[
\left|\{\partial\mcL(\vect\theta)-\partial\mcL(\vect\theta^*)\}_J
-\vect H_{J\times J}(\vect\theta-\vect\theta^*)_J\right|_2
\leq C\{\sqrt{Da_s}+a_s+D^2\}.
\]

Intersect this event with $\mcE_n$ and write $J=\widehat J$. Define $\vect u_t=\{\partial\mcL(\vect\theta^{(t)})-\partial\mcL(\vect\theta^*)\}_J-\vect H_{J\times J}\vect e_t$. Since $\eta\leq M_H^{-1}$, the matrix $\vect A_J=\mbI-\eta\vect H_{J\times J}$ satisfies $\|\vect A_J\|\leq1-\eta m_H$. When an update retains $J$, its error follows $\vect e_{t+1}=\vect A_J\vect e_t-\eta\vect g_J-\eta\vect u_t$. Young's inequality gives $C\sqrt{Da_s}\leq m_HD/4+C^2a_s/m_H$. For sufficiently large $n$, $D\leq\epsilon_3$ also gives $CD^2\leq m_HD/4$. Hence, for a constant $C_u$,
\[
|\vect e_{t+1}|_2
\leq(1-\eta m_H/2)|\vect e_t|_2+
\eta\{|\vect g_J|_2+C_u a_s\}.
\]
Choose $C_3\geq4(C_g+C_u)/m_H$. Since $a_s\leq1$ eventually, the forcing term is at most $\eta m_H\epsilon_3/4$. Every current error of norm at most $\epsilon_3$ therefore produces an error of norm at most $(1-\eta m_H/4)\epsilon_3$. The region assumption places this next iterate in $\Theta$, completing the containment induction. Iterating the displayed recursion proves the first bound.

Containment also gives $|\vect u_t|_2\leq Ca_s^{3/4}$. Apply the recursion to $\vect e_t+\vect H_{J\times J}^{-1}\vect g_J$. The geometric kernel has total mass at most $\eta/\{1-(1-\eta m_H)\}=m_H^{-1}$, proving the expansion bound.

\end{proof}

\phantomsection\label{app:maximum_norm_contraction}

A positive step size with $q=\|\mbI-\eta\vect H\|_\infty<1$ exists if and only if $H_{ii}>\sum_{j\ne i}|H_{ij}|$ for every $i$. The admissible step sizes are exactly $0<\eta<\min_i 2/(H_{ii}+\sum_{j\ne i}|H_{ij}|)$. This follows from $q=\max_i\{|1-\eta H_{ii}|+\eta\sum_{j\ne i}|H_{ij}|\}$. For $0<\eta\leq1/\max_iH_{ii}$, we have $q=1-\eta\min_i(H_{ii}-\sum_{j\ne i}|H_{ij}|)$ and $\|\vect H^{-1}\|_\infty\leq\{\min_i(H_{ii}-\sum_{j\ne i}|H_{ij}|)\}^{-1}$. A lower bound on the row margins then gives $q\leq q_0<1$. For the inverse bound, choose a coordinate with $|v_i|=|\vect v|_\infty$ and use $|\vect H\vect v|_\infty\geq\{H_{ii}-\sum_{j\ne i}|H_{ij}|\}|\vect v|_\infty$.

\begin{corollary}\label{cor:iht_warm_start}
Let $m=\tilde s$, $1\leq s<m\leq p$, $p>n\geq2$, and $a_s=m\log p/n\to0$. Consider the residual objective under the assumptions of Theorem \ref{thm:hdLip_subgrad}(2). Let $\vect\theta^*$ be an interior stationary point of the expected objective, let $S=\supp(\vect\theta^*)$ have size $s$, and suppose $\mbB(\vect\theta^*,R)\subseteq\Theta$ for $R>0$. Assume that $0<\mu_0\leq L_0$ and, for every index set $I\supseteq S$ with $|I|\leq2m+s$,
\[
\mu_0\mbI_I\preceq\nabla^2\bar{\mcL}(\vect\theta)_{I\times I}\preceq L_0\mbI_I
\]
whenever $\supp(\vect\theta)\subseteq I$ and $|\vect\theta-\vect\theta^*|_2\leq R$. Put $\kappa_0=L_0/\mu_0$, suppose $m-s\geq64\kappa_0^2s$, and run IHT with working sparsity $m$ and step size $\eta=L_0^{-1}$. There exist $C,\gamma,K,c>0$ such that, for sufficiently large $n$, on one event of probability at least $1-Cp^{-\gamma}$, every trajectory initialized at an $m$-sparse $\vect\theta^{(0)}$ with $|\vect\theta^{(0)}-\vect\theta^*|_2\leq R/2$ remains in $\mbB(\vect\theta^*,R)$ and satisfies
\[
|\vect\theta^{(t)}-\vect\theta^*|_2
\leq\left(1-\frac1{2\kappa_0}\right)^t|\vect\theta^{(0)}-\vect\theta^*|_2+K\sqrt{a_s},\qquad t\geq0.
\]
In particular, $|\vect\theta^{(t)}-\vect\theta^*|_2\leq2K\sqrt{a_s}$ for every $t\geq\lceil c\kappa_0\log n\rceil$.
\end{corollary}

\begin{proof}
Work on the increment and pointwise score event in Theorem \ref{thm:hdLip_subgrad}. Its probability is at least $1-Cp^{-\gamma}$. Write $\vect g=\partial\mcL$, $\vect G=\nabla\bar{\mcL}$, and
$
\Gamma=|\vect g(\vect\theta^*)|_{(2m+s)}.
$
Stationarity gives $\vect G(\vect\theta^*)=\vect0$. The pointwise maximum-norm score bound therefore yields $\Gamma\leq C_g\sqrt{a_s}$. If an index set has more than $2m$ coordinates but at most $2m+s\leq3m$, split it into two sets of size at most $2m$. The restricted increment bound then gives, for every $m$-sparse $\vect\theta\in\Theta$ and every such index set $I$,
\[
\big|\{\vect g(\vect\theta)-\vect g(\vect\theta^*)-\vect G(\vect\theta)+\vect G(\vect\theta^*)\}_I\big|_2
\leq C\{\sqrt{Da_s}+a_s\},\qquad D=|\vect\theta-\vect\theta^*|_2.
\]
The endpoint pair has joint support of size at most $m+s\leq2m$, as required by that theorem. Both bounds hold simultaneously for the decisions and index sets selected by the algorithm.

We use the projection mechanism of \citet[Lemma~1]{jain_etal.2014nips} and give the needed distance estimate. For an arbitrary vector $\vect z$, let $J$ be a retained set of its $m$ largest coordinates in magnitude, let $\vect y=\vect z_J$ extended by zero, and put $A=S\setminus J$, $B=J\setminus S$, and $r=\sqrt{s/(m-s)}$. The ordering of the retained coordinates gives $|\vect z_A|_2\leq r|\vect z_B|_2$. With $\vect e=\vect z-\vect\theta^*$ on $J\cup S$ and $E=|\vect e_{J\cup S}|_2$, we have $\vect e_B=\vect z_B$ and
\[
\begin{aligned}
|\vect y-\vect\theta^*|_2^2
&=|\vect e_J|_2^2+|\vect e_A-\vect z_A|_2^2\\
&\leq E^2+r^2|\vect e_B|_2^2+2r|\vect e_A|_2|\vect e_B|_2
\leq(1+r+r^2)E^2\leq(1+r)^2E^2.
\end{aligned}
\]

Suppose that the current iterate $\vect\theta$ has distance $D\leq R$, set $\vect z=\vect\theta-\eta\vect g(\vect\theta)$, and let $J$ be its next retained set. Set $I=\supp(\vect\theta)\cup J\cup S$. Then $|I|\leq2m+s$. The segment from $\vect\theta^*$ to $\vect\theta$ stays in the assumed coordinate ball. Integration of the population Hessian along this segment gives
\[
\big|\{\vect\theta-\vect\theta^*-\eta[\vect G(\vect\theta)-\vect G(\vect\theta^*)]\}_I\big|_2
\leq\left(1-\frac1{\kappa_0}\right)D.
\]
Combining this inequality with the increment bound and the projection estimate yields
\[
D^+\leq(1+r)\left[\left(1-\frac1{\kappa_0}\right)D+\eta\{\Gamma+C\sqrt{Da_s}+Ca_s\}\right].
\]
Young's inequality gives $C\sqrt{Da_s}\leq\mu_0D/8+2C^2a_s/\mu_0$. Put $C_1=C+2C^2/\mu_0$. Since $r\leq1/(8\kappa_0)$,
\[
(1+r)\left(1-\frac7{8\kappa_0}\right)
\leq1-\frac3{4\kappa_0}\leq Q:=1-\frac1{2\kappa_0}.
\]
Consequently
\[
D^+\leq QD+\eta(1+r)(\Gamma+C_1a_s),\qquad
\frac{\eta(1+r)(\Gamma+C_1a_s)}{1-Q}
\leq\frac4{\mu_0}(\Gamma+C_1a_s)=:\mathcal R_n.
\]
Choose $K=4(C_g+C_1)/\mu_0$. For $a_s\leq1$, $\mathcal R_n\leq K\sqrt{a_s}$, which is at most $R/2$ for sufficiently large $n$. For $D\leq R$, the one-step bound is at most $QR+(1-Q)R/2<R$. Starting with $D_0\leq R/2$, induction proves containment and
\[
D_t\leq Q^tD_0+(1-Q^t)\mathcal R_n.
\]
Since $Q^t\leq\exp\{-t/(2\kappa_0)\}$ and $a_s\geq n^{-1}$ eventually, increasing $c$ gives the stated entry count.

A bounded conditional density derivative gives Lipschitz bounds on the population Hessian on these coordinate subspaces by the proof of Corollary \ref{thm:hdH_smooth}. Projections with up to $2m+s$ coordinates satisfy the required moment bounds by splitting them into two projections with at most $2m$ coordinates and applying the triangle inequality for the sub-Gaussian norm. Under this density derivative condition, the spectral bounds $m_H\mbI\preceq\nabla^2\bar{\mcL}(\vect\theta^*)\preceq M_H\mbI$ imply the displayed local Hessian bounds for sufficiently small $R$, with $\mu_0=m_H/2$ and $L_0=M_H+m_H/2$.
\end{proof}

\subsection{Stopping rules and sampling distribution}

For a proximal or IHT update from $\vect\theta$ to $\vect\theta^+$ with step size $\eta>0$, write $k=|\vect\theta-\vect\theta^+|_2$ and $G=k/\eta+L\max\{k,\epsilon_1\}$.

\begin{proposition}\label{prop:restricted_stopping}
For parts \textup{(i)} and \textup{(ii)}, suppose $\mcL$ is convex and satisfies Conditions \ref{cond:hdL_smooth} and \ref{cond:hdmu_convex}.
\begin{enumerate}[label=\textup{(\roman*)},leftmargin=*]
\item Let $\vect\theta^+$ solve \eqref{eq:pgd_iterate} with a proper convex penalty $r$, and let $\hat{\vect\theta}\in\argmin{\vect\theta\in\Theta}F(\vect\theta)$. Suppose $\vect\theta\in\Thetar$ and $\vect\theta^+,\hat{\vect\theta}\in\Theta_{\mathrm{sc}}\cap\Thetar$. Then
\[
\begin{aligned}
|\vect\theta^+-\hat{\vect\theta}|_2
&\leq\max\{4G/\mu,\epsilon_2\},\\
0\leq F(\vect\theta^+)-F(\hat{\vect\theta})
&\leq\max\{G^2/\mu,G\epsilon_2\}.
\end{aligned}
\]
\item Let $\vect\theta^+$ be an IHT update from $\vect\theta$. Suppose $\vect\theta\in\Thetar$, $\vect\theta^+,\vect\theta^*\in\Theta_{\mathrm{sc}}\cap\Thetar$, and $S\subseteq\supp(\vect\theta^+)$. Then
\[
\begin{aligned}
|\vect\theta^+-\vect\theta^*|_2
&\leq\max\{2(G+\sqrt{\tilde s}b_n)/\mu,\epsilon_2\},\\
\mcL(\vect\theta^+)-\mcL(\vect\theta^*)
&\leq\max\{G^2/\mu,G\epsilon_2\}.
\end{aligned}
\]
\item Suppose Condition \ref{cond:hdH_smooth} holds and $\vect\theta\in\Thetar$ satisfies $|\vect\theta-\vect\theta^*|_2\leq\epsilon_3$. For $d>0$, either of the following conditions gives $|\vect\theta-\vect\theta^*|_\infty\leq d$,
\[
\begin{aligned}
\frac{|\vect\theta^+-\vect\theta|_\infty}{\eta}+b_n+r_n+\lambda
&\leq\frac{d}{\|\vect H^{-1}\|_\infty}&&\text{for $\ell_1$-penalized PGD},\\
|\{\partial\mcL(\vect\theta)\}_J|_\infty+b_n+r_n
&\leq\frac{d}{\|\vect H_{J\times J}^{-1}\|_\infty}&&\text{if $\supp(\vect\theta)\subseteq J\supseteq S$}.
\end{aligned}
\]
In the second condition, assume $|J|\leq\tilde s$.
\end{enumerate}
\end{proposition}

\begin{proof}
For part (i), proximal optimality gives $\vect\xi^+\in\partial r(\vect\theta^+)$ with $\partial\mcL(\vect\theta)+\lambda\vect\xi^+=(\vect\theta-\vect\theta^+)/\eta$. Thus $\vect v=\partial\mcL(\vect\theta^+)+\lambda\vect\xi^+\in\partial F(\vect\theta^+)$, and Condition \ref{cond:hdL_smooth} gives $|\vect v|_{(2\tilde s)}\leq k/\eta+L\max\{k,\epsilon_1\}=G$. Write $\vect\delta=\vect\theta^+-\hat{\vect\theta}$ and $D=|\vect\delta|_2$. Since $|\vect\delta|_0\leq2\tilde s$, convexity and optimality of $\hat{\vect\theta}$ give $0\leq F(\vect\theta^+)-F(\hat{\vect\theta})\leq\langle\vect v,\vect\delta\rangle\leq GD$. If $D\geq\epsilon_2$, Condition \ref{cond:hdmu_convex} for $(\hat{\vect\theta},\vect\theta^+)$ and convexity of $r$ give
\[
0\leq F(\vect\theta^+)-F(\hat{\vect\theta})\leq GD-\frac\mu4D^2\leq\frac{G^2}{\mu},
\]
which also implies $D\leq4G/\mu$. If $D<\epsilon_2$, the convexity bound gives $F(\vect\theta^+)-F(\hat{\vect\theta})\leq G\epsilon_2$, proving part (i).

For part (ii), put $J^+=\supp(\vect\theta^+)$, $\vect\delta=\vect\theta^+-\vect\theta^*$, and $D=|\vect\delta|_2$. The inclusion $S\subseteq J^+$ gives $\supp(\vect\delta)\subseteq J^+$ and $|J^+|\leq\tilde s$. Coordinatewise optimality gives $\{\partial\mcL(\vect\theta)\}_{J^+}=\{\vect\theta-\vect\theta^+\}_{J^+}/\eta$. Condition \ref{cond:hdL_smooth} therefore yields $|\{\partial\mcL(\vect\theta^+)\}_{J^+}|_2\leq G$, while $|\{\partial\mcL(\vect\theta^*)\}_{J^+}|_2\leq\sqrt{\tilde s}b_n$. Convexity gives $\mcL(\vect\theta^+)-\mcL(\vect\theta^*)\leq GD$. If $D\geq\epsilon_2$, adding Condition \ref{cond:hdmu_convex} for both orientations gives
\[
\frac\mu2D^2\leq\langle\partial\mcL(\vect\theta^+)-\partial\mcL(\vect\theta^*),\vect\delta\rangle\leq(G+\sqrt{\tilde s}b_n)D.
\]
Hence $D\leq2(G+\sqrt{\tilde s}b_n)/\mu$. The same condition for $(\vect\theta^*,\vect\theta^+)$ gives $\mcL(\vect\theta^+)-\mcL(\vect\theta^*)\leq GD-\mu D^2/4\leq G^2/\mu$. If $D<\epsilon_2$, convexity gives $\mcL(\vect\theta^+)-\mcL(\vect\theta^*)\leq G\epsilon_2$, proving part (ii).

Adding $k$ to the distance bounds in parts \textup{(i)} and \textup{(ii)} gives the current-iterate stopping tests in Section \ref{sec:hd_stopping}. Choose $\varrho>\epsilon_2$, with $\varrho>4L\epsilon_1/\mu$ for PGD or $\varrho>2(L\epsilon_1+\sqrt{\tilde s}b_n)/\mu$ for IHT, to place the stopping radius above the approximation and score terms at $k=0$.

For part \textup{(iii)}, Condition \ref{cond:hdH_smooth} gives $|\vect H(\vect\theta-\vect\theta^*)|_\infty\leq|\partial\mcL(\vect\theta)|_\infty+b_n+r_n$. For $\ell_1$-penalized PGD, proximal optimality gives $(\vect\theta-\vect\theta^+)/\eta=\partial\mcL(\vect\theta)+\lambda\vect\xi$, with $\vect\xi\in\partial|\vect\theta^+|_1$ and $|\vect\xi|_\infty\leq1$. Thus $|\partial\mcL(\vect\theta)|_\infty\leq|\vect\theta^+-\vect\theta|_\infty/\eta+\lambda$, and multiplication by $\vect H^{-1}$ proves the first criterion. For the second, $\supp(\vect\theta-\vect\theta^*)\subseteq J$ gives $|\vect H_{J\times J}(\vect\theta-\vect\theta^*)_J|_\infty\leq|\{\partial\mcL(\vect\theta)\}_J|_\infty+b_n+r_n$. Multiplication by $\vect H_{J\times J}^{-1}$ proves the second criterion.
\end{proof}

\begin{proof}[Proof of Corollary \ref{cor:hd_normality}]
Restrict the residual objective to the deterministic coordinate set $J$. The conditional density of $Y$ given $\vect Z_J$ is $\rho_J(y\mid\vect Z_J)=\mbE\{\rho(y\mid\vect Z)\mid\vect Z_J\}$ and has the same bound. Every unit projection of $\vect Z_J$ has bounded sub-Gaussian norm, and the restricted region is convex and bounded. Theorem \ref{thm:Lip_subgrad} therefore applies with dimension $|J|$ and scale $a_J=|J|\log n/n\leq a_s$. Its uniform increment event has probability at least $1-Cn^{-\gamma|J|}$. Intersect it with the localization event and, for IHT, with $\{\widehat J=J\}$. Their intersection has probability tending to one. Put $\vect e_t=\{\vect\theta^{(t)}-\vect\theta^*\}_J$ and $\vect g_J=\{\partial\mcL(\vect\theta^*)\}_J$. Independence and stationarity give
\[
\mbE|\vect g_J|_2^2=\frac1n\mbE|\vect\psi_{i,J}|_2^2\leq C|J|/n.
\]

The PGD iterates have support $S$ from time 1. Each IHT update retaining $J$ copies the unthresholded update on $J$. Thus, for $1\leq t<T$,
\[
\vect e_{t+1}=(\mbI-\eta\vect H_{J\times J})\vect e_t-\eta\vect g_J-\eta\vect u_t,
\qquad
|\vect u_t|_2\leq C\{\sqrt{|\vect e_t|_2a_J}+a_J+|\vect e_t|_2^2\}.
\]
Under Theorem \ref{thm:pgd_phases}(iii) or Proposition \ref{prop:stable_iht_inference}\ref{part:stable_iht_expansion}, the same recursion coordinatewise gives $|\vect e_{t+1}|_\infty\leq q|\vect e_t|_\infty+\eta(b_n+r_n)\leq d$, proving containment by induction. Theorem \ref{thm:iht_phases}(iv) gives Euclidean containment for the other IHT case. Hence $|\vect e_t|_2\leq C_3\sqrt{a_s}=o(1)$ throughout the window. Let $q_J=\|\mbI-\eta\vect H_{J\times J}\|$ and $\alpha_J=1-q_J$. The bound $q\leq q_0<1$ gives $q_J\leq q_0$ and $\eta m_H\leq1+q_0$, while $\eta\leq M_H^{-1}$ gives $\alpha_J\geq\eta m_H$. In both cases $\eta/\alpha_J\leq C$. Young's inequality and the vanishing containment radius yield
\[
|\vect e_{t+1}|_2\leq(1-\alpha_J/2)|\vect e_t|_2+\eta\{|\vect g_J|_2+Ca_J\}.
\]
The geometric kernel has mass at most $2\eta/\alpha_J\leq C$. For $t_0=\lfloor T/2\rfloor$, the stated iteration counts therefore give
\[
\sup_{t_0\leq t\leq T}|\vect e_t|_2=O_P(\sqrt{|J|/n}+a_J),\qquad
\sup_{t_0\leq t<T}|\vect u_t|_2=O_P\{(|J|/n)^{3/4}\sqrt{\log n}+a_J\}.
\]
Apply the linear recursion from $t_0$ to $T$ to $\vect e_t+\vect H_{J\times J}^{-1}\vect g_J$. Its kernel has mass at most $\eta/\alpha_J\leq C$, and its transient is $O_P(n^{-1})$ after increasing $c$. This proves the expansion. Each deterministic coordinate of $\vect H_{J\times J}^{-1}\vect\psi_{i,J}$ has bounded third absolute moment. The variance lower bound gives Lyapunov's condition, and the dimension condition makes both remainder terms $o(n^{-1/2})$. Slutsky's theorem proves the normal limit. The complement of the joint event has vanishing probability and does not affect either conclusion.
\end{proof}

\begin{proposition}\label{prop:stable_iht_inference}\label{app:stable_iht_inference}
\begin{enumerate}[label=\textup{(\roman*)},ref=(\roman*),leftmargin=*]
\item\label{part:stable_iht_expansion} Suppose Condition \ref{cond:hdH_smooth} holds, $\tilde s\geq s$, $\mcB_d\subseteq\Theta$, $\sqrt{\tilde s}d\leq\epsilon_3$, $q<1$, $\vect\theta^{(0)}\in\mcB_d$, and $\eta(b_n+r_n)\leq(1-q)d$. For every $J\supseteq S$ with $|J|=\tilde s$, if $\supp(\vect\theta^{(0)})\subseteq J$ and the next $T$ IHT updates retain $J$, then every iterate belongs to $\mcB_d$ and
\[
\begin{aligned}
\{\vect\theta^{(T)}-\vect\theta^*\}_J
={}&-\vect H_{J\times J}^{-1}\{\partial\mcL(\vect\theta^*)\}_J\\
&+(\mbI-\eta\vect H_{J\times J})^T\left[\{\vect\theta^{(0)}-\vect\theta^*\}_J+\vect H_{J\times J}^{-1}\{\partial\mcL(\vect\theta^*)\}_J\right]+\vect R_{J,T},
\end{aligned}
\]
where $|\vect R_{J,T}|_\infty\leq\eta(1-q^T)r_n/(1-q)$.

\item\label{part:stable_iht_score_intervals} Consider the residual objective under either set of assumptions in Theorem \ref{thm:hdLip_subgrad}, with $\nabla\bar{\mcL}(\vect\theta^*)=\vect0$. Let $\widehat J\supseteq S$ be a random retained set of size $\tilde s$, and suppose the conditions of part \ref{part:stable_iht_expansion} hold on an event $\mcE_n$ of probability at least $1-\delta_n$, where $\delta_n\to0$. For every $0<\alpha<1$, let $B_{n,\alpha}=C\sqrt{\log(2p/\alpha)/n}$. Then
\[
\Pr\left\{|\theta_j^{(T)}-\theta_j^*|\leq q^Td+\frac{\eta(1-q^T)}{1-q}(B_{n,\alpha}+r_n)\text{ for every }j\in\widehat J\right\}\geq1-\alpha-\delta_n.
\]
If $q\leq q_0<1$, $r_n=C_4a_s^{3/4}$, and $T\geq\lceil c\log n\rceil$ for sufficiently large $c$, the radius is $O\{\sqrt{\log p/n}+a_s^{3/4}\}$ and tends to zero throughout $a_s\to0$.

\item\label{part:stable_iht_generic_normality} Let $J=J_n\supseteq S$ be deterministic. For IHT, suppose the conditions of part \ref{part:stable_iht_expansion} hold with probability tending to one for a random retained set $\widehat J$, and $\Pr(\widehat J=J)\to1$. For PGD, take $J=S$ and suppose the conditions of Theorem \ref{thm:pgd_phases}(iii) hold with probability tending to one. Assume $\vect H$ is deterministic, $q\leq q_0<1$, and $\partial\mcL(\vect\theta^*)=n^{-1}\sum_i\vect\psi_i$ with i.i.d. mean-zero scores. For a deterministic $j\in J$, write
\[
\xi_{i,J,j}=\vect e_{j,J}^{\tp}\vect H_{J\times J}^{-1}\vect\psi_{i,J},\qquad
\sigma_{J,j}^2=\operatorname{Var}(\xi_{i,J,j}).
\]
Suppose $\sigma_{J,j}\geq c>0$, $\mbE|\xi_{i,J,j}|^{2+\nu}\leq C\sigma_{J,j}^{2+\nu}$ for some $\nu>0$, and $\sqrt n\,r_n=o_P(1)$. If
\[
\sqrt n\,q^T\left(d+\frac{\eta b_n}{1-q}\right)=o_P(1)\quad\text{for IHT},
\qquad
\sqrt n\,q^{T-1}\left(d+\frac{\eta b_n}{1-q}\right)=o_P(1)\quad\text{for PGD},
\]
then
\[
\frac{\sqrt n(\theta_j^{(T)}-\theta_j^*)}{\sigma_{J,j}}\xrightarrow{d}\mcN(0,1).
\]
For the residual objective, Corollary \ref{cor:hd_normality} gives the sharp expansion and normal limit.

\item\label{part:stable_iht_gaussian} Consider the residual objective under the assumptions of Theorem \ref{thm:hdLip_subgrad}(2), with $\nabla\bar{\mcL}(\vect\theta^*)=\vect0$ and $\vect\psi_i=\vect b-\vect Z_i f'_+(Y_i-\vect Z_i^{\tp}\vect\theta^*)$. Let $\widehat J\supseteq S$ be a random retained set of size $\tilde s$. Suppose the conditions of part \ref{part:stable_iht_expansion} hold on an event $\mcE_n$ of probability at least $1-\delta_n$, where $\delta_n\to0$, and assume $\vect H$ is deterministic and $q\leq q_0<1$. Use $\xi_{i,J,j},\sigma_{J,j}$ from part \ref{part:stable_iht_generic_normality}. Take a nonempty deterministic family $\mathfrak J_n$ of nonempty retained sets of size $\tilde s$, with $\Pr(\widehat J\in\mathfrak J_n)\to1$. Let $\mathcal I_n=\{(J,j):J\in\mathfrak J_n,\ j\in J\}$, $D_n=|\mathcal I_n|$, and $\ell_n=\log(2nD_n)$. Assume $\sigma_{J,j}\geq c>0$ uniformly over $\mathcal I_n$ and
\[
\ell_n^5=o(n),\qquad
\sqrt{n\ell_n}\left[q^T\left(d+\frac{\eta b_n}{1-q}\right)+\frac{\eta r_n}{1-q}\right]=o_P(1).
\]
Let $c_{n,1-\alpha}$ be the $(1-\alpha)$-quantile of $\max_{(J,j)\in\mathcal I_n}|G_{J,j}|$, where $\vect G$ is mean-zero Gaussian with the covariance of $(\xi_{i,J,j}/\sigma_{J,j})_{(J,j)\in\mathcal I_n}$. For every $0<\alpha<1$,
\[
\liminf_{n\to\infty}\Pr\left\{\theta_j^*\in\left[\theta_j^{(T)}\pm\frac{\sigma_{\widehat J,j}c_{n,1-\alpha}}{\sqrt n}\right]\text{ for every }j\in\widehat J\right\}\geq1-\alpha.
\]
The critical value can also be $\Phi^{-1}\{1-\alpha/(2D_n)\}$. For the family of all retained sets of size $\tilde s$, with $r_n=C_4a_s^{3/4}$ and $T\geq\lceil c\log n\rceil$ for sufficiently large $c$, the regime $\{\tilde s\log p+\log n\}^5=o(n)$ suffices for the two displayed rate conditions.
\end{enumerate}
\end{proposition}

\begin{proof}
For part \ref{part:stable_iht_expansion}, write $\vect e_t=\{\vect\theta^{(t)}-\vect\theta^*\}_J$, $\vect g_J=\{\partial\mcL(\vect\theta^*)\}_J$, and $\vect A_J=\mbI-\eta\vect H_{J\times J}$. The initial error is supported on $J$, and every retained coordinate copies the unthresholded update. Therefore
\[
\vect e_{t+1}=\vect A_J\vect e_t-\eta\vect g_J-\eta\vect u_t,\qquad
\vect u_t=\{\partial\mcL(\vect\theta^{(t)})-\partial\mcL(\vect\theta^*)\}_J-\vect H_{J\times J}\vect e_t.
\]
If $\vect\theta^{(t)}\in\mcB_d$, its error has Euclidean norm at most $\sqrt{\tilde s}d\leq\epsilon_3$, and Condition \ref{cond:hdH_smooth} gives $|\vect u_t|_\infty\leq r_n$. Since $\|\vect A_J\|_\infty\leq q$, the next error has maximum norm at most $qd+\eta(b_n+r_n)\leq d$. Its support is in $J$, and $\mcB_d\subseteq\Theta$ gives the induction. Iterating the recursion proves the expansion and the displayed remainder bound. The same Neumann series gives
\[
\|\vect H_{J\times J}^{-1}\|_\infty\leq\frac{\eta}{1-q},\qquad
\left|\vect e_T+\vect H_{J\times J}^{-1}\vect g_J\right|_\infty
\leq q^T\left(d+\frac{\eta b_n}{1-q}\right)+\frac{\eta r_n}{1-q}.
\]

For part \ref{part:stable_iht_score_intervals}, each score coordinate has a uniformly bounded sub-Gaussian norm, since its random part is bounded by $C_f|Z_j|$ and its affine term subtracts its expectation. A coordinatewise tail bound and a union bound give $\Pr\{b_n>B_{n,\alpha}\}\leq\alpha$. On $\mcE_n$, the recursion in part \ref{part:stable_iht_expansion} gives $|\vect\theta^{(T)}-\vect\theta^*|_\infty\leq q^Td+\eta(1-q^T)(b_n+r_n)/(1-q)$. Intersecting these events proves the confidence bound. Boundedness of $\Theta$ and $\mcB_d\subseteq\Theta$ bound $d$, and a sufficiently large $c$ makes the transient negligible, proving the stated radius.

For part \ref{part:stable_iht_generic_normality}, the localization event, intersected with $\{\widehat J=J\}$ for IHT, has probability tending to one. Part \ref{part:stable_iht_expansion} gives the IHT expansion, and Theorem \ref{thm:pgd_phases}(iii) gives the PGD expansion. The respective transient conditions and $\sqrt n\,r_n=o_P(1)$ imply $\sqrt n(\theta_j^{(T)}-\theta_j^*)/\sigma_{J,j}=-(\sqrt n\,\sigma_{J,j})^{-1}\sum_{i=1}^n\xi_{i,J,j}+o_P(1)$. The moment assumption bounds the Lyapunov ratio by $Cn^{-\nu/2}$, and the Lyapunov central limit theorem and Slutsky's theorem prove the normal limit.

For part \ref{part:stable_iht_gaussian}, extend $\vect H_{J\times J}^{-1}\vect e_{j,J}$ by zeros outside $J$. This vector is $\tilde s$-sparse with Euclidean norm at most $m_H^{-1}$. Hence $\xi_{i,J,j}/\sigma_{J,j}$ has a uniformly bounded sub-Gaussian norm. Its exponential and fourth-moment bounds and its unit variance verify the assumptions of \citet[Theorem~2.1]{chernozhukov_etal.2022aos}, which gives rectangle Gaussian approximation error $C(\ell_n^5/n)^{1/4}$ for the standardized score array. Nazarov's inequality \citep{nazarov.2003}, applied to $(\vect G,-\vect G)$, bounds the probability that $\max_{(J,j)\in\mathcal I_n}|G_{J,j}|$ lies in any interval of width $\delta>0$ by $C\delta\sqrt{\log(2D_n)}$. The second rate condition gives a deterministic $\varepsilon_n=o(\ell_n^{-1/2})$ for which the selected standardized expansion remainder is at most $\varepsilon_n$ with probability tending to one. On this event, bounding every standardized influence coordinate in absolute value by $c_{n,1-\alpha}-\varepsilon_n$ implies simultaneous coverage, and anti-concentration proves the bound. The Gaussian union bound permits the normal critical value. For the full retained-set family, $\ell_n\leq C\{\tilde s\log p+\log n\}$, and
\[
\sqrt{n\ell_n}a_s^{3/4}\leq C\left[\frac{\{\tilde s\log p+\log n\}^5}{n}\right]^{1/4}.
\]
A sufficiently large $c$ also makes the transient negligible, proving the sufficient regime.
\end{proof}

\subsection{Index objectives}\label{app:sparse_index}

\begin{corollary}\label{cor:sparse_index}
Consider the index objective \eqref{eq:index_loss} with i.i.d. observations and $p>n\geq2$. Impose Assumptions \ref{cond:discont}, \ref{cond:theta}, and \ref{cond:density}, and write $\nu=\diff f'_+$.
\begin{itemize}
\item[1)] If $|\vect X|_\infty\leq C_X$ almost surely and $\Lambda_{\max}\{\mbE(\vect X\vect X^{\tp})\}=O(1)$, then \eqref{eq:hdlip_subgrad} holds with probability at least $1-Cp^{-\gamma\tilde s}$.

\item[2)] If $\sup_{|\vect v|_2=1,\,|\vect v|_0\leq2\tilde s}\|\vect v^{\tp}\vect X\|_{\psi_2}\leq C_X$ and either $\nu$ has compact support or $\rho_X$ has a bounded continuous first derivative, then \eqref{eq:hdlip_subgrad} holds with $|\cdot|_\infty$ replaced by $|\cdot|_{(2\tilde s)}$, with the same probability bound.
\end{itemize}
Both increment bounds hold for some $C,\gamma>0$, uniformly over $\vect\theta_1,\vect\theta_2\in\Theta$ with $|\supp(\vect\theta_1)\cup\supp(\vect\theta_2)|\leq2\tilde s$, for every selection of scenario subgradients. Under either set of assumptions, at $\vect\theta^*\in\Theta$, the bound $|\partial\mcL(\vect\theta^*)-\nabla\bar{\mcL}(\vect\theta^*)|_\infty\leq C\sqrt{\log p/n}$ holds with probability at least $1-Cp^{-\gamma}$. The increment and pointwise bounds therefore hold jointly with probability at least $1-Cp^{-\gamma}$.
Under the population strong convexity and Taylor assumptions of Corollary \ref{thm:hdH_smooth}, its conclusions hold for this objective. Compact support of $\nu$ and a bounded continuous density derivative give the Taylor bound in $|\cdot|_{(2\tilde s)}$. At a stationary population solution, use the score $\vect\psi_i=\vect b+\vect X_i f'_+(\vect X_i^{\tp}\vect\theta^*)$. Under their respective hypotheses, Corollary \ref{cor:hd_normality}, Proposition \ref{prop:stable_iht_inference}, Theorem \ref{thm:iht_phases}(iv), and Corollary \ref{cor:iht_warm_start} hold with this objective and score.
\end{corollary}

\begin{proof}
Write $m=\tilde s$, $g=f'_+$, $\nu=\diff g$, and $a_s=m\log p/n$. Use $a_{\vect\theta}(c)$, $\vect x_{\vect\theta}(c)$, and $\rho_{\vect\theta}(c\mid\vect U)$ from the proof of Proposition \ref{prop:index_objectives}, with $\vect U=\vect X_{-1}$. Convexity of $\Theta$ and $|\theta_1|\geq c_0$ fix the sign of its first coefficient. Every decision support contains coordinate 1, and a segment between decisions with joint support of size at most $2m$ stays in that subspace and in $\Theta$.

For part 1, the conditional distribution of $X_1$ is supported on $[-C_X,C_X]$. Thus $\rho_{\vect\theta}(c\mid\vect U)>0$ implies $|a_{\vect\theta}(c)|\leq C_X$, and every coordinate of $\vect x_{\vect\theta}(c)$ is bounded by $C_X$ on this event. For every unit vector $\vect u$, the second-moment bound gives, uniformly in $c$,
\[
\mbE[\rho_{\vect\theta}(c\mid\vect U)|\vect u^{\tp}\vect x_{\vect\theta}(c)|]
\leq C\{|u_1|C_X+\mbE|\vect u_{-1}^{\tp}\vect U|\}\leq C.
\]
Apply the crossing formula \eqref{eq:svm_all_level_crossing} along the segment between $\vect\theta_0$ and $\vect\theta_1$. Since $\nu(\mbR)\leq2C_f$, the coordinate bound on each section gives, for $d=|\vect\theta_1-\vect\theta_0|_2$,
\[
\mbE\left|X_j\{g(\vect X^{\tp}\vect\theta_1)-g(\vect X^{\tp}\vect\theta_0)\}\right|^k
\leq d\,k!C^k,\qquad k\geq1.
\]
The conditional density bound also gives the weighted slab moments at every level, and integration against $\nu$ gives the strict-bracket moments. Bernstein's inequality therefore gives deviation $C\{\sqrt{dt/n}+t/n\}$ for a fixed decision pair and coordinate.

Take $p^{-12}$-nets of $\Theta$ on coordinate subspaces of dimension at most $2m$. Their combined logarithmic cardinality is $O(m\log p)$. A decision and its representative have index difference at most $w=C_X\sqrt{2m}\,p^{-12}$. The strict bracket $g(\vect X^{\tp}\tilde{\vect\theta}+2w)-f'_-(\vect X^{\tp}\tilde{\vect\theta}-2w)$ has weighted moments bounded by $w k!C^k$. Applying Bernstein's inequality over the nets and coordinates bounds interpolation by $C(w+a_s)\leq Ca_s$. Applying the pairwise bound with $t=Am\log p$ gives part 1, with failure probability $Cp^{-\gamma m}$, for every subgradient selection.

For part 2, the sparse projection assumption supplies the section moments used in the proof of Proposition \ref{prop:index_objectives}.

For every unit $2m$-sparse vector $\vect v$ and every $2m$-sparse $\vect\theta\in\Theta$,
\[
\vect v^{\tp}\vect x_{\vect\theta}(c)
=\vect v^{\tp}\vect X-\frac{v_1}{\theta_1}\vect\theta^{\tp}\vect X+\frac{v_1}{\theta_1}c.
\]
The triangle inequality for $\|\cdot\|_{\psi_2}$, applied separately to the projections along $\vect v$ and $\vect\theta$, bounds the sub-Gaussian norm of the random part by $C$. Hence all section moments used in \eqref{eq:svm_section_moments} hold for sparse test vectors on compact level intervals. When the conditional density derivative is bounded, the proof of \eqref{eq:svm_section_density_tail} also applies, since $\vect X^{\tp}\vect\theta$ is sub-Gaussian for these sparse decisions. Cauchy--Schwarz then gives the same section moments at every level. In particular, for a displacement direction $\vect u$ and a test vector $\vect v$, each of sparsity at most $2m$, the mixed section moments are bounded by $k!C^k$ without a factor depending on $m$.

Apply \eqref{eq:svm_all_level_crossing} along the preceding sparse segment. For $d=|\vect\theta_1-\vect\theta_0|_2$, integration against the finite measure $\nu$ gives
\[
\mbE\left|\vect v^{\tp}\vect X\{g(\vect X^{\tp}\vect\theta_1)-g(\vect X^{\tp}\vect\theta_0)\}\right|^k
\leq d\,k!C^k,\qquad k\geq2.
\]
The first moment is at most $Cd$. The slab bound \eqref{eq:svm_all_level_slabs}, integrated against $\nu$, gives the strict-bracket moments. When $\nu$ has compact support, only a fixed neighborhood of the support of $\nu$ is used, and its slab widths are eventually smaller than one. When the density derivative is bounded, the bounds hold at all levels.

Take $p^{-12}$-nets of $\Theta$ on each coordinate subspace of dimension at most $2m$, and $1/4$-nets of the unit spheres on all test supports of that size. The logarithm of their combined cardinality is $O(m\log p)$. All decision net points are chosen in $\Theta$, and hence retain the coefficient bound on coordinate 1. The sub-Gaussian bound on sparse projections gives
\[
\max_{i\leq n}|\vect X_i|_{(2m)}\leq C\sqrt{m\log p}
\]
with failure probability $Cp^{-\gamma m}$. For a decision and a representative on the same subspace, their index values differ by at most $w=Cp^{-12}\sqrt{m\log p}$ on this event. Monotonicity brackets every subgradient selection by
\[
g(\vect X^{\tp}\tilde{\vect\theta}+2w)-f'_-(\vect X^{\tp}\tilde{\vect\theta}-2w).
\]
The slab moments bound each weighted bracket moment by $w k!C^k$. The population contribution outside the displayed sample event is at most $Cp^{-10}$ by the sparse projection moments. Bernstein's inequality over the decision and sphere nets therefore bounds interpolation by $Ca_s$. Apply the pairwise bound with $t=Am\log p$ over the net pairs and test vectors. The sphere-net inequality gives the restricted increment bound with failure probability $Cp^{-\gamma m}$. The strict brackets cover every subgradient selection.

Under either set of assumptions, $|g|\leq C_f$ makes each pointwise score coordinate bounded or sub-Gaussian. A coordinate union bound gives the pointwise deviation with failure probability $Cp^{-\gamma}$. Intersecting it with the increment event proves the joint probability bound.

Under part 2 of this corollary, conditional-density integration gives, for unit $2m$-sparse $\vect u,\vect v$,
\[
P|\vect v^{\tp}\vect X\{f'_+(\vect X^{\tp}(\vect\theta+d\vect u))-f'_+(\vect X^{\tp}\vect\theta)\}|\leq Cd.
\]
Consequently, the population gradient increment has $|\cdot|_{(2m)}$-norm at most $Cd$. Combining this bound with the increment bound above, separately for $d\geq a_s$ and $d<a_s$, proves part 1 of Corollary \ref{thm:hdH_smooth} for the index objective.

For restricted strong convexity, let $D=|\vect\theta_1-\vect\theta_2|_2$ and $\vect\theta_t=\vect\theta_2+t(\vect\theta_1-\vect\theta_2)$. This segment stays in $\Theta_{\rm sc}$, and every pair on it has joint support of size at most $2m$. The increment bound proved above therefore controls the empirical subgradient increment minus its expectation at $\vect\theta_t$ and $\vect\theta_2$ on the same event. Pairing it with the $2m$-sparse displacement and integrating gives
\[
\mcL(\vect\theta_1)-\mcL(\vect\theta_2)-\langle\partial\mcL(\vect\theta_2),\vect\theta_1-\vect\theta_2\rangle \geq\frac{\mu_0}{2}D^2-C\sqrt{a_s}\,D^{3/2}-Ca_sD \geq\frac{\mu_0}{4}D^2
\]
when $D\geq C_2a_s$. This proves the stated restricted strong convexity with $\mu=\mu_0$.

For the local approximation, $|\cdot|_\infty\leq|\cdot|_{(2m)}$ and the assumed population gradient expansion give, for $D=|\vect\theta-\vect\theta^*|_2\leq C_3\sqrt{a_s}$,
\[
\big|\partial\mcL(\vect\theta)-\partial\mcL(\vect\theta^*)-\vect H(\vect\theta-\vect\theta^*)\big|_\infty
\leq C(\sqrt{Da_s}+a_s+D^2)\leq C_4a_s^{3/4}.
\]
This proves the local condition with the radius and remainder.

Suppose now that $\nu$ has compact support and the density derivative is bounded. On its support, for two decisions with joint support of size at most $2m$,
\[
\|a_{\vect\theta}(c)-a_{\vect\theta'}(c)\|_{\psi_2}
\leq C|\vect\theta-\vect\theta'|_2.
\]
Subtracting their defining fractions and applying the sparse projection bound separately to $\vect\theta$, $\vect\theta'$, and their displacement gives this inequality. Conditional integration gives $\vect H(\vect\theta)=\mbE\int\vect x_{\vect\theta}(c)\vect x_{\vect\theta}(c)^{\tp}\rho_{\vect\theta}(c\mid\vect U)\nu(\diff c)$. The density derivative bound and H\"older's inequality therefore give
\[
|\vect v^{\tp}\{\vect H(\vect\theta)-\vect H(\vect\theta')\}\vect u|
\leq C|\vect\theta-\vect\theta'|_2
\]
for unit $2m$-sparse $\vect u,\vect v$. Integrating this bound along the segment to $\vect\theta^*$ proves \eqref{eq:hd_taylor}, and also its version in $|\cdot|_{(2m)}$. Combining that Taylor bound with the increment event proves the local subgradient approximation in Corollary \ref{thm:hdH_smooth}. When $\nu$ is not compactly supported, the stated Taylor assumption supplies this step.

Under the bounded-coordinate assumptions, the first-moment crossing calculation gives a population maximum-norm increment at most $Cd$. Conversion to $|\cdot|_{(2m)}$ multiplies the smoothness constant by $\sqrt{2m}$. In the strong convexity calculation the scalar empirical remainder acquires the same factor, and $D\geq C_2ma_s$ suffices. The maximum-norm local approximation still has remainder $C_4a_s^{3/4}$.

For a deterministic $J\supseteq S$, coordinate 1 belongs to $J$. The conditional density of $X_1$ given $\vect X_{J\setminus\{1\}}$ is the conditional average of $\rho_X(x_1\mid\vect X_{-1})$ over the omitted coordinates. Its density and derivative bounds are preserved by conditional averaging and dominated difference quotients. Every unit projection of $\vect X_J$ has bounded sub-Gaussian norm. The restricted region remains convex and bounded, with $|\theta_1|\geq c_0$. Applying Proposition \ref{prop:index_objectives} to the restriction to $J$ gives increment scale $|J|\log n/n$. Independence gives $|\{\partial\mcL(\vect\theta^*)\}_J|_2=O_P(\sqrt{|J|/n})$. Together with the Taylor bound above, these estimates give the inputs used in the recursion and normal limit for a deterministic retained set in Corollary \ref{cor:hd_normality}. The uniform increment and pointwise score bounds proved above supply the events used for data-dependent retained sets and for the warm-start recursion.
\end{proof}

\bibliographystyle{plainnat-nourl}
\bibliography{subGD_ER}

@inproceedings{jain_etal.2014nips,
  author    = {Jain, Prateek and Tewari, Ambuj and Kar, Purushottam},
  title     = {On iterative hard thresholding methods for high-dimensional {M}-estimation},
  booktitle = {Advances in Neural Information Processing Systems},
  year      = {2014},
}

@article{blumensath.2009acha,
  author  = {Blumensath, Thomas and Davies, Mike E.},
  title   = {Iterative hard thresholding for compressed sensing},
  journal = {Applied and Computational Harmonic Analysis},
  year    = {2009},
  volume  = {27},
  number  = {3},
  pages   = {265--274},
}

@article{wang_liu_zhang.2014,
  author    = {Wang, Zhaoran and Liu, Han and Zhang, Tong},
  title     = {Optimal computational and statistical rates of convergence for sparse nonconvex learning problems},
  journal   = {The Annals of Statistics},
  year      = {2014},
  volume    = {42},
  number    = {6},
  pages     = {2164--2201},
  publisher = {Institute of Mathematical Statistics},
}

@article{bunea_etal.2007aos,
  author    = {Bunea, Florentina and Tsybakov, Alexandre B. and Wegkamp, Marten H.},
  title     = {Aggregation for {Gaussian} regression},
  journal   = {The Annals of Statistics},
  year      = {2007},
  volume    = {35},
  number    = {4},
  pages     = {1674--1697},
  publisher = {Institute of Mathematical Statistics},
}

@inproceedings{negahban_etal.2009nips,
  author    = {Negahban, Sahand and Yu, Bin and Wainwright, Martin J. and Ravikumar, Pradeep},
  title     = {A unified framework for high-dimensional analysis of {M}-estimators with decomposable regularizers},
  booktitle = {Advances in Neural Information Processing Systems},
  year      = {2009},
}

@techreport{devolder_etal.2013,
  author      = {Devolder, Olivier and Glineur, Fran{\c{c}}ois and Nesterov, Yurii},
  title       = {First-order methods with inexact oracle: the strongly convex case},
  year        = {2013},
  number      = {2013/16},
  institution = {CORE},
  type        = {CORE Discussion Paper},
}

@article{chen_lee.2023joe,
  author  = {Chen, Le-Yu and Lee, Sokbae},
  title   = {Sparse quantile regression},
  journal = {Journal of Econometrics},
  year    = {2023},
  volume  = {235},
  number  = {2},
  pages   = {2195--2217},
}

@article{chang_etal.2026or,
  author  = {Chang, Jinyuan and Yang, Lin and Zhang, Yichen and Zhou, Wen-Xin},
  title   = {Feature-rich, data-private: a sparse learning framework for the high-dimensional newsvendor},
  journal = {Operations Research},
  year    = {2026},
  doi     = {10.1287/opre.2025.2370},
  note    = {Articles in Advance},
}

@article{Dontchev_Rockafellar.2004,
  author  = {Dontchev, Asen L. and Rockafellar, R. Tyrrell},
  title   = {Regularity and conditioning of solution mappings in variational analysis},
  journal = {Set-Valued Analysis},
  year    = {2004},
  volume  = {12},
  pages   = {79--109},
}

@misc{julien_etal.2012arXiv,
  author        = {Lacoste-Julien, Simon and Schmidt, Mark and Bach, Francis},
  title         = {A simpler approach to obtaining an {O}(1/t) convergence rate for the projected stochastic subgradient method},
  year          = {2012},
  eprint        = {1212.2002},
  archiveprefix = {arXiv},
}

@article{beck_teboulle.2012siopt,
  author  = {Beck, Amir and Teboulle, Marc},
  title   = {Smoothing and first order methods: a unified framework},
  journal = {SIAM Journal on Optimization},
  year    = {2012},
  volume  = {22},
  number  = {2},
  pages   = {557--580},
}

@article{donoghue_candes.2015focm,
  author  = {O'Donoghue, Brendan and Cand{\`e}s, Emmanuel J.},
  title   = {Adaptive restart for accelerated gradient schemes},
  journal = {Foundations of Computational Mathematics},
  year    = {2015},
  volume  = {15},
  pages   = {715--732},
}

@article{Jensen_etal.2012bitnm,
  author  = {Jensen, Tobias Lindstr{\o}m and J{\o}rgensen, Jakob Heide and Hansen, Per Christian and Jensen, S{\o}ren Holdt},
  title   = {Implementation of an optimal first-order method for strongly convex total variation regularization},
  journal = {BIT Numerical Mathematics},
  year    = {2012},
  volume  = {52},
  pages   = {329--356},
}

@article{diakonikolas_guzman.2026siopt,
  author  = {Diakonikolas, Jelena and Guzm{\'a}n, Crist{\'o}bal},
  title   = {Optimization on a finer scale: bounded local subgradient variation perspective},
  journal = {SIAM Journal on Optimization},
  year    = {2026},
  volume  = {36},
  number  = {1},
  pages   = {152--184},
  doi     = {10.1137/24M1659510},
  url     = {https://doi.org/10.1137/24M1659510},
}

@article{devolder_etal.2014mp,
  author  = {Devolder, Olivier and Glineur, Fran{\c{c}}ois and Nesterov, Yurii},
  title   = {First-order methods of smooth convex optimization with inexact oracle},
  journal = {Mathematical Programming},
  year    = {2014},
  volume  = {146},
  number  = {1--2},
  pages   = {37--75},
  doi     = {10.1007/s10107-013-0677-5},
  url     = {https://doi.org/10.1007/s10107-013-0677-5},
}

@article{shen_li_etal.2025aos,
  author    = {Shen, Yinan and Li, Jingyang and Cai, Jian-Feng and Xia, Dong},
  title     = {Computationally efficient and statistically optimal robust high-dimensional linear regression},
  journal   = {The Annals of Statistics},
  year      = {2025},
  volume    = {53},
  number    = {1},
  pages     = {374--399},
  publisher = {Institute of Mathematical Statistics},
}

@article{davis_drusvyatskiy.2022mor,
  author  = {Davis, Damek and Drusvyatskiy, Dmitriy},
  title   = {Graphical convergence of subgradients in nonconvex optimization and learning},
  journal = {Mathematics of Operations Research},
  year    = {2022},
  volume  = {47},
  number  = {1},
  pages   = {209--231},
  doi     = {10.1287/moor.2021.1126},
  url     = {https://doi.org/10.1287/moor.2021.1126},
}

@article{ruan.2025mor,
  author  = {Ruan, Feng},
  title   = {On the uniform convergence of subdifferentials in stochastic optimization and learning},
  journal = {Mathematics of Operations Research},
  year    = {2025},
  volume  = {51},
  number  = {3},
  pages   = {2323--2353},
  doi     = {10.1287/moor.2024.0533},
}

@article{Wang_He.2024et,
  author  = {Wang, Lan and He, Xuming},
  title   = {Analysis of global and local optima of regularized quantile regression in high dimensions: a subgradient approach},
  journal = {Econometric Theory},
  year    = {2024},
  volume  = {40},
  number  = {2},
  pages   = {233--277},
}

@article{davis_etal.2018jota,
  author  = {Davis, Damek and Drusvyatskiy, Dmitriy and MacPhee, Kellie J. and Paquette, Courtney},
  title   = {Subgradient methods for sharp weakly convex functions},
  journal = {Journal of Optimization Theory and Applications},
  year    = {2018},
  volume  = {179},
  pages   = {962--982},
}

@book{rockafellar1997convex,
  author    = {Rockafellar, R. Tyrrell},
  title     = {Convex analysis},
  year      = {1997},
  volume    = {28},
  publisher = {Princeton University Press},
}

@book{beck.2017,
  author    = {Beck, Amir},
  title     = {First-order methods in optimization},
  year      = {2017},
  publisher = {Society for Industrial and Applied Mathematics},
  address   = {Philadelphia, PA},
  doi       = {10.1137/1.9781611974997},
}

@article{horowitz.1998ecomc,
  author    = {Horowitz, Joel L.},
  title     = {Bootstrap methods for median regression models},
  journal   = {Econometrica},
  year      = {1998},
  volume    = {66},
  number    = {6},
  pages     = {1327--1351},
  publisher = {[Wiley, Econometric Society]},
  urldate   = {2026-06-08},
}

@article{fernandes_etal.2021jbes,
  author    = {Fernandes, Marcelo and Guerre, Emmanuel and Horta, Eduardo},
  title     = {Smoothing quantile regressions},
  journal   = {Journal of Business \& Economic Statistics},
  year      = {2021},
  volume    = {39},
  number    = {1},
  pages     = {338--357},
  publisher = {Taylor \& Francis},
}

@inproceedings{xu_yan_etal.2016nips,
  author    = {Xu, Yi and Yan, Yan and Lin, Qihang and Yang, Tianbao},
  title     = {Homotopy smoothing for non-smooth problems with lower complexity than {O}(1/$\epsilon$)},
  booktitle = {Advances in Neural Information Processing Systems},
  year      = {2016},
}

@article{xiao_zhang.2013siamjo,
  author  = {Xiao, Lin and Zhang, Tong},
  title   = {A proximal-gradient homotopy method for the sparse least-squares problem},
  journal = {SIAM Journal on Optimization},
  year    = {2013},
  volume  = {23},
  number  = {2},
  pages   = {1062--1091},
}

@article{polyak1992acceleration,
  author    = {Polyak, Boris T. and Juditsky, Anatoli B.},
  title     = {Acceleration of stochastic approximation by averaging},
  journal   = {SIAM Journal on Control and Optimization},
  year      = {1992},
  volume    = {30},
  number    = {4},
  pages     = {838--855},
  publisher = {SIAM},
}

@article{nesterov1983method,
  author    = {Nesterov, Yurii},
  title     = {A method for solving the convex programming problem with convergence rate $O (1/k^2)$},
  journal   = {Dokl. Akad. Nauk {SSSR}},
  year      = {1983},
  volume    = {269},
  number    = {3},
  pages     = {543--547},
}

@article{he_pan_etal.2023joe,
  author  = {He, Xuming and Pan, Xiaoou and Tan, Kean Ming and Zhou, Wen-Xin},
  title   = {Smoothed quantile regression with large-scale inference},
  journal = {Journal of Econometrics},
  year    = {2023},
  volume  = {232},
  number  = {2},
  pages   = {367--388},
}

@article{nesterov2005smooth,
  author  = {Nesterov, Yurii},
  title   = {Smooth minimization of non-smooth functions},
  journal = {Mathematical Programming},
  year    = {2005},
  volume  = {103},
  number  = {1},
  pages   = {127--152},
  doi     = {10.1007/s10107-004-0552-5},
}

@article{moreau1965proximite,
  author  = {Moreau, J. J.},
  title   = {Proximit{\'e} et dualit{\'e} dans un espace hilbertien},
  journal = {Bulletin de la Soci{\'e}t{\'e} Math{\'e}matique de France},
  year    = {1965},
  volume  = {93},
  pages   = {273--299},
}

@article{duchi_etal.2012siamjo,
  author  = {Duchi, John C. and Bartlett, Peter L. and Wainwright, Martin J.},
  title   = {Randomized smoothing for stochastic optimization},
  journal = {SIAM Journal on Optimization},
  year    = {2012},
  volume  = {22},
  number  = {2},
  pages   = {674--701},
}

@article{wang_peng_etal.2020jasa,
  author    = {Wang, Lan and Peng, Bo and Bradic, Jelena and Li, Runze and Wu, Yunan},
  title     = {A tuning-free robust and efficient approach to high-dimensional regression},
  journal   = {Journal of the American Statistical Association},
  year      = {2020},
  volume    = {115},
  number    = {532},
  pages     = {1700--1714},
  publisher = {Taylor \& Francis},
}

@article{koo_etal.2008jmlr,
  author  = {Koo, Ja-Yong and Lee, Yoonkyung and Kim, Yuwon and Park, Changyi},
  title   = {A {Bahadur} representation of the linear support vector machine},
  journal = {Journal of Machine Learning Research},
  year    = {2008},
  volume  = {9},
  number  = {44},
  pages   = {1343--1368},
}

@book{koenker.2005,
  author     = {Koenker, Roger},
  title      = {Quantile regression},
  year       = {2005},
  publisher  = {Cambridge University Press},
  place      = {Cambridge},
  collection = {Econometric Society Monographs},
}

@article{chen_etal.2019aos,
  author    = {Chen, Xi and Liu, Weidong and Zhang, Yichen},
  title     = {Quantile regression under memory constraint},
  journal   = {The Annals of Statistics},
  year      = {2019},
  volume    = {47},
  number    = {6},
  pages     = {3244--3273},
  publisher = {Institute of Mathematical Statistics},
}

@article{jaeckel.1972aoms,
  author    = {Jaeckel, Louis A.},
  title     = {Estimating regression coefficients by minimizing the dispersion of the residuals},
  journal   = {The Annals of Mathematical Statistics},
  year      = {1972},
  volume    = {43},
  number    = {5},
  pages     = {1449--1458},
  publisher = {Institute of Mathematical Statistics},
  urldate   = {2025-08-19},
}

@article{ding_wang.2023arXiv,
  author  = {Ding, Lijun and Wang, Alex L.},
  title   = {Sharpness and conditioning of nonsmooth convex formulations in statistical signal recovery},
  journal = {SIAM Journal on Optimization},
  year    = {2026},
  volume  = {36},
  number  = {3},
  pages   = {1269--1296},
  doi     = {10.1137/24M1678891},
}

@article{ban_rudin.2019or,
 author={Ban, Gah-Yi and Rudin, Cynthia},
 title={The Big Data Newsvendor: Practical Insights from Machine Learning},
 journal={Operations Research},
 year={2019},
 volume={67},
 number={1},
 pages={90--108},
 doi={10.1287/opre.2018.1757}
}

@article{rockafellar_uryasev.2000jor,
 author={Rockafellar, R. Tyrrell and Uryasev, Stanislav},
 title={Optimization of Conditional Value-at-Risk},
 journal={The Journal of Risk},
 year={2000},
 volume={2},
 number={3},
 pages={21--41},
 doi={10.21314/JOR.2000.038}
}

@book{shapiro_etal.2021,
 author={Shapiro, Alexander and Dentcheva, Darinka and Ruszczy{\'n}ski, Andrzej},
 title={Lectures on Stochastic Programming: Modeling and Theory},
 edition={3rd},
 publisher={Society for Industrial and Applied Mathematics},
 year={2021}
}

@article{polyak.1969,
  author = {Polyak, Boris T.},
  title = {Minimization of unsmooth functionals},
  journal = {USSR Computational Mathematics and Mathematical Physics},
  year = {1969},
  volume = {9},
  number = {3},
  pages = {14--29},
  doi = {10.1016/0041-5553(69)90061-5},
  url = {https://doi.org/10.1016/0041-5553(69)90061-5}
}

@techreport{ruppert.1988,
  author = {Ruppert, D.},
  title = {Efficient estimations from a slowly convergent {Robbins--Monro} process},
  institution = {School of Operations Research and Industrial Engineering, Cornell University},
  number = {781},
  year = {1988},
  month = feb,
  url = {https://hdl.handle.net/1813/8664}
}

@article{davis_drusvyatskiy_jiang.2024aos,
  author = {Davis, Damek and Drusvyatskiy, Dmitriy and Jiang, Liwei},
  title = {Asymptotic normality and optimality in nonsmooth stochastic approximation},
  journal = {The Annals of Statistics},
  year = {2024},
  volume = {52},
  number = {4},
  pages = {1485--1508},
  doi = {10.1214/24-AOS2401},
  url = {https://doi.org/10.1214/24-AOS2401}
}

@article{king_rockafellar.1993mor,
  author = {King, Alan J. and Rockafellar, R. Tyrrell},
  title = {Asymptotic theory for solutions in statistical estimation and stochastic programming},
  journal = {Mathematics of Operations Research},
  year = {1993},
  volume = {18},
  number = {1},
  pages = {148--162},
  doi = {10.1287/moor.18.1.148},
  url = {https://doi.org/10.1287/moor.18.1.148}
}

@article{shapiro.1993mor,
  author = {Shapiro, Alexander},
  title = {Asymptotic Behavior of Optimal Solutions in Stochastic Programming},
  journal = {Mathematics of Operations Research},
  year = {1993},
  volume = {18},
  number = {4},
  pages = {829--845},
  doi = {10.1287/moor.18.4.829}
}

@book{vershynin.2018hdp,
  author = {Vershynin, Roman},
  title = {High-Dimensional Probability: An Introduction with Applications in Data Science},
  publisher = {Cambridge University Press},
  year = {2018},
  series = {Cambridge Series in Statistical and Probabilistic Mathematics},
  volume = {47},
  doi = {10.1017/9781108231596}
}

@article{pasupathy.2010or,
  author  = {Pasupathy, Raghu},
  title   = {On choosing parameters in retrospective-approximation algorithms for stochastic root finding and simulation optimization},
  journal = {Operations Research},
  year    = {2010},
  volume  = {58},
  number  = {4},
  pages   = {889--901},
}

@article{royset_szechtman.2013or,
  author  = {Royset, Johannes O. and Szechtman, Roberto},
  title   = {Optimal budget allocation for sample average approximation},
  journal = {Operations Research},
  year    = {2013},
  volume  = {61},
  number  = {3},
  pages   = {762--776},
}

@article{nesterov.2015mp,
  author  = {Nesterov, Yurii},
  title   = {Universal gradient methods for convex optimization problems},
  journal = {Mathematical Programming},
  year    = {2015},
  volume  = {152},
  number  = {1--2},
  pages   = {381--404},
}

@article{agarwal_etal.2012aos,
  author  = {Agarwal, Alekh and Negahban, Sahand and Wainwright, Martin J.},
  title   = {Fast global convergence of gradient methods for high-dimensional statistical recovery},
  journal = {The Annals of Statistics},
  year    = {2012},
  volume  = {40},
  number  = {5},
  pages   = {2452--2482},
}

@article{he_shao.1996aos,
  author  = {He, Xuming and Shao, Qi-Man},
  title   = {A general {B}ahadur representation of {M}-estimators and its application to linear regression with nonstochastic designs},
  journal = {The Annals of Statistics},
  year    = {1996},
  volume  = {24},
  number  = {6},
  pages   = {2608--2630},
}

@article{tan_wang_zhou.2022jrssb,
  author = {Tan, Kean Ming and Wang, Lan and Zhou, Wen-Xin},
  title = {High-dimensional quantile regression: convolution smoothing and concave regularization},
  journal = {Journal of the Royal Statistical Society: Series B},
  year = {2022},
  volume = {84},
  number = {1},
  pages = {205--233},
  doi = {10.1111/rssb.12485}
}

@article{oliveira_thompson.2023mp1,
  author = {Oliveira, Roberto I. and Thompson, Philip},
  title = {Sample average approximation with heavier tails {I}: non-asymptotic bounds with weak assumptions and stochastic constraints},
  journal = {Mathematical Programming},
  year = {2023},
  volume = {199},
  number = {1--2},
  pages = {1--48},
  doi = {10.1007/s10107-022-01810-x}
}

@article{oliveira_thompson.2023mp2,
  author = {Oliveira, Roberto I. and Thompson, Philip},
  title = {Sample average approximation with heavier tails {II}: localization in stochastic convex optimization and persistence results for the {Lasso}},
  journal = {Mathematical Programming},
  year = {2023},
  volume = {199},
  number = {1--2},
  pages = {49--86},
  doi = {10.1007/s10107-023-01940-w}
}

@article{jin_scheinberg_xie.2024siopt,
  author = {Jin, Billy and Scheinberg, Katya and Xie, Miaolan},
  title = {High probability complexity bounds for adaptive step search based on stochastic oracles},
  journal = {SIAM Journal on Optimization},
  year = {2024},
  volume = {34},
  number = {3},
  pages = {2411--2439},
  doi = {10.1137/22M1512764}
}

@article{he_tan_zhou.2023jrssb,
  author = {He, Xuming and Tan, Kean Ming and Zhou, Wen-Xin},
  title = {Robust estimation and inference for expected shortfall regression with many regressors},
  journal = {Journal of the Royal Statistical Society: Series B},
  year = {2023},
  volume = {85},
  number = {4},
  pages = {1223--1246},
  doi = {10.1093/jrsssb/qkad063}
}

@article{lam_wang_wu_zhang.2022arxiv,
  author = {Lam, Henry and Wang, Kaizheng and Wu, Yuhang and Zhang, Yichen},
  title = {Adaptive data fusion for multi-task non-smooth optimization},
  journal = {arXiv preprint arXiv:2210.12334},
  year = {2022},
  doi = {10.48550/arXiv.2210.12334}
}

@article{chernozhukov_etal.2022aos,
  author = {Chernozhukov, Victor and Chetverikov, Denis and Kato, Kengo and Koike, Yuta},
  title = {Improved central limit theorem and bootstrap approximations in high dimensions},
  journal = {The Annals of Statistics},
  year = {2022},
  volume = {50},
  number = {5},
  pages = {2562--2586},
  doi = {10.1214/22-AOS2193}
}

@incollection{nazarov.2003,
  author = {Nazarov, Fedor},
  title = {On the maximal perimeter of a convex set in {$\mathbb{R}^n$} with respect to a {Gaussian} measure},
  booktitle = {Geometric Aspects of Functional Analysis},
  editor = {Milman, Vitali D. and Schechtman, Gideon},
  series = {Lecture Notes in Mathematics},
  volume = {1807},
  pages = {169--187},
  publisher = {Springer},
  year = {2003},
  doi = {10.1007/978-3-540-36428-3_15}
}

\end{document}